%% file: main.tex
\documentclass[11pt,a4paper]{article}
\usepackage{graphicx}
\usepackage{url}
\usepackage{amssymb}
\usepackage[centertags]{amsmath}
\usepackage{multirow}
\usepackage{subfigure}
\usepackage[affil-it]{authblk}
\usepackage{cite}
\usepackage{notoccite}
\usepackage{tocbibind}
\usepackage[colorlinks,linkcolor=blue,anchorcolor=blue,citecolor=blue,urlcolor=blue]{hyperref}

\graphicspath{
{./} {Calibration/} {CentralDetector/} {ExpOverview/} {Installation/} {Introduction/}
{OfflineComputing/} {SiPM/} {TriggerDAQDCS/} {VetoShielding/}
}

\makeatletter
\newcommand\blfootnote[1]{%
  \begin{NoHyper}
  \renewcommand\thefootnote{}\footnote{#1}%
  \addtocounter{footnote}{-1}%
  \end{NoHyper}
}
\renewcommand{\thefigure}{\ifnum \c@section>\z@ \thesection-\fi \@arabic\c@figure}
\renewcommand{\thetable}{\ifnum \c@section>\z@ \thesection-\fi \@arabic\c@table}
\@addtoreset{equation}{section}
\@addtoreset{table}{section}
\@addtoreset{figure}{section}
\@addtoreset{footnote}{section}
\makeatother

\begin{document}
\pagestyle{empty}
\begin{center}
\mbox{  }
\vskip 1.6in
{\Huge\bf  TAO Conceptual Design Report}
\vskip 0.4in
{\Large A Precision Measurement of the Reactor Antineutrino Spectrum with Sub-percent Energy Resolution}
\vskip 0.6in
{\LARGE  The JUNO Collaboration}
\vskip 0.4in
{\LARGE May 17, 2020}

\vskip 1in
\input{TAOabstract.tex}
\end{center}

\clearpage
\pagestyle{plain}

\title{The JUNO Collaboration}
\input{TAOauthors.tex}
\date{}
\maketitle
\clearpage

\input{Introduction/ExecutiveSummary.tex}
\clearpage

\setcounter{section}{0}
\tableofcontents
\clearpage
\listoffigures
\clearpage
\listoftables
\clearpage


\input{Introduction/section.tex}

\clearpage
\input{ExpOverview/section.tex}
\clearpage
\input{CentralDetector/section.tex}

\clearpage
\input{Calibration/section.tex}

\clearpage
\input{VetoShielding/section.tex}

\clearpage
\input{SiPM/section.tex}

\clearpage
\input{TriggerDAQDCS/section.tex}
\clearpage
\input{OfflineComputing/section.tex}

\clearpage
\input{Installation/section.tex}

\clearpage
\input{TAOacknowledgement.tex}

\bibliographystyle{unsrt}
\bibliography{references}

\end{document}

%% file: TAOabstract.tex
\begin{abstract}

The Taishan Antineutrino Observatory (TAO, also known as JUNO-TAO) is a satellite experiment of the Jiangmen Underground Neutrino Observatory (JUNO). A ton-level liquid scintillator detector will be placed at $\sim 30$~m from a core of the Taishan Nuclear Power Plant. The reactor antineutrino spectrum will be measured with sub-percent energy resolution, to provide a reference spectrum for future reactor neutrino experiments, and to provide a benchmark measurement to test nuclear databases. A spherical acrylic vessel containing 2.8~ton gadolinium-doped liquid scintillator will be viewed by 10~m$^2$ Silicon Photomultipliers (SiPMs) of $>50$\% photon detection efficiency with almost full coverage. The photoelectron yield is about 4500 per MeV, an order higher than any existing large-scale liquid scintillator detectors. The detector operates at -50$^\circ$C to lower the dark noise of SiPMs to an acceptable level. The detector will measure about 2000 reactor antineutrinos per day, and is designed to be well shielded from cosmogenic backgrounds and ambient radioactivities to have a $\sim 10$\% background-to-signal ratio. The experiment is expected to start operation in 2022.

\end{abstract}

%% file: TAOauthors.tex
\author[5]{Angel Abusleme}
\author[45]{Thomas Adam}
\author[67]{Shakeel Ahmad}
\author[55]{Sebastiano Aiello}
\author[67]{Muhammad Akram}
\author[67]{Nawab Ali}
\author[29]{Fengpeng An}
\author[10]{Guangpeng An}
\author[22]{Qi An}
\author[55]{Giuseppe Andronico}
% ...
\affil[5]{Pontificia Universidad Cat\'{o}lica de Chile, Santiago, Chile}
% ...
\affil[10]{Institute of High Energy Physics, Beijing, China}
% ...
\affil[22]{University of Science and Technology of China, Hefei, China}
% ...
\affil[29]{East China University of Science and Technology, Shanghai, China}
% ...
\affil[45]{IPHC, Universit\'{e} de Strasbourg, CNRS/IN2P3, F-67037 Strasbourg, France}
% ...
\affil[55]{INFN Catania and Dipartimento di Fisica e Astronomia dell Universit\`{a} di Catania, Catania, Italy}
% ...
\affil[67]{Pakistan Institute of Nuclear Science and Technology, Islamabad, Pakistan}
% ...

%% file: Introduction/ExecutiveSummary.tex
\begin{center}
{\Large\bf Executive Summary}
\end{center}

The Taishan Antineutrino Observatory (TAO, also known as JUNO-TAO) is a satellite experiment of the Jiangmen Underground Neutrino Observatory (JUNO)~\cite{Djurcic:2015vqa}. TAO consists of a ton-level liquid scintillator (LS) detector at $\sim30$ meters from a reactor core of the Taishan Nuclear Power Plant in Guangdong, China. About 4500 photoelectrons per MeV could be observed by instrumenting with almost full coverage ($\sim10$~m$^2$) of Silicon Photomultipliers (SiPMs) of  $>50$\% photon detection efficiency, resulting in an unprecedented energy resolution approaching to the limit of LS detectors. The detector operates at -50$^\circ$C to lower the dark noise of SiPM to an acceptable level. The TAO experiment is expected to start operation in 2022.

The main purposes of the TAO experiment are 1) to provide a reference spectrum for JUNO, eliminating the possible model dependence due to fine structure in the reactor antineutrino spectrum in determining the neutrino mass ordering~\cite{An:2015jdp}; 2) to provide a benchmark measurement to test nuclear databases, by comparing the measurement with the predictions of the summation method; 3) to provide increased reliability in measured isotopic antineutrino yields due to a larger sampled range of fission fractions; 4) to provide an opportunity to improve nuclear physics knowledge of neutron-rich isotopes~\cite{INDC-NDS-0786}; 5) to search for light sterile neutrinos with a mass scale around 1~eV; 6) to provide increased reliability and verification of the technology for reactor monitoring and safeguard.

\begin{figure}[htb]
    \centering
    \includegraphics[width=0.6\columnwidth]{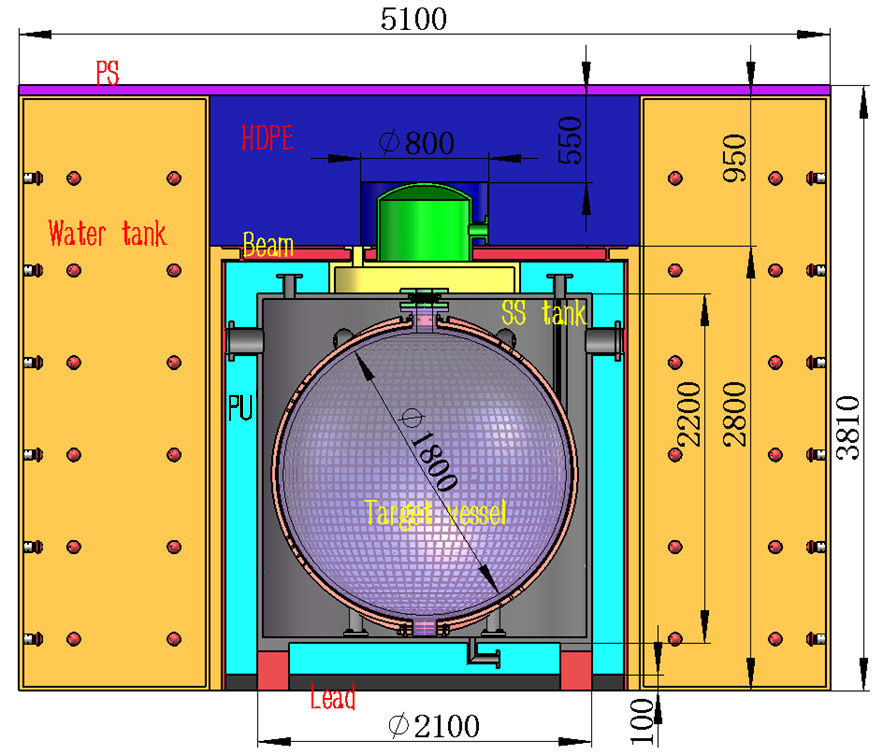}
    \caption{Schematic view of the TAO detector, which consists of a Central Detector (CD) and an outer shielding and veto system. The CD consists of 2.8~ton gadolinium-doped LS filled in a spherical acrylic vessel and viewed by 10~m$^2$ SiPMs, a spherical copper shell that supports the SiPMs, 3.45~ton buffer liquid, and a cylindrical stainless steel tank insulated with 20~cm thick Polyurethane (PU). The outer shielding includes 1.2~m thick water in the surrounding tanks, 1~m High Density Polyethylene (HDPE) on the top, and 10~cm lead at the bottom. The water tanks, instrumented with Photomultipliers (shown by red circles), and the Plastic Scintillator (PS) on the top comprise the active muon veto system. The dimensions are displayed in mm.     \label{fig:cdscheme}}
\end{figure}

The schematic drawing of the TAO detector is shown in Figure~\ref{fig:cdscheme}. The Central Detector (CD) detects reactor antineutrinos with 2.8~ton Gadolinium-doped LS (GdLS) contained in a spherical acrylic vessel of 1.8~m in inner diameter. To fully contain the energy deposition of gammas from the Inverse Beta Decay (IBD) positron annihilation, a 25-cm selection cut will be applied for positron vertex from the acrylic vessel, resulting in 1 ton fiducial mass. The IBD event rate in the fiducial volume will be about 2000 (4000) events per day with (without) the detection efficiency taken into account.
SiPM tiles are installed on the inner surface of a spherical copper shell of 1.882~m in inner diameter. The gap between the SiPM surface and the acrylic vessel is about 2~cm. The copper shell is installed in a cylindric stainless steel tank of an outer diameter of 2.1~m and a height of 2.2~m. The stainless steel tank is filled with Linear Alkylbenzene (LAB), also the solvent of the GdLS, which serves as the buffer liquid to shield the radioactivity of the outer tank, to stabilize the temperature, and to couple optically the acrylic and the SiPM surfaces. The stainless steel tank is insulated with 20~cm thick Polyurethane (PU) to operate at -50$^\circ$C to reduce the dark noise of SiPMs to $\sim$~100~Hz/mm$^2$. The central detector is surrounded by 1.2~m thick water tanks on the sides and 1~m High Density Polyethylene (HDPE) on the top to shield the ambient radioactivity and cosmogenic neutrons. Cosmic muons will be detected by the water tanks with PMTs instrumented and by Plastic Scintillator (PS) on the top.

\begin{figure}[htb]
    \centering
    \subfigure{\includegraphics[width=0.48\columnwidth]{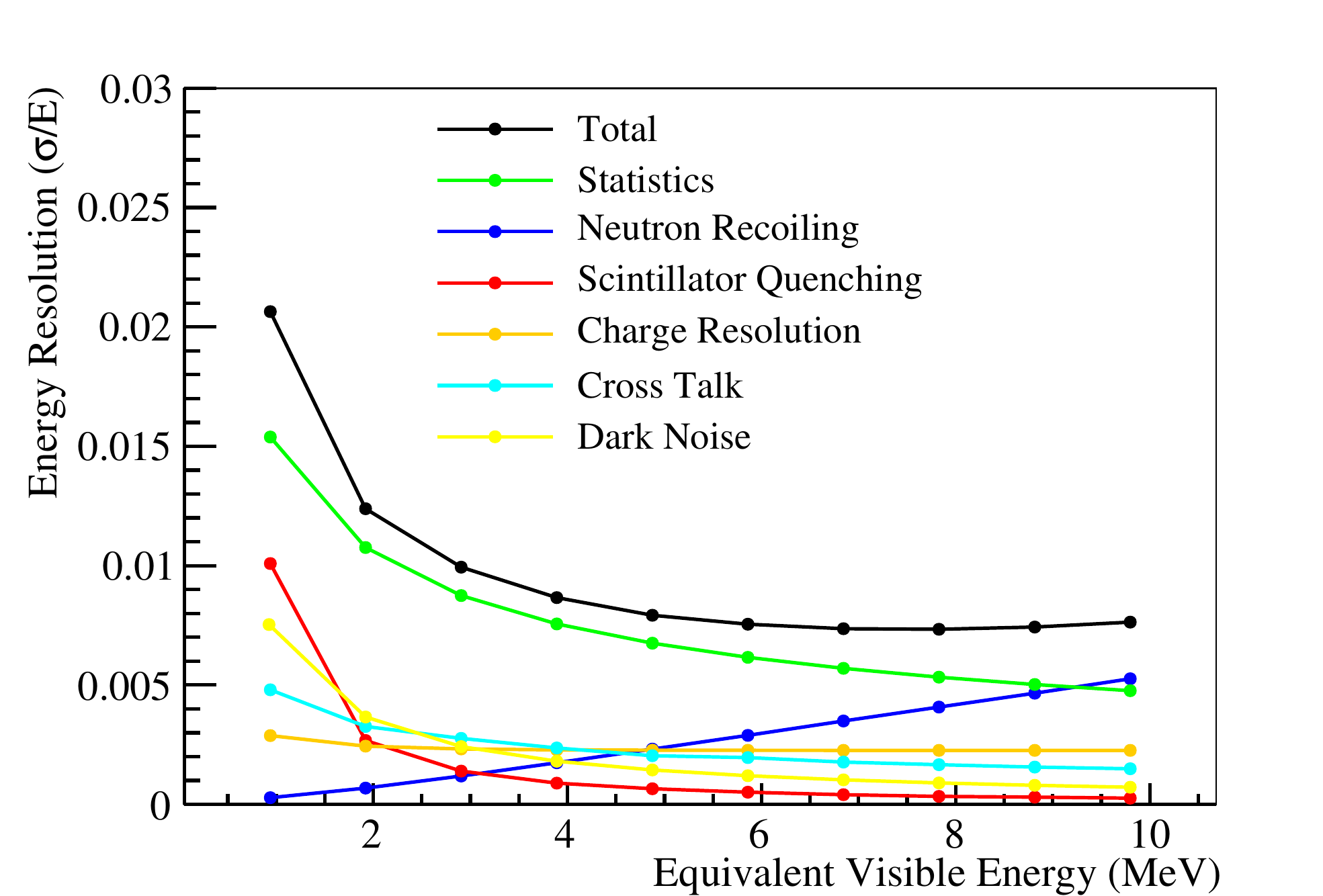}}
    \subfigure{\includegraphics[width=0.455\columnwidth]{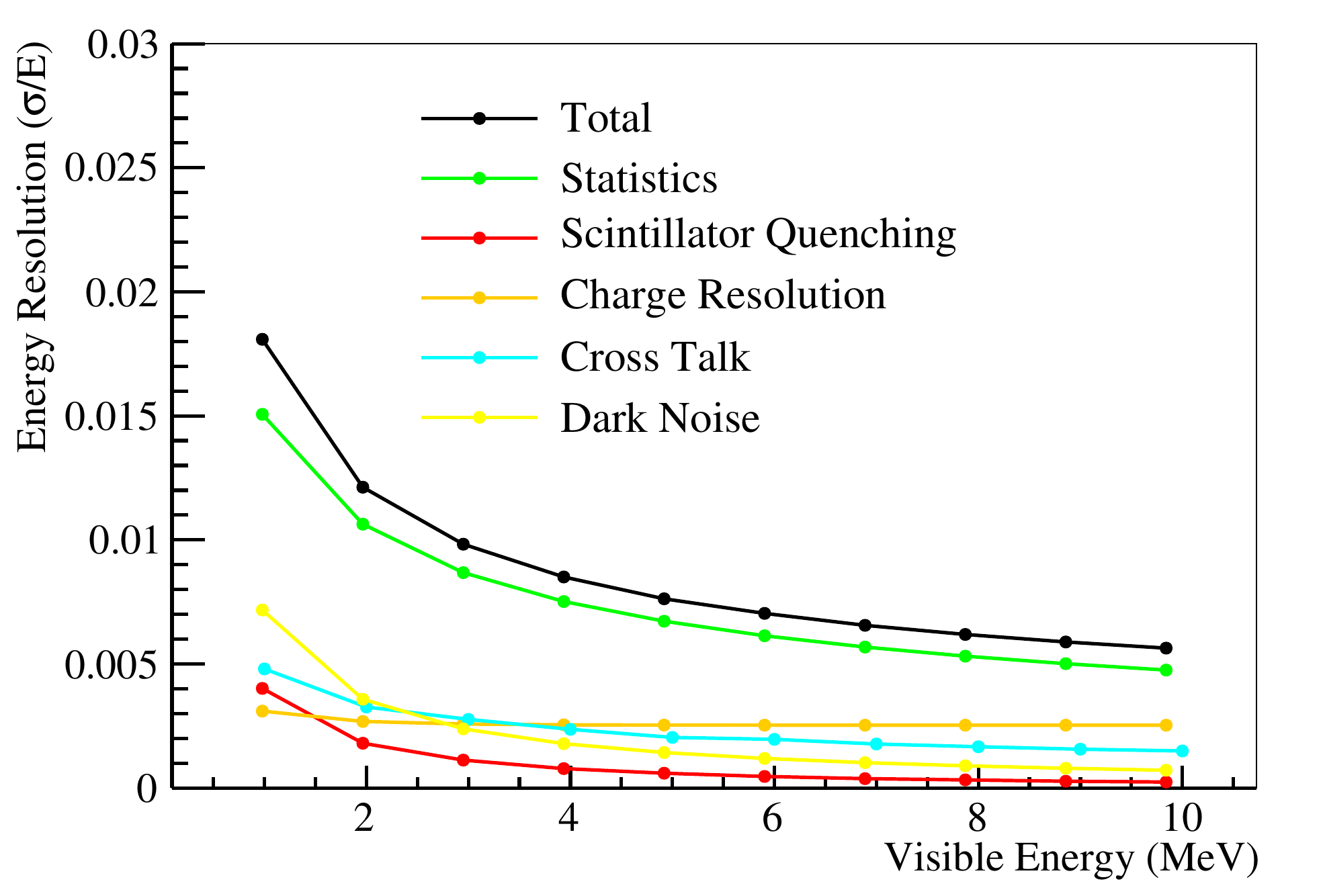}}
    \caption{Expected energy resolution of the TAO detector. Left: energy resolution for reactor antineutrinos versus equivalent visible energy defined as the reconstructed neutrino energy minus a constant shift of 0.78~MeV due to the IBD reaction kinetics. It includes the effect from the spread of the neutron kinetic energy in the IBD reaction. Right: energy resolution for electrons without the effect of neutron kinetic energy. Details are described in Section~\ref{sec:Expoverview}. \label{fig:Eres}}
\end{figure}

Although $3\%/\sqrt{E[\rm{MeV}]}$ energy resolution ($E$ is the visible energy) will be enough for TAO to serve as a reference detector of JUNO, the energy resolution should be as high as possible to study the fine structure of the reactor antineutrino spectrum and create a highly resolved benchmark to test nuclear databases. New findings in the measurement of the reactor antineutrino spectrum might be achieved with a state-of-the-art detector. A photoelectron yield of about 4500 photoelectrons per MeV are expected for TAO from simulations, corresponding to an energy resolution of $1.5\%/\sqrt{E[{\rm MeV}]}$ in photoelectron statistics. However, when approaching to the limit of the energy resolution of LS detectors, non-stochastic effects become prominent. At low energies, the contribution from the LS quenching effect might be quite large, although not very well understood thus model dependent. At high energies, the smearing from neutron recoil of IBD becomes dominant. Taking into account the projected dark noise, cross talk, and charge resolution of the SiPMs, the expected energy resolution of TAO is shown in Figure~\ref{fig:Eres}. The usual $1/\sqrt{E}$ behavior is not valid here. In most of the energy region of interest, the energy resolution of TAO will be sub-percent.

Taishan Nuclear Power Plant locates in Chixi town of Taishan city in Guangdong province, 53~km from the JUNO experiment. It has two cores currently in operation. Another two cores might be built later. All reactors are European Pressurised Reactor (EPR) with 4.6~GW thermal power. At $\sim30$~meter baseline, the far core contributes about 1.5\% to the total reactor antineutrino rate in the TAO detector. The Taishan Neutrino Laboratory for the TAO detector is in a basement at 9.6~m underground, outside of the concrete containment shell of the reactor core. Muon rate and cosmogenic neutron rate are measured to be 1/3 of those on the ground. Simulations show that the cosmogenic fast neutron background, accidental background, and cosmogenic $^8$He/$^9$Li background can be well controlled to $<10$\% of the signal with proper shielding and muon veto.
The expected rates of IBD signal and the residual backgrounds passing the IBD selection cuts (see details in Section~\ref{sec:Expoverview}) are summarized in Table~1.

\begin{table}[htb]
\setlength{\belowcaptionskip}{5pt}
\begin{center}
\caption{Summary of the event rates of the IBD signal and backgrounds.\label{tab:summarySB}}
\begin{tabular}{r  l}
  \hline\hline
  IBD signal & \mbox{ }\mbox{ } 2000~events/day \\
  Muon rate & \mbox{ }\mbox{ } 70~Hz/m$^2$ \\
  Singles from radioactivity & $< 100$~Hz \\
  Fast neutron background after veto & $< 200$~events/day \\
  Accidental background rate & $< 190$~events/day \\
  $^{8}$He/$^{9}$Li background rate & $\sim$  54~events/day \\
  \hline
\end{tabular}
\end{center}
\end{table}

The detector R\&D started in 2018. A GdLS recipe has been developed and showed good transparency and light yield at -50$^\circ$C. The SiPMs and the readout electronics have been preliminarily tested at the same temperature. A prototype detector is currently being tested at -50$^\circ$C.

%% file: Introduction/section.tex
\section{Physics Goals}
\label{sec:introduction}
\blfootnote{Editors: Stefano Mari (smari@os.uniroma3.it) and Liang Zhan (zhanl@ihep.ac.cn)}
\blfootnote{Major contributors: Jun Cao, Jianrun Hu, Bed\v{r}ich Roskovec}

\begin{figure}[htb]
    \centering
    \includegraphics[width = 0.6\columnwidth]{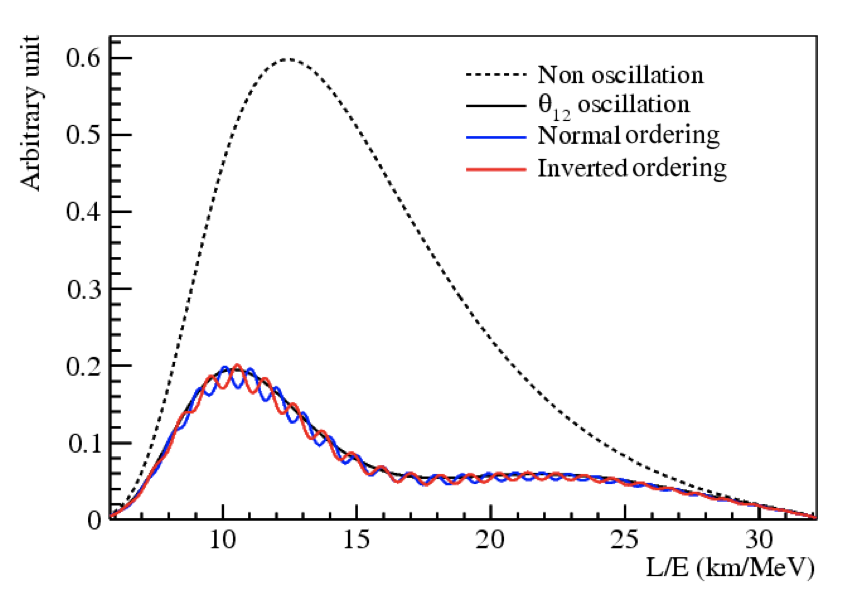}
    \caption{The reactor antineutrino $L/E$ spectra for different mass ordering. $L$ is the baseline and $E$ is the antineutrino energy. Figure is taken from Ref.~\cite{An:2015jdp}.}
    \label{fig:osc_spec}
\end{figure}

The three-neutrino oscillation framework has been well supported by the
observations from solar neutrinos, atmospheric neutrinos, accelerator
neutrinos and reactor antineutrinos.
The neutrino mixing matrix, relating the mass eigenstates ($\nu_1, \nu_2,
\nu_3$) and the flavor eigenstates ($\nu_e, \nu_{\mu}, \nu_{\tau}$), is commonly
expressed as the Pontecorvo-Maki-Nakagawa-Sakata (PMNS) matrix~\cite{Pontecorvo:1957cp, Pontecorvo:1967fh, Maki:1962mu}.
The neutrino oscillations can be described by six parameters: three mixing
angles, $\theta_{12}$, $\theta_{13}$, $\theta_{23}$, and two independent
mass splittings, $\Delta m^2_{21} \equiv m^2_2 - m^2_1$, $\Delta
m^2_{32} \equiv m^2_3 - m^2_2$ (or $\Delta m^2_{31}$, where $m_1, m_2$, and
$m_3$ are masses of the mass eigenstates), and one CP-violation phase
$\delta_{CP}$.
Today a very large set of oscillation results obtained with an amazing variety
of experimental configurations and techniques can be interpreted in the
three-neutrino framework.
The mixing angles and the mass splittings have been measured with
precision below 10\%~\cite{Tanabashi:2018oca}. The unknown CP-violation phase and the neutrino mass ordering (i.e.\ the sign of $|\Delta m^2_{32}|$) are the major goals of the next generation neutrino experiments.
The neutrino mass ordering (NMO) has two possibilities: normal ordering ($m_1 <
m_2 < m_3$) (NO) and inverted ordering ($m_3 < m_1 < m_2$) (IO).
The mass ordering can be determined by reactor antineutrino experiments at medium baselines (a few tens of kilometers) via the interplay effects between the short- and long-wavelength oscillations~\cite{PETCOV200294}.
The JUNO experiment aims to determine the neutrino mass ordering and to improve the uncertainties of three oscillation parameters to below 1\% by a precise measurement of the reactor
neutrino energy spectrum with an energy resolution of
$3\%/\sqrt{E[\textrm{MeV}]}$~\cite{An:2015jdp}.
In Figure~\ref{fig:osc_spec} the two energy spectra corresponding to NO and IO are reported.

In commercial reactors like the Taishan reactors, electron antineutrinos ($\bar\nu_e$) are generated from
thousands of beta decay branches of the fission products from four major
isotopes, $^{235}$U, $^{238}$U, $^{239}$Pu and $^{241}$Pu.
When detecting reactor antineutrinos via IBD reaction, the expected antineutrino energy spectrum in a detector at a given time $t$ is
calculated as
\begin{equation}\label{equ_generic}
    S_d(E_{\nu}, t)  = N_d\epsilon_d\sigma(E_{\nu})\sum_r \frac{P_{ee}(E_{\nu},L_{rd})}{ 4\pi L^2_{rd}}\phi_r(E_{\nu},t),
\end{equation}
where $E_{\nu}$ is the $\bar\nu_e$ energy, $d$ is the detector index, $r$ is the
reactor index, $N_d$ is the number of free protons in detector target, $\epsilon_d$ is the detection efficiency, $L_{rd}$ is the distance from detector $d$ to reactor $r$,
$P_{ee}(E_{\nu},L_{rd})$ is the $\bar{\nu}_e$ survival probability, $\sigma(E_{\nu})$ is the IBD cross section, and $\phi_r(E_{\nu},t)$ is the energy spectrum of antineutrinos from reactor $r$ which can be calculated as
\begin{equation}\label{equ_prediction}
\phi_r(E_{\nu},t) = \frac{W_r(t)}{\sum_i f_{ir}(t) e_i}\sum_i f_{ir}(t)s_i(E_{\nu}),
\end{equation}
where $W_r(t)$ is the thermal power of reactor $r$, $e_i$ is the mean energy released
per fission for isotope $i$, $f_{ir}(t)$ is the fission fraction, $s_i(E_{\nu})$
is the $\bar\nu_e$ energy spectrum per fission for each isotope.

The impacts of the uncertainties of the thermal power and fission fractions on the antineutrino flux are expected to be at sub-percent level according to the uncertainty evaluation in Daya Bay experience~\cite{An:2016srz}.
The energy spectrum per fission for each isotope has been estimated in literatures~\cite{PhysRev.113.280,PhysRevC.83.054615,Huber:PhysRevC84, VonFeilitzsch:1982jw,Schreckenbach:1985ep,HAHN1989365,PhysRevLett.112.202501,PhysRevC.24.1543,PhysRevLett.109.202504} by two main approaches.
One is the summation method~\cite{PhysRevLett.112.202501,PhysRevC.24.1543,PhysRevLett.109.202504} which
sums all the antineutrino energy spectra corresponding to thousands of beta
decay branches for about 1000 isotopes in the fission products, with information in nuclear databases.
This method results in an overall 10\%--20\% energy dependent uncertainty in the energy spectrum due to inadequate decay information and lack of relevant uncertainties in nuclear structures and fission yields.
The other method is the beta conversion method~\cite{PhysRev.113.280,PhysRevC.83.054615,Huber:PhysRevC84, VonFeilitzsch:1982jw,Schreckenbach:1985ep,HAHN1989365}
which converts the measured $\beta$ energy spectra from the individual fission
isotopes $^{235}$U, $^{239}$Pu, and $^{241}$Pu
to the corresponding antineutrino energy spectra using a set of virtual beta spectra.

\begin{figure}[htb]
\centering
\includegraphics[width=0.45\columnwidth]{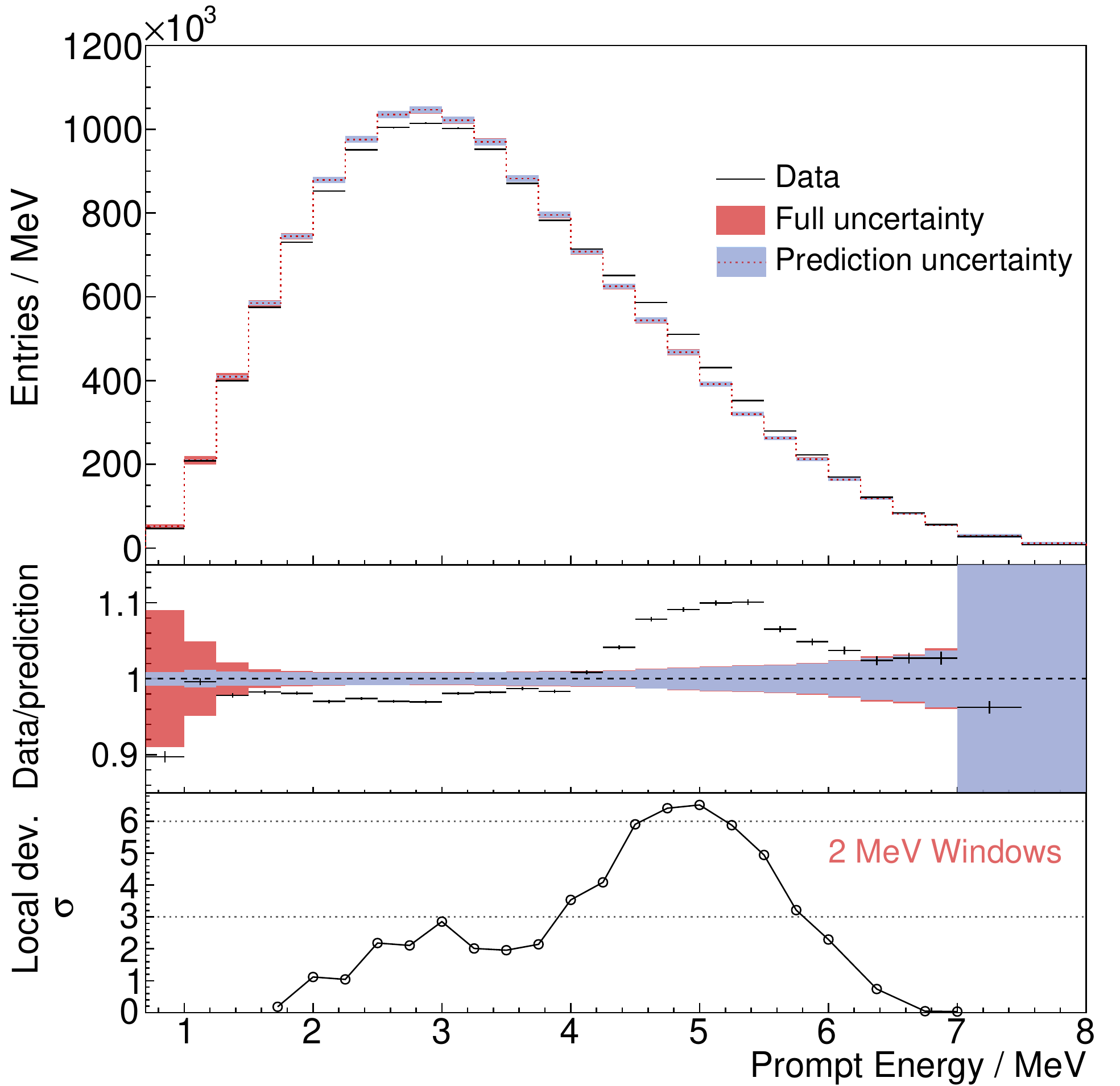}
\includegraphics[width=0.45\columnwidth]{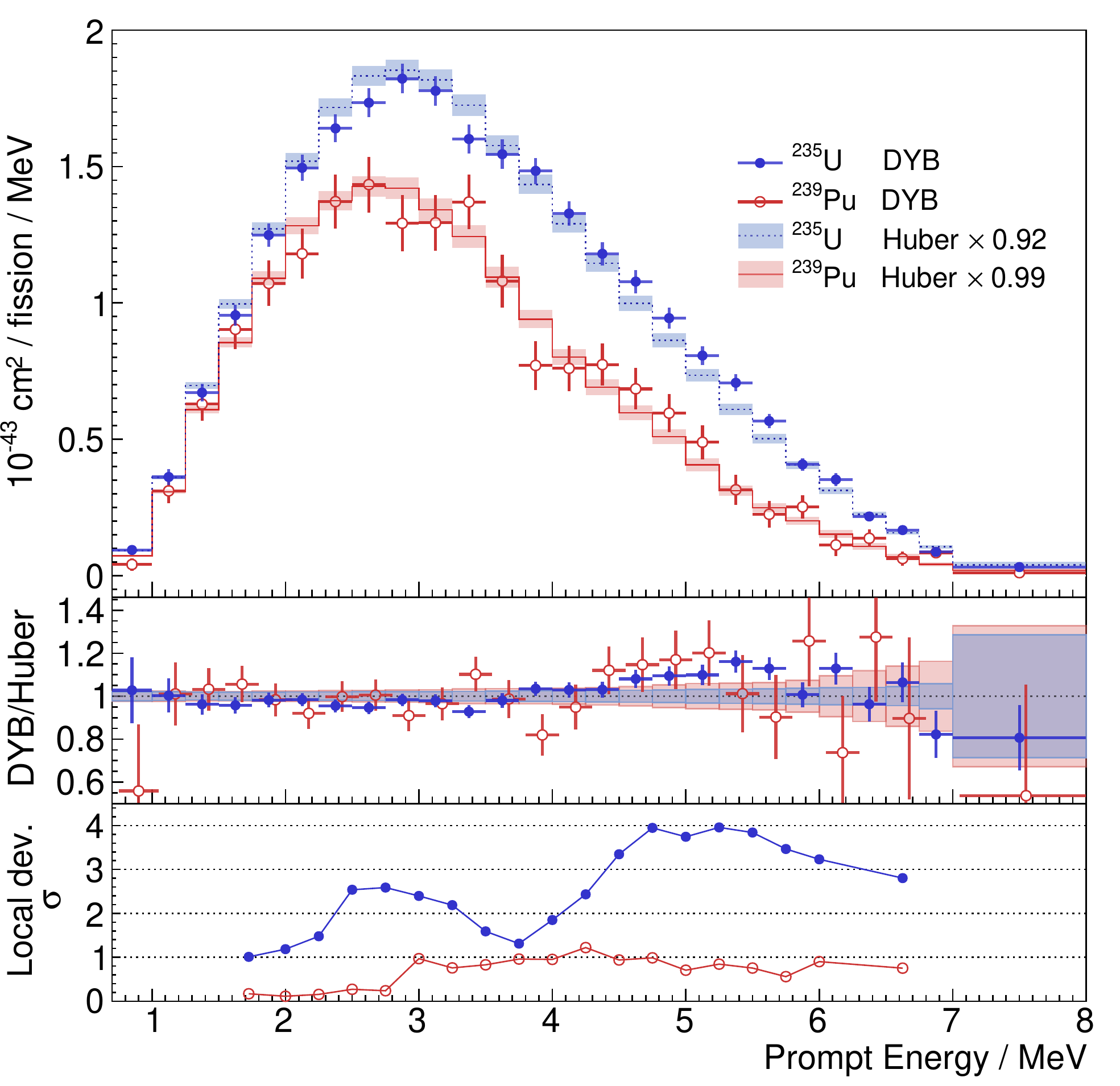}
\caption{ Left: Predicted and measured prompt energy spectra.
  The prediction is based on the Huber-Mueller model and is   normalized to the number of measured events.
  Right: Comparison of the extracted $^{235}$U and $^{239}$Pu spectra and the corresponding Huber-Mueller model predictions with the best-fit normalization factors 0.92 and 0.99, respectively.
  Ratio of the extracted spectra to the predicted spectra and the significance of local deviations are also shown in the middle and bottom panels, respectively. Figures are taken from Daya Bay~\cite{Adey:2019ywk}.
\label{fig:spectracomp}}
\end{figure}

The observed antineutrino yield per fission shows a deficit compared with the model predictions, namely, the reactor antineutrino anomaly~\cite{Mention:2011rk}.
The recent reactor antineutrino experiments, Daya Bay~\cite{An:2015nua}, RENO~\cite{Bak:2018ydk}, Double Chooz~\cite{DoubleChooz:2019qbj}, NEOS~\cite{Ko:2016owz}, and others confirmed the reactor antineutrino anomaly and observed a new discrepancy in the Huber-Mueller~\cite{Huber:PhysRevC84,PhysRevC.83.054615} model predictions.
The observed antineutrino energy spectrum shows an excess around 5~MeV
compared with the model predictions.
Figure~\ref{fig:spectracomp} shows the prompt energy spectrum compared with the model predictions at the Daya Bay experiment.
The variation of the energy spectrum versus the fission fractions is also studied and two major components, $^{235}$U and $^{239}$Pu, are extracted and compared with model predictions.
Those observations of the total energy spectrum and the extracted isotopic energy spectra at Daya Bay disagree with the model predictions. Although the Huber-Mueller model was used in these comparisons, summation models (e.g. Ref.~\cite{Estienne:2019ujo}) show a similar deficit and bump.

To reduce the impact from the uncertainties of reactor flux models,
reactor antineutrino experiments often deploy near detectors to provide the reference spectrum.
Daya Bay, RENO and Double Chooz experiments have shown the success of such relative measurements.
A precise measurement of the reactor antineutrino spectrum with an energy
resolution of $3\%/\sqrt{E[\textrm{MeV}]}$ at JUNO provides the sensitivity on
neutrino mass ordering when comparing it with the spectra predicted under the hypotheses
of normal mass ordering and inverted mass ordering, respectively.
The current reactor antineutrino experiments, such as Daya Bay, can provide a reference spectrum for JUNO to correct the problems of reactor flux anomaly and the 5-MeV bump.
However, the energy resolution is not sufficient to constrain the fine structure spectrum (details in later sections).

The TAO experiment will deliver a precise antineutrino energy spectrum measurement with sub-percent energy resolution in most of the energy region of interest, providing new and important data in addition to current reactor neutrino experiments.
With this new data, TAO can achieve several physics goals:
\begin{itemize}
    \item Measurement of a high-resolution antineutrino energy spectrum, which serves as a benchmark to test nuclear databases, provides increased reliability in measured isotopic antineutrino yields, and gives an opportunity to improve nuclear physics knowledge of neutron-rich isotopes;
    \item Providing the reference spectrum for JUNO to reduce the model dependence on the reactor antineutrino spectrum;
    \item Searching for light sterile neutrinos with a mass scale around 1~eV;
    \item Verification of the detector technology for reactor monitoring and safeguard applications.
\end{itemize}
Details of the above physics goals will be described in the following sections.

\subsection{Fine structure measurement}
\begin{figure}[htb]
    \centering
    \includegraphics[width = 0.6\columnwidth]{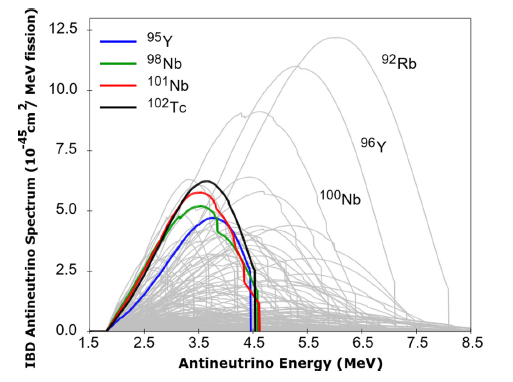}
    \caption{Calculated antineutrino energy spectra from many fission products in a commercial reactor. Figure is taken from Ref.~\cite{Sonzogni:2017voo}.}
    \label{fig:branch}
\end{figure}

In a reactor, the antineutrino energy spectrum is composed of spectra from
thousands of beta decay branches.
For each individual decay branch, the Coulomb correction produces a sharp edge at the
end point of the individual antineutrino spectrum.
As a result, the antineutrino energy spectrum has fine structure due to the
discontinuities at the edge of each decay branch.
A demonstration of percent-level fine structure in a spectrum calculated with the summation method is given e.g. in Ref.~\cite{PhysRevLett.114.012502}.
The popular Huber-Mueller model does not show fine structure because it uses about 30 virtual beta spectra without detailed structure to convert the $\beta$ spectra to antineutrino spectra.
Figure~\ref{fig:branch} shows an example of the summation calculation of antineutrino spectra from many fission products in Ref.~\cite{Sonzogni:2017voo}.
The cutoff at the edge of each decay branch is clearly visible.
However, the exact shape, amplitude and uncertainty of the fine structure is determined by thousands of beta-decay branches and thus it is hard to be quantified due to lack of information.
The measurement of the fine structure at TAO experiment will provide a benchmark to test nuclear databases, by comparing the measurement with the predictions of the summation method.
The measurement will also provide an opportunity to improve nuclear physics knowledge of neutron-rich isotopes~\cite{INDC-NDS-0786} in reactors.

\begin{figure}[htb]
    \centering
    \includegraphics[width = 0.6\columnwidth]{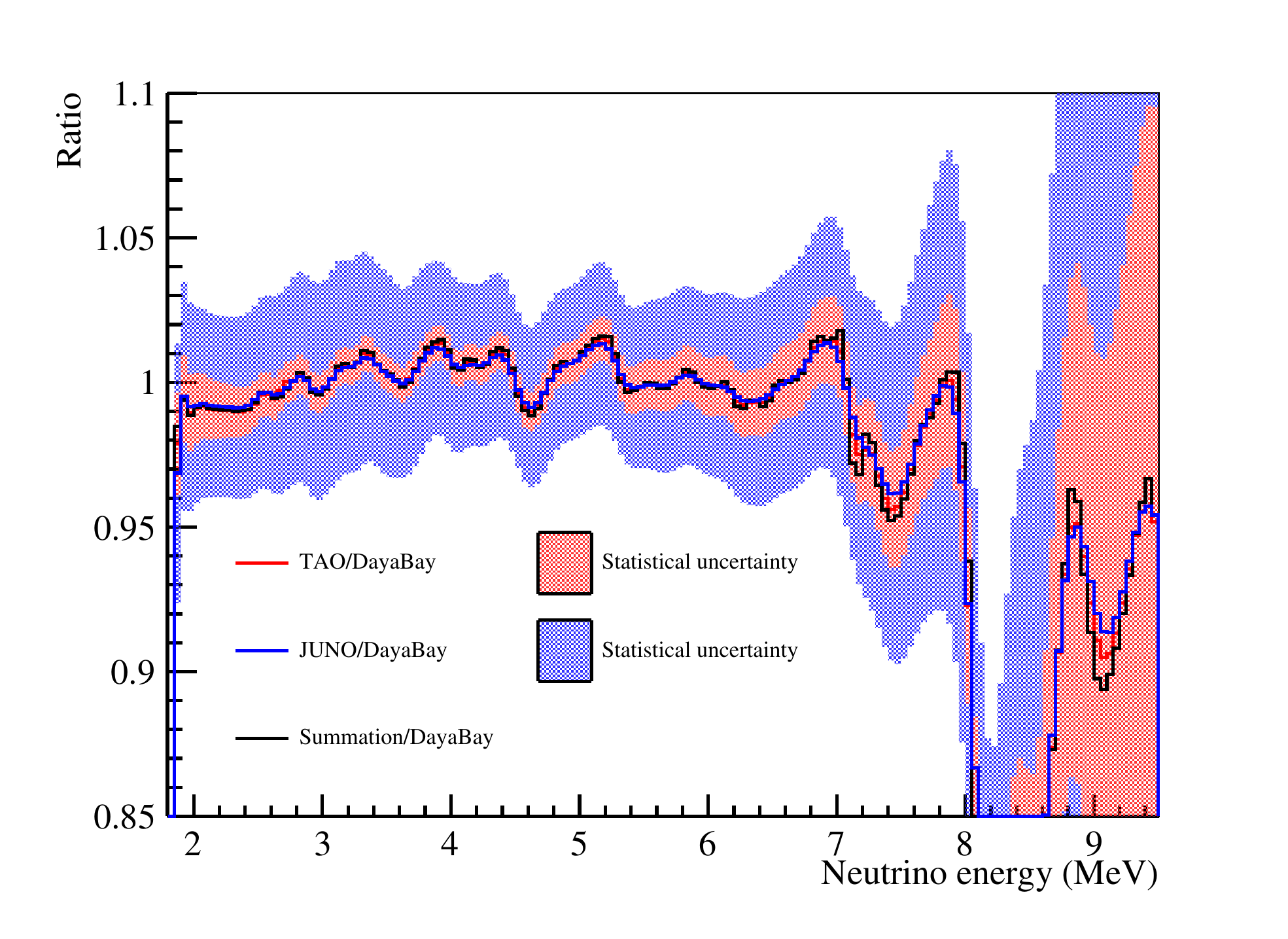}
    \caption{Comparison of the summation spectrum~\cite{Estienne:2019ujo} and three convoluted energy spectra with respective energy resolutions of TAO, JUNO and Daya Bay. Ratio of the other three spectra to the Daya Bay convoluted spectrum shows the difference is about 2\%. The TAO and JUNO spectra reproduces similar structures as the summation spectrum to less than 1\%. The bin width is set to be 50~keV.}
    \label{fig:finestructure}
\end{figure}

The Daya Bay neutrino experiment will eventually collect more than $5\times 10^6$
reactor antineutrino signals, the largest-ever sample.
This performance enables precise reactor antineutrino spectral shape measurement with sub-percent uncertainties around 3~MeV energy~\cite{Adey:2019ywk}.
However, the energy resolution of $8\%/\sqrt{E[\textrm{MeV}]}$ in Daya Bay~\cite{An:2016srz} is not sufficient to measure the fine structure of the energy spectrum.
Other reactor antineutrino experiments, such as RENO, Double Chooz, and NEOS are also limited by energy resolution ($5\%-7\%/\sqrt{E[\textrm{MeV}]}$).
TAO is designed to have a sub-percent energy resolution,
better than the JUNO experiment ($3\%/\sqrt{E[\textrm{MeV}]}$).
The impact of the energy resolution in the measurement of the fine structure is demonstrated by toy MC calculations.
The authors of Ref.~\cite{Estienne:2019ujo} provide a summation spectrum with sufficient energy bins to include the fine structure from the end points of each decay branches.
The summation spectrum is convoluted with different energy resolutions, corresponding to the designed values of TAO, JUNO and the actual value of Daya Bay.
Figure~\ref{fig:finestructure} shows the comparison of the summation spectrum and three convoluted energy spectra.
As shown in the figure, TAO and JUNO can reproduce the summation spectrum better than 1\% including fine structure, while Daya Bay is different at 2\% level.

Currently the summation method calculations have an uncertainty of about 10\%--20\% due to insufficient nuclear data.
For the fine structure, no reliable calculations are available with well presented uncertainties.
A precise measurement of the reactor antineutrino spectrum from the TAO experiment will provide a benchmark to validate the summation spectrum calculation.
With three years of data taking, TAO will collect about two million antineutrino events.
A statistical uncertainty below 1\% in the energy range of 2.5--6~MeV can constrain the fine structure to better than 1\%, providing a reference spectrum for JUNO with a bin width of about 30~keV~\cite{An:2015jdp}.

\subsection{Reference spectrum for JUNO}
The energy resolution is essential for JUNO to distinguish the multiple
oscillation pattern driven by $\Delta m^2_{31}$ and $\Delta m^2_{32}$ in the
hypotheses of normal mass ordering or inverted mass ordering.
The uncertainty of the fine structure in the antineutrino energy spectrum has
impact on the sensitivity of mass ordering.
Due to insufficient decay information and lack of uncertainties in the nuclear structures and
fission yields in nuclear databases, the summation method has an uncertainty at
the 10\% level.
Current predicted antineutrino spectrum from reactor flux models, including both
summation method and conversion method, disagrees with the measured spectrum at Daya Bay experiment and other reactor antineutrino experiments.
Thus, the current reactor flux models cannot provide a reliable reference spectrum
including fine structure for JUNO as an input of the neutrino mass ordering identification.

TAO will provide a precise reference spectrum for JUNO with sub-percent energy resolution, and the event rate will be 33 times higher than JUNO.
With the input spectrum from TAO, the predicted antineutrino energy spectrum for
JUNO without oscillations can be expressed as
\begin{equation}
    S_{\mathrm{JUNO}}(E_{\nu}) =  S_{\mathrm{TAO}}(E_{\nu}) + \sum_i \Delta f_iS_i(E_{\nu}),
    \label{eq:flux}
\end{equation}
where $S_{\mathrm{TAO}}(E_{\nu})$ is the reference antineutrino energy spectrum
from TAO, $\Delta f_i$ is the possible difference of fission fractions for four
major isotopes, and $S_i(E_{\nu})$ is the antineutrino spectrum for each
isotope.
If TAO has the same components of reactor antineutrino flux as JUNO, it will be an ideal near detector for JUNO to cancel all the antineutrino shape uncertainty.
However, since TAO detects mainly the antineutrinos produced by one of the Taishan reactor cores, it could measure a different flux with respect to the one seen by JUNO with possible different running time periods.
JUNO mainly receives the reactor antineutrinos from two Taishan reactors and six Yangjiang reactors.
Taishan and Yangjiang reactors are different types of reactors, with 4.6~GW and 2.9~GW thermal power respectively.
The difference of the fission fractions for four major isotopes, $^{235}$U, $^{238}$U, $^{239}$Pu and $^{241}$Pu are considered in the term related to $\Delta f_i$ in Eq.~\ref{eq:flux}.

\begin{figure}[htb]
    \centering
    \includegraphics[width = 0.6\columnwidth]{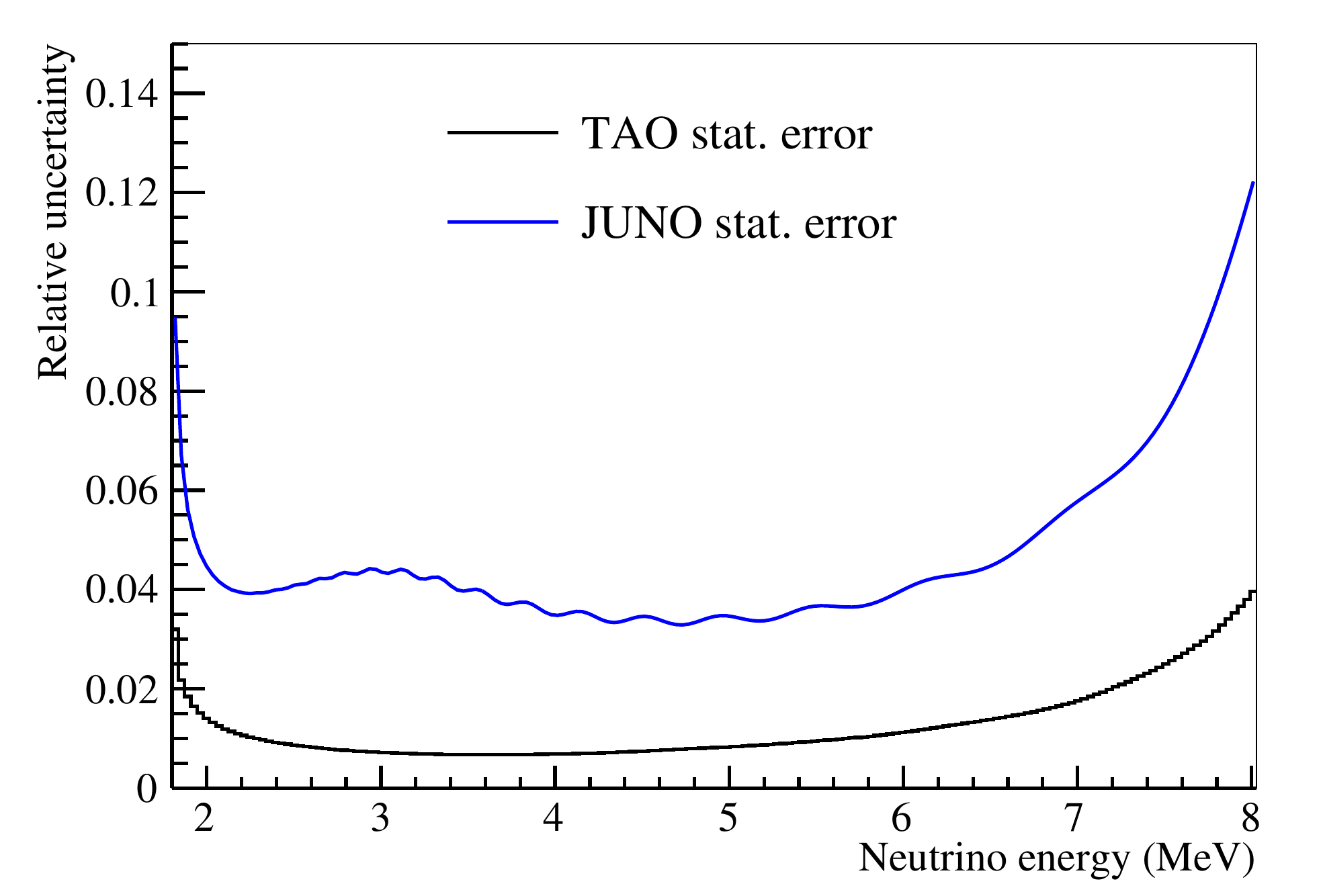}
    \caption{The statistical uncertainty of TAO (JUNO) with three (six) year's data taking.  }
    \label{fig:statistics}
\end{figure}

When using TAO as reference spectrum for JUNO, the statistical uncertainty of TAO will be propagated to JUNO as an input of the bin-to-bin spectral shape uncertainty.
Figure~\ref{fig:statistics} shows the statistical uncertainty of TAO with three years of data taking, and the statistical uncertainty of JUNO with six years of data taking.
The expected antineutrino event sample at TAO is nearly 20 times of JUNO.
It shows the statistical uncertainty of TAO is better than 1\% in most of the energy region of interest.

\begin{figure}[htb]
    \centering
    \includegraphics[width = 0.6\columnwidth]{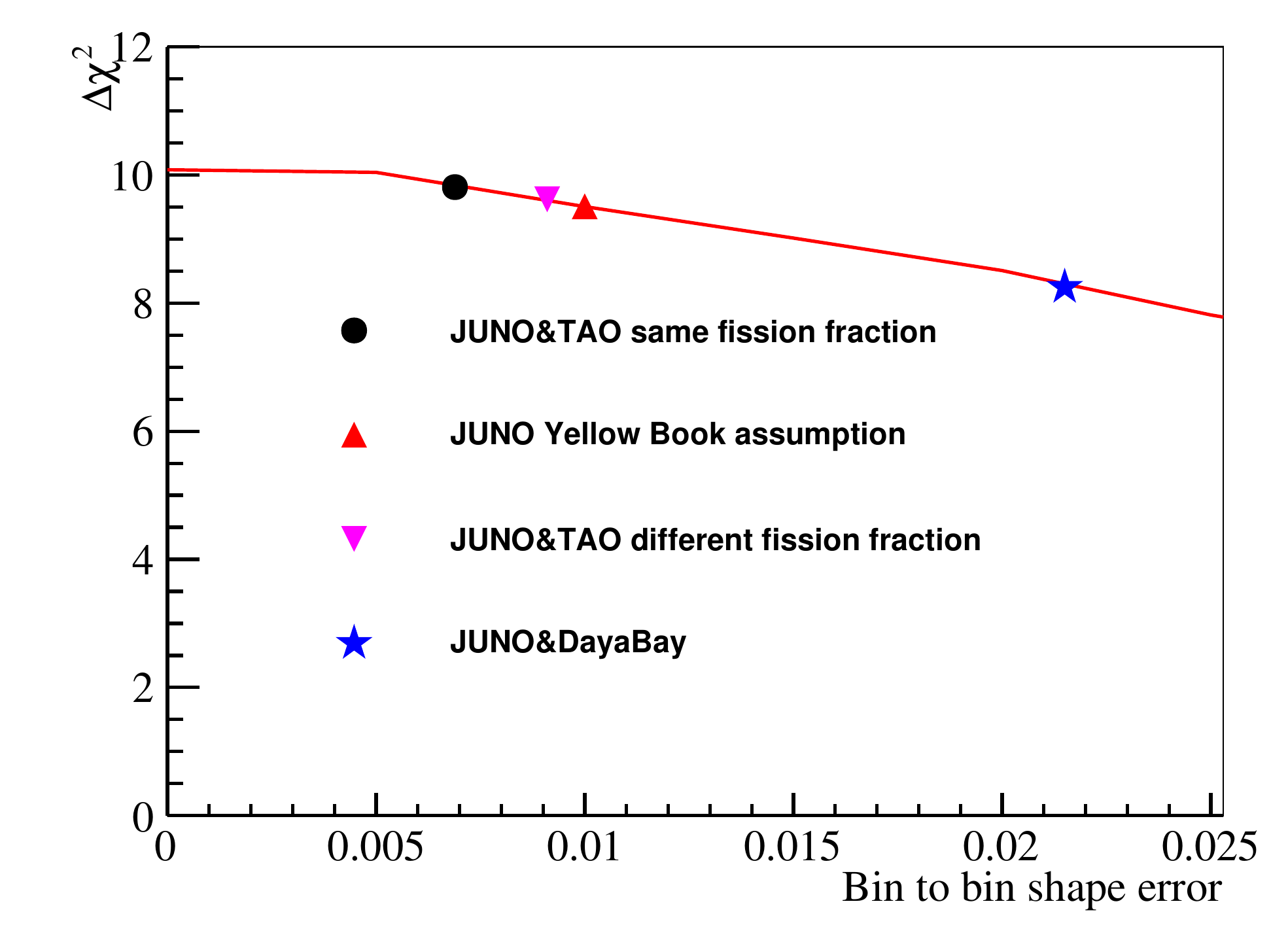}
    \caption{The neutrino mass ordering sensitivity of JUNO with the inputs in
    Ref.~\cite{An:2015jdp} as a function of the input bin-to-bin shape uncertainty.
    The $\Delta\chi^2$ represents the mass ordering determination sensitivity~\cite{An:2015jdp}, defined as the $\Delta\chi^2 = |\chi^2_{NO} - \chi^2_{IO}|$, where $\chi^2_{NO}$ ($\chi^2_{IO}$) is the standard $\chi^2$ of fitting the expected data in the hypothesis of normal ordering (inverted ordering) to the simulated data.
    Several cases of using TAO or Daya Bay constraint uncertainties and the JUNO yellow book~\cite{An:2015jdp} result are shown as markers.
    The markers show the cases for different reference spectra as inputs for JUNO.}
    \label{fig:MO}
\end{figure}

In Ref.~\cite{An:2015jdp}, a 1\% bin-to-bin spectral shape uncertainty is assumed for JUNO.
The bin-to-bin uncertainty could be as large as 10\% based on the input of the uncertainty of the summation spectrum.
Even with the constraint from the spectrum measurement of the Daya Bay experiment, the bin-to-bin uncertainty is at the level of 2\% as indicated in Figure~\ref{fig:finestructure} due to insufficient energy resolution of Daya Bay experiment.
With the constraint from TAO experiment, the bin-to-bin uncertainty can be reduced to below 1\% level.
With the assumption of 10\% difference on fission fractions for TAO and JUNO, the bin-to-bin uncertainty from the reference spectrum is about 1\%.
Figure~\ref{fig:MO} shows the mass ordering sensitivity of JUNO as a function of the input bin-to-bin shape uncertainty.
The markers show the cases for different reference spectra as inputs for JUNO.
The mass ordering sensitivity ($\Delta \chi^2$) is improved with the input of TAO by $\sim$1.5 compared with the case using the Daya Bay reference spectrum, and is slightly better than the result in Ref.~\cite{An:2015jdp} with assumption of 1\% bin-to-bin spectral shape uncertainty.

Another method to use TAO reference spectrum in JUNO instead of using Eq.~\ref{eq:flux} is to perform a combined analysis of TAO and JUNO spectra.
In this method, the correlation coefficients between TAO and JUNO data is crucial.
The constraint from the TAO spectrum is naturally implemented in the combined analysis.
A preliminary result of combined analysis obtains consistent results as shown in Figure~\ref{fig:MO}

\subsection{Search for light sterile neutrinos}
The majority of experimental data using accelerator, atmospheric, reactor and solar neutrinos can be explained by nowadays well-established three-flavor neutrino mixing, parameterized by the PMNS matrix~\cite{Pontecorvo:1957cp,Pontecorvo:1967fh,Maki:1962mu}. However, some observed phenomena are in tension with this three-flavor paradigm, when attempting to explain them by neutrino oscillations, which are a natural consequence of neutrino mixing. Those are: so-called Reactor Antineutrino Anomaly (RAA)~\cite{Mention:2011rk}, an observed reactor $\bar\nu_e$ deficit with respect to the state-of-the-art prediction models; anomalous $\bar\nu_e$ appearance in the $\bar\nu_\mu$ beam at the LSND~\cite{Aguilar:2001ty} and MiniBooNE~\cite{Aguilar-Arevalo:2013pmq,Aguilar-Arevalo:2018gpe} experiments; and deficit in number of $\nu_e$'s from radioactive calibration source in gallium experiments~\cite{KOSTENSALO2019542}. All those can be accommodated if we extend our model by an additional fourth neutrino with a mass splitting of approximate 1~eV$^2$. Corresponding flavor state would not participate in the weak interactions, since there are only three light active neutrinos~\cite{ALEPH:2005ab}, and thus it is called 'sterile'. Nevertheless, it can still mix with the active ones and demonstrates its presence via neutrino oscillations.

We investigate the TAO sterile neutrino sensitivity in the framework of a 3+1 model, which contains an additional sterile state and corresponding new mass state on top of the three known flavor and mass states. Taking into account TAO's short oscillation baseline, the oscillation probability for $\bar\nu_e$ disappearance can be approximated as:
\begin{equation}
    P_{\bar\nu_e\rightarrow\bar\nu_e}(L,E)=1-4\sum_{i=1}^{3}{\vert U_{ei}\vert^2\vert U_{e4}\vert^2\sin^2\frac{\Delta m_{4i}^2L}{4E}},
\end{equation}
where mass splittings are defined as $\Delta m_{ij}^2\equiv m^2_i-m^2_j$, $m_i$ being the mass of the i-th neutrino state, and
$\vert U_{ei}\vert$ are elements of the extended $4\times4$ unitary mixing matrix.
Using the parametrization of Ref.~\cite{Palazzo:2013bsa}, $U_{ei}$ can be expressed in terms of the neutrino mixing angles $\theta_{12}$, $\theta_{13}$ and $\theta_{14}$.

We assume a simple geometry of a cylindrical reactor with height of 3~m and radius of 2~m, where antineutrinos are produced uniformly in this volume. The TAO detector is spherical. Its center is placed at 10~m below the reactor center and 30~m far on the horizontal level to match the possible location of the experimental hall. Due to the proximity of the reactor and detector, both cannot be treated as point-like in this study and their dimensions are thus taken into account in the oscillation probability calculation.

The reactor antineutrinos are detected via IBD reaction on free protons. As a nominal setting, we assume 3~years of data taking with 80\% reactor time on and 50\% IBD detection efficiency. We use Huber-Mueller model ~\cite{Huber:PhysRevC84,PhysRevC.83.054615} to calculate the antineutrino spectra, however, with an inflated bin-to-bin uncorrelated shape uncertainty of 5\% (for 50~keV bin width), which is set to be a conservative estimate to the one in Ref.~\cite{An:2016srz}. The choice of the default spectrum shape has a negligible impact on the sensitivity.

We take into account major sources of background: accidental coincidences, decays of unstable muon spallation products, i.e.\ $^9$Li and {}$^8$He decays, and fast neutrons. Rates and spectrum shapes are determined from the TAO simulation (see Section~\ref{sec:Expoverview}). We assume uniformly distributed background when further dividing the fiducial volume into virtual segments.

In order to quantify the difference of expected spectra with the prediction, we define the $\chi^2$ as:
\begin{equation}
    \label{eq:chi2_steriles}
\begin{split}
    \chi^2_{min}=
    & \min_{\alpha's}\sum_i^{\text{segments}}\sum_{j}^{\text{bins}}\left(\frac{M_{ij}-P_{ij}}{\sqrt{M_{ij}}}\right)^2+\\
    & +\left(\frac{\alpha_{Acc}}{\sigma_{Acc}}\right)^2+\left(\frac{\alpha_{Li}}{\sigma_{Li}}\right)^2+
    \left(\frac{\alpha_{FN}}{\sigma_{FN}}\right)^2+\sum_i^{bins}\left(\frac{\alpha_i}{1}\right)^2,\\
\end{split}
\end{equation}
where $M_{ij}$ is the expected number of events, $P_{ij}$ is the prediction for $i$-th virtual segment of the fiducial volume and $j$-th energy bin. The prediction is given as:
\begin{equation}
    \label{eq:prediction}
    P_{ij}=(1+\alpha_R)R_{ij}+(1+\alpha_{Acc})A_{ij}+(1+\alpha_{Li})L_{ij}+(1+\alpha_{FN})F_{ij}+(1+\alpha_j)\sigma^{sh}_{ij},
\end{equation}
where $R_{ij}$ is the reactor antineutrino spectrum, function of sterile oscillation parameters $\sin^22\theta_{14}$ and $\Delta m^2_{41}$. $A_{ij}$, $L_{ij}$ and $F_{ij}$ are the accidentals, {}$^9$Li/{}$^8$He and fast neutron backgrounds respectively for i-th segment and j-th energy bin. Each of the components has a corresponding rate nuisance parameter~$\alpha$ and a relative uncertainty of 0.1\%, 10\% and 10\% for the backgrounds respectively. The antineutrino rate nuisance parameter is unconstrained. The last term in Eq.~\ref{eq:prediction} represents the spectra shape uncertainty and is defined as:
\begin{equation}
    \sigma_{ij}^{sh}=\sqrt{\left(\sigma_{R}^{sh}\times R_{ij}\right)^2+\left(\sigma_{Li}^{sh}\times L_{ij}\right)^2+\left(\sigma_{FN}^{sh}\times F_{ij}\right)^2}
\end{equation}
with corresponding nuisance parameters $\alpha_j$ fully correlated among segments and energy bin-to-bin uncorrelated. We use a 5\%, 10\% and 3\% bin-to-bin uncorrelated relative shape uncertainty (for 50~keV bin width) for reactor antineutrinos,  {}$^9$Li/{}$^8$He and fast neutrons, respectively. The accidentals spectrum is assumed to be known without uncertainty. We minimize over all nuisance parameters in Eq.~\ref{eq:chi2_steriles}.

We use the CL$_s$ statistical method~\cite{Read:2002hq,Junk1999435} to determine TAO sterile neutrino sensitivity, where we assume measured data to follow the classical three-neutrino model. The CL$_s$ method compares two hypotheses, in our case classical ($3\nu$) and alternative sterile neutrino ($4\nu$) scenarios. In order to further reduce the computational demands,  we employ the so-called Gaussian CL$_s$ method~\cite{CLsMethod}, which approximates parent distributions with normal ones.

As a nominal setting, we assume 3 years of data taking. Increased statistics improves the sensitivity only a little. We assume conservative 5\% (for 50 keV bin width) relative bin-to-bin uncorrelated reactor antineutrino shape uncertainty. This is a major systematic uncertainty and its improvement will result in stringent sterile neutrino limits. We can achieve two times better limits with a 2\% uncertainty. We use four virtual segments of the TAO detector, which will improve the sensitivity for $\Delta m^2_{41}\gtrsim0.3\text{ eV}^2$ utilizing their relative comparison.

\begin{figure}[htb]
    \centering
    \includegraphics[width = 0.6\columnwidth]{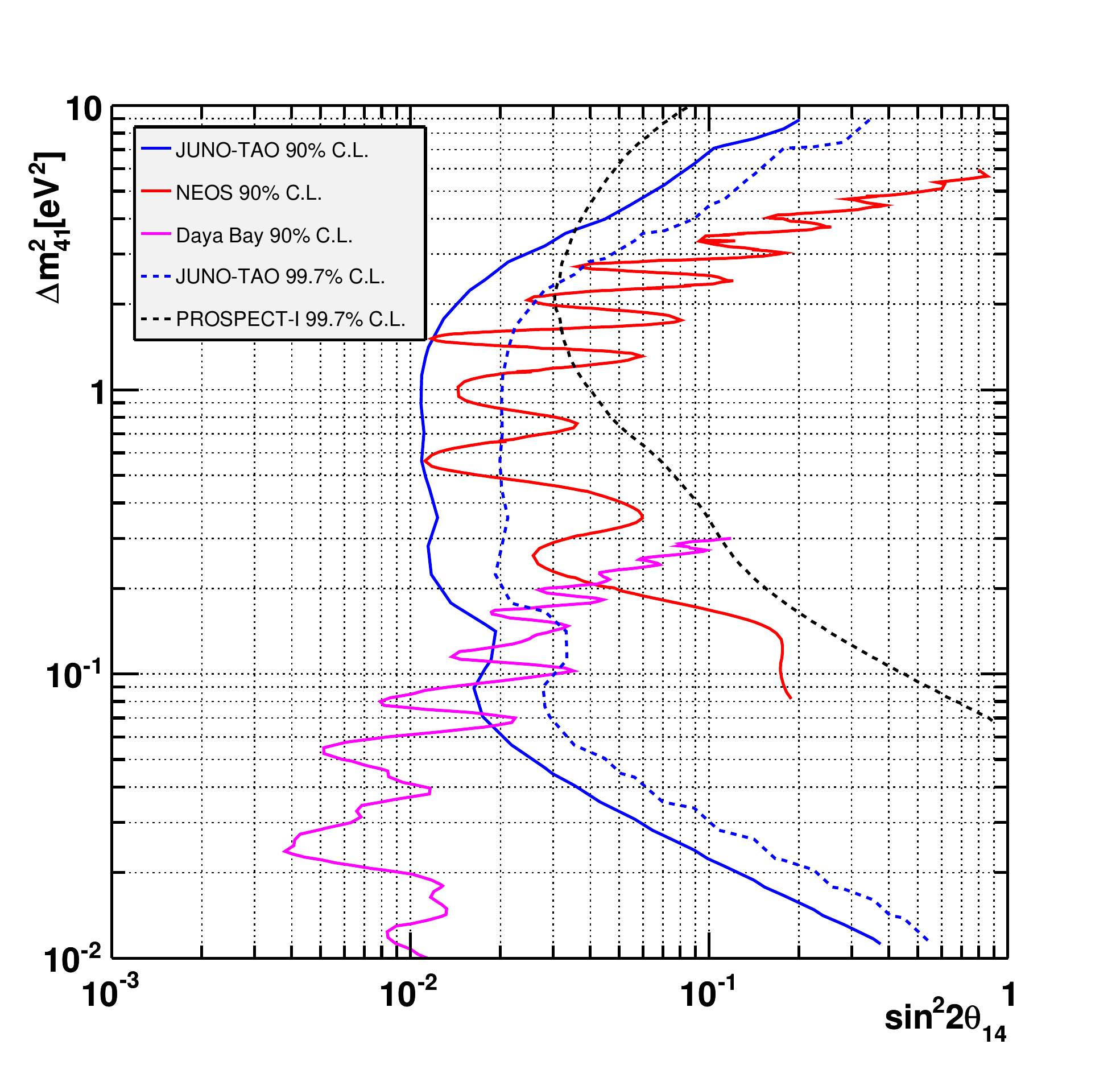}
    \caption{The comparison of TAO's sterile neutrino sensitivity using nominal settings with the Daya Bay~\cite{Adamson:2020jvo}, NEOS~\cite{Ko:2016owz} experiments at 90\% C.L. (solid lines) and PROSPECT phase-I~\cite{Ashenfelter:2015uxt} at 99.7\% C.L. (dashed lines). Parameter space to the right from the curves is excluded on more than the respective confidence level.}
    \label{fig:sterile_sensitivity_comparison}
\end{figure}

The search for sterile neutrinos via reactor antineutrino oscillations is in the scope of several experiments. The Daya Bay experiment used eight detectors placed at the baselines $\gtrsim300$~m to set the most stringent limit on the sterile neutrino mixing for $\Delta m^2_{41}\leq0.2\text{ eV}^2$~\cite{Adamson:2020jvo}. Experiments such as PROSPECT~\cite{Ashenfelter:2018iov}, STEREO~\cite{AlmazanMolina:2019qul}, DANSS~\cite{Alekseev:2018efk}, look for the oscillation signature at very short baselines $\sim$10~m covering large values of $\Delta m^2_{41}$ from $\sim$0.2~eV$^2$ to $\sim$20~eV$^2$. The intermediate distances of $\sim$30~m were explored by Bugey-3~\cite{Declais:1994su} and NEOS~\cite{Ko:2016owz} experiments, covering $\Delta m^2_{41}$ from approximately 3$\times10^{-2}$~eV$^2$ to about 5~eV$^2$. The TAO sterile neutrino sensitivity to the new mixing angle $\theta_{14}$ as a function of new mass splitting $\Delta m^2_{41}$ is shown together with a representative experiment of each baseline range in Figure~\ref{fig:sterile_sensitivity_comparison}. TAO is complementary to Daya Bay and those very short baseline experiments, demonstrated as an expected sensitivity of the PROSPECT phase-I~\cite{Ashenfelter:2015uxt} while it is competitive and eventually leading experiments at $\sim$30~m distances, here represented by NEOS. Bugey-3 has a similar sensitivity. Furthermore, the TAO experiment is likely to set the best sterile neutrino limits around $\Delta m^2_{41}=0.5\text{ eV}^2$ with a future improvement of the reactor antineutrino spectrum uncertainty expected from Daya Bay.

\subsection{Reactor monitoring and safeguard}
Antineutrino detectors have proven the ability to monitor in real time the nuclear reactor power~\cite{Bowden2009,Boireau:2015dda} and in the longer time scales the fuel composition~\cite{An:2017osx}.
This provided a complementary way of reactor monitoring with respect to the standard methods.
Moreover, such capability offers an interesting tool as a safeguard against undeclared and/or independent verification of the declared reactor power and fissile inventory.
The effort of developing such a monitoring tool promoted by the International Atomic Energy Agency (IAEA) is ongoing across the globe.
TAO is an ideal detector to greatly contribute to this effort.

\begin{figure}
    \centering
    \includegraphics[width = 0.6\columnwidth]{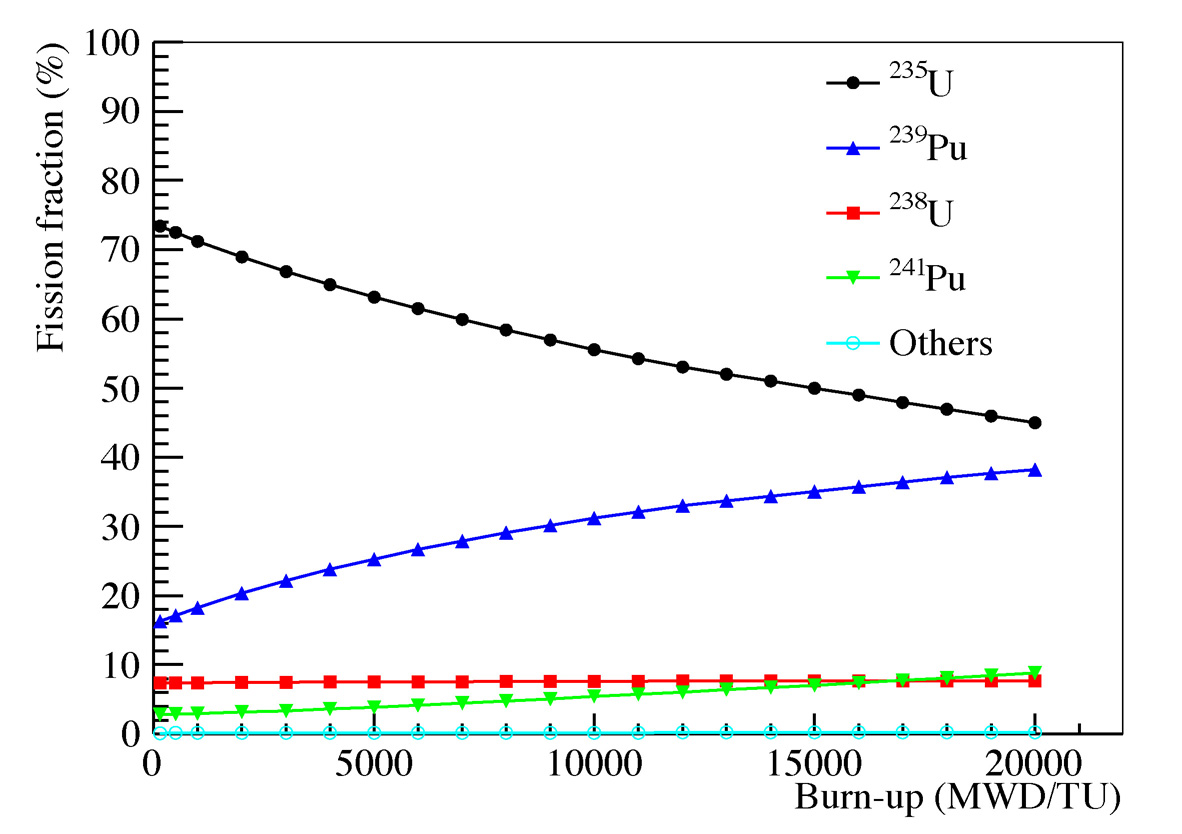}
    \caption{The fission fraction evolution for a typical running cycle of one Daya Bay reactor~\cite{An:2017osx}.}
    \label{fig:fissionfractions}
\end{figure}

More than 99.7\% of antineutrinos from a typical nuclear reactor come from decays of fission daughters of four major isotopes: ${}^{235}$U, ${}^{238}$U, ${}^{239}$Pu and ${}^{241}$Pu. The number of emitted neutrinos is in the first approximation proportional to the reactor power and enables the real time reactor power monitoring. However, in more detail, antineutrino flux and energy spectrum change with nuclear fuel composition evolution as ${}^{235}$U in the reactor fuel is consumed and ${}^{239}$Pu and ${}^{241}$Pu are produced during the operation of a commercial reactor. Figure~\ref{fig:fissionfractions} shows an example of evolution of the fission fractions, relative contributions of each isotope to the total number of fissions, for the four major isotopes during a running cycle of one Daya Bay reactor~\cite{An:2017osx}. The cycle between nuclear fuel replacement is usually few months long.

Naturally breed plutonium could be nonetheless subject of interest for military purposes, namely building nuclear weapons. To prevent such a proliferation, IAEA representatives would like to monitor the reactors operation and the fissile inventory. However, checks and sharing the operation information is not always available or inspections might not be infallible since it is not performed constantly. Neutrino detectors could provide such a missing information or independently verify their truthfulness, from reactor operation activity on daily bases, see e.g.~\cite{Bowden2009,Boireau:2015dda} to the fissile inventory in case of undeclared refueling and/or fuel processing, see e.g.~\cite{Christensen2015,Stewart2019}.

The main aim of the safeguard is to determine the amount of plutonium produced in the reactor and reveal its eventual removal by fuel reprocessing. This, as well as reactor monitoring in general, can be done from the overall neutrino flux and/or antineutrino energy spectrum measurements. Each of the four isotopes has a unique antineutrino yield and produces a unique energy spectrum. The observed neutrino flux and spectrum are linear combinations of four isotopes with contributions proportional to their fission fractions. The change of the fuel composition with burn-up leads to the neutrino flux and spectrum evolution as it was demonstrated e.g. in~\cite{An:2017osx}. Measuring these quantities with suitable detectors will allow to monitor the reactor performance and determine the amount of plutonium produced. The necessary input for such an analysis is of course knowing precisely the isotopic antineutrino yield and energy spectrum. Their accurate measurement is in demand since recent antineutrino experiments revealed discrepancy from theoretical predictions~\cite{An:2015nua,Ko:2016owz,Adey:2018qct,Bak:2018ydk,DoubleChooz:2019qbj}.

\begin{figure}
    \centering
    \includegraphics[width = 0.6\columnwidth]{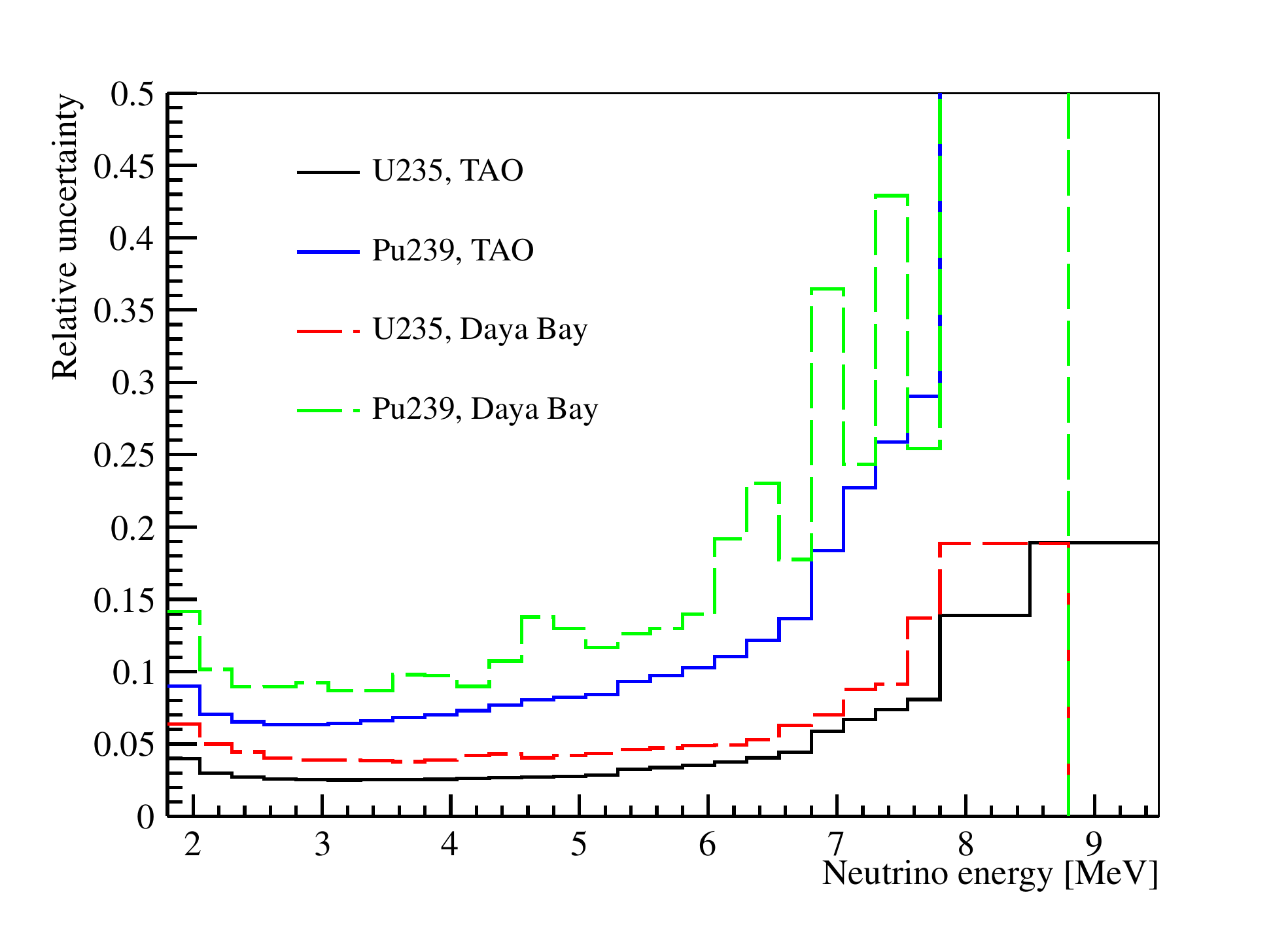}
    \caption{The expected relative spectrum shape uncertainty for the extracted ${}^{235}$U and ${}^{239}$Pu antineutrino spectra for three years of data taking of TAO and their comparison with the Daya Bay experiment results~\cite{Adey:2019ywk}.}
    \label{fig:decomposition}
\end{figure}

TAO will bring significant improvement in the precision of both flux and spectrum measurements. Based on the variation of the reactor antineutrino spectrum as a function of the fission fractions, the individual isotope spectra can be extracted in the experiment and later used by  other experiments as an input. Using the same method performed by the Daya Bay experiment~\cite{Adey:2019ywk}, which extracted the spectra for individual isotopes from the commercial reactor for the first time, antineutrino spectra can be acquired from the TAO data as well. The expected relative uncertainty of the extracted $^{235}$U and ${}^{239}$Pu antineutrino spectra for three years of data taking is shown in Figure~\ref{fig:decomposition}. The uncertainty for TAO is smaller than in the Daya Bay result due to the advantage of monitoring a single reactor as opposed to six, thus having among others a larger fission fraction variation. TAO will also provide fine structure shape due to its superb energy resolution $<$2\%~at~1~MeV.

TAO will join the global effort towards the nuclear reactor monitoring using reactor antineutrinos. Among current and proposed experiments, it is envisaged to provide the most precise measurement of the $^{235}$U and ${}^{239}$Pu antineutrino spectra from commercial reactors. In addition, spectra will be measured with unprecedented fine structure resolution. The TAO measurement can serve as an input for other reactor monitoring and safeguard studies.

%% file: ExpOverview/section.tex
\section{Experiment Overview}
\label{sec:Expoverview}
\blfootnote{Editors: Liang Zhan (zhanl@ihep.ac.cn)}
\blfootnote{Major contributors: Marco Grassi, Wei Wang, Lianghong Wei, Chengzhuo Yuan}

\subsection{The TAO detector}
The reactor antineutrinos detected by JUNO come mainly from six Yangjiang reactors and two Taishan reactors.
Six Yangjiang reactors have a full thermal power of 2.9~GW each.
Currently, the Taishan nuclear power plant has two reactors with a full thermal power of 4.6~GW each.
A candidate location for the TAO experiment is in a basement about 30~m from the center of one Taishan reactor core. With an overburden of several meters-water-equivalent, the measured cosmic muon flux is one third of that at the ground level.

The detector design will be described in detail in Section~\ref{sec:cd} and Section~\ref{sec:veto}.
Figure~\ref{fig:cdscheme} shows a sketch of the TAO detector.
It consists of a central detector (CD), including a cryostat in order to keep
the operating temperature at -50$^\circ$C, a water Cherenkov detector, and a
passive shield. The central detector is a liquid scintillator (LS) detector with a spherical acrylic vessel in a diameter of 1.8~m to contain the LS. A preliminary fiducial volume cut rejects the outer 0.25~m layer of LS, yielding 1~ton fiducial mass in a radius of 0.65~m. The LS mixture is based on Linear Alkylbenzene (LAB) because of its excellent transparency, high flash point, low chemical reactivity, and good light yield. The LS is loaded with 0.1\% gadolinium to reject effectively the accidental backgrounds with a delayed IBD neutron capture signal of $\sim 8$~MeV, much higher than the natural radioactivity backgrounds.
The liquid scintillator also consists of 2 g/L 2,5-diphenyloxazole (PPO) as the
fluor and 1 mg/L p-bis-(o-methylstyryl)-benzene (bis-MSB) as the wavelength
shifter. A small amount of ethanol (0.1\%) is added
in order to maintain the optical properties of the mixture at low temperature. The density of the GdLS is 0.916 g/ml at -50$^\circ$C. The light yield is about 12000 photons per MeV.
The liquid scintillator is contained in an acrylic vessel, which is submerged in a liquid buffer in a cylindrical cryostat with a radius of about 2 m, which preserves the temperature.
The cryostat is filled with non-scintillating LAB  as a liquid buffer in order to maintain good thermal performance. The scintillation light produced in the LS is detected by about 4100 Silicon Photomultiplier (SiPM) tiles with a total area of $\sim 10$ m$^2$, as described in Section~\ref{sec:sipm}, which ensure a high photo-detection efficiency and $>95\%$ photo-coverage.
To reduce the dark noise to a manageable level, the SiPM tiles need be cooled down to a low temperature (-50$^\circ$C). A copper sphere encloses the acrylic vessel providing the mechanical support to keep the SiPM tiles pointing to the center of the detector. The outer surface of the copper sphere is used to instrument the readout electronics and support the cooling pipes.

The central detector is surrounded by water tanks to shield environmental radioactivity from the rock and air.
The water tanks are equipped with 3-inch Photomultiplier Tubes (PMTs) to detect the Cherenkov light from cosmic muons, acting as a veto detector, with an expected efficiency of $>90$\%.
In addition to the water tanks, layers of High Density Polyethylene (HDPE) are placed on top of the TAO detector in order to provide a passive shield, mainly against the neutrons produced by the cosmic muons and the radioactivity from the materials outside the detector.
The HDPE shielding is covered by a plastic scintillator layer on top for tagging the cosmic muons.
Lead bricks are laid at the bottom of the detector, acting as a passive shield against the radioactivity.

\subsection{Signal}
\begin{figure}[htb]
    \centering
    \includegraphics[width = 0.6\columnwidth]{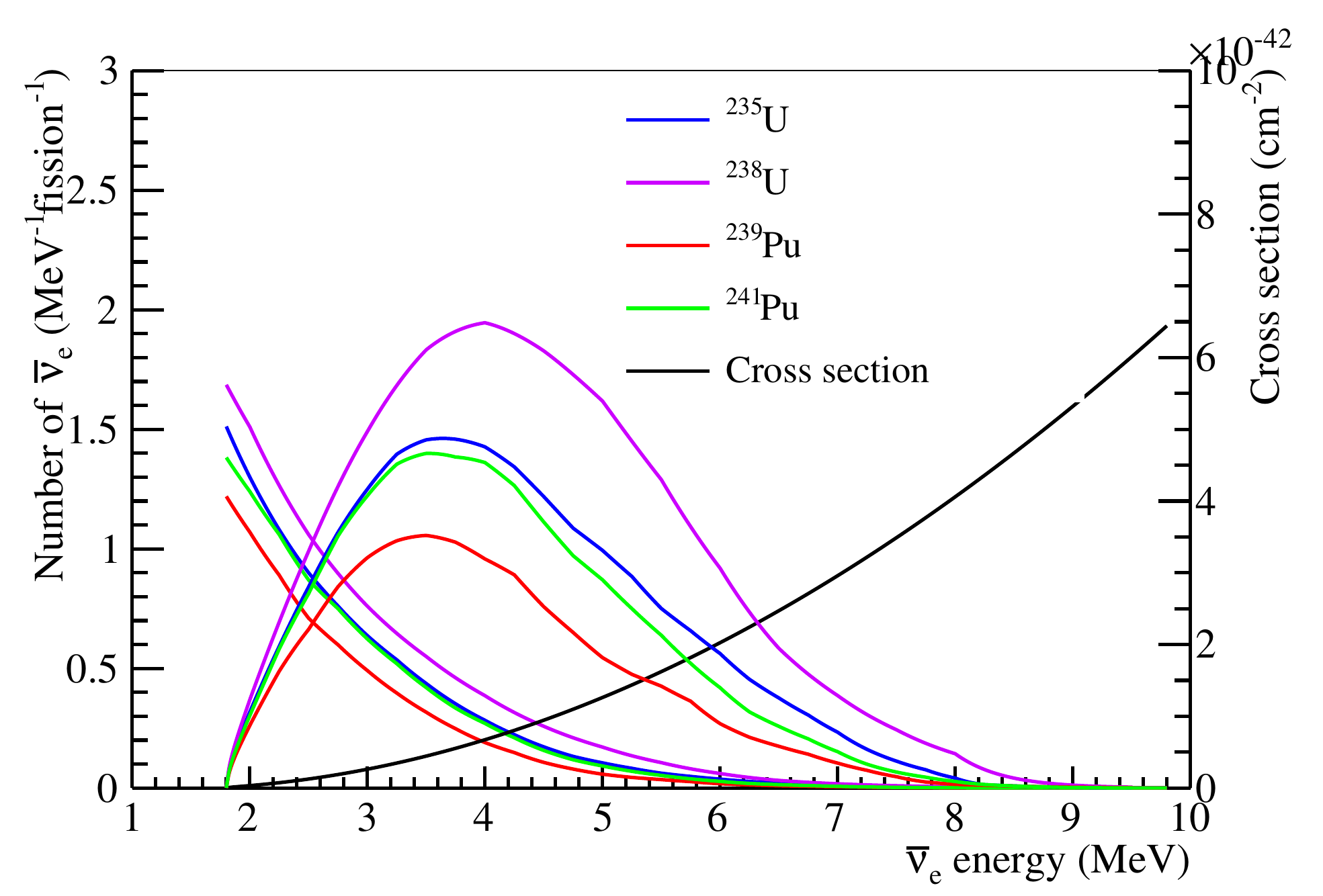}
    \caption{Antineutrino energy spectrum per fission in a commercial reactor weighted by the inverse beta decay cross section, from Ref.~\cite{An:2016srz}.}
    \label{fig:nu_spec}
\end{figure}

Reactor antineutrinos ($\bar\nu_e$) are generated from the fission products of four major isotopes, $^{235}$U, $^{238}$U, $^{239}$Pu  and $^{241}$Pu.
The $\bar\nu_e$ energy spectrum is measured via the inverse $\beta$-decay
(IBD) reaction, $\bar\nu_e \ + \ p \rightarrow e^+ \ + \ n $, in the gadolinium-doped liquid scintillator.
The IBD cross section increases steadily for energies above its 1.8 MeV threshold.
The antineutrino spectrum from a commercial reactor decreases with increasing energy, therefore the resulting IBD spectrum has a bell shape with the maximum around 3.5-4.0 MeV, as reported
in Figure~\ref{fig:nu_spec} from Ref.~\cite{An:2016srz}.

The coincidence of the prompt scintillation generated by the $e^+$ with the delayed neutron capture on Gd provides a distinctive  $\bar\nu_e$ signature.
The IBD neutrons are predominantly captured by hydrogen emitting one 2.2~MeV gamma or by gadolinium emitting several gammas with a total energy of about 8~MeV. The average capture time is about 30 $\mu$s with 0.1\% loaded Gd by mass.
The gammas produced by the gadolinium capture are clear signatures of the IBD events above the energy of natural radioactivities.

The antineutrino energy $E_{\bar\nu_e}$ is correlated to the detected prompt energy from the
positron, $E_{e^+}$, as $E_{\bar\nu_e} \approx E_{e^+} + (m_n - m_p - m_e)$. The kinetic energy of the outgoing neutron is less than tens of keV, which can be ignored in a first-order approximation.
The energy deposited by the positron in the scintillator converts to light, and the energy resolution is in a first-order approximation determined by the photon counting statistics.
The light yield of the LS is larger than 12000 photons per MeV, and about 4500 photoelectrons can be collected, corresponding to an energy resolution of $1.5\%/\sqrt{E [\mathrm{MeV}]}$.
TAO is designed to provide a photon detection efficiency of $\sim50$\%.
This requirement can be satisfied by using the SiPMs as the photosensors.

The expected antineutrino energy spectrum in the TAO detector ignoring the neutrino oscillation is expressed as
\begin{equation}\label{equ_nuspec}
    S(E_{\nu})  = \frac{N_p\epsilon\sigma(E_{\nu})}{ 4\pi L^2}\phi(E_{\nu}),
\end{equation}
where $E_{\nu}$ is the $\bar\nu_e$ energy, $N_p$ is the target proton number, $\epsilon$ is the detection efficiency, $L$ is the distance from detector to the reactor,
$\sigma(E_{\nu})$ is the IBD cross section, and $\phi(E_{\nu})$ is the reactor antineutrino flux integrated over time. The $N_p$ is about $7.2 \times 10^{28}$ per ton of LS assuming a 12\% hydrogen mass fraction.
The baseline is about 30~m. The reactor antineutrino flux from the Taishan reactor core with 4.6~GW thermal power is calculated with a nominal fission fraction of 0.561, 0.076, 0.307 and 0.056 for $^{235}$U, $^{238}$U, $^{239}$Pu and $^{241}$Pu, respectively.

\begin{figure}[htb]
    \centering
    \includegraphics[width = 0.6\columnwidth]{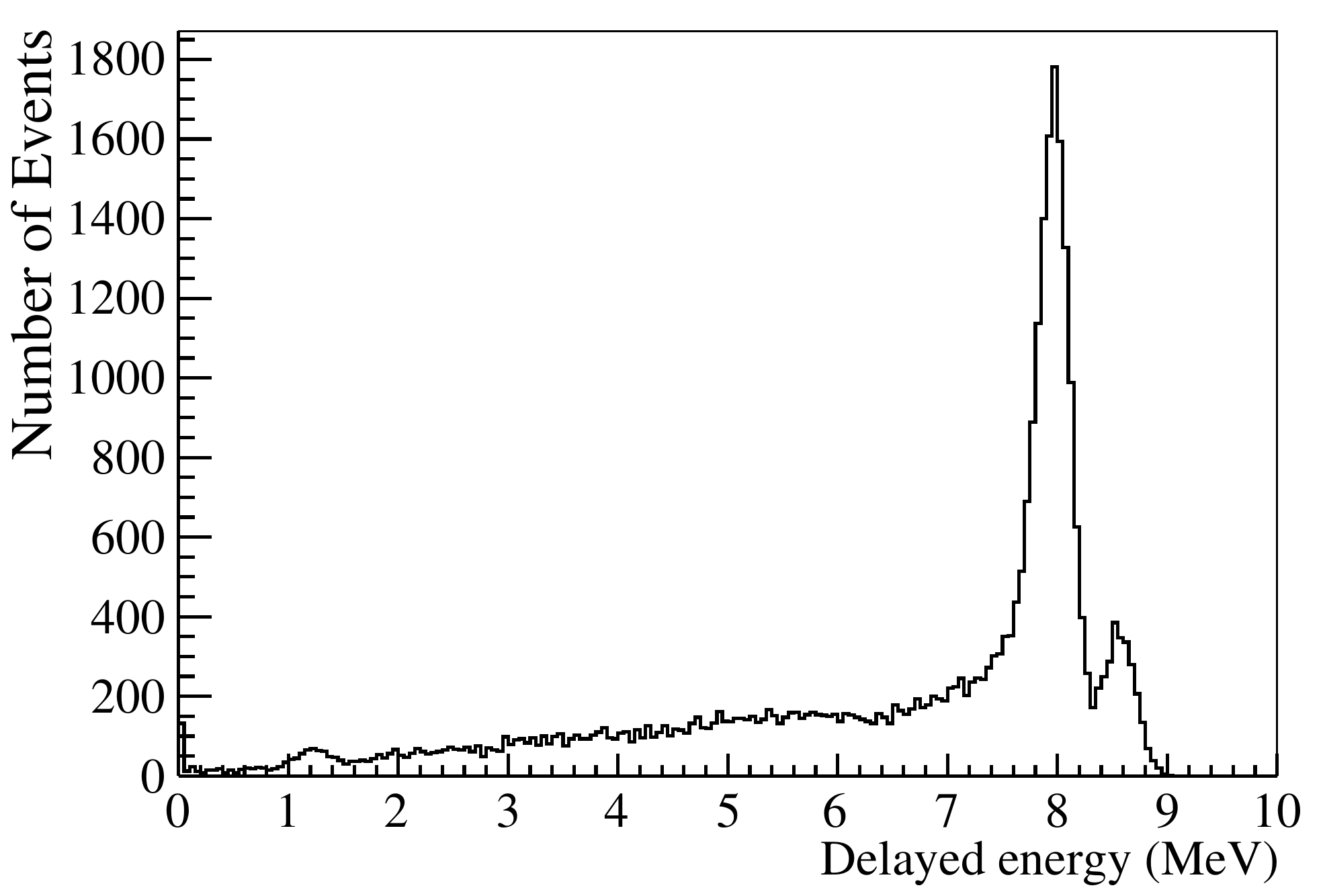}
    \caption{The delayed energy distribution in the detector for the Gd-capture events in the fiducial volume. Two peaks around 8~MeV correspond to captures on two natural isotopes of Gd, 7.94~MeV for $^{157}$Gd and 8.54~MeV for $^{155}$Gd. }
    \label{fig:delayE}
\end{figure}

The IBD signals are selected based on the tagging of the $\sim 8$~MeV signal of neutron capture on Gd as the delayed signal. The H capture signal with a lower energy of 2.2~MeV is not considered in the IBD selection, otherwise the background rate will increase by about one order of magnitude. The overall detection efficiency with a preliminary set of cuts is about 50\% in the 1~ton fiducial volume from the Geant4 simulation.
The IBD detection efficiency can be broken down to:
\begin{enumerate}
  \item The IBD neutrons are mostly captured by Gd and H. The Gd capture fraction is 87\%, which is the main detection channel. The H capture fraction is 13\%, and the C capture fraction is less than 0.1\%.
  \item To select the $\sim 8$~MeV delayed signal of neutron capture on Gd, a 7--9~MeV energy cut can be applied and this yields a 59\% detection efficiency. A simulation of the delayed energy for the neutron capture on Gd events is shown in Figure~\ref{fig:delayE}.
  \item The efficiency for the prompt energy cut ($> 0.9$~MeV) is 99.8\%.
  \item The efficiency for the prompt-delayed coincidence time cut ($1~\mathrm{\mu s}<\Delta T < 100~\mathrm{\mu s} $) is about 97\%.
\end{enumerate}

Integrating the $\bar\nu_e$ energy in Eq.~\ref{equ_nuspec}, the TAO detector will detect about 2000 IBD events per day in the fiducial volume based on the preliminary selection cuts. The selection cuts are still in the progress of optimization with the consideration of the detector design and the backgrounds.
The observed number of IBD events is further reduced by the decrease of live time due to the application of cosmic-ray muon veto. The expected muon veto efficiency is larger than 90\%.

\subsection{Backgrounds}
A background for IBDs consists of two events which pass the selection criteria but are not caused by the reactor antineutrinos.
These two events, prompt and delayed, can be correlated or uncorrelated in time.
Two uncorrelated but randomly close events passing the energy cuts form a so-called accidental background.
Natural radioactivity is one of the major sources of the prompt events because it easily passes the $> 0.9$~MeV prompt energy cut but is impossible to pass the $> 7$~MeV delayed energy cut.
This is also one of the reasons to load Gd in liquid scintillator for producing delayed events with energy $\sim 8$~MeV.
The neutrons in the environment or produced by cosmic muons are the major sources of delayed candidate if they are captured on Gd.
Another type of background is correlated background if the prompt and delayed signals are correlated in space and time, such as the fast neutron background and $^8$He/$^9$Li background.
They are produced by the cosmic muons as spallation products.

\begin{figure}[htb]
    \centering
    \includegraphics[width = 0.6\columnwidth]{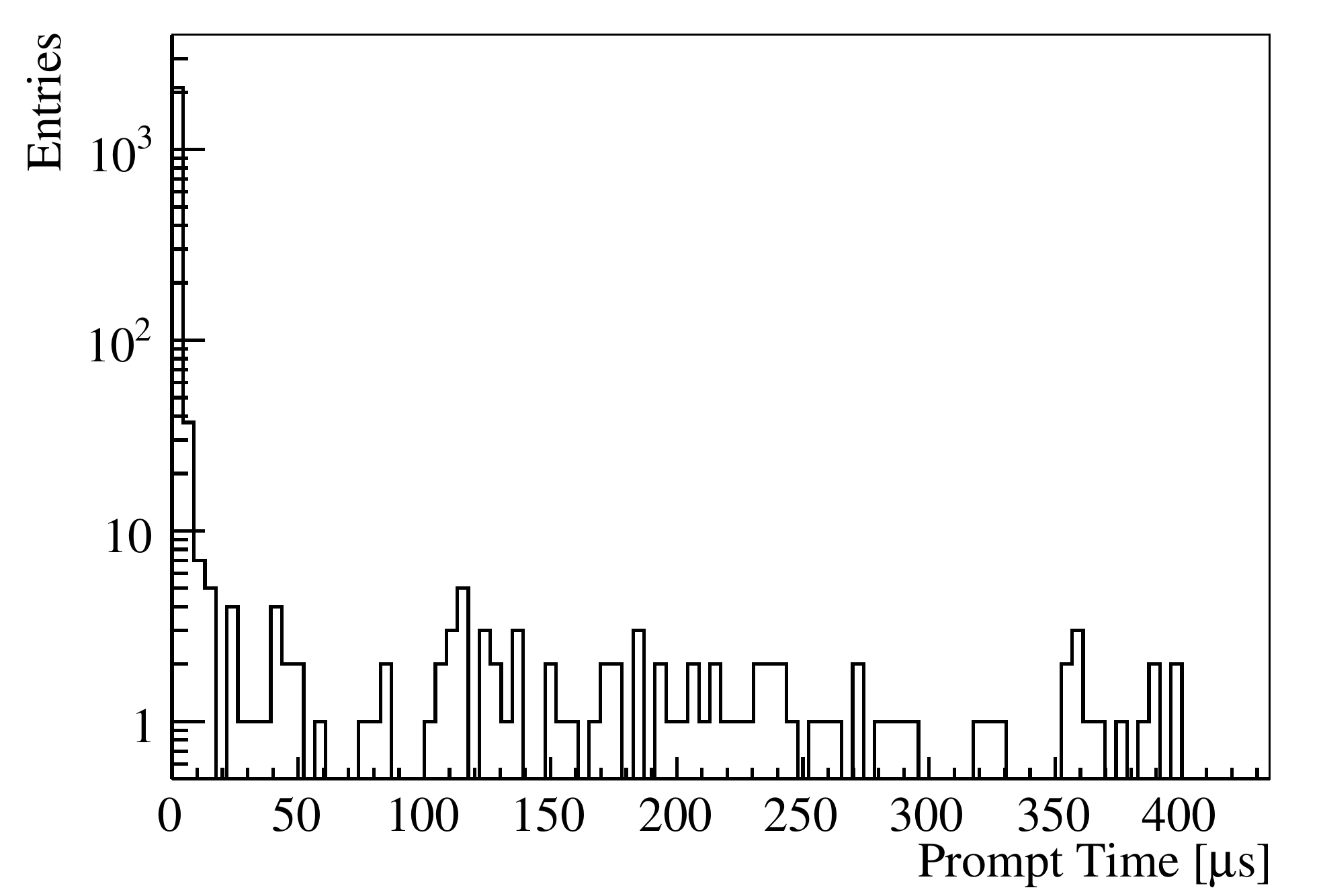}
    \caption{The time interval between the prompt signal of the selected fast neutron background and the preceding muon event.}
    \label{fig:promptT}
\end{figure}

\begin{figure}[htb]
    \centering
    \subfigure{\includegraphics[width=0.48\columnwidth]{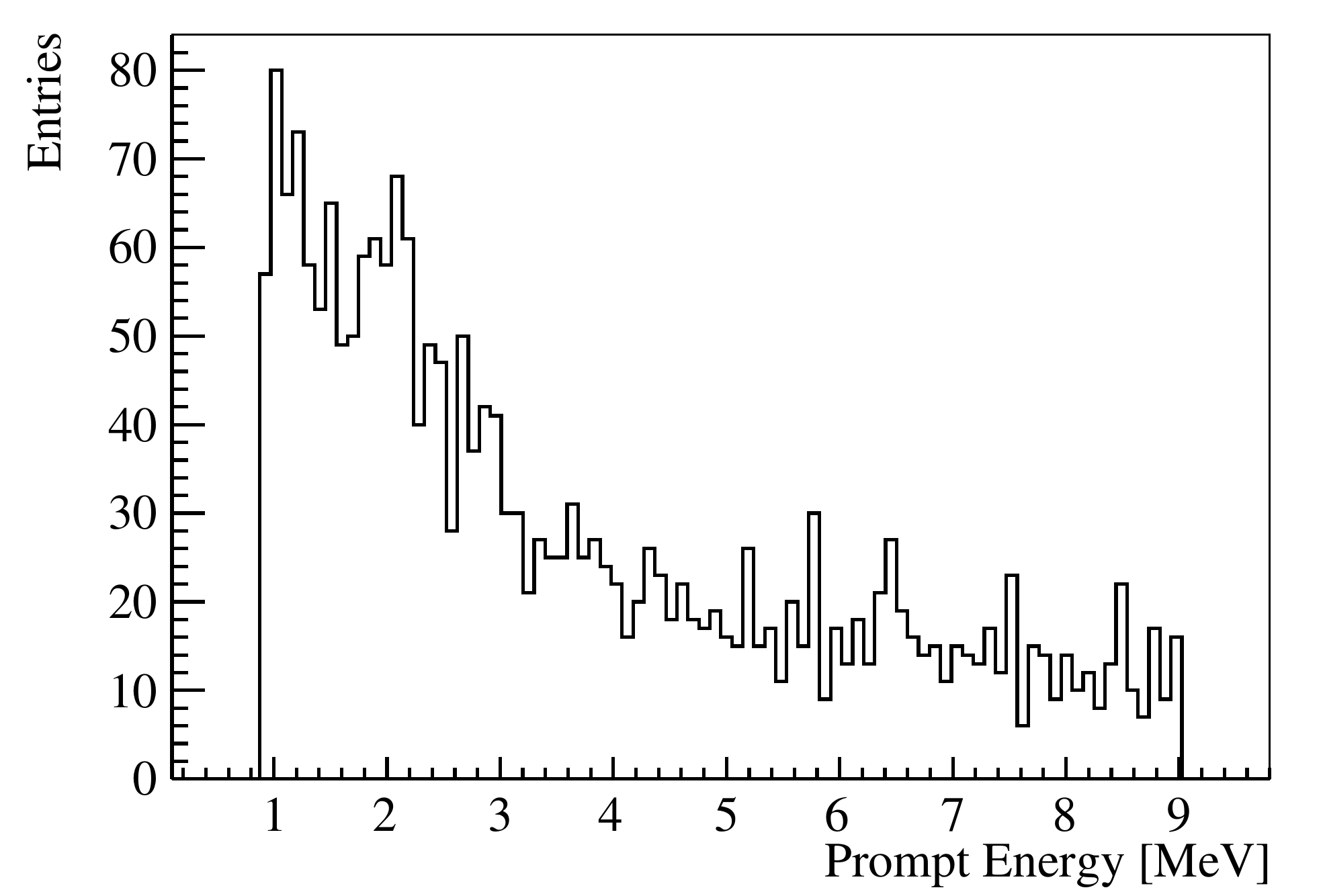}}
    \subfigure{\includegraphics[width=0.48\columnwidth]{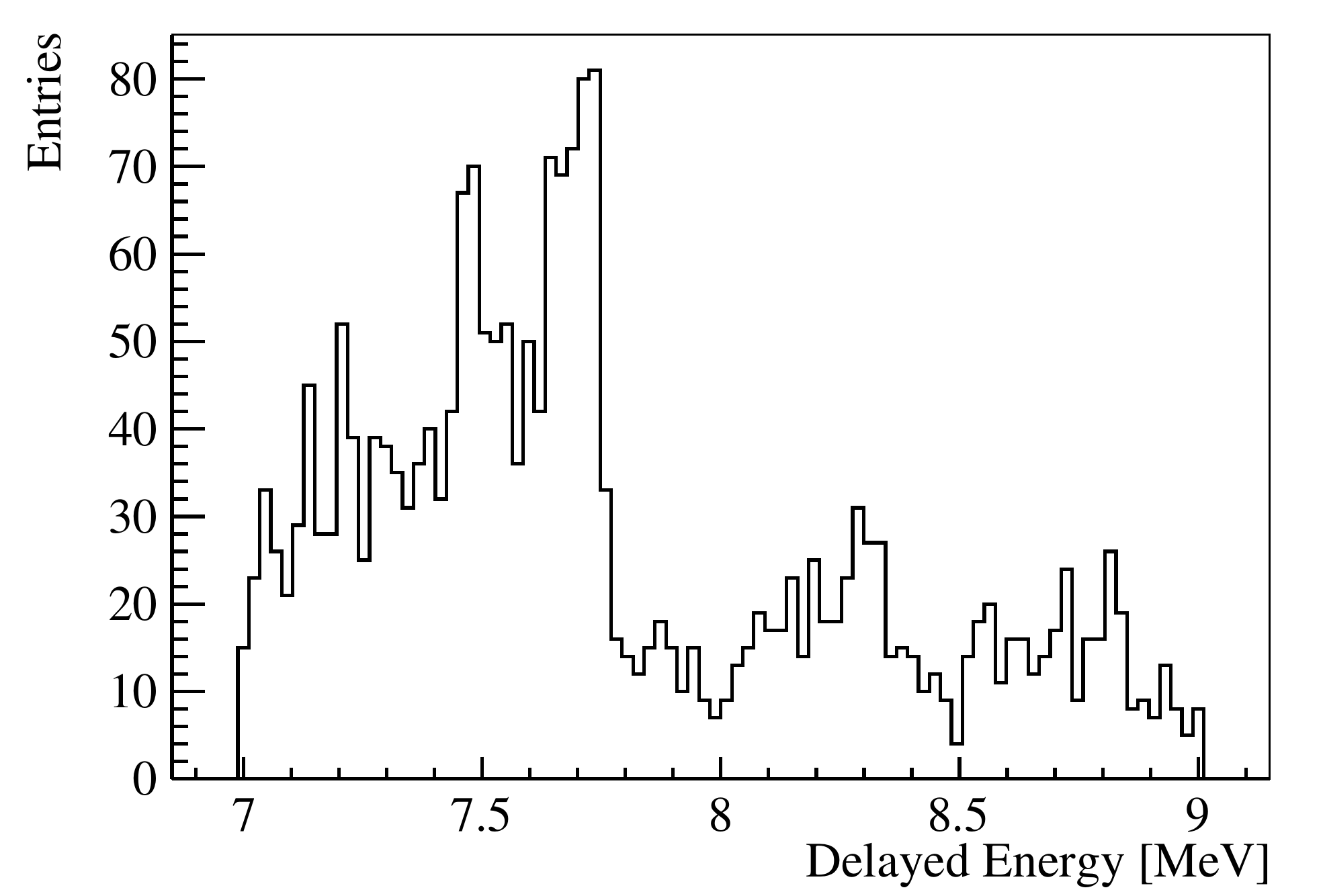}}
    \caption{Prompt (left) and delayed (right) energy spectra of the fast neutron background.}
    \label{fig:FNenergy}
\end{figure}

The fast neutron background, arising from cosmic muon spallation in the materials surrounding the detector, is the leading background component due to relatively small overburden of the TAO experiment.
Fast neutron interactions are characterized by a prompt energy deposit due
to recoiled proton, and by a delayed deposit due to the neutron capture upon
thermalization.
The muon rate in the experimental hall is about 70~Hz/m$^2$, namely one third of the rate on the surface.
A muon event generator is constructed to generate muons on the ground and a Geant4 simulation package is used to propagate the muon to the detector.
The detector geometry, shown in Figure~\ref{fig:cdscheme}, is implemented in the simulation.
The fast neutron background is selected with the same selection criteria as the IBD selection for the muon simulation sample.
The fast neutron background rate is estimated to be 1880~events/day assuming no muon veto applied, similar to the IBD signal rate.
Figure~\ref{fig:promptT} shows time interval between the prompt signal in the selected fast neutron background and the preceding muon event.
The proton recoils by energetic neutrons should complete within microsecond.
Some prompt events with time interval larger than 10~$\mu$s could be formed by the neutron capture signals when multiple neutrons are produced by the muons.
Figure~\ref{fig:FNenergy} shows the energy spectrum of the prompt and delayed signals of the fast neutron backgrounds.
The 8-MeV Gd-neutron-capture peak is clearly seen in the delayed energy spectrum.

To reduce the fast neutron background, the muon veto will be applied.
The muon event rate, tagged by the veto detector, either by the water tank or the plastic scintillator, is about 4000~Hz. The veto time window cannot be as long as the criteria used in the Daya Bay (600~$\mu$s)~\cite{An:2012eh} or JUNO experiment (1.5~ms)~\cite{An:2015jdp}. That would result in a 100\% dead time. A preliminary 20~$\mu$s veto window is envisaged, which will introduce a less than 10\% dead time.
The veto time is sufficient to cut most of the fast neutron background.
After applying the muon veto, the fast neutron background is reduced to less than 200~events/day.
The muon veto detector is also functioning as shielding material to reject the neutron penetrating to the central detector.
The shielding helps to reduce the rate of the fast neutron background produced outside of the veto detector without muon tagging, which contributes about 1/3 of the fast neutron background.
The muon veto cut cannot reject some of the delayed signal because the neutron capture time is about 30~$\mu$s in average.
Those signals in the energy range of 7--9~MeV without coincident prompt signal is called delayed-like signal which has a rate of 0.22~Hz and could form the delayed signal of an accidental background.

Despite the short distance between the detector and the reactor core, the
largest source of $\gamma$-rays is natural radioactivity, namely the
concrete of the walls of the experimental hall, and the Printed
Circuit Boards (PCBs) hosting the photosensors and the readout electronics.
$^{40}$K emission dominates the concrete induced $\gamma$-ray yield, while
$^{238}$U and $^{232}$Th are responsible for most of the PCB-induced
$\gamma$-ray flux. The former can be reduced by using a passive water
shielding, while the latter needs to be controlled through careful material
selection. We simulated events due to the radioactive elements contained in the concrete, PCB, stainless steel vessel, water, acrylic, GdLS and HDPE, and found the total event rate is below 100~Hz above 0.9~MeV with 1.2~m thick water shielding.
The accidental background rate can be calculated by $R_d(1-exp(-R_p\Delta T))$, where $R_d$ is the rate of the delayed events, $R_p$ is the rate of the prompt events, and the $\Delta T$ is the coincidence time.
With the inputs of $R_d = 0.22$~Hz, $R_p = 100$~Hz, and $\Delta T = 100~\mu$s, the expected accidental event rate is 190/day, at a similar level of the fast neutron background.

In the TAO liquid scintillator, the cosmic muons can interact with $^{12}$C and produce radioactive isotopes.
Among them, $^9$Li and $^8$He with half-lives of 0.178~s and 0.119~s, respectively, are the most serious correlated background source to IBD signals, because they can decay by emitting both an electron and a neutron which form the prompt and delayed coincident signals, respectively.
The $^9$Li and $^8$He is often modeled empirically as being proportional to $E_{\mu}^{0.74}$, where $E_\mu$ is the average energy of the muons at the detector~\cite{Wang:2001fq}.
The production yield has been measured in the KamLAND detector~\cite{Abe:2009aa}.
At TAO experiment, the production yield can be extrapolated considering the average muon energy is 260~GeV for KamLAND and 8.4~GeV for TAO.
The veto time window of 20~$\mu$s for fast neutron background is difficult to reject the $^9$Li and $^8$He because of their long half-lives.
The production rates of $^9$Li and $^8$He in the fiducial volume of TAO detector is 45 and 9 per day, respectively.
Table~\ref{tab:background} summarizes some important results of the singles and background simulation.

\begin{table}
\setlength{\belowcaptionskip}{5pt}
\begin{center}
\caption{Summary of the IBD signal and background simulation results. \label{tab:background}}
\begin{tabular}{r l}
  \hline\hline
   IBD signal & \mbox{ }\mbox{ } 2000~events/day \\
  Muon rate & \mbox{ }\mbox{ } 70~Hz/m$^2$ \\
  Fast neutron background before veto & \mbox{ }\mbox{ } 1880~events/day \\
  Fast neutron background after veto & $< 200$~events/day \\
  Singles from radioactivity & $< 100$~Hz \\
  Accidental background rate & $< 190$~events/day \\
  $^{8}$He/$^{9}$Li background rate & $\sim 54$~events/day \\
  \hline
\end{tabular}
\end{center}
\end{table}

\subsection{Energy resolution}
\label{sec:energyresolution}

The energy resolution is a key parameter for the TAO experiment.
It is predominantly determined by the statistics of the collected photoelectrons (p.e.).
Compared with 1200~p.e./MeV in the JUNO experiment~\cite{An:2015jdp}, the photoelectron yield of 4500~p.e./MeV is expected considering the following improvements.
\begin{itemize}
  \item The coverage of photon sensors is improved to $\sim 95$\% from 75\% in JUNO.
  \item The photon detection efficiency is improved to $\sim 50$\% using SiPMs, while it is $\sim 27$\% with PMTs in JUNO.
  \item Smaller dimension of the TAO detector increases the photoelectron statistics by 40\% due to less photons absorbed in the liquid scintillator.
  \item Low temperature at -50$^\circ$C could increase the photon yield of LS by $\sim 25$\%~\cite{Xia:2014cca}.
\end{itemize}
The expected energy resolution as a function of energy obtained from the TAO detector simulation is shown in Figure~\ref{fig:Eres}.
It takes into account several detector effects:
\begin{itemize}
  \item Statistics:
  Preliminary Monte Carlo study shows that a photoelectron yield of about 4500 photoelectrons per MeV can be reached, providing the energy resolution required by the TAO physics goals.
  \item Scintillator quenching: The quenched energy is simulated step by step in Geant4 using Birks' law~\cite{Chou:1952jqv} as
\begin{equation}
\Delta E_q = \frac{\Delta E}{1+k_B\frac{dE}{dx}+C(\frac{dE}{dx})^2},
\end{equation}
  where $k_B = 6.5\times 10^{-3}~\mathrm{g/cm^2/MeV}$ and $C = 1.5\times 10^{-6}~\mathrm{g^2/cm^4/MeV^2}$ are Birks' constants, $\frac{dE}{dx}$ is the stopping power, $\Delta E$ is the deposited energy before quenching, and $\Delta E_q$ is the quenched energy which is used to determine the mean value of the number of scintillation photons to be generated.
  The interaction processes for a particle in LS is random and the particle energy and deposited energy in each step fluctuate event-by-event.
  As a result, the fluctuation of the total quenched energy presents even for monoenergetic particles.
  \item Charge resolution: The charge resolution of one SiPM channel is assumed to be 16\% in the simulation. The number of the SiPM channels is about 4100. The energy resolution due to the SiPM charge resolution is $0.16/\sqrt{Nhit}$ where $Nhit$ is the number of fired channels. $Nhit$ varies from 2800 to 4100 as a function of energy in the range of 1-10~MeV calculated from a toy MC.
  \item Cross talk:  Due to optical cross talk effect, a SiPM can generate multiple photoelectrons when there is only one real photoelectron. The number of generated superfluous photoelectrons (n) follows the following distribution,
  \begin{equation}
  P(n|c) = \frac{(cn)^{n-1}e^{-cn}}{n!}
  \end{equation}
  where the cross talk probability is $c$ as a parameter of the SiPM properties.
  In simulation, the number of cross-talk photoelectrons fluctuates in each event.
  We subtract the average number of cross-talk photoelectrons while its fluctuation contributes to the energy resolution.   The cross talk probability is assumed to be 10\% based on the studies on the SiPMs described in Section~\ref{sec:sipm}.
  \item Dark noise: Dark noise rate is about 250~kHz/mm$^2$ for a typical SiPM at room temperature. At -50$^\circ$C, it could be reduced by three orders of magnitudes to 100~Hz/mm$^2$. The total area of the SiPMs is about 10~m$^2$ and the readout time window is set to be 1~$\mu$s here. The fluctuation of the number of dark noise photoelectrons is then $\sqrt{1000}$. This fluctuation affects the energy resolution in the energy reconstruction. It contributes a constant term to the energy resolution independent of the visible energy of physical events.
  \item Neutron recoil: The kinetic energy of the neutron in the IBD reaction final state shares a part of the initial energy of the incident antineutrino, which introduces an energy smearing of the positron in the final state. Based on the kinetics of the IBD reaction, the energy spread of the positron is
  \begin{equation}
  \Delta E = 2(E_{\nu}-\Delta_{np})E_{\nu}/M_p,
  \end{equation}
  where $\Delta_{np}$ is the mass difference of the neutron and the proton, $M_p$ is the proton mass, and $E_{\nu}$ is the energy of the incident antineutrino.
  The resulting energy resolution is $\Delta E/\sqrt{12}$, as an approximation of the standard deviation of a uniform distribution.
  However, a small fraction of neutron kinetic energy is detectable.
  The recoiled protons in the neutron thermalization process produce a small amount of scintillation photons that can mix with the photons produced by the positron.
  Considering the quenching effect of the recoiled proton, the energy resolution of neutron recoiling effect becomes $(1-Q_f)\Delta E/\sqrt{12}$, where $Q_f$ is the average quenching factor for the IBD neutron, which is about 0.29 determined from simulation.
  This effect emerges during the IBD reaction and is relevant to the determination of the antineutrino energy instead of the detection of the visible energy.
\end{itemize}

\subsection{Systematic uncertainties}
The goal of the TAO experiment is to precisely measure the reactor antineutrino energy spectrum.
The systematic uncertainties of the expected number of IBD events could be a few percent based on the experience of previous reactor antineutrino experiments~\cite{An:2012eh}.
The detection efficiency uncertainty dominates the uncertainty of the number of events.
The uncertainties (rate uncertainty) independent on the antineutrino energy, which affect overall antineutrino rate, do not have an impact on the precision of the
measured antineutrino spectral shape.

\begin{figure}[htb]
    \centering
    \includegraphics[width = 0.6\columnwidth]{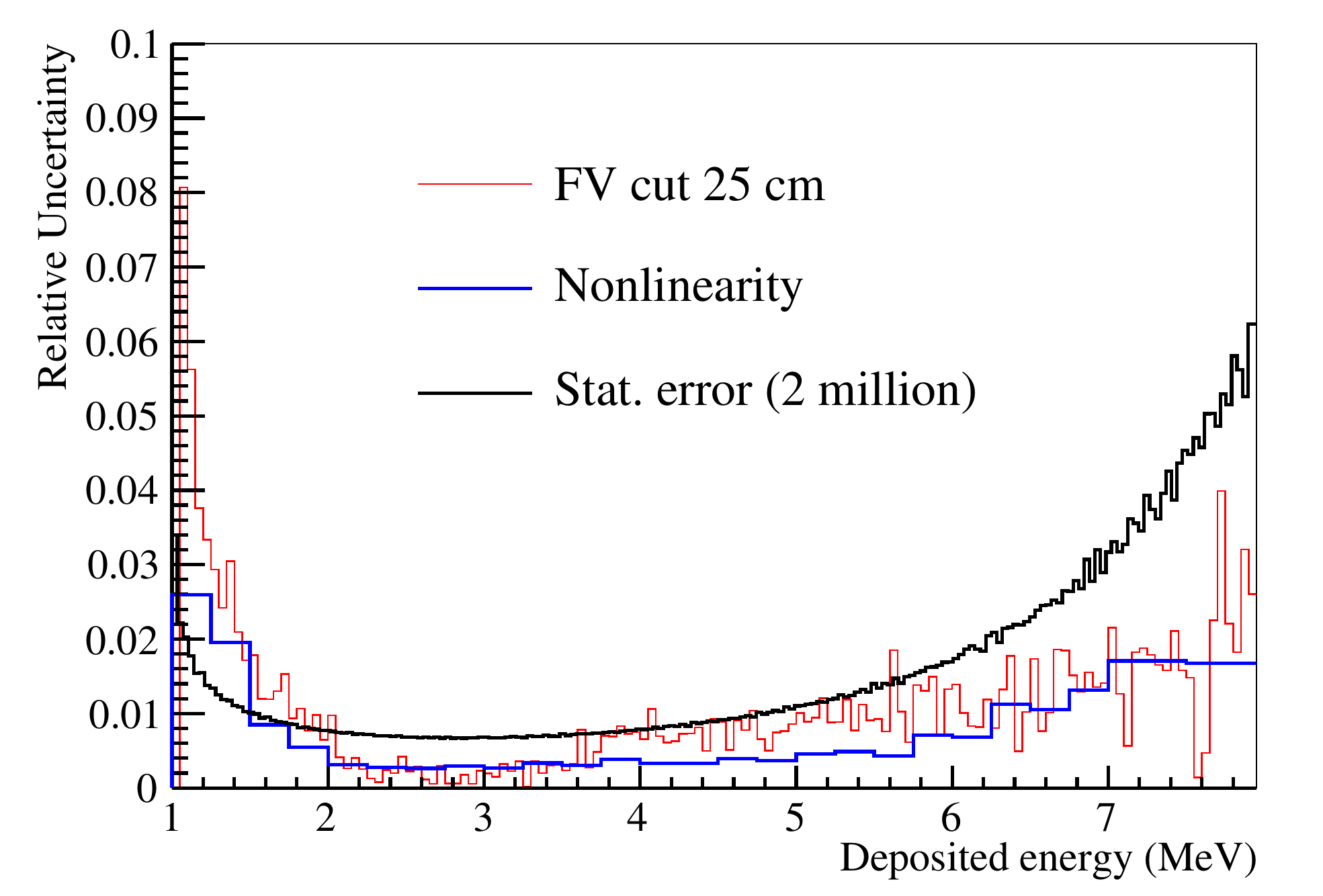}
    \caption{Three main sources of the spectral shape uncertainties. The uncertainty of the nonlinear energy scale is taken from Daya Bay~\cite{Adey:2019zfo}.}
    \label{fig:shapeError}
\end{figure}

The precision of the spectral shape measurement is driven by three main uncertainties.
The first is the statistical uncertainty, uncorrelated between energy bins.
The second is the energy scale uncertainty, determined by the uncertainties of the parameters in the energy scale model.
The third uncertainty is induced by the fiducial volume cut, which distorts the energy spectrum due to the energy leakage.
Figure~\ref{fig:shapeError} shows the impacts of the three major shape
systematic uncertainties on the reactor antineutrino spectrum measurement.

The IBD signal event rate is about 2000/day.
Two (four) million events will be collected in three (six) years data taking.
The statistical uncertainty from TAO is one source of the bin-to-bin shape uncertainty for JUNO when propagating the uncertainty of the reference spectrum from TAO to JUNO.
It is smaller than 1\% in the energy range of 2--5~MeV using the same bin width (35~keV) that used in the mass ordering analysis at JUNO~\cite{An:2015jdp}.
The nonlinearity uncertainty is taken from Daya Bay~\cite{Adey:2019zfo} under the assumption of similar performance between TAO and Daya Bay using quite similar liquid scintillator.
Due to the relatively small size of the TAO detector, a fraction of IBD events have energy leakage even with the 25~cm fiducial volume cut.
The energy leakage is simulated using Geant4 and can also be investigated by comparing the events produced by the calibration sources placed at the edge of the detector in simulation.
Different fiducial volume cuts correspond to different energy leakage and distorted energy spectra.
A preliminary reconstruction algorithm shows $<5$~cm vertex resolution and $<5$~cm bias in radial direction can be obtained for the IBD events.
With the assumption of 5~cm vertex resolution and bias, a toy MC is performed to propagate the uncertainty of the vertex to the simulated energy spectrum for events surviving the fiducial volume cut.
The shape uncertainty due to the fiducial volume cut is estimated.
As shown in Figure~\ref{fig:shapeError}, the statistical uncertainty dominates in majority of the energy range.

%% file: CentralDetector/section.tex
\section{Central Detector}
\label{sec:cd}
\blfootnote{Editors: Jun Cao (caoj@ihep.ac.cn) and Yuguang Xie (ygxie@ihep.ac.cn)}
\blfootnote{Major contributors: Xiaoyan Ma, Liting Sun, Xilei Sun, Menfen Wang, Zhangquan Xie, Huan Yang, Chengzhuo Yuan}

\subsection{Overview}
The TAO experiment has two primary physics goals,
\begin{itemize}
\item to provide a model-independent reference spectrum for JUNO,
\item to provide a benchmark to test nuclear databases by measuring the fine structure of the spectrum,
\end{itemize}
together with other goals described in Sec.~\ref{sec:introduction}. The first goal requires $<$$3\%/\sqrt{E({\rm MeV})}$ energy resolution and $>$10 statistics of JUNO, and is relatively easy to achieve. The second one has not specific requirement but prefers as high as possible energy resolution and statistics.

The Central Detector (CD) of TAO is designed to have a fiducial mass of one ton Gadolinium-doped Liquid Scintillator (GdLS). The reactor antineutrino event rate is more than 30 times higher than that of JUNO with selection efficiency included. The layout and support of the Silicon Photomultipliers (SiPMs) will be optimized to have as high as possible coverage. Three options are under consideration, with 94\%, 95.5\%, and 96.9\% coverage, respectively. About 4500 p.e./MeV could be reached.

TAO central detector is a two-layer detector, with an inner layer of 2.8 ton GdLS as the antineutrino target contained in a spherical acrylic vessel, and an outer layer of 3.45 ton Linear Alkylbenzene (LAB) as buffer liquid contained in a cylindrical Stainless Steel Tank (SST). SiPM Photosensor will be installed on a spherical copper shell wrapping the acrylic vessel, with 18~mm distance between the SiPM surface and the acrylic vessel. To reduce the dark noise of the SiPMs, the whole CD will be operated at -50$^\circ$C.

GdLS for TAO is adapted from the Daya Bay GdLS~\cite{Beriguete:2014gua} by reducing the fluor concentration and adding co-solvent to avoid fluor precipitation at low temperature. The carboxylic gadolinium complex is dissolved into LAB with a Gd mass fraction of 0.1\%. The concentration of the fluor PPO is 2~g/L and that of wavelength shifter bis-MSB is 1~mg/L. To improve the solubility of the fluor and wavelength shifter, a co-solvent of 0.05\% ethanol is added into GdLS. LAB can be used as non-scintillating buffer liquid and has no problem to work at -50$^\circ$C if the water content is removed carefully.

The mechanical structures, cryostat design, and R\&D of the low temperature GdLS will be described in the following.

\subsection{Mechanical structures}

\begin{figure}[htb]
\begin{center}
\includegraphics[width=0.8\textwidth]{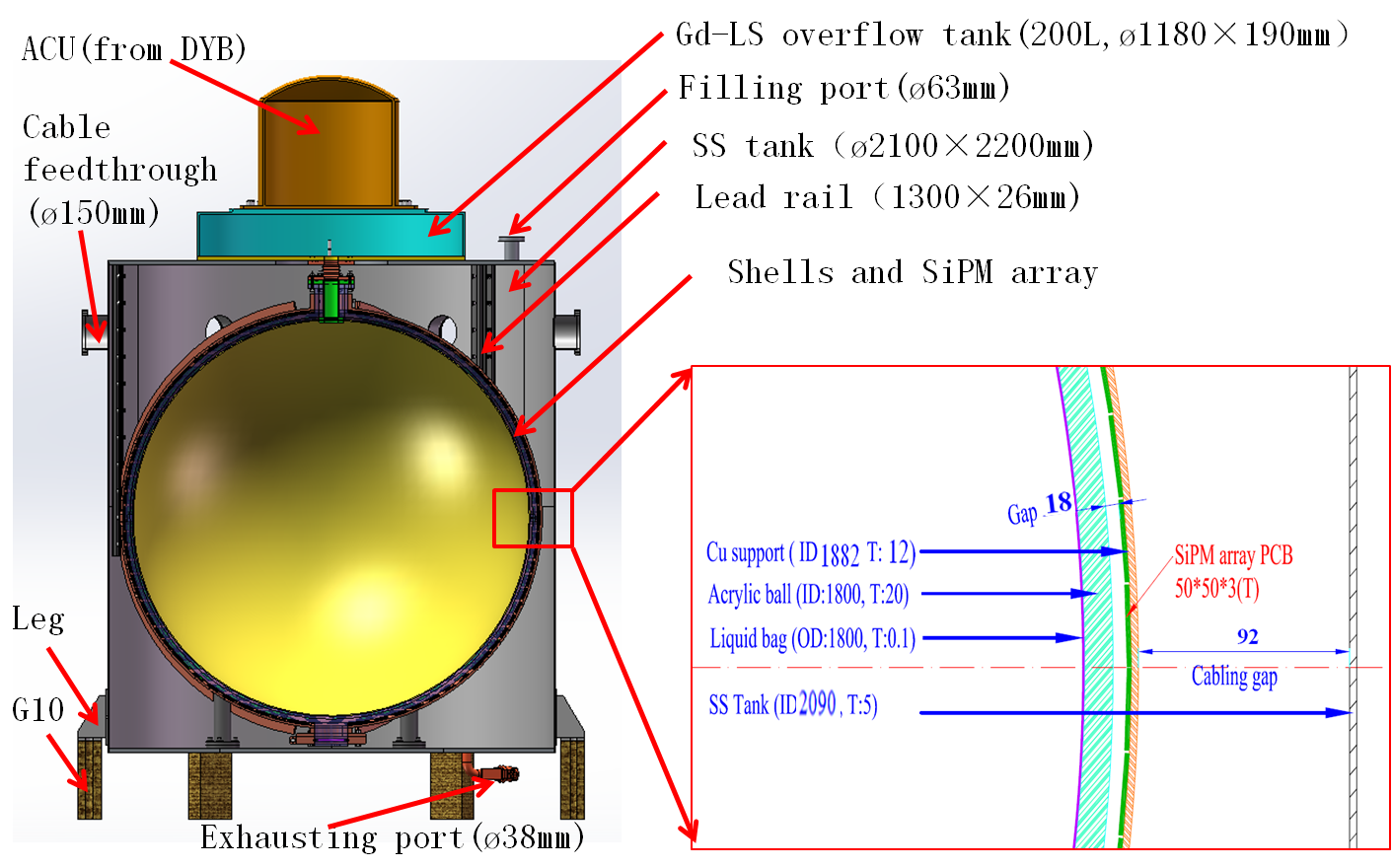}
\caption{The mechanical structure of the TAO central detector. The zoom on the right picture shows in detail the dimension chain of the detector.
\label{fig:cdmechanics}}
\end{center}
\end{figure}

The scheme of the central detector is shown in Figure~\ref{fig:cdmechanics}. The 2.8~ton GdLS is filled in a 1.80~m diameter spherical acrylic vessel. After a 25-cm fiducial volume cut to reduce the energy spectrum distortion due to the gamma energy leakage into buffer liquid, the fiducial mass is 1 ton. The thickness of the acrylic vessel is 20~mm. A 12-mm thick spherical copper shell of an inner diameter of 1.882~m wraps the acrylic vessel and provides the mechanical support for the SiPM tiles and their electronics readout. About 4100 SiPM tiles, each of $50\times50$ mm$^2$ in dimension, are installed on the inner surface of the copper shell. Each tile consists of $8\times8$ array of 6~mm pixels, or $5\times5$ array of 10~mm pixels of SiPM cells mounted on a low background Printed Circuit Board (PCB). The Frontend Electronics (FEE) is in another PCB on the outer side of the copper shell. The copper shell is submerged in buffer liquid (LAB) contained in a cylindrical stainless steel tank of an inner diameter of 2.10~m and inner height of 2.20~m. The tank is wrapped with 20-cm insulation material, Polyurethane (PU) foam, to limit a heat leakage to be smaller than 500~W.

\begin{table}[htb]
\setlength{\belowcaptionskip}{5pt}
\begin{center}
\caption{Dimension chain of the TAO central detector. Both Inner Diameter (ID) and Outer Diameter (OD) are listed for the acrylic vessel, the copper shell, and the SST. The buffer liquid will be filled to 2100~mm level while the SST height is 2200~mm. The lid and bottom panel of SST are 10~mm thick. The buffer liquid inside and outside the copper shell are summed over. \label{tab:cd:dimension}}
\begin{tabular}{r c c c c c c}
\hline\hline
        & Diameter & Thickness & Height & Volume  & Density & Weight \\
        & (mm) & (mm) & (mm) & (m$^3$) & (g/cm$^3$) & (ton) \\\hline
GdLS    &      & 900  &      & 3.504  &  0.916     & {2.8} \\\hline
Liquid Bag (optional) & 1800 & 0.1 & && & 0.001 \\\hline
Acrylic Vessel ID & {1800} & {20} & & 0.208 & 1.2 & 0.25 \\\hline
Acrylic Vessel OD & 1840 &    &    &   &       &       \\\hline
Gap w/ Buffer Liquid & & {18} & & 0.195 & & \\\hline
SiPM & 1876 & 1 & & 0.011 & 2.34 & 0.026 \\\hline
SiPM-PCB & 1878 & 2 & & 0.022 & 1.8 & 0.04 \\\hline
Copper Shell ID & 1882 & 12 & & 0.135 & 8.9 & {1.2} \\\hline
Copper Shell OD & 1906 &    & &       &       &   \\\hline
Buffer Liquid   & 2090  &  & 2100 & 3.579 & 0.914 & {3.45} \\\hline
SST ID & 2090 & 5 & 2200 & 0.142 & 7.93 & 1.13 \\\hline
SST OD & 2100 &   & 2220 &      &       &       \\\hline
SST Insulation OD & 2500 & 200 & 2620 & 5.172 & 0.05 & 0.26 \\\hline
SST Insulation support OD & 2506 & 3 & 2626 & 0.091 & 7.93 & 0.725 \\\hline
Total Central Detector & {2506} & & {2626} & 12.6 & & {9.9} \\
\hline
\end{tabular}
\end{center}
\end{table}

The dimension chain of the CD is listed in Table~\ref{tab:cd:dimension}. From the center of the CD to the outmost insulation support panel, the materials and structures are GdLS, liquid bag (optional), 20-mm thick acrylic vessel, an 18~mm gap filled with buffer liquid, 3-mm thick SiPM on PCB, 12-mm thick copper shell, buffer liquid, 5-mm thick SST, 200-mm thick thermal insulation layer, and a 3-mm thick stainless steel panel that contain the insulation layer. The copper shell is spherical and the SST is cylindrical, therefore heights are shown starting from the buffer liquid. The height of the SST is 2200~mm, but the buffer liquid will be filled to 2100~mm level, leaving 100~mm space for liquid overflow and nitrogen cover. The buffer liquid inside and outside the copper shell is connected, so its weight is summed in one row, totaled 3.45 ton. The diameter of the CD is 2.506~m. The height is 2.626~m. And the total weight is about 9.9 ton, without flanges, possible reinforcing ribs, and other minor parts.

The mechanical structure of the CD includes the SST, the support of the SiPM (i.e.\ the copper shell), and the acrylic vessel. They are described in the following together with the nitrogen system, cabling and piping, and assembly.

\subsubsection{Stainless steel tank}

The stainless steel tank is the outer vessel of the antineutrino detector of TAO. It provides mechanical support for all components inside the SST, as well as Automatic Calibration Unit (ACU, see Section~\ref{sec:calibration}) and overflow tank on the top of the lid. It also provides an air-tight environment for the liquid scintillator. Inside the SST, the temperature will be maintained at -50$^\circ$C, while outside the SST, it is at room temperature, so a layer of insulation is required.

The SST is a cylindrical vessel made of SS 304. The dimension and requirements of SST are listed in Table~\ref{tab:cd:sst}. Constrained by the transportation passageways, especially the elevator to the underground laboratory in the basement of the building, which has a dimension of $1990~({\rm Depth)}\times 1390~({\rm Width}) \times 1990~({\rm Height})$ mm, the SST has to be shipped in parts and welded together in the laboratory.

\begin{table}[htb]
\setlength{\belowcaptionskip}{5pt}
\begin{center}
\caption{Dimension and requirements of the Stainless Steel Tank.\label{tab:cd:sst}}
\begin{tabular}{r l}
\hline\hline
Height & 2200 mm \\
Outer Diameter & 2100 mm \\
Thickness of the wall & 5 mm \\
Thickness of the lid and bottom & 10 mm \\
Overflow tank height & 200 mm \\
Leakage & $\leq 10^{-6}$ mbar$\cdot$L/s \\
\hline
\end{tabular}
\end{center}
\end{table}

\begin{figure}[htb]
    \centering
    \includegraphics[width=0.35\textwidth]{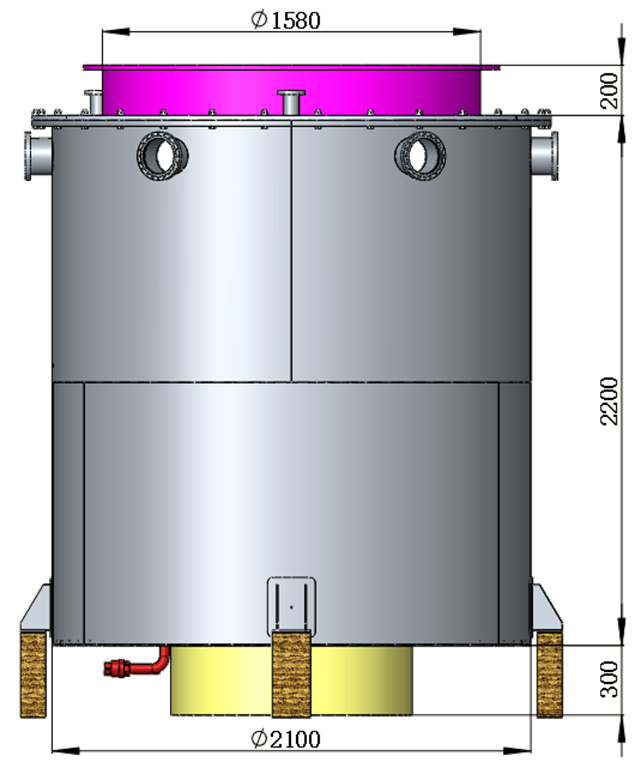}
    \caption{Structure of the stainless steel tank.
    \label{fig:sst}}
\end{figure}

The SST consists of the lid, the barrel, the bottom and the support. The lid and the bottom are both divided into 3 pieces. The barrel is divided into 6 pieces, as shown in Figure~\ref{fig:sst}. These parts will be welded together in the underground laboratory.

On the lid of the SST, there are an overflow tank, the buffer liquid inlet, and the coolant inlets/outlets. The overflow tank has a flange serving as the interface to the automatic calibration unit on top of it.

On the wall of the SST, the exact number and size of the cable feedthroughs depend on the SiPM readout scheme, which has a few options to be determined with further R\&D. Preliminarily we have designed six DN160 flanges for mounting feedthroughs on the wall, which keep the possibility to lead 4100 channels out for all options. The support legs are welded at the bottom of the cylinder and will sit on blocks made from Fiber Reinforced Plastics (FRP) or titanium alloy to keep a good insulation.

\begin{figure}[htb]
\begin{center}
\includegraphics[width=0.9\textwidth]{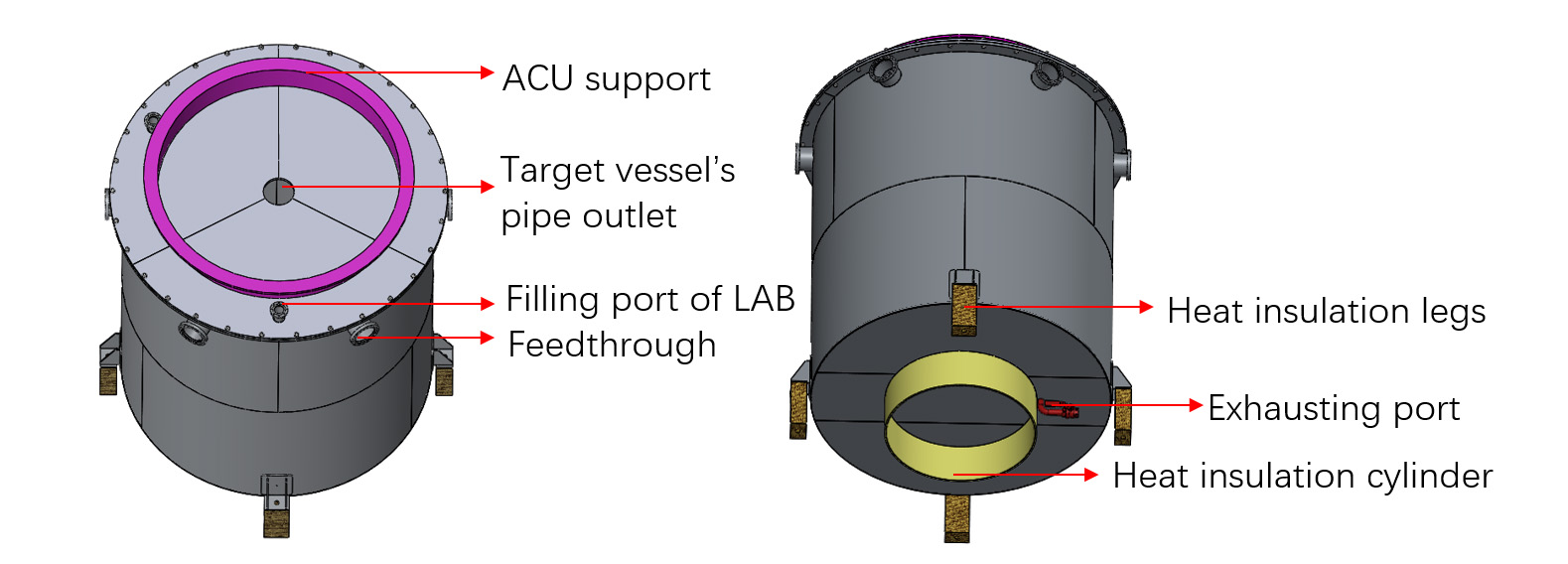}
\caption{Top view (left) and bottom view (right) of the stainless steel tank.
\label{fig:sstlid}}
\end{center}
\end{figure}

At the bottom of the SST, there are outlets for cleaning, and outlets for the buffer liquid and GdLS. An FRP cylinder is designed to support and reinforce the bottom panel. Another option is to use a support frame linking with the legs. The structure is shown in Figure~\ref{fig:sstlid}.

The big flange between the barrel and the lid will be challenging, since it has to be tailored due to the limitation of the transportation passageway (elevator). The flange will be welded onsite, with tools to control the deformation of the flange and the lid. The grooves will be polished after welding. Double O-ring is designed for the air-tight sealing. A couple of companies had been identified with existing experience. The structure of the lid is shown in Figure~\ref{fig:sstlid}.

The strength of the SST has been analyzed with the Finite Element Analysis (FEA) for all possible working conditions, such as empty tank, overturn, and during filling, at both room temperature and low temperature. Due to limited height of the laboratory, the copper shell and the acrylic vessel inside have to be installed into SST horizontally. Then, the assembled SST needs an overturn from horizontal position to vertical position. The strength of the SST is good enough, and the only attention needs be paid is the overturn operation.

\subsubsection{SiPM support}
In order to have a photosensor coverage close to 100\%, design of the layout and support of the SiPMs is challenging. The support should have a good thermal conductivity since the heat produced by the readout as well as the SiPMs themselves need be transferred smoothly, so the working temperature of the SiPMs keeps stable. The support should be made with very low background material as it is only several centimeters from the GdLS. The support material should be compatible with the buffer liquid and should have good a mechanical strength to avoid deformation and damage to the SiPM tiles during the assembly, overturn (see Section~\ref{sec:assembly}), and lifting.

\begin{figure}[htb]
\begin{center}
\includegraphics[width=0.3\textwidth]{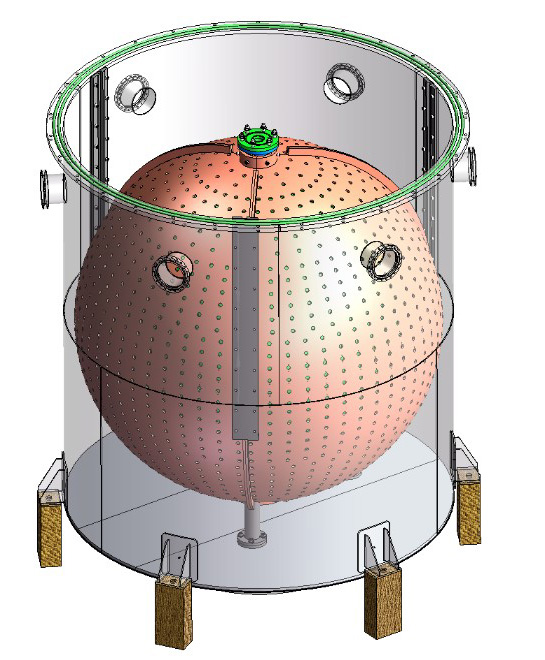}
\includegraphics[width=0.3\textwidth]{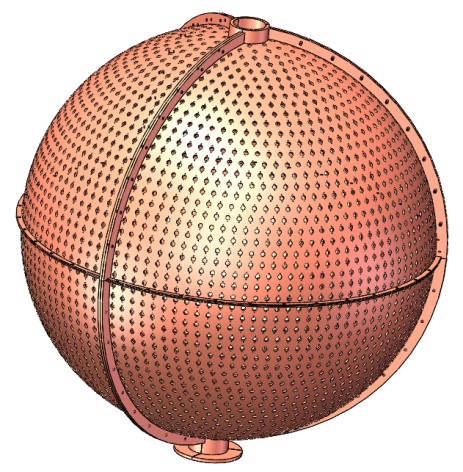}
\includegraphics[width=0.3\textwidth]{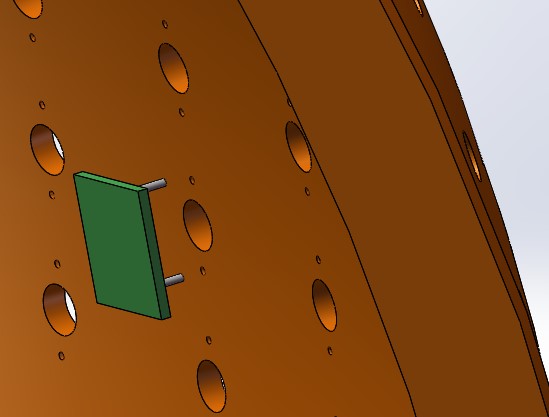}
\caption{Left: the spherical copper shell as the SiPM support structure. Middle: the sphere is divided into 6 pieces. Right: the local drawing of the copper shell with one $50\times50$ mm SiPM tile and its fixture.
\label{fig:cusupport}}
\end{center}
\end{figure}

Design of the support structure is shown in Figure~\ref{fig:cusupport}. It is a shell structure made of about 1200~kg oxygen-free copper. The copper shell also supports and fixes the acrylic vessel through only two flanges on the top and at the bottom of the acrylic vessel. The 12-mm thickness is determined by both the mechanical strength and the thermal capacity to stabilize the temperature. The sphere is divided into 6 pieces. The upper and lower half sphere each consists of 3 pieces. They are bolted together through the reinforcing ribs on the edge of each piece. The tolerance of the diameter is designed to be $\pm2.5$ mm. The deformation of the support structure is estimated to be about 2-4~mm for all working conditions.

\begin{figure}[htb]
	\begin{center}
		\includegraphics[width=0.8\textwidth]{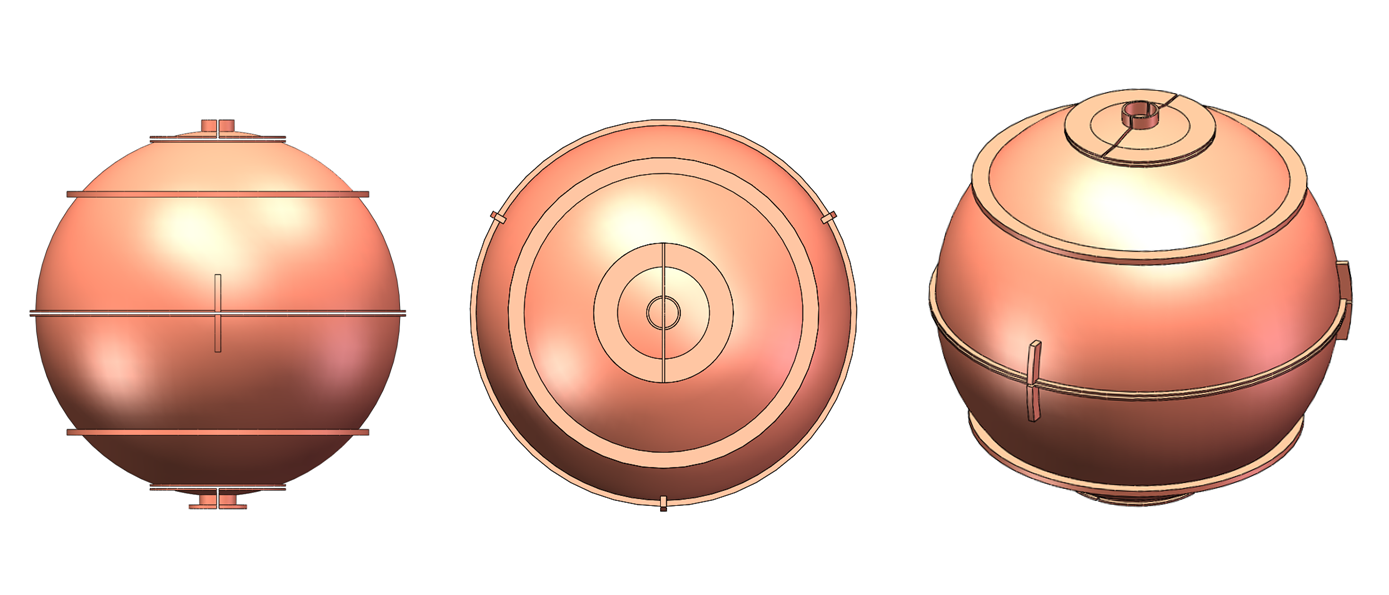}
		\caption{Latitudinal division option of the copper shell.
			\label{fig:cs2}}
	\end{center}
\end{figure}

It is also possible to divide the copper shell latitudinally rather than longitudinally. In this case, eight pieces are required. Four big pieces are in ring shape and make up about 80\% of the sphere. The other four small pieces combine as the flange cover for the top and the bottom, as shown in Figure~\ref{fig:cs2}. Manufacture of the copper rings will be more difficult but higher SiPM coverage could be achieved with this option.

\begin{figure}[htb]
	\begin{center}
		\includegraphics[width=0.8\textwidth]{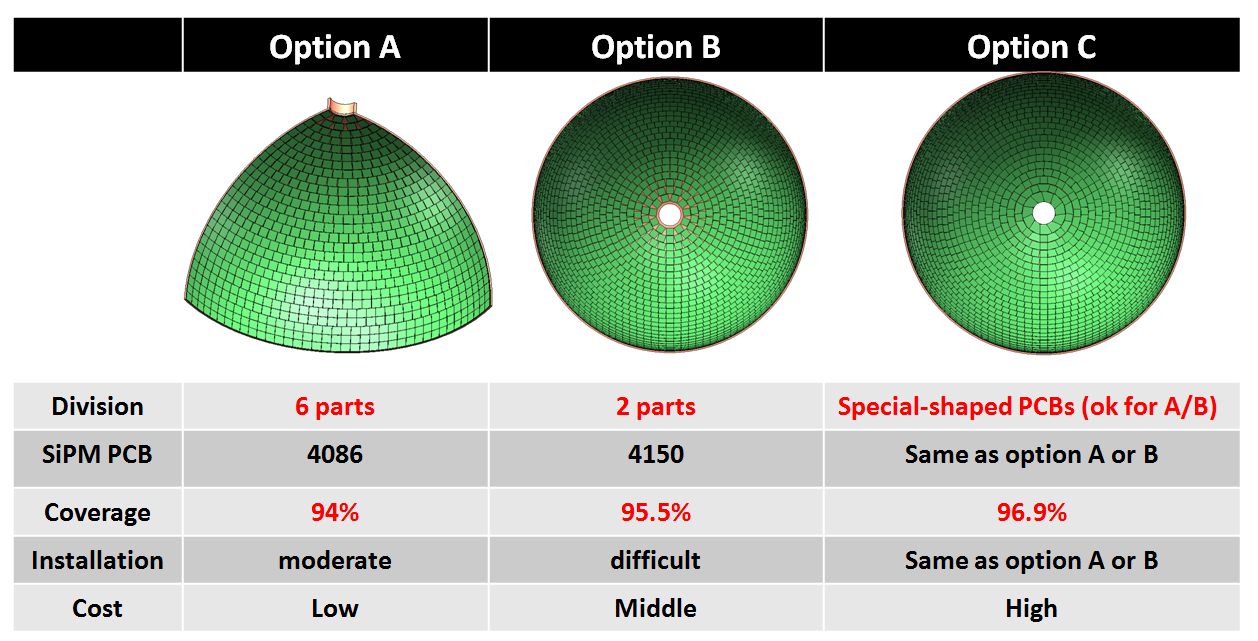}
		\caption{Comparison of three options of SiPM layout.
			\label{fig:coveragecom}}
	\end{center}
\end{figure}

Several options of SiPM layout have been considered. Due to the mismatch of the square SiPM tiles and the spherical support structure, the default layout with 6 pieces copper shell (option A in Figure~\ref{fig:coveragecom}) is the simplest and could reach a photo-sensor coverage of 94\%. The layout with latitudinal division (option B) could reach a coverage of 95.5\%. With multiple sizes of SiPM tiles (Option C), i.e.\ with several types of non-square tiles, the coverage could be improved further to 96.9\%. Option C works for both latitudinal and longitudinal divisions but needs tuning of the SiPM tile shape.

The PCB of each SiPM tile is fixed onto the copper shell with two bolts. The position tolerance of the bolt holes, expressed in angular precision, is 0.01$^\circ$.

Due to the height limitation of the laboratory, the assembled copper shell (with SiPM tiles installed) has to be installed into the SST horizontally, both laid flat. Three guide rails are designed on the inner wall of the SST. After installation, the guide rails also fix the X and Y positions of the copper shell.

A potential problem is the compatibility of the buffer LAB and the copper shell. Copper easily develops a green patina. Our compatibility test shows that copper can pollute the candidate buffer liquid LAB, especially when the copper surface is not clean enough. We also find that passivation of copper will apparently improve the compatibility. Further R\&D is necessary to determine if the pollution is acceptable, if the passivation could be damaged during assembly, if there are better buffer liquids, etc.

Deformation of the detector components when cooling down from room temperature to -50$^\circ$C must be seriously studied. The PCB of the SiPM tile contracts more than copper. The gap between adjacent PCBs will increase, thus no interference will happen. However, the connection between PCB and bolts may break since the largest contraction of PCBs will be 0.26 mm as the finite element analysis shown. A prototype test of a local model will be done to check the design and the assembly procedure of SiPM tiles.

\subsubsection{Acrylic vessel}
\label{sec:acrylicvessel}

\begin{figure}[htb]
\begin{center}
\includegraphics[width=0.45\textwidth]{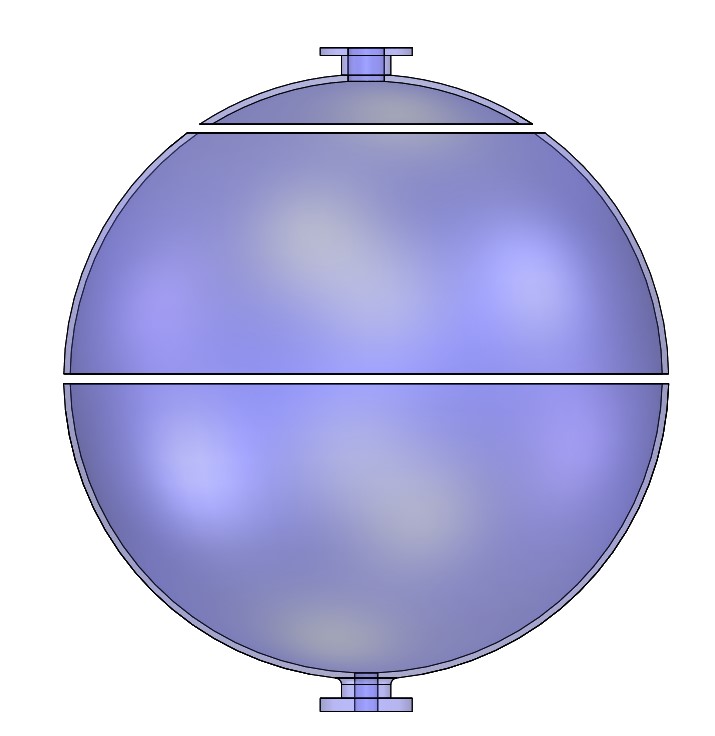}
\includegraphics[width=0.45\textwidth]{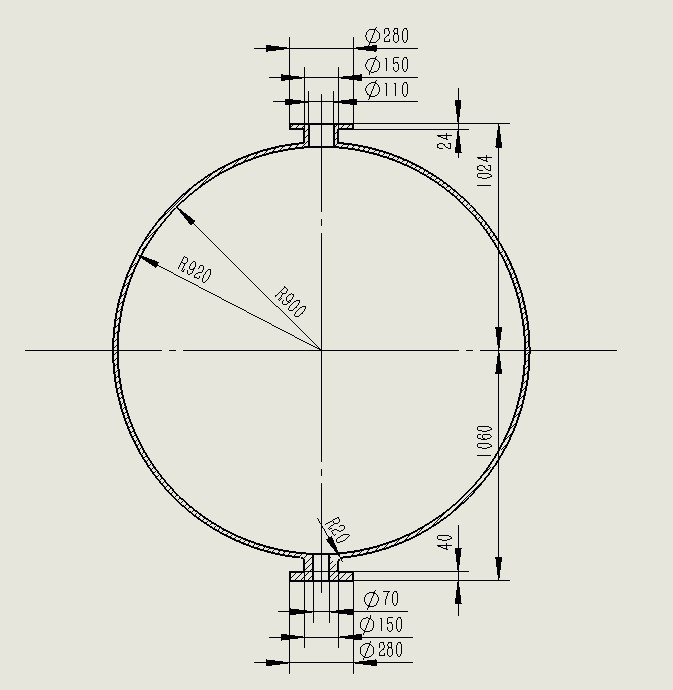}
\caption{Left: Partition of the acrylic vessel. Right: Drawing of the acrylic vessel (dimensions in mm).
\label{fig:acrylicvessel}}
\end{center}
\end{figure}

The reactor antineutrino target, GdLS, will be contained in an acrylic vessel of an inner diameter of 1800 mm and thickness of 20~mm. The thickness is possible to reduce by about 5~mm while the strenth is still enough. The acrylic vessel will be divided into 3 pieces, as shown in Figure~\ref{fig:acrylicvessel}.

There is an upper chimney on the top of the acrylic vessel, connecting to the overflow tank by flanges and a bellow. The upper chimney is fixed in the chimney of the copper shell via jack bolts during installation, but is loosened during running. The bottom chimney connects to the GdLS outlet by a flange, and is fixed onto the copper shell via a clamp.

\begin{figure}[htb]
\begin{center}
\includegraphics[width=0.6\textwidth]{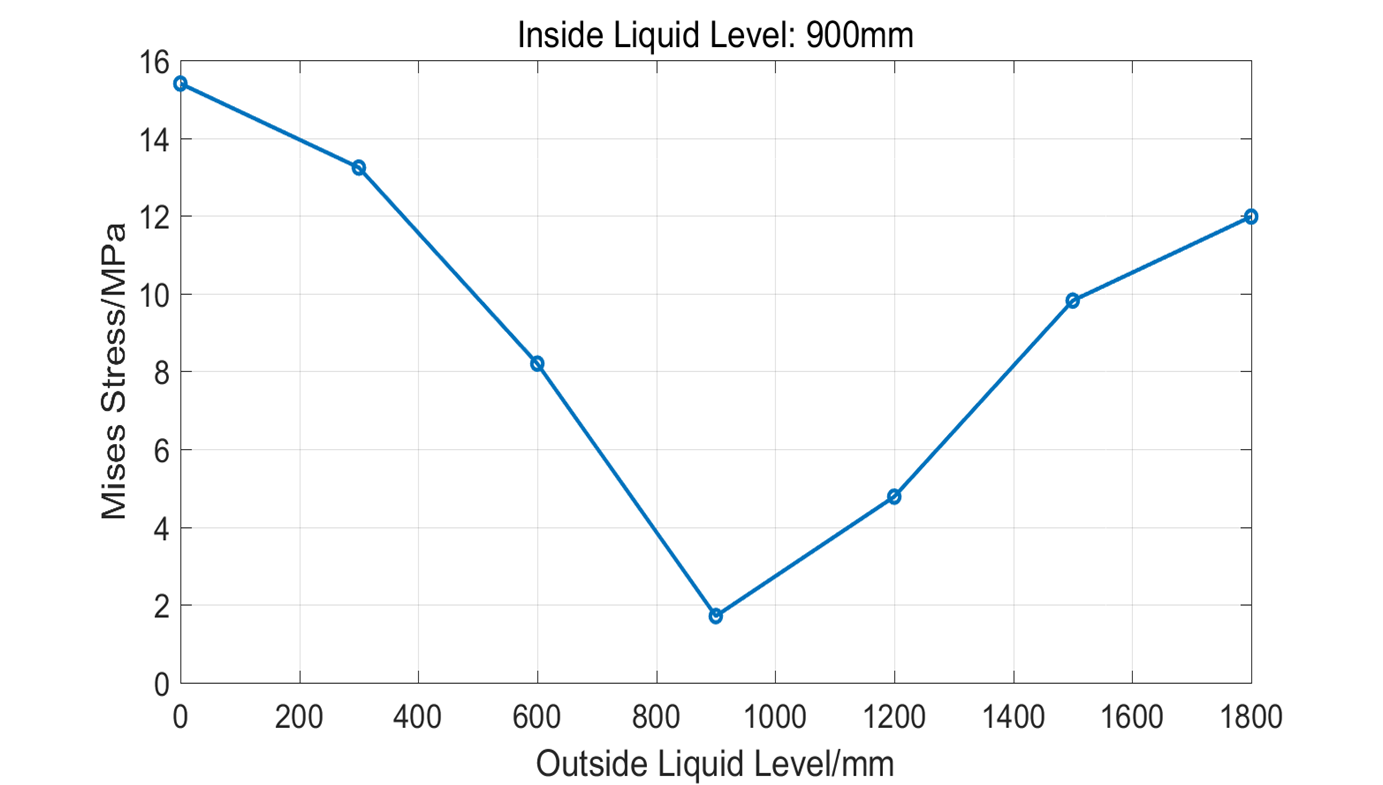}
\caption{Stress of the acrylic vessel with 20-mm thickness as a function of LAB level when the LS level is at the equator (i.e.\ 900 mm).
\label{fig:acrylicvesselstress} }
\end{center}
\end{figure}

The stress of the acrylic vessel has been analyzed with FEA. Acrylic material has been studied extensively by JUNO to construct its 35.4-m diameter acrylic vessel~\cite{Djurcic:2015vqa}. The maximum stress of the acrylic vessel should be less than 5~MPa for long term use to avoid crazing and cold flow, and could be relaxed to 10~MPa for short term (days). For the TAO acrylic vessel, the stress is small when empty and after filled. The maximum stress is found to be around the central supporting leg at the very bottom during filling with unequal liquid level inside and outside. The worst case happens when the inside (GdLS) liquid level is at the equator, which is 900~mm as we define 0~mm at the bottom of the acrylic vessel. For this case, the maximum stress of the acrylic vessel is shown in Figure~\ref{fig:acrylicvesselstress} for different outside liquid levels. When no buffer liquid is filled, the stress is close to 16~MPa. When the buffer is fully filled (while the GdLS is still half filled), the stress is 12~MPa. To keep the stress being less than 10~MPa during the filling, the inner and outer liquid level difference should be controlled to be less than 400~mm.

\begin{figure}[htb]
	\begin{center}
		\includegraphics[width=0.9\textwidth]{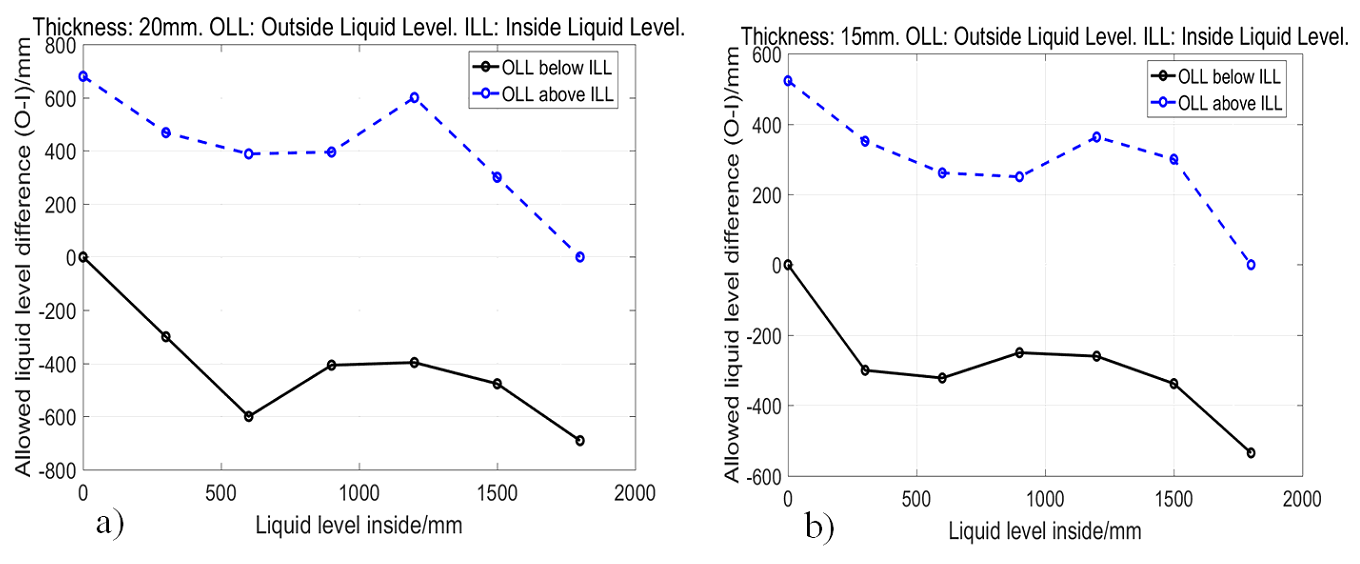}
		\caption{Allowed liquid level difference (Outside Liquid Level (OLL) minus Inside Liquid Level (ILL)) during the filling process for 20-mm (a) and 15-mm (b) thick acrylic vessels.
			\label{fig:acrylicfilling} }
	\end{center}
\end{figure}

More FEAs have been done in order to optimize the thickness of the acrylic vessel. Both 20-mm and 15-mm thick acrylic vessels are analyzed to investigate the allowed liquid level difference during filling. Both options are safe enough during installation, cooling, and running, taking 5~MPa as the allowed stress. The filling process is considered as the worst case. Taking 10~MPa as the allowed stress, Figure~\ref{fig:acrylicfilling}(a) shows the allowed liquid level difference for GdLS at different level for 20-mm thick acrylic. The maximum allowed value is about 400~mm no matter the outside liquid level is above or below the inside one. If the thickness is reduced to 15 mm, the vessel is still safe when the liquid level difference is controlled to be less than 250~mm,  as shown in Figure~\ref{fig:acrylicfilling}(b).

The three pieces of acrylic components will be bonded together to form the spherical vessel in the laboratory. The bonding uses about 5 kg flammable liquid MMA monomer and the bonding process takes about 2 days, followed by annealing and polishing procedures. Such operation might carry certain fire risk. An alternative option has been considered with the acrylic pieces connected together without bonding, in case the bonding is not allowed in the laboratory close to the reactor core. In that case a liquid bag is required to be placed inside the acrylic vessel to contain the liquid. The acrylic vessel will provide mechanical support to the liquid bag. A similar scheme has been studied for JUNO and believed to be feasible after a prototype test and technical reviews~\cite{Djurcic:2015vqa}, although it may carry larger risk than the current acrylic scheme of JUNO. Small adaption will be needed for TAO, mainly the flange connecting the liquid bag and the chimney of acrylic vessel.

When using the alternative liquid bag option, the buffer liquid has to have a matched density with the GdLS. The density of the GdLS is 0.86~kg/L at room temperature. The buffer liquid should have density slightly lower than GdLS. LAB is a natural choice although it might be aggressive to detector materials. The bag will be made of PA (Nylon) or PE/PA (Polyethylene/Nylon) composite film of a thickness of 100~$\mu$m. The light transmission is $>$96\%.

\subsubsection{Nitrogen system}
Water content in GdLS and buffer liquid (LAB) needs be reduced to very low level. Laboratory tests show that water in GdLS and buffer liquid should be less than 10~ppm and 5~ppm, respectively, in order to maintain transparency at -50$^\circ$C. When exposing to air, water vapor in air will be absorbed into GdLS and buffer liquid. Therefore, nitrogen protection and bubbling before filling is necessary to remove water. Nitrogen bubbling also helps to improve the light yield of the GdLS by about 13\% and improve the pulse shape discrimination power by purging the oxygen. Radon in air of the laboratory could permeate into the detector, dissolve into the liquid scintillator, and produce backgrounds. Nitrogen cover will help to reduce the radon permeation. However, experiences from the Daya Bay experiment show that even moderate precaution is enough to reduce the backgrounds from radon to acceptable level, due to the high event rate at such a short baseline to reactor. During operation, the liquid has to be covered by flushing nitrogen to isolate the detector from water vapor and radon in air.

One consideration is to use a circular pipe with many holes installed at the bottom of the SST to purge the water content in the buffer liquid, while the GdLS will be purged before filling into the acrylic vessel through an air-tight filling system. The inlet and outlet of nitrogen are designed on the lid of the SST and sealed with double O-ring flanges.

Nitrogen will be provided either by liquid nitrogen bottles or with a small nitrogen generator. For the first option, the bottles need be shipped every two weeks during the experiment data taking. For the second option, maintenance of the generator might increase the workload in the power plant. Again, commercial purity of liquid nitrogen and moderate precaution is enough for TAO to reach acceptable level of radon background.

\subsubsection{Piping, cabling, and backend electronics box}

Cooling pipes are attached on the surface of the copper shell to effectively remove the heat generated by SiPMs and the readout electronics, so the temperature keep stable for the whole volume of GdLS in the acrylic vessel, with minimum convection of the buffer liquid. The pipes made of copper will be fixed on the copper shell by bolts with a certain pressure. They penetrate through the sidewall of SST, go down to the ground along the outside of SST, and finally link to the refrigerator. The layout and routing of the cooling pipes will be elaborated later, and could be further optimized with a prototype test.

The SiPM tiles and the FEE boards are mounted on the inner and outer surfaces of the copper shell respectively, and both are connected by a high density connector (such as the Samtec connector). FEE and SiPM tile can be connected with the two PCBs parallel or orthogonal to each other. The readout cables of the FEE are routed also along the outer surface of the copper shell and connect to the Field Programmable Gate Array (FPGA) boards, which are installed in the upper part of the SST. The signals from the FPGA boards are read out by the Data Acquisition (DAQ) system via optical links through the feed-through on the sidewall of the SST.

To reduce the number of cables penetrating the SST wall, it is preferred to keep the backend electronics in the tank. The signal cables of the 4100 SiPM tiles will be routed into three backend electronics boxes near the feedthroughs on the SST wall. The backend electronics including the FPGA usually produce a lot of heat. These boxes are designed to be thermally insulated, e.g. made of Teflon. Some branches of the cooling pipes will go through the insulated boxes to take the heat away directly. With these backend electronics boxes, the cable plugs can be accommodated in three flanges, although the readout scheme, discrete or ASIC, is to be determined.

\subsubsection{Assembly}
\label{sec:assembly}

The TAO detector will be pre-assembled at the Institute of High Energy Physics (IHEP) in Beijing, then disassembled and shipped to the Taishan power plant, and assembled again there. Due to the transportation limit, all components can not be larger than $1.99 \times 1.39 \times 1.99$ m in dimension. The SST and the acrylic vessel used for preassembly at IHEP will not be re-used since the cutting process will cause large distortion. So a new SST and a new acrylic vessel will be made for the final detector, while all other components will be reused.

The onsite assembly should be as simple as possible to avoid logistic difficulties. The SST will be welded onsite from 6 pieces with a mould to control the deformation during welding and ensure the dimension precision. After the lid is sealed and the cabling and piping outside the SST are finished, the insulation layer wrapping the SST will be made onsite. A thin (about 3 mm) steel shell surrounds the SST and keeps 20-cm gap in-between, then the polyurethane is filled in the gap and then foams to the shell shape. The insulation layer for the bottom and lid of the SST can be made separately, and the bottom part should be ready before the SST is moved to the targeted position. More details can be found in Section~\ref{sec:installation}.

\subsection{Low temperature control}
The low temperature system will lower the temperature inside the SST to -50$^\circ$C and keep it stable during data taking. The heat sources include the heat leakage from outside the SST (500 W), the heat produced by the SiPMs and the FEE readout (500-1000 W), and the heat from the backend electronics ($<1000$ W). The heat generated inside the SST could be revised when the readout scheme is decided and tested. The light yield of the GdLS is a function of temperature, increasing about 0.35\% per degree as temperature decreases~\cite{Xia:2014cca}. To have a stable detector energy scale, the temperature fluctuation should be controlled within $\pm0.5^\circ$C for GdLS in the acrylic vessel.

The low temperature environment is realized with a cryostat and a cryogenerator. The cryostat includes the SST with insulation and the coiled pipe for coolant. The design goal is:
\begin{itemize}
\item The temperature inside the SST is uniformly -50$^\circ$C, while keeping the ability of the cryogenerator to cool the SST down to -70$^\circ$C. The heat load inside the SST is $<2.5$ kW. Stirrer is not allowed, and the disturbance of the liquid due to convection should be small.
\item The key requirement on the temperature uniformity is to keep $\pm0.5^\circ$C inside the acrylic vessel (for GdLS).
\item The material, fabrication, and assembly should satisfy the cleanness requirements.
\item The cooling process from room temperature to -50$^\circ$C is about 2 weeks, which could be fine tuned by balancing the experimental requirements and the cost.
\end{itemize}

A single layer SST with insulation, instead of a double-layer vacuum vessel, will be adopted since the cryostat has to be welded together from several pieces in the laboratory due to transportation limitation. 20-cm thick Polyurethane (PU) is chosen as the insulation material wrapping the SST. A layer of steel shell is required to form a mould outside of the SST and keep a 20-cm gap between them. PU foaming material will be injected into the gap to form the insulation layer.

\subsubsection{Heat production and cooling design}

There are three major sources of heat in the central detector, including
\begin{itemize}
\item the heat leakage from environment. After insulation with 20-cm polyurethane for the tank and a polyethylene hat for the calibration device (ACU), the heat leakage is estimated to be 460~W;
\item the heat generated by SiPMs and the readout electronics on the copper shell, which is estimated to be 500~W (ASIC option) or 1000~W (discrete option);
\item the heat of about 1000~W from the backend electronics, mainly from FPGA, if they are installed inside SST.
\end{itemize}

The thermal conductivity of Polyurethane is 0.03~W/(m$\cdot$K). A simple calculation of the bulk heat leakage from environment at 25$^\circ$C to inside SST at -50$^\circ$C through 20-cm PU insulation is 225~W. A detailed thermal simulation for a SST model including the support legs and cable and pipe penetrations yields a heat leakage of 460~W.

Inside the detector, the heat sources include SiPMs, the frontend electronics, and the backend electronics (see Sections~\ref{sec:ASIC} and \ref{sec:discrete}). The heat generation is expected to be stable and uniform on the surface of the supporting structure, the copper shell. Since the temperature variation has apparent impacts to the light yield of the GdLS, the core requirement of the cryogenic system is to keep the temperature of the GdLS stable and uniform to $\pm0.5^\circ$C. The convection of thick liquid at low temperature may have impacts to the light transmission. The heat source and the cooling source should be as close as possible to take the heat away immediately. Therefore, corresponding to the three heat sources, the coolant coiled pipes are designed to have three groups. One group is attached to the SST wall, lid, and bottom to take away the heat leaking from environment. Another group is attached on the surface of the copper shell to take away the heat generated by the FEE. The third group will pass the insulated boxes that contain the backend electronics.

\begin{figure}[htb]
\begin{center}
\includegraphics[width=0.4\textwidth]{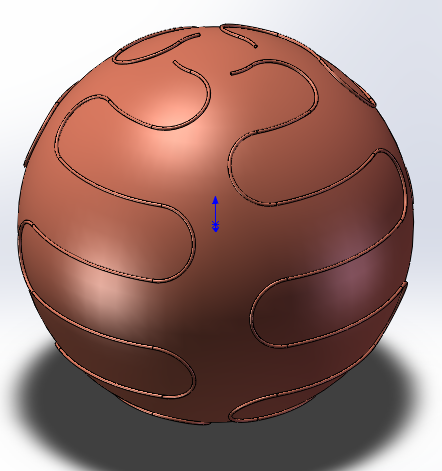}
\caption{The layout of the cooling pipes on the copper shell.
\label{fig:coolantcoiled}}
\end{center}
\end{figure}

The copper shell is divided into several pieces and shipped into the underground laboratory. The cooling pipes need to follow this design. To avoid interference with the frontend electronics mounted on the copper shell, the maximum diameter of the cooling pipes should be less than 20~mm. The layout of the cooling pipes on the copper shell is shown in Figure~\ref{fig:coolantcoiled}.

The whole central detector will operate at -50$^\circ$C. We have evaluated other cooling options but considered them as less feasible. One of those options is to keep the copper shell and the SiPMs at -50$^\circ$C but GdLS at 20$^\circ$C. We would not need to worry about the water content and fluor solubility in GdLS then. The thermal conductivity of acrylic is 0.19~W/(m$\cdot$K). Suppose we use 10-cm thick acrylic vessel to separate the GdLS and the SiPMs with good thermal insulation, the heat flow passing the acrylic is 1330~W. This heat capacity should be compensated either by circulating the GdLS with heat feeding, or by adding an interlayer with liquid flow to feed the heat. Detailed analyses were done with ANSYS Steady State Thermal and Fluent software. The thermal conductivity of LAB at 23$^\circ$C is measured to be 0.1426~W/(m$\cdot$K)~\cite{Wu:2019gzo}. Another measurement has been done for low temperatures, which shows 0.139~W/(m$\cdot$K) at -50$^\circ$C as listed in Table~\ref{tab:viscos}. GdLS properties in the simulation have been taken as the same as LAB. It is found that the temperature of GdLS is hard to be uniform in the acrylic vessel. The temperature differences are 5.5~degree and 3~degree for the above two options, respectively. A hybrid option to keep the GdLS at -30$^\circ$C is possible to satisfy the uniformity requirement of $\pm0.5^\circ$C, but still too complex to be feasible and reliable. Therefore, the option to have the whole detector working at -50$^\circ$C is chosen, while the R\&D efforts are put on developing a stable low temperature GdLS.

\subsubsection{Copper shell vs stainless steel shell}

A 12-mm thick spherical copper shell provides the mechanical support for the SiPM tiles and their electronics readout. It also serves as a temperature stabilizer to provide stable and uniform temperature for GdLS inside the acrylic vessel and SiPMs on the inner surface of the shell structure. We require the uniformity of temperature inside the acrylic vessel and working temperature of the SiPM both to be $\pm0.5$ degree to get uniform energy scale of the detector.

\begin{figure}[htb]
\begin{center}
\includegraphics[width=0.8\textwidth]{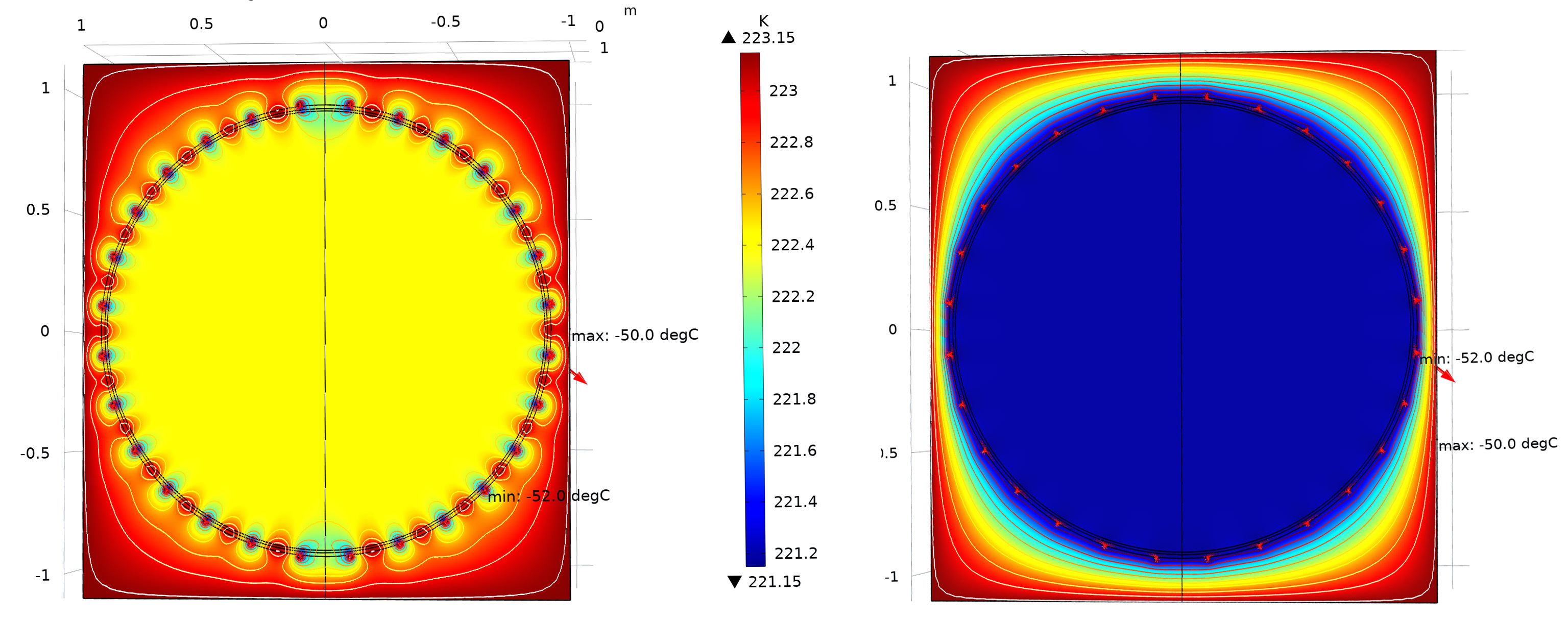}
\caption{The temperature field distribution for the stainless steel design (left) and the copper shell design (right). On the shell there are 14 layers of coiled coolant pipe, which appear as cold spots on the cross section figure.
\label{fig:coppertf}}
\end{center}
\end{figure}

Stainless steel shell is also considered as its strength is better than copper and manufacturing is easier. However, its heat conductivity is about 15~W/(m$\cdot$K, more than 20 times smaller than that of copper. The temperature field is analyzed for steady state. A heat of 50~W/m$^2$ is uniformly loaded on the copper or stainless steel shell. The coiled coolant pipe has 14 layers, attached on the outer surface of the shell. The coolant temperature is set to be -52$^\circ$C. As shown in the left of Figure~\ref{fig:coppertf}, the temperature of the bulk GdLS is -50.8$^\circ$C, while the temperature range is $-50.8\pm0.5^\circ$C when the distance is 50~mm from the stainless steel shell of a thickness of 15~mm. On the edge of the acrylic vessel, some of the GdLS has larger temperature variation than required. Although events in this area will be rejected by the fiducial volume cut, the temperature variation might have certain impacts to physics. Also, temperature difference is undesirable as it may cause GdLS convection, whose impact is unknown, especially for the stick liquid at low temperature. In the right of Figure~\ref{fig:coppertf}, the temperature field for the 12~mm thick copper shell with the same setting is shown. The temperature is very uniform and all GdLS has a temperature of -52$^\circ$C.

A special transient state analysis has also been done to simulate the non-uniform heat load, e.g. some electronics generate more heat or even short-circuited. A heat of 500~W/m$^2$ is loaded in an area of 0.5~m$^2$ on the shell, as an extreme case. Large temperature variation is observed for the stainless steel shell design, and it takes 10~hours to reach a steady state. For the copper shell design, the temperature difference is only 0.4 degree, and it takes 10~minutes to reach a steady state.

\subsubsection{Refrigerator system}

\begin{figure}[htb]
\begin{center}
\includegraphics[width=0.6\textwidth]{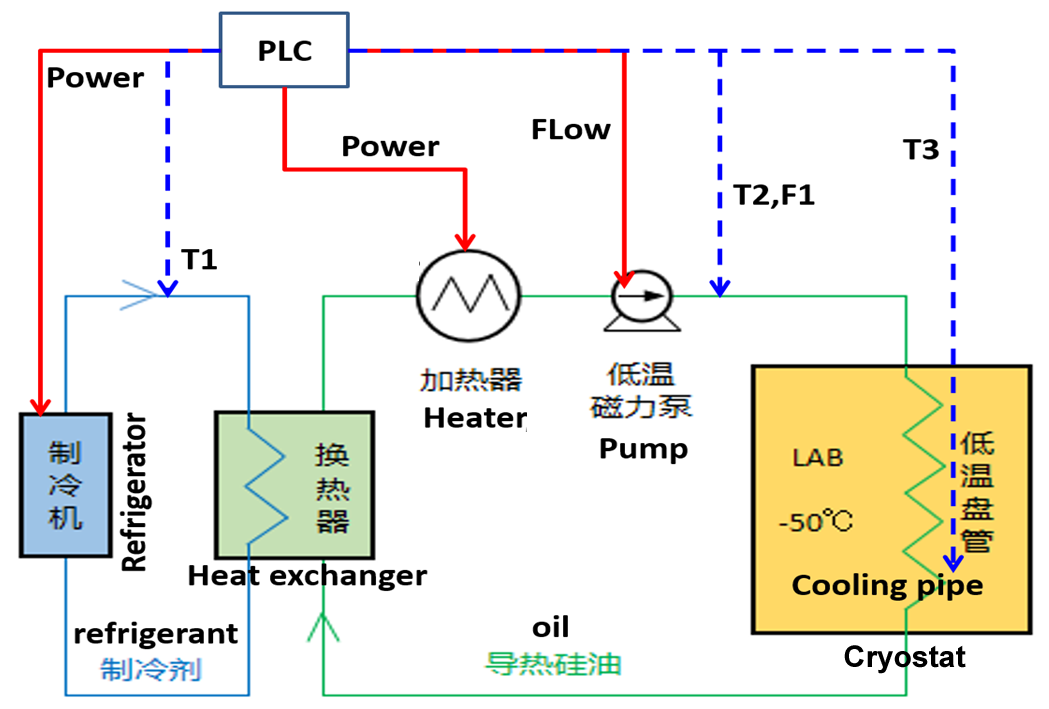}
\caption{The scheme of the refrigerator system.
\label{fig:refrischeme}}
\end{center}
\end{figure}

The scheme of the refrigerator system is shown in Figure~\ref{fig:refrischeme}, including the refrigerator, heat exchanger, coiled coolant pipe, heater, meters, Programmable Logic Controller (PLC), and the cryostat (the detector). The coolant is cooled by the heat exchanger, and driven by a magnetic drive pump. To control the temperature of the coolant, an electric heater is designed before the heat exchanger. The temperature control is realized by changing the power of the heater. Silicone oil is cooled in the heat exchanger and takes away the heat in the detector. The cycling pipes connecting with the SST and the refrigerator are vacuum cooling pipes made of stainless steel, so the diameter of the pipe could be as small as 50~mm. Otherwise, normal pipe must be wrapped with insulation layer up to about 10~cm. All coolant pipes outside the detector need be well insulated and avoid water condensation. The heat exchanger is made of AISI-316 stainless steel. The electric heater is controlled with a Proportional Integral Derivative (PID) controller.

The secondary refrigerant is silicone oil, which has a broad work temperature from -80$^\circ$C to 195$^\circ$C. The silicone oil has good temperature stability, good thermal conductivity, and is chemical inert, non-toxic, and environmentally friendly. The cooling capacity is taken with 1.5 times margin, about 3.5~kW at -50$^\circ$C. The cooling capacity of the refrigerator chosen for the prototype experiment is shown in Table~\ref{tab:coolingcap}.

\begin{table}[htb]
\setlength{\belowcaptionskip}{5pt}
\begin{center}
\caption{Cooling capacity of the LT-80A1WN refrigerator.\label{tab:coolingcap}}
\begin{tabular} {c c}
\hline\hline
Temperature ($^\circ$C) & Cooling capacity (kW) \\ \hline
-40 & 4.8 \\
-60 & 4.0 \\
-75 & 2.0 \\
\hline
\end{tabular}
\end{center}
\end{table}

\subsubsection{Temperature monitoring}

The temperature in the detector must be measured and monitored during the whole process of the experiment running. The temperature data in the SST is crucial for the control of the refrigerator system, and temperature data outside the SST is also important to estimate and monitor the heat leakage in different areas. The temperature sensors will be set uniformly on the outer surface of copper shell, SST and PU insulation to monitor the full detector system. Number and locations of the temperature sensors are still under optimization, and will be verified with the prototype detector.

The performance of the cryogenic system has been tested in a prototype for JUNO~\cite{Zhang:2019rfl}. After implemented with 20-cm insulation layer on the outer tank, installed cooling pipes and a refrigerator system, replaced the LS and water to TAO GdLS and buffer LAB, and instrumented with 10 temperature sensors, the prototype has been adapted as a low temperature liquid scintillator detector of 70~L GdLS and about 6~ton buffer LAB. The prototype has been successfully cooled down from 20$^\circ$C to -50$^\circ$C. The heat load (basically from the heat leakage of the insulation) is found to be about 500~W as expected. The temperature non-uniformity in the 1.8-m cylinder space has reached 1.3$^\circ$C and the temperature stability is 0.02$^\circ$C in 20~days.

\subsection{Liquid scintillator}
Gadolinium-doped liquid scintillator (GdLS) will be used as the neutrino target of TAO to register a clean delayed signal of IBD from the neutron capture on Gd and thus reduce the accidental background. To lower the dark noise of SiPM to 100~Hz/mm$^2$ as required by TAO, the GdLS and the buffer liquid should work at -50$^\circ$C or lower. The GdLS for TAO should have good
\begin{itemize}
\item transparency at -50$^\circ$C;
\item light yield at -50$^\circ$C;
\item chemical stability for several years;
\item safety (fire risk, toxicity, environmental safety, etc.).
\end{itemize}

Linear Alkylbenzene is used in Daya Bay and JUNO as the solvent of the liquid scintillator. It has advantages on all above requirements by TAO GdLS. Especially it has a high flash point, $>130^\circ$C, thus very suitable for using near the reactor. Commercially available LAB is a mixture of carbon number in the linear chain ranging from 9 to 14 in general. It has a freezing point lower than -60$^\circ$C, but its water content and the fluor in LAB-based LS may precipitate at low temperature. By tuning the recipe of the Daya Bay GdLS and adding co-solvent, we have developed the GdLS (and undoped LS) that works well at low temperature.

\subsubsection{Recipe}
LAB has a freezing point lower than -60$^\circ$C, although it depends on particular sample given it is a mixture. The viscosity, density, specific heat capacity, and thermal conductivity of a typical LAB sample used for JUNO have been measured at different temperatures as shown in Table~\ref{tab:viscos}.

\begin{table}[htb]
\setlength{\belowcaptionskip}{5pt}
\begin{center}
\caption{Viscosity, density, specific heat capacity, and thermal conductivity of a typical LAB at different temperatures. \label{tab:viscos}}
\begin{tabular}{r c c c c }
\hline\hline
  Temperature & -20$^\circ$C &  -30$^\circ$C &  -40$^\circ$C &  -50$^\circ$C \\ \hline
  Viscosity (mm$^2$/s) & 54.2 & 114.7 & 283.4 & 802.5 \\ \hline
  Density (g/mL) & 0.896 & 0.902 & 0.908 & 0.914 \\ \hline
  Specific heat capacity (J/(g$\cdot$K)) & 1.784 & 1.761 & 1.740 & 1.727 \\ \hline
  Thermal conductivity (W/(m$\cdot$K)) & 0.143 & 0.142 & 0.140 & 0.139 \\
\hline
\end{tabular}
\end{center}
\end{table}

\begin{figure}[htb]
\begin{center}
\includegraphics[width=0.7\textwidth]{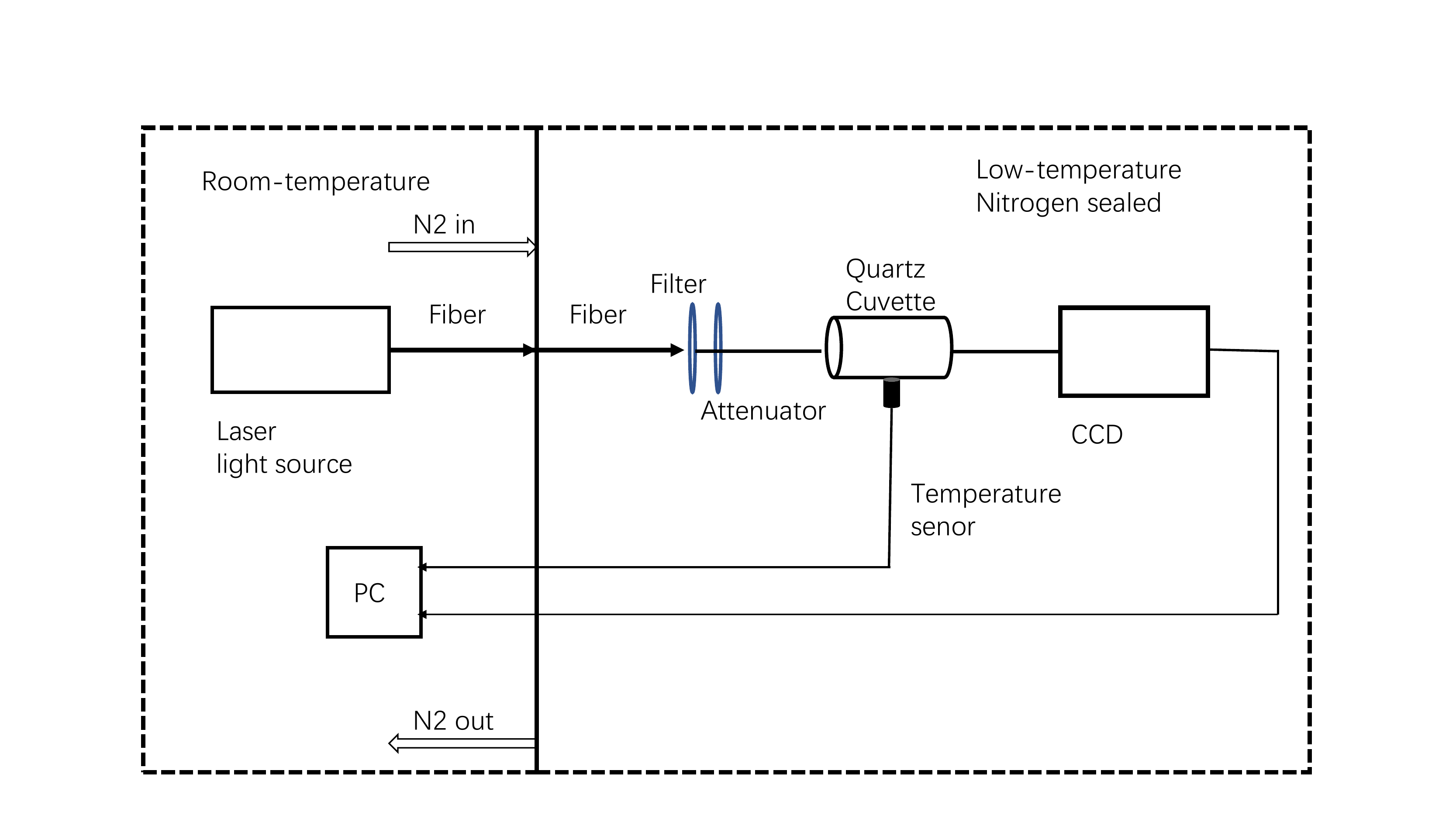}
\caption{Scheme of the apparatus measuring the transparency of liquid at low temperature.
\label{fig:lsccd}}
\end{center}
\end{figure}

Although still in liquid state, normal LAB may turn cloudy at low temperature due to the water content in it. At room temperature, the water content in LAB is often at $\sim40$ ppm, while the saturated content is $\sim200$ ppm. The transparency of the liquid sample at low temperature was measured with a customized apparatus shown in Figure~\ref{fig:lsccd}. When the precipitate appears, the light received by Charge Coupled Device (CCD) will reduce due to scattering of the incident light. Furthermore, the edge of the light spot on CCD will turn fuzzy. We found that LAB will keep clear if the water content is less than 5 ppm. Water in LAB can be removed by bubbling with dry nitrogen.

Di-isopropylnaphthalene (DIN) as another solvent candidate also has high flash point. The NEOS experiment adds 10\% DIN into LAB-based liquid scintillator and gets better pulse shape discrimination~\cite{Ko:2016owz}, which is attractive for an experiment at shallow overburden to reject fast neutron background. However, DIN itself turns cloudy starting from -20$^\circ$C. Adding 10\% DIN into LAB also degrade the transparency. Further R\&D will be done on this option for lower fraction of DIN. Another solvent candidate, pseudocumene, has a flashing point of about 40$^\circ$C (inflammable) and a freezing point of -43.78$^\circ$C, not suitable for TAO.

\begin{figure}[htb]
	\centering
	\includegraphics[width=0.5\textwidth]{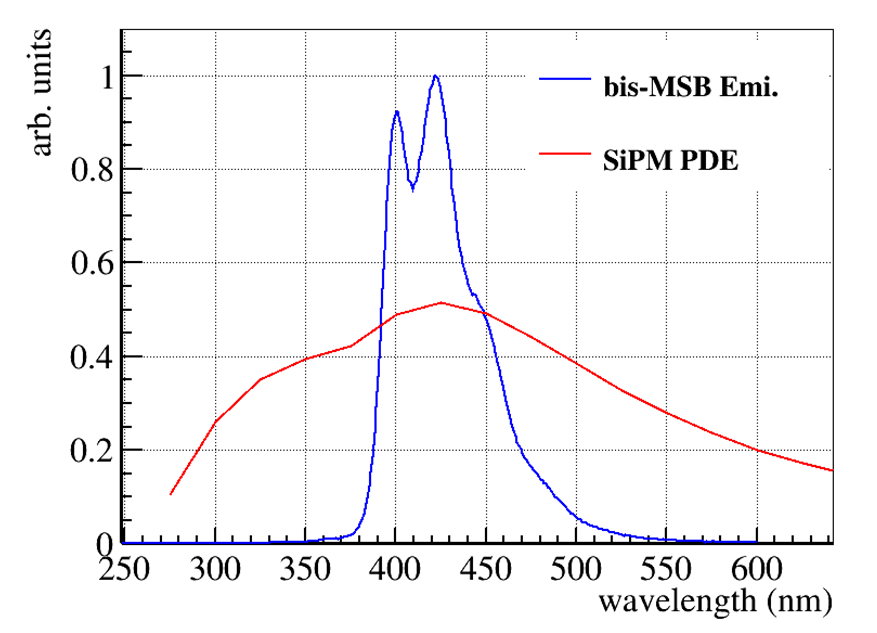}
	\caption{The bis-MSB emission spectrum and the typical SiPM PDE as a function of wavelength.
		\label{fig:sipmppo}}
\end{figure}

Daya Bay liquid scintillator has 3~g/L PPO as fluor and 15~mg/L bis-MSB as wavelength shifter~\cite{Beriguete:2014gua}. New R\&D for JUNO using one Daya Bay detector~\cite{DYBJUNOLS} shows that 1 to 3~mg/L of bis-MSB will have the highest light output for JUNO. For most detectors of various size, 3~mg/L bis-MSB is enough, slightly depending on the purity of the solvent. Higher concentration of bis-MSB may slightly reduce the light yield if the solvent has very little absorption like JUNO, but may have advantages by quickly converting the light to longer wavelength, if the solvent or other components in LS have absorption at the emission wavelength of PPO, which also can improve the photon collection efficiency. The bis-MSB emission spectrum and the typical SiPM Photon Detection Efficiency (PDE) as a function of wavelength overlaps well, as shown in Figure~\ref{fig:sipmppo}.

\begin{figure}[htb]
    \centering
    \includegraphics[width=0.6\textwidth]{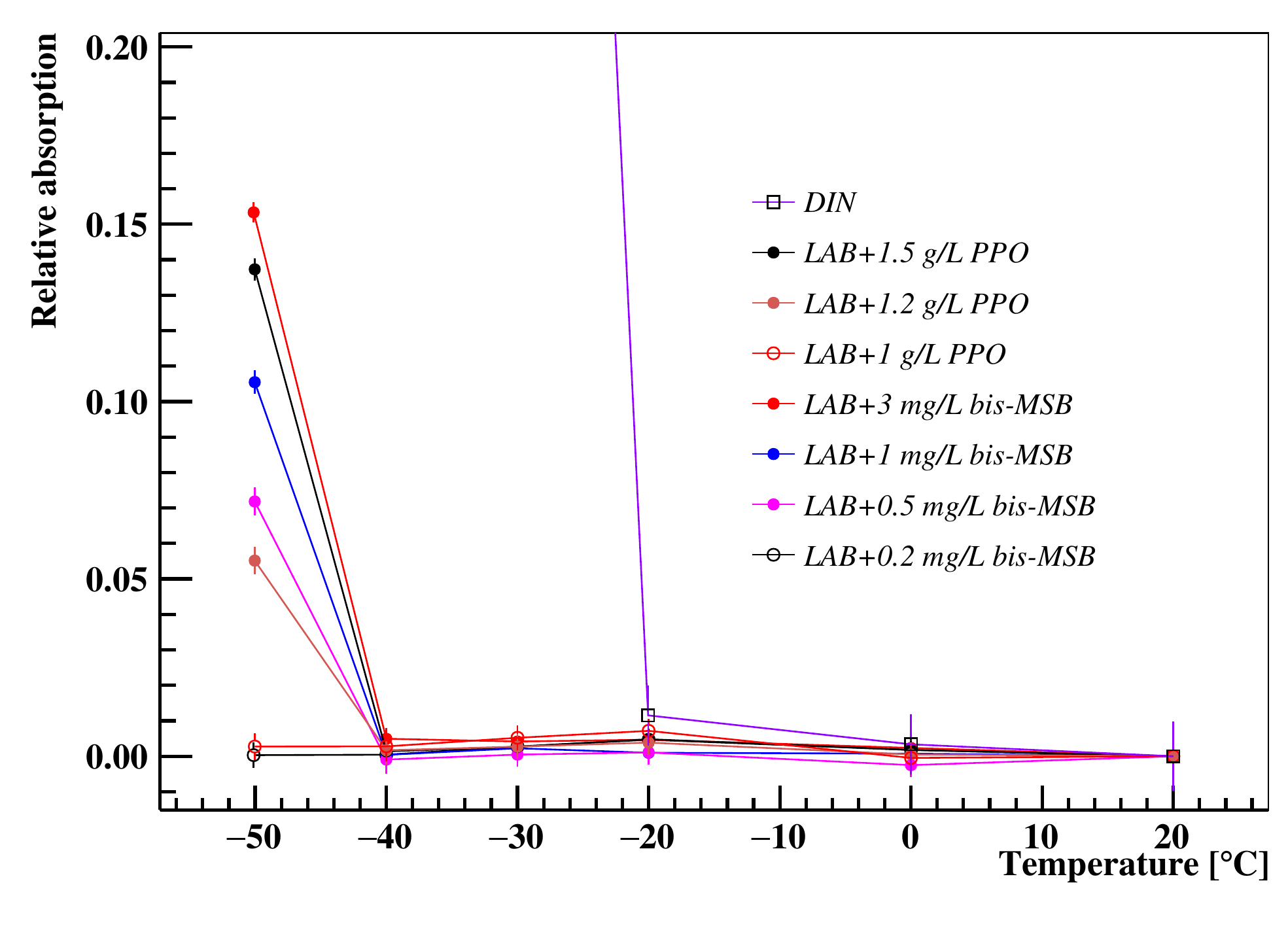}
    \caption{Solubility testing of PPO and bis-MSB in LAB at low temperatures.
    \label{fig:lsppo}}
\end{figure}

At low temperature, the solute PPO and bis-MSB in LS may precipitate. We found that the solubility of PPO in LAB at -50$^\circ$C is between 1 and 1.2~g/L. The absorption of the solution suddenly increases when the temperature decreases from -40$^\circ$C to -50$^\circ$C, as shown in Figure~\ref{fig:lsppo}. The light spot on CCD also turns fuzzy, illustrating the light scattering due to the precipitation of PPO. Similarly, the solubility of bis-MSB in LAB is between 0.2 and 0.5~mg/L.

\begin{figure}[htb]
\begin{center}
  \includegraphics[width=0.6\textwidth]{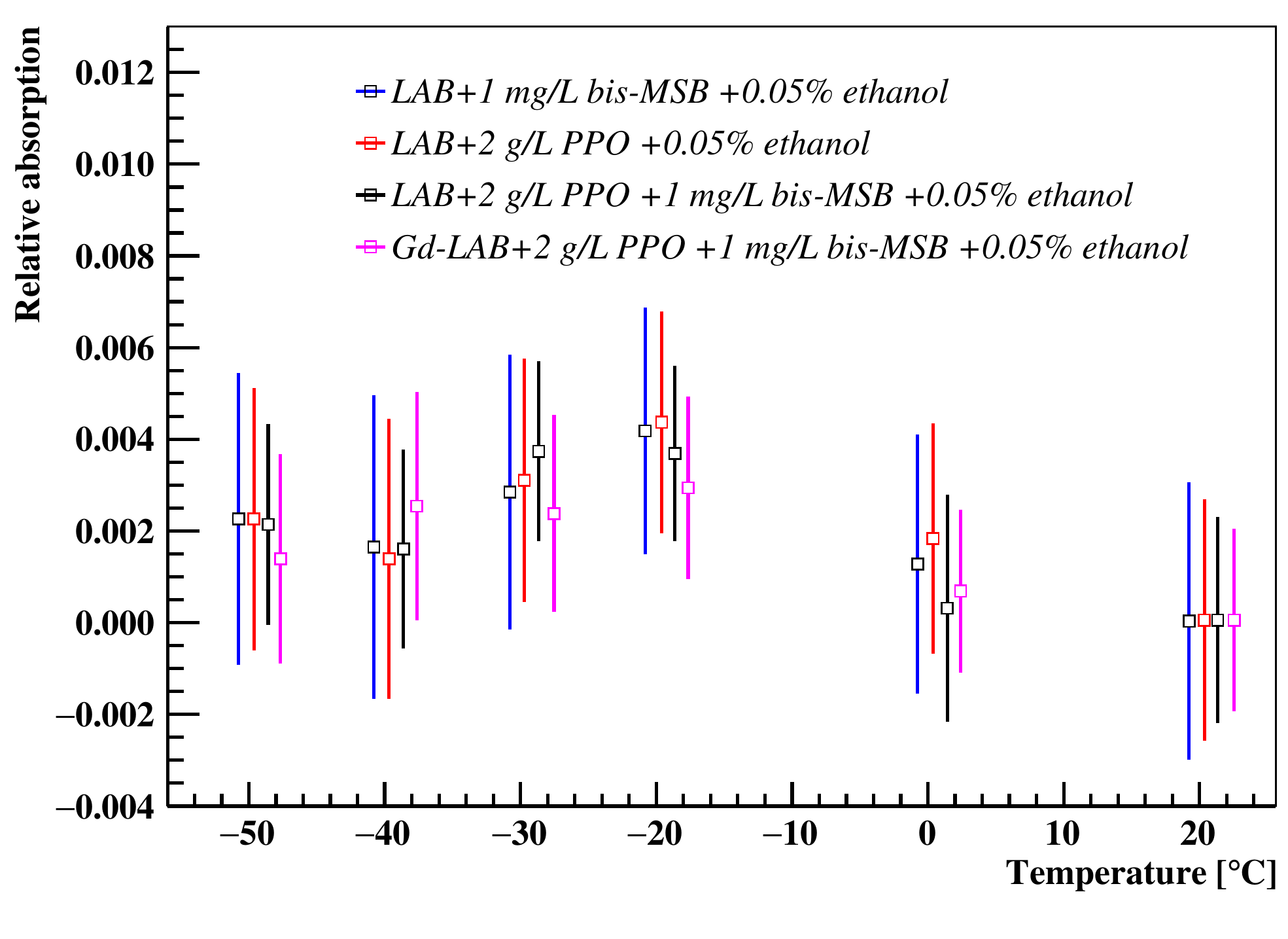}	
  \caption{Absorption of the liquid scintillator and GdLS with co-solvent.
\label{fig:lslowtemp}}
\end{center}
\end{figure}

However, Liquid scintillator with 1 g/L PPO and 0.2 mg/L bis-MSB has significantly lower light yield comparing to the optimal recipe for JUNO with 2.5~g/L PPO and 1-3 mg/L bis-MSB. Alcohol has low freezing point and is lipophilic. Addition of a small fraction of alcohol into LS will increase the solubility of PPO and bis-MSB, thus cure the light yield problem at low temperature, as shown in Figure~\ref{fig:lslowtemp}. Besides ethanol, less volatile pentanol is also studied but found to be not compatible with the reagent of the gadolinium.

As a conclusion, the recipe of the TAO GdLS is chosen to be LAB plus 2~g/L PPO and 1~mg/L bis-MSB, with 0.1\% mass fraction of gadolinium using Daya Bay's Gd-complex~\cite{Beriguete:2014gua}, and $\sim0.05$\% ethanol as co-solvent. Since the GdLS needs the water to be removed by bubbling with nitrogen and ethanol is volatile, we will add more ethanol, e.g. 0.5\%, at the beginning and will monitor the ethanol content during operation with a sensor in the overflow tank of the detector. The Gd-complex, a complex of GdCl$_3$ and carboxylic acid, contains a fraction of water molecule in the complex. The water content in GdLS can be higher than in LS without Gd-doping to keep clear at low temperature. We require that the water content should be lower than 10 ppm in TAO GdLS, and lower than 5 ppm in LAB as buffer liquid. The absorption is shown in Figure~\ref{fig:lslowtemp}, which shows that the GdLS keeps clear at -50$^\circ$C.

Adding alcohol will lower the flash point of GdLS and increase the fire risk. With 0.05\% ethanol, the flash point of GdLS is still $>100^\circ$C. However, if the ethanol fraction is 0.5\%, the flash point will be only 40$^\circ$C.

\subsubsection{Liquid scintillator light yield}
\begin{figure}[htb]
\begin{center}
  \includegraphics[width=0.6\textwidth] {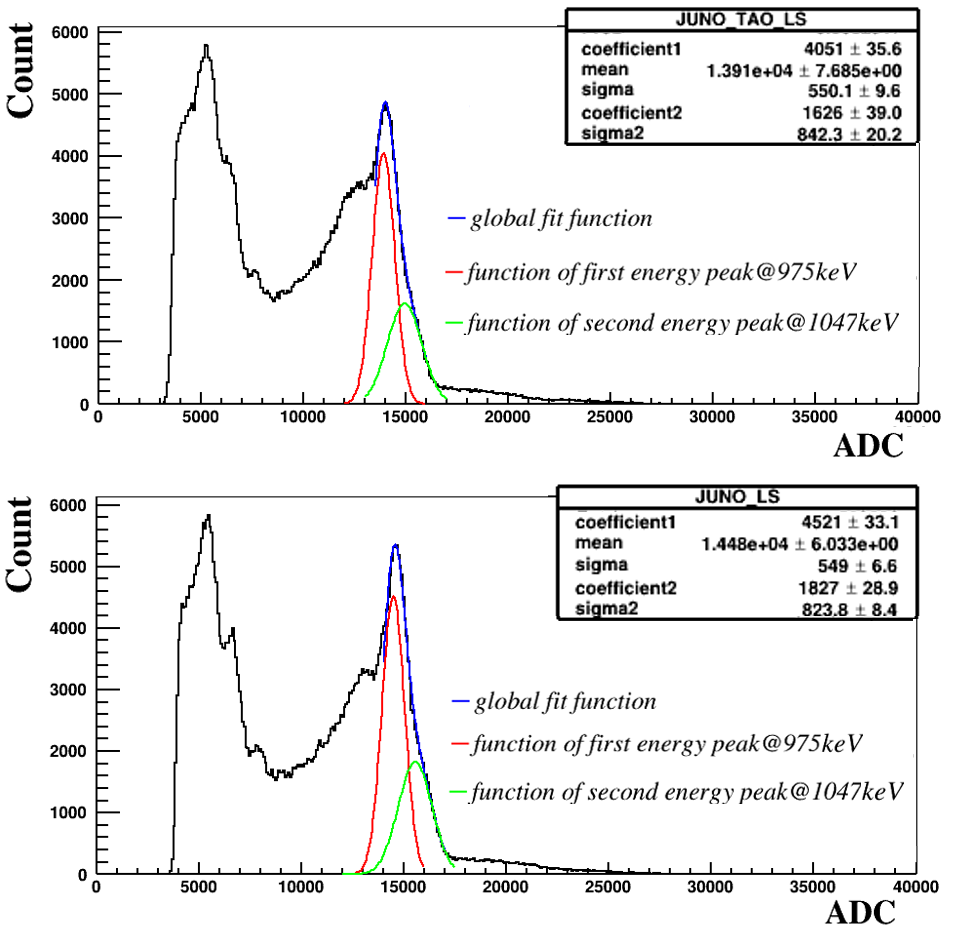}
  \caption{Light yield of the ``JUNO-TAO LS" and the JUNO LS at room temperature.
\label{fig:lslightyield}}
\end{center}
\end{figure}

Adding co-solvent may have impact to the light yield. We measured the light yield of a LS referred as ``JUNO-TAO-LS" (LAB + 2 g/L PPO + 1 mg/L bis-MSB + 0.05 \% ethanol), which has the same recipe as the TAO GdLS but without Gd, relative to the JUNO LS (LAB + 2.5 g/L PPO + 3 mg/L bis-MSB) with a customized apparatus, in which six PMTs are coupled to a cubic acrylic vessel of 2-cm dimension. Coincidence of 6 PMTs reduces the noise significantly. The internal conversion electron of a $^{207}$Bi source is used to measure the light yield, as shown in Figure~\ref{fig:lslightyield}. The gain of PMTs at low temperature has been calibrated. We find that the relative light yield of the TAO LS is 96\% of that of JUNO at room temperature.

The light yield of liquid scintillator usually increases at lower temperature due to less thermal quenching~\cite{Xia:2014cca}.
For low temperature, the light yield of TAO GdLS is still under study. The non-linear energy response of TAO GdLS will be studied in laboratory with a similar apparatus in Ref.~\cite{Zhang:2014iza} but placed in a low temperature cryostat.

\subsubsection{Production and filling}

The production of TAO GdLS will be similar to the Daya Bay GdLS~\cite{Beriguete:2014gua}. However, we need control the water content and ethanol content in GdLS, which is actually challenging since ethanol is volatile, and an exposure to air will increase the water content in GdLS. The production and filling thus need special caution.

\begin{figure}[htb]
\begin{center}
  \includegraphics[width=0.6\textwidth]{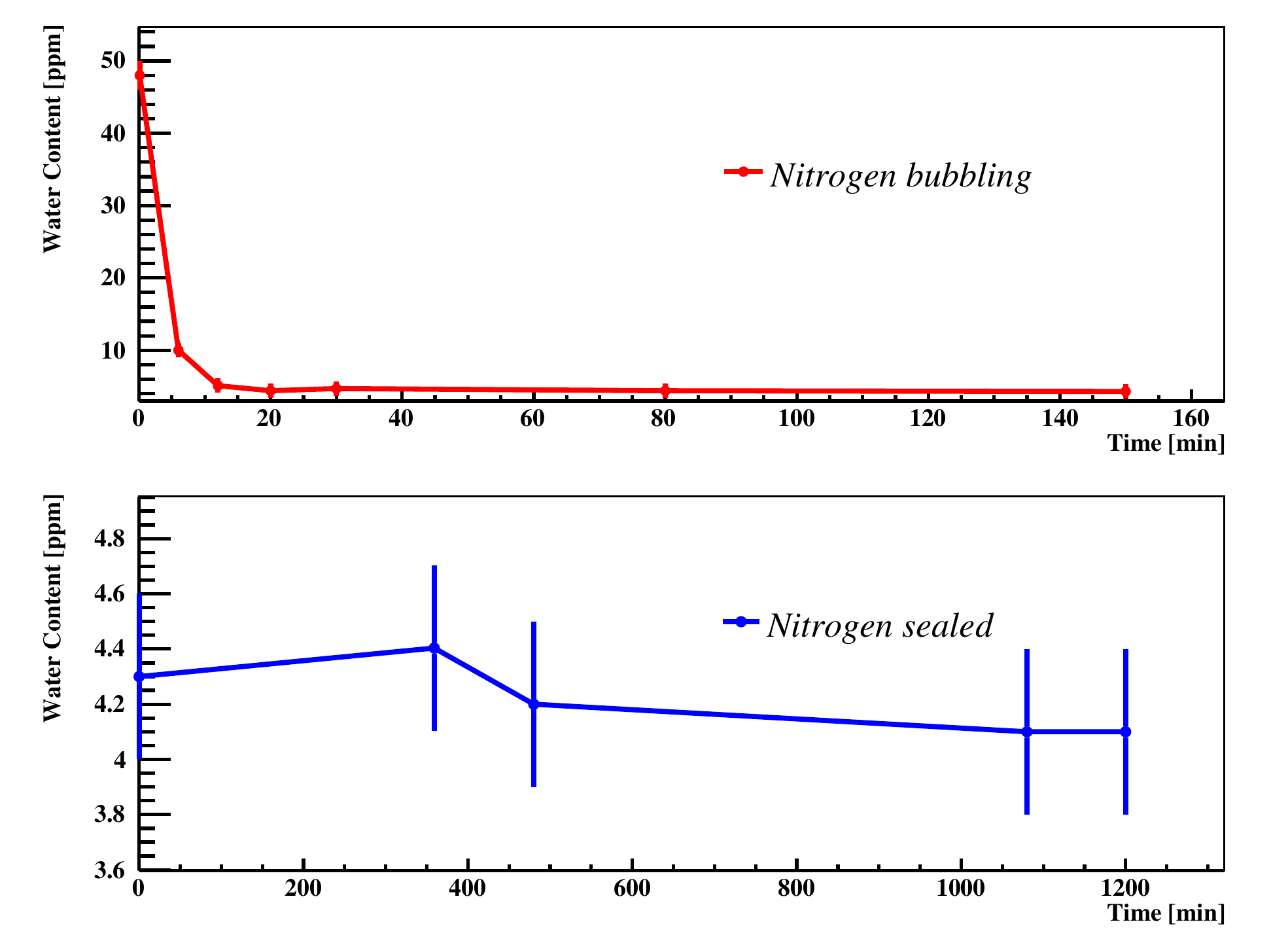}
  \caption{Top: Water content in LS as a function of bubbling time with high purity dry nitrogen. Bottom: Water content in LS as a function of time after bubbling is done and LS is sealed with nitrogen atmosphere.
  \label{fig:watercontent}}
\end{center}
\end{figure}

Several methods to remove water have been studied. Nitrogen bubbling is the preferred method, without risks to pollute the very transparent GdLS. A sample of 1 L LS was bubbled with high purity dry nitrogen of 2 L/min. The water content is measured with a Karl Fisher equipment, as shown in Figure~\ref{fig:watercontent}. After two hours of bubbling, the water content is stable at 5 ppm, and keeps stable after being sealed and covered with nitrogen.

The ethanol content in GdLS needs be monitored during operation since the flow nitrogen cover may take the ethanol away. An ethanol sensor will be installed in the overflow tank for that purpose. With the partition coefficient of ethanol content in air and in GdLS, the measurement in nitrogen cover can be converted to the ethanol content in GdLS. Details will be worked out later, such as the selection of the sensor and the measurement of partition coefficient at low temperature, etc.

GdLS is compatible with acrylic vessel and relevant materials used for TAO, given the experience from Daya Bay~\cite{An:2015qga}. The compatibility of LAB as the buffer liquid and detector materials needs be studied later.

%% file: Calibration/section.tex
\section{Calibration System}
\label{sec:calibration}
\blfootnote{Editors: Zhimin Wang (wangzhm@ihep.ac.cn) and Liang Zhan (zhanl@ihep.ac.cn)}

\subsection{Requirements}
The precise measurement of reactor antineutrino spectrum in the TAO experiment requires extreme care in the characterization of the detector properties as well as frequent monitoring of the detector performance.
The requirements for the calibration of the detector response are listed in Table~\ref{tab:calib:requirements}.
These can be met by a comprehensive automated calibration program with LED light sources and radioactive sources deployed into the detector volume at regular intervals (i.e., daily or weekly). Radioactive isotopes and spallation neutrons produced by cosmic muons in TAO provide additional full-volume calibration data. One Automatic Calibration Unit (ACU) of the Daya Bay experiment will be reused in TAO after the Daya Bay decommissioning. Similar calibration procedures will be employed in TAO and the energy scale calibration is expected to reach similar precision as Daya Bay ($<1.0\%$)~\cite{Adey:2019zfo}. The spatial uniformity has impact on the energy resolution in the energy reconstruction. To achieve a sub-percent energy resolution in most of the energy region, a 0.5\% spatial uniformity is required.

\begin{table}[htb]
\begin{center}
\caption{Requirements for the detector calibration. \label{tab:calib:requirements}}
\begin{tabular}{c c c}
    \hline\hline
    Requirements & Description & Proposed Solutions \\ \hline
         SiPM gains	& gains vs. time for all channels & LED, dark noise \\ \hline
         SiPM timing	&  $\sim 1$~ns timing calibration	& LED, gamma sources  \\ \hline
         SiPM PDE & relative PDE for each SiPM tile & LED, gamma sources \\ \hline
         SiPM cross talk &  optical cross talk and afterpulsing & p.e.\ spectrum \\ \hline
         Energy scale & p.e.\ yield vs. energy at center, $<1\%$& gamma sources, neutron capture \\ \hline
         Spatial uniformity	& spatial uniformity for p.e.\ yield, $<0.5\%$ & gamma sources, neutron capture \\ \hline
         Charge pattern & charge pattern vs. vertex & LED, gamma sources, neutron \\ \hline
\end{tabular}
\end{center}
\end{table}

\subsection{Radioactive sources}
The main goal of TAO is to precisely measure the antineutrino energy spectrum via inverse beta-decay reaction (IBD).
The prompt signal, as a proxy for the antineutrino energy, is produced by positron in the IBD reaction.
The positron ionization and subsequent annihilation gammas produce signal in the energy region of 1--10~MeV.
The delayed signal is produced by the gammas emitted from neutron captures on Gd.
No monoenergetic positron calibration source is available and a set of gamma sources are used to characterize the detector response.

It is necessary to characterize the detector properties carefully before data taking and monitor the stability of the detectors during the whole data taking period. Calibration sources must be deployed regularly throughout the active volume of the detector to simulate and monitor the detector response to positrons, neutron capture gammas and gammas from the environment.
The sources available for calibrations in TAO are listed in Table~\ref{tab:calib:sources}.
These sources cover the energy range from about 0.5~MeV to 10~MeV and thus can be used for a thorough energy calibration.
The positron produced by the $^{22}$Na and $^{68}$Ge source will be absorbed by the enclosure and only the two annihilation gammas will be released.
Dissolving the positron source in a small scintillator volume filled in a transparent vessel is also in consideration.
The detector response depends on the particle species.
Conversion of the gamma response to the positron response can be determined from the simulation. Similar conversion procedure based on the experience in Ref.~\cite{Adey:2019zfo} will be performed for TAO and the related uncertainty will be investigated.

\begin{table}[htb]
\setlength{\belowcaptionskip}{5pt}
\begin{center}
\caption{Possible calibration sources.  \label{tab:calib:sources} }
\begin{tabular}{r l}
\hline\hline
    Event Type & Calibrations \\ \hline
    {\bf Neutron sources:}  & Neutron response,     \\
    Am-Be, $^{252}$Cf, Pu(C) &  detection efficiency    \\ \hline
    {\bf Positron sources:} $^{22}$Na, $^{68}$Ge & \begin{tabular}{l}Positron response, energy scale,\\ trigger threshold\end{tabular}   \\ \hline
    \begin{tabular}{r} {\bf Gamma sources:} \\ \\ \\  $^{137}$Cs  \\  $^{54}$Mn  \\ $^{65}$Zn \\$^{40}$K  \\Neutron capture on H \\  $^{22}$Na \\ $^{60}$Co  \\  Am-Be  \\  $^{238}$Pu-$^{13}$C \\ Neutron capture on Gd  \end{tabular}  & \begin{tabular}{l}Energy scale, stability, \\ energy resolution, \\spatial and temporal variations \\ 0.662 MeV \\ 0.835 MeV \\ 1.351 MeV \\  1.461 MeV  \\ 2.223 MeV  \\ annihilation + 1.275 MeV  \\ 1.173 + 1.333 MeV  \\  4.43 MeV \\6.13 MeV \\ $\sim$ 8~MeV \end{tabular}     \\ \hline
\end{tabular}
\end{center}
\end{table}

Neutron sources simulates the delayed signal of IBD reactions at the calibrated points.
The Am-Be source can be used to calibrate the detection efficiency of neutron capture signal by detecting the 4.43 MeV gamma in coincidence with the neutron. The $^{238}$Pu-$^{13}$C source will similarly provide a 6.13 MeV gamma as a prompt signal in coincidence with the delayed neutron. The neutron detection efficiency can be determined with neutron sources by measurement of the energy spectrum for neutron capture signals. In addition, neutron sources allow us to determine the appropriate selection cut for IBD coincidence time by measurement of the neutron capture time. The positron annihilation signal in IBD reactions can be simulated by positron sources. Radioactive sources must be encapsulated in small containers to prevent any possible contamination of the ultra-pure liquid scintillator. They can be regularly deployed to the whole active volume in Z axis of the detector. Calibration data are used to benchmark the Monte Carlo (MC), then MC will be used to predict the IBD responses.

\begin{figure}[htb]
    \centering
    \includegraphics[width = 0.6\columnwidth]{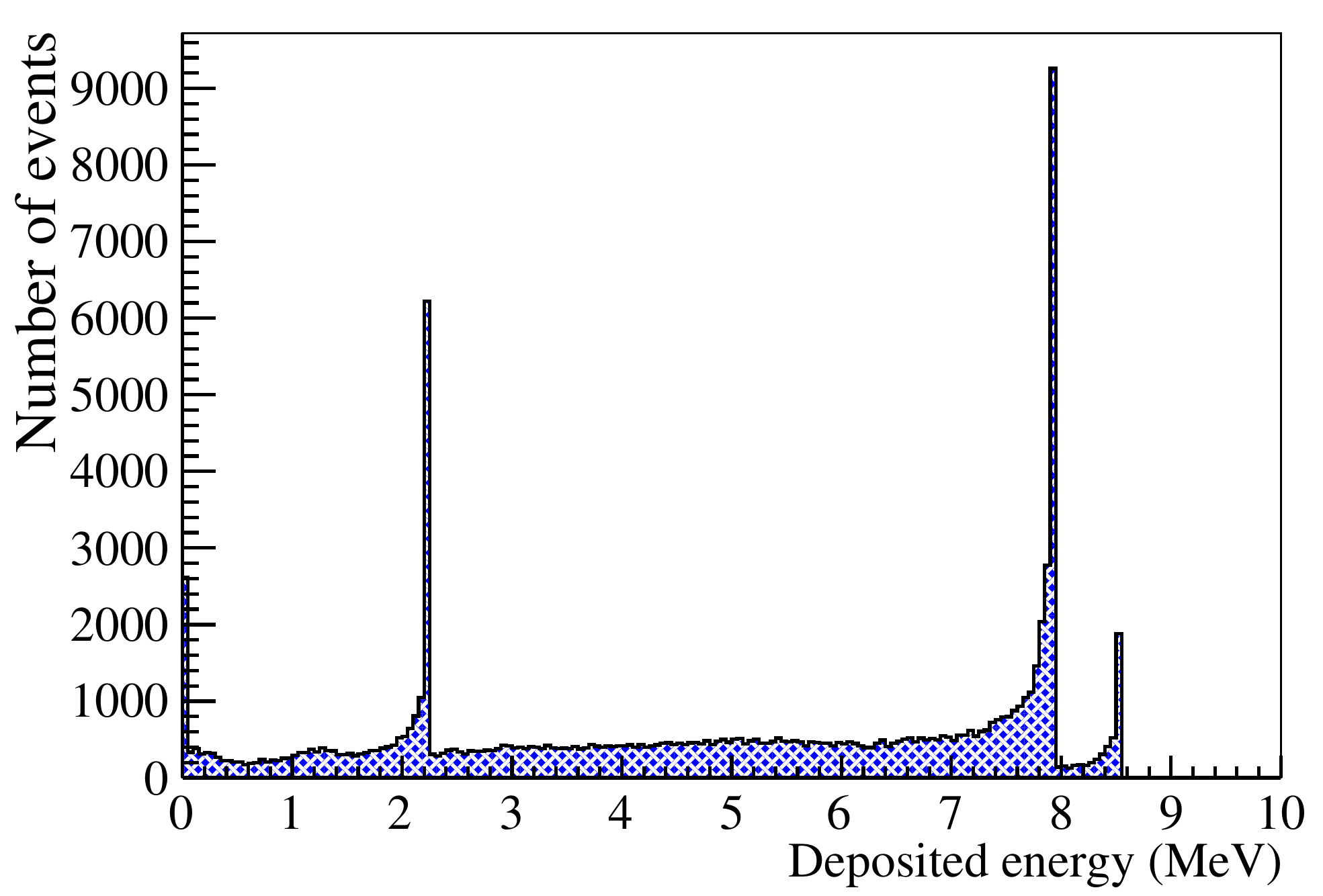}
    \caption{Energy deposit of spallation neutrons captured by Gd and H. The tails below the energy peaks are caused by the gammas escaping the LS volume.}
    \label{fig:calib:spallation-nGd}
\end{figure}

As an example, the simulated detector response (deposited energy) for the 8~MeV n-Gd capture signals throughout the detector volume is shown in Figure~\ref{fig:calib:spallation-nGd}.
Regular monitoring of the full-volume response for these events, compared with the regular automated source deployments, will provide precise information on the stability of the detector, particularly the optical properties .
With the help of MC simulations, this comparison can be used to assess the detection efficiency and its stability.

Cosmic muons passing through the detector produces short-lived radioactive isotopes and spallation neutrons. These events will follow the muon signal (detected in the muon system as well as the central detector) and will be uniformly distributed throughout the detector volume. Therefore, they provide very useful information in the full detector volume, which is complementary to the information obtained by deploying point sources.
Such events are used by KamLAND and Daya Bay to study the energy and position reconstruction as well as to check the uniformity. TAO will use spallation neutron and $^{12}B$ for the energy scale calibration.
The beta-alpha correlated decays in $^{238}$U and $^{232}$Th decay chains can be used to calibrate the full volume response. The $^{214}$Bi--$^{214}$Po--$^{210}$Pb ($^{212}$Bi--$^{212}$Po--$^{208}$Pb) beta-alpha chain produces a 7.686~MeV (8.87~MeV) alpha with a 164~$\mu$s (0.3~$\mu$s) half-life correlated with the previous beta. Considering the scintillator quenching effect, the alpha produces $\sim 1$~MeV visible energy at the emission point without position smearing as gammas. The alpha rate, depending on the actual GdLS radioactivity, is estimated assuming that the radioactivity in TAO is similar to Daya Bay. The rates of these events for TAO are given in Table~\ref{tab:calib:muon-rate}.

\begin{table}[htb]
\setlength{\belowcaptionskip}{5pt}
\begin{center}
\caption{Neutron, $^{12}$B and $\alpha$ rates in TAO. \label{tab:calib:muon-rate}}
\begin{tabular}{r l}
\hline\hline
 \mbox{  }\mbox{  }   Event Type \mbox{  } & \mbox{  } Rates (/day) \mbox{  }\mbox{  }\\ \hline
         Neutron \mbox{  } & \mbox{  } $\sim 259200$   \\
      $^{12}$B \mbox{  } & \mbox{  } $\sim 6000$  \\
            alpha \mbox{  }  & \mbox{  } $\sim 1000$  \\ \hline
\end{tabular}
\end{center}
\end{table}

\begin{figure}[htb]
    \centering
    \includegraphics[width = 0.7\columnwidth]{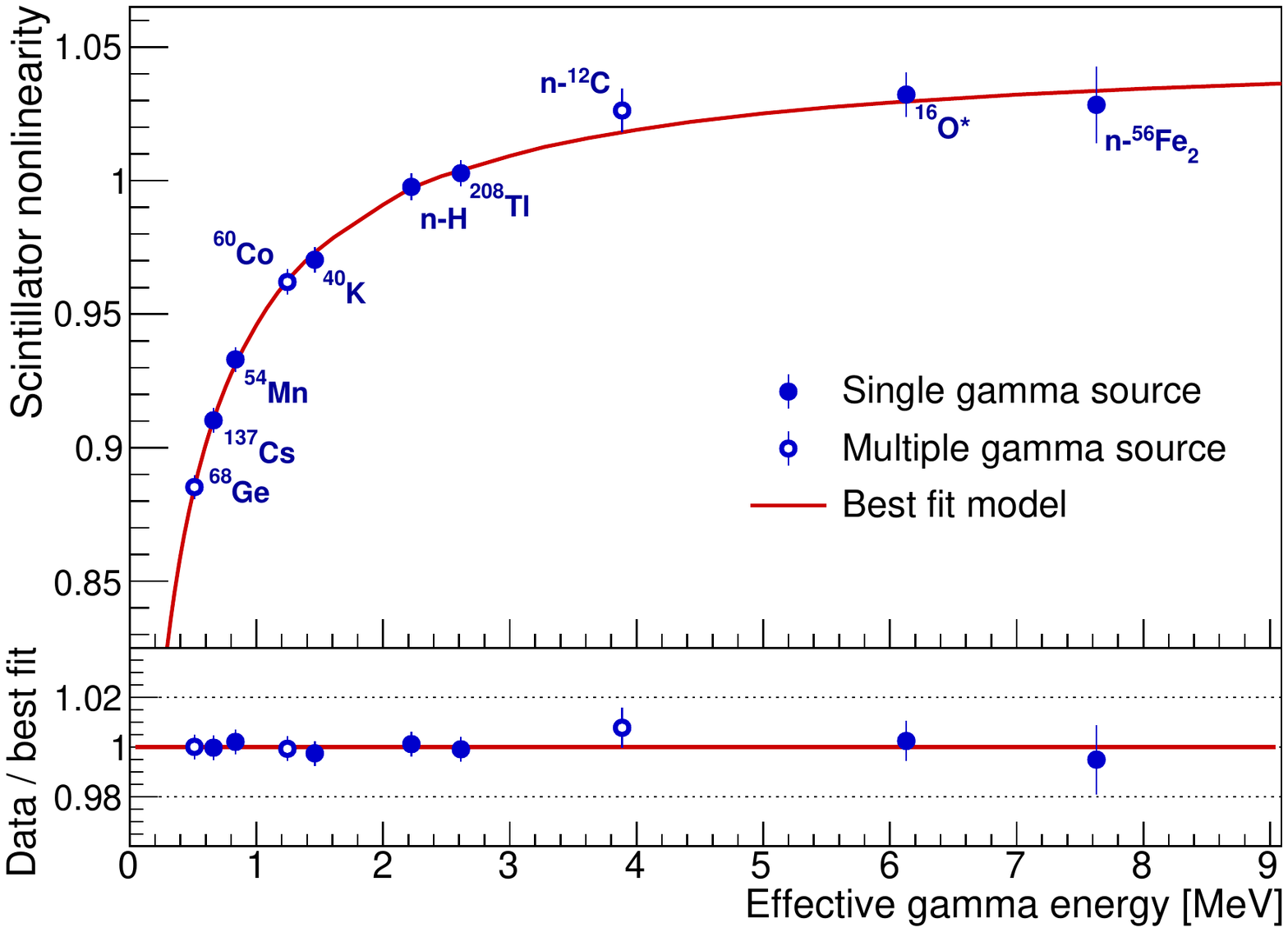}
    \caption{Figure taken from the Daya Bay experiment~\cite{Adey:2019zfo} with the estimated LS nonlinearity~(red line), and the measured energy peak from gamma sources.}
    \label{fig:LSNL}
\end{figure}

Figure~\ref{fig:LSNL} shows the measured scintillator nonlinearity in the energy scale calibration from the Daya Bay experiment~\cite{Adey:2019zfo}. Not all of the sources in Table~\ref{tab:calib:sources} can be deployed regularly by ACU.
To deploy other sources not installed in the ACU in some intensive calibration campaigns, it requires special operation to shutdown the cooling system and open the thermal insulation. Simulation in TAO is being carried out to study what sources are necessary to achieve a similar precision to Daya Bay.

\subsection{LED source}
LEDs have proven to be reliable and stable light sources that can generate fast pulses down to nano-second widths at similar wavelengths (420~nm) to the scintillation light in liquid scintillator. Especially, photons are emitted from the LED promptly, comparing to an exponential decay nature of the scintillation light. Therefore, the LED light source is usually an essential component of the detector calibration system.

The most important function of the LED light source is the timing alignment of all SiPM readout channels. High intensity light pulses from the LED diffuser ball at the detector center can be used to get the alignment constants of all channels. Timing is usually a function of the charge, depending on both the design of the electronics and the time resolution of the SiPMs. In addition to the time alignment constants, the relationship of the timing and charge, the so-called time walk curve, can be obtained by varying the light intensity of the LED.

The gain of the SiPMs can be  monitored by regularly deploying the LED diffuser ball at the detector center, and generating low intensity light pulses corresponding single photoelectron (p.e.) signals at each readout channel. Since SiPMs have very good charge resolution, multiple p.e.\ signals from the calibration or physical events can also be used to calibrate the gain of each channel, as shown in Figure~\ref{charge_spec} and Figure~\ref{fig:FEBfinger}.

\begin{figure}[htb]
    \centering
    \includegraphics[width=0.9\columnwidth]{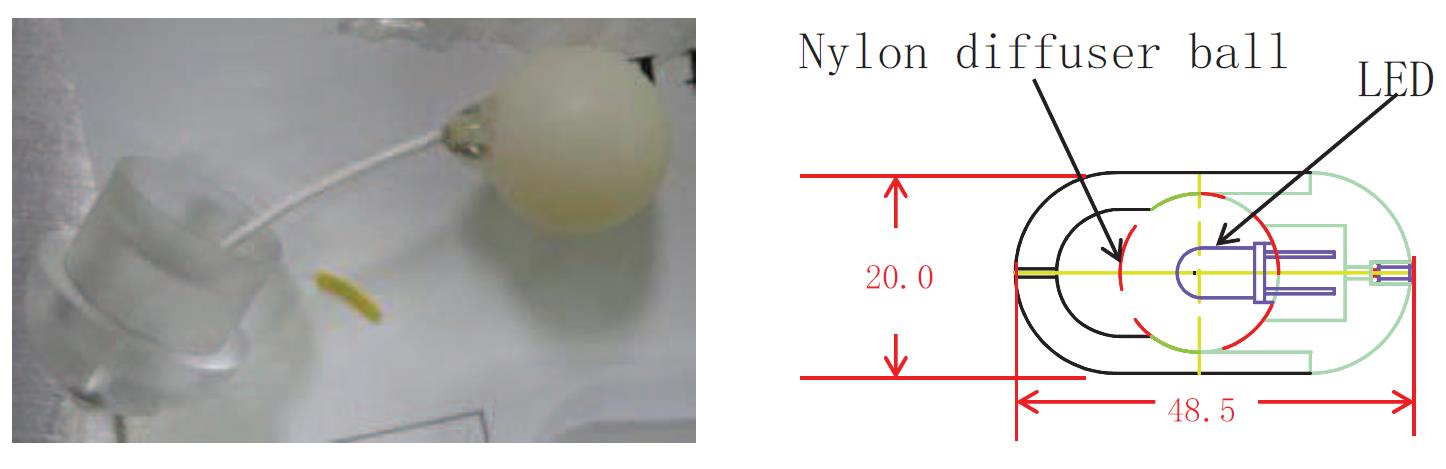}
    \caption{Left: picture of the LED potted in the nylon diffuser ball; right: conceptual diagram of the LED in Daya Bay.}
    \label{fig:calib:ACU-LED}
\end{figure}

An example of the LED source from Daya Bay~\cite{Liu:2013ava} is showing in Figure~\ref{fig:calib:ACU-LED}.
The LED is potted in a nylon diffuser ball which is then encapsulated in an acrylic enclosure, to avoid contamination of the liquid scintillator.

A blue laser diode is under consideration as an alternative of the LED to achieve better time calibration. The deployment cable has to be changed from electrical to optical fiber comparing to the Daya Bay practice, and the reliability at low temperature need further R\&D.

\subsection{Automated calibration unit}
\label{sec:acu}

The detector will be instrumented with an Automated Calibration Unit (ACU) recycled from the Daya Bay experiment~\cite{Liu:2013ava}. The automated deployment system will be used to monitor the detector on a routine basis, perhaps daily or weekly, and will allow the full z-axis access inside the detector.

\begin{figure}[htb]
    \centering
    \includegraphics[width=0.55\columnwidth]{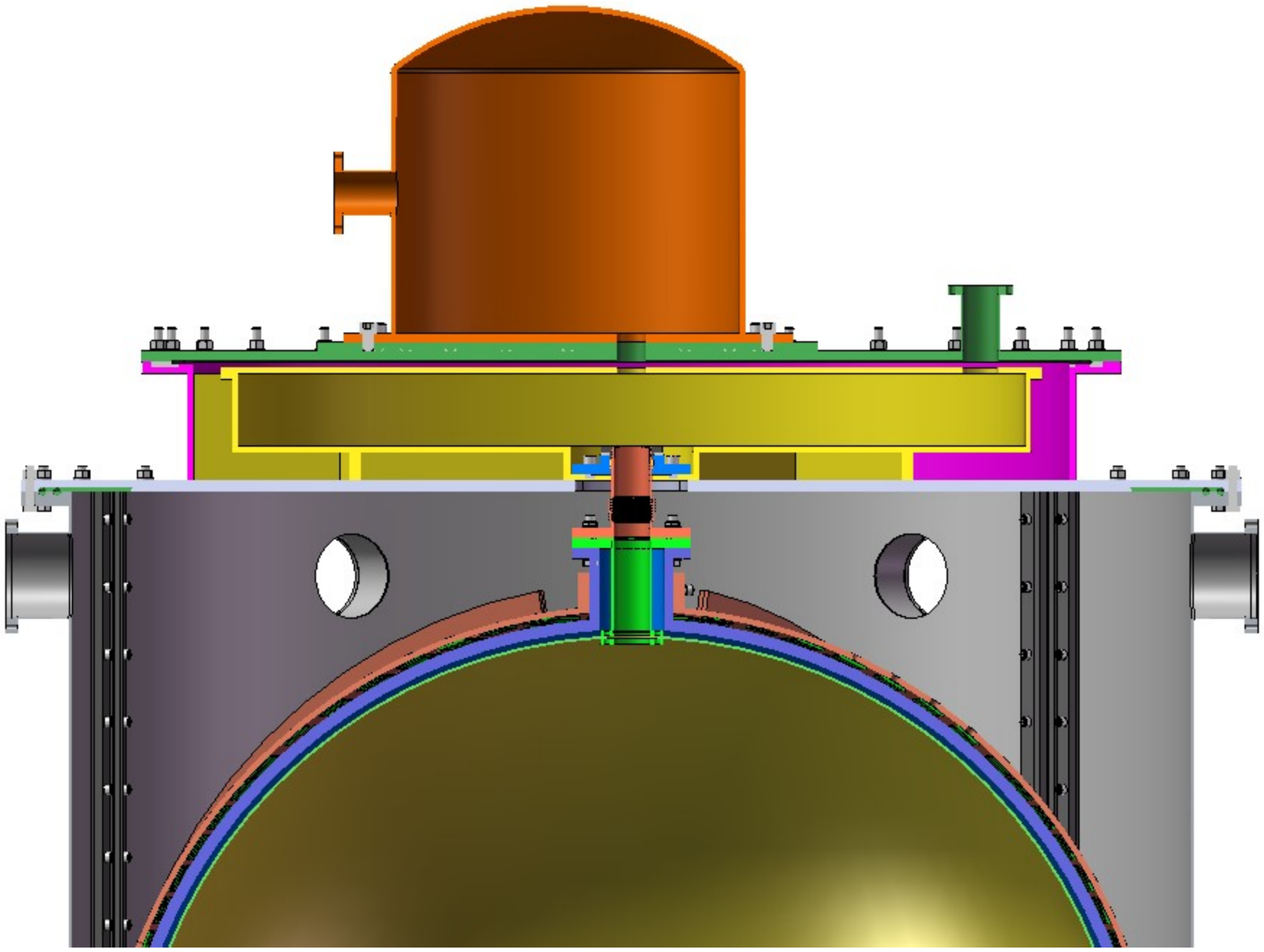}
    \caption{ACU installed on top of the TAO central detector.}
    \label{fig:calib:ACU-assembly}
\end{figure}

\begin{figure}[htb]
    \centering
    \subfigure[Side view]{\includegraphics[width=0.5\columnwidth]{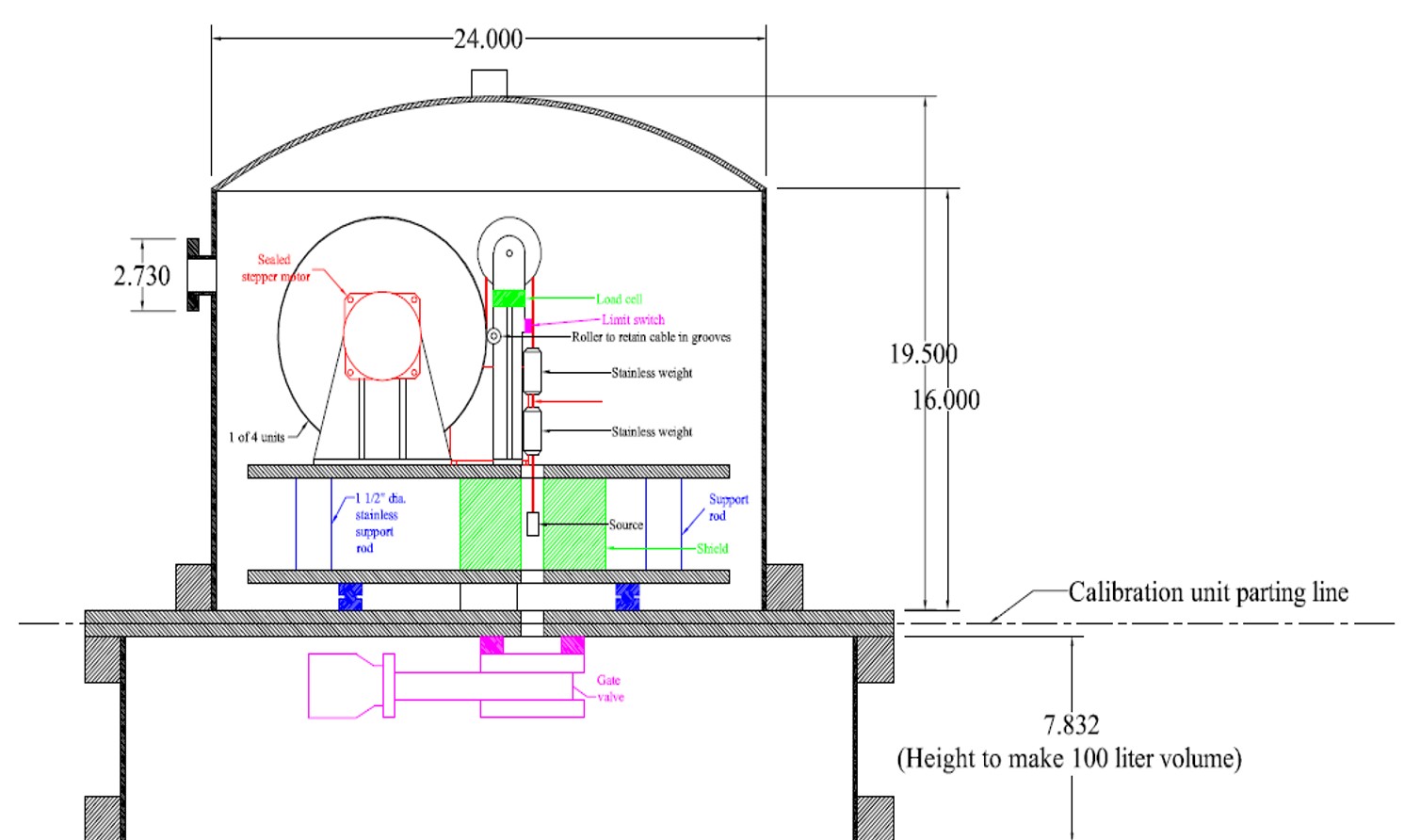}}
    \subfigure[Bird view]{\includegraphics[width=0.35\columnwidth]{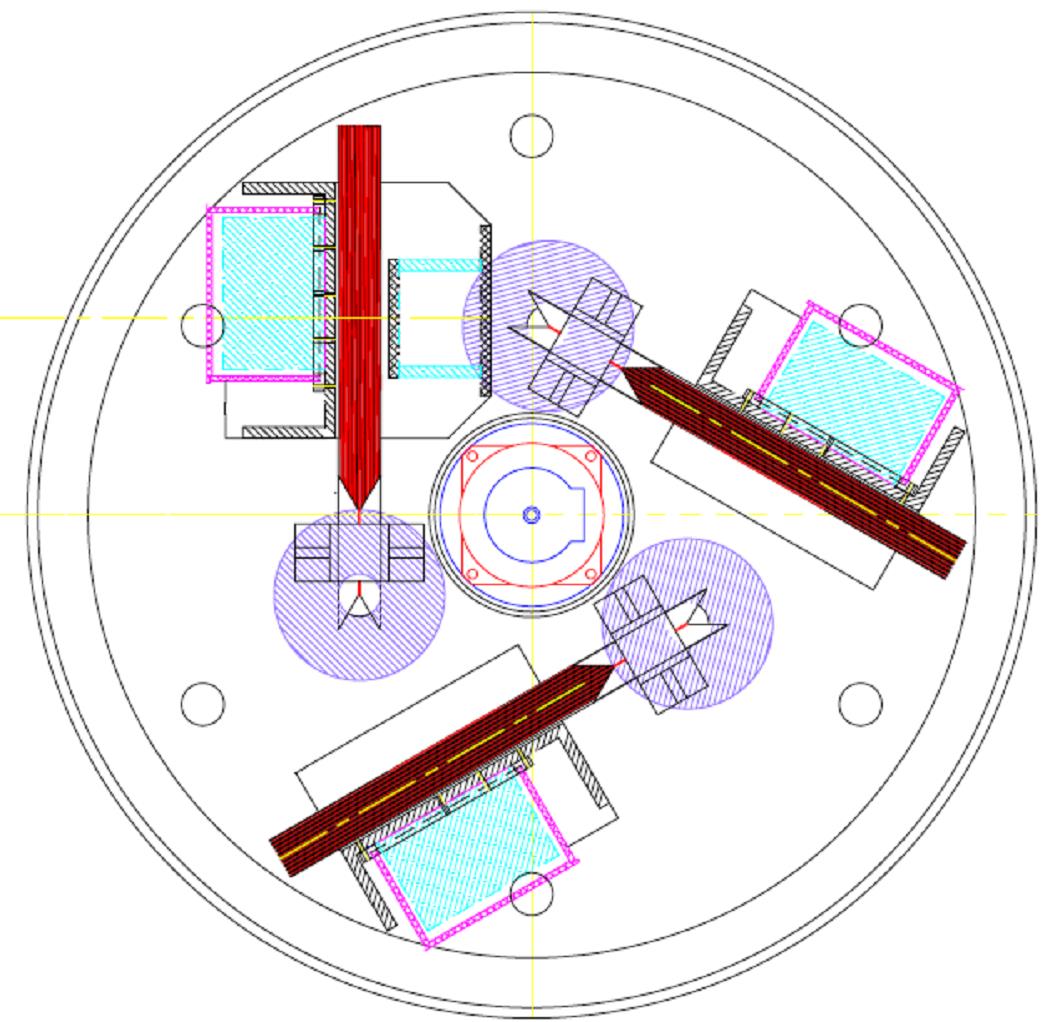}}
    \caption{ACU schematic drawing.}
    \label{fig:calib:ACU-schematic}
\end{figure}

\begin{figure}[htb]
    \centering
    \includegraphics[width=0.6\columnwidth]{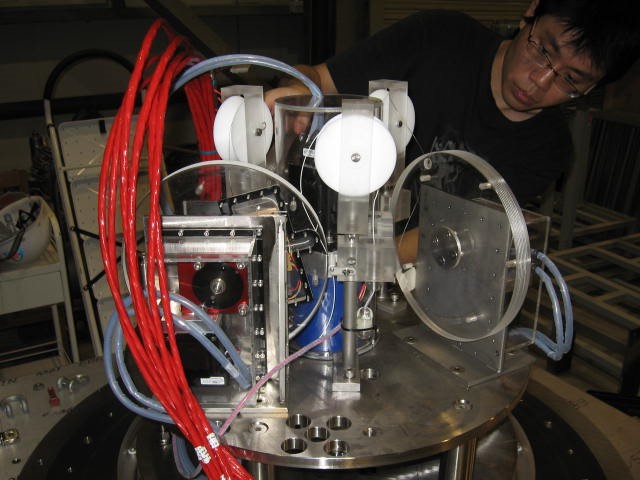}
    \caption{Picture of the inner structure of a Daya Bay ACU.}
    \label{fig:calib:ACU-photo}
\end{figure}

The ACU system will be located above a single port on the top of the detector vessel, as shown in Figure~\ref{fig:calib:ACU-assembly}. Three source assemblies (potentially containing several isotopes in each) can be lowered into liquid scintillator along the central axis of the detector volume, one assembly at a time. This will be facilitated by 3 independent stepping-motor driven source deployment units all mounted on a common turntable.
The turntable and deployment units will all be enclosed in a sealed stainless steel bell jar. The schematic drawing of ACU is shown in Figure~\ref{fig:calib:ACU-schematic} and a photo of the inside turntable is shown in Figure~\ref{fig:calib:ACU-photo}. All internal components have been certified to be compatible with the liquid scintillator.
The deployment systems will be operated under computer-automated control in coordination with the data acquisition system (to facilitate separation of source monitoring data from physics data). Each source can be withdrawn into a shielded enclosure on the turntable for storage. The source assemblies can be deployed at a speed $\sim 7$~mm/s and the deployed source position will be known to better than 2~mm.

We plan to include a LED with diffuser ball on the deployment axis 1 in ACU, a $^{68}$Ge (positron) in the axis 2, and a combined source of $^{60}$Co ($\gamma$) and neutron ($^{252}$Cf, Am-Be, or $^{238}$Pu-$^{13}$C) in axis 3 in regular calibration, similar to Daya Bay.
These sources can be deployed in sequence into the detector. During automated calibration or monitoring periods, only one source (or a combined source) would be deployed in detector at a time.
Simulation studies indicate that we can use these regular automated source deployments to track and compensate for changes in 1) average gain of the detector (photoelectron yield per MeV); 2) Number of operative SiPMs; 3) scintillation light attenuation length, as well as other optical properties of the detector system.
Simulation studies are in progress to determine if more sources could be combined into the three source assemblies. Making use of the excellent energy resolution of TAO, difference sources in one assembly might be able to calibrate the detector simultaneously with negligible impacts to the lower energy peak from the tail of higher energy peaks.
The minimal number of locations necessary to sufficiently characterize the detector spatial uniformity is also under investigation.

The automated deployment of sources will be scheduled in sequence, and is expected to take 1-2 hours for each calibration run. Physical data acquisition will be suspended during this period, and these data runs will be designated as calibration runs.
For some special calibration runs, we can configure the ACU to change the regular sources.
The ACU control system will need to communicate and coordinate with the data acquisition system during these calibration runs so that all the data are properly recorded and labeled.

The ACU is originally designed to be operating at room temperature. In TAO experiment, the ACU has to be operating at -50$^{circ}$C. It is known that the O-ring material is not suitable to work at low temperature and will be replaced. We will test the ACU system in a cryogenic box with a volume of 1~m$^3$, and then in a prototype detector.

\subsection{Calibration strategy}
The selection of the radioactive sources, the calibration points for spatial uniformity, the activity of radioactive sources, and the calibration run time are being studied with simulations.

\subsubsection{SiPM parameters}
Although SiPM parameters will be measured in bench test as described in Section~\ref{sec:sipm}, the calibration in detector is important for redundancy and monitoring.
The SiPM gains for each channel can be calibrated with the single p.e.\ spectrum using LED or SiPM dark noises.
To avoid the contamination of multiple p.e.\ signal in the single p.e.\ spectrum, LED can be configured at sub-Poissonian intensity level, such as an average of 0.1 p.e.\ for each channel.
Gain stability will be monitored at regular intervals during calibration runs.
An alternative method is to use dark noises which is usually single p.e.\ signals. The dark noises can be accumulated during the normal physics running without interruption by the regular calibration.

The SiPM timing calibration requires fast LED light pulse with a width better than 1~ns level, as already achieved in Daya Bay. The time alignment constants are mainly determined by the readout electronics and the cable length and remains stable over long periods. High intensity light pulse is necessary to ensure all readout channel are fired simultaneously after correcting the time of flight. This calibration can be done at a very low frequency.

The relative PDE might be able to be monitored from the p.e.\ yield of each SiPM, if the uniformity of the diffuse ball is good enough. Isotopic light from the diffuse ball in the central axis then could provide ``identical'' photon inputs for SiPM.
Calibration sources at the detector center is an alternative choice if the isotropy of the LED diffuse ball cannot meet the requirement.

The optical cross talk of each SiPM can be determined by measurement of the p.e.\ spectrum.
The number of p.e.\ of each SiPM should follow the Poisson distribution from a pure statistics.
The deviation from the Poisson distribution will provide the calibration of the cross talk.

\subsubsection{Spatial uniformity}
The spatial uniformity is determined by the detector geometry and its optical properties and is studied with a preliminary Geant4 simulation.
The scintillation light is collected by the array of the SiPM tiles as shown in Figure~\ref{sim_cd}. The detector is filled with GdLS in a spherical acrylic vessel of a diameter of 1.80~m and a thickness of 2~cm.  The acrylic vessel is surrounded by fully covered SiPM array and dipped in LAB buffer, where the SiPM surface has 15~mm distance to the acrylic. Both at the top and bottom of the GdLS sphere, the SiPM is removed because of a 15~cm diameter calibration chimney located at the top of the detector and an acrylic supporting leg of 10~cm diameter at the bottom. Due to the geometry mismatch of the spherical vessel and the rectangle SiPM tiles, the photo-coverage around the poles is much less than that at the equator.
The chimney and the supporting leg as well as the photo-coverage non-uniformity affect the optical asymmetry of the detector.

\begin{figure}[htb]
    \centering
    \includegraphics[width=0.6\columnwidth]{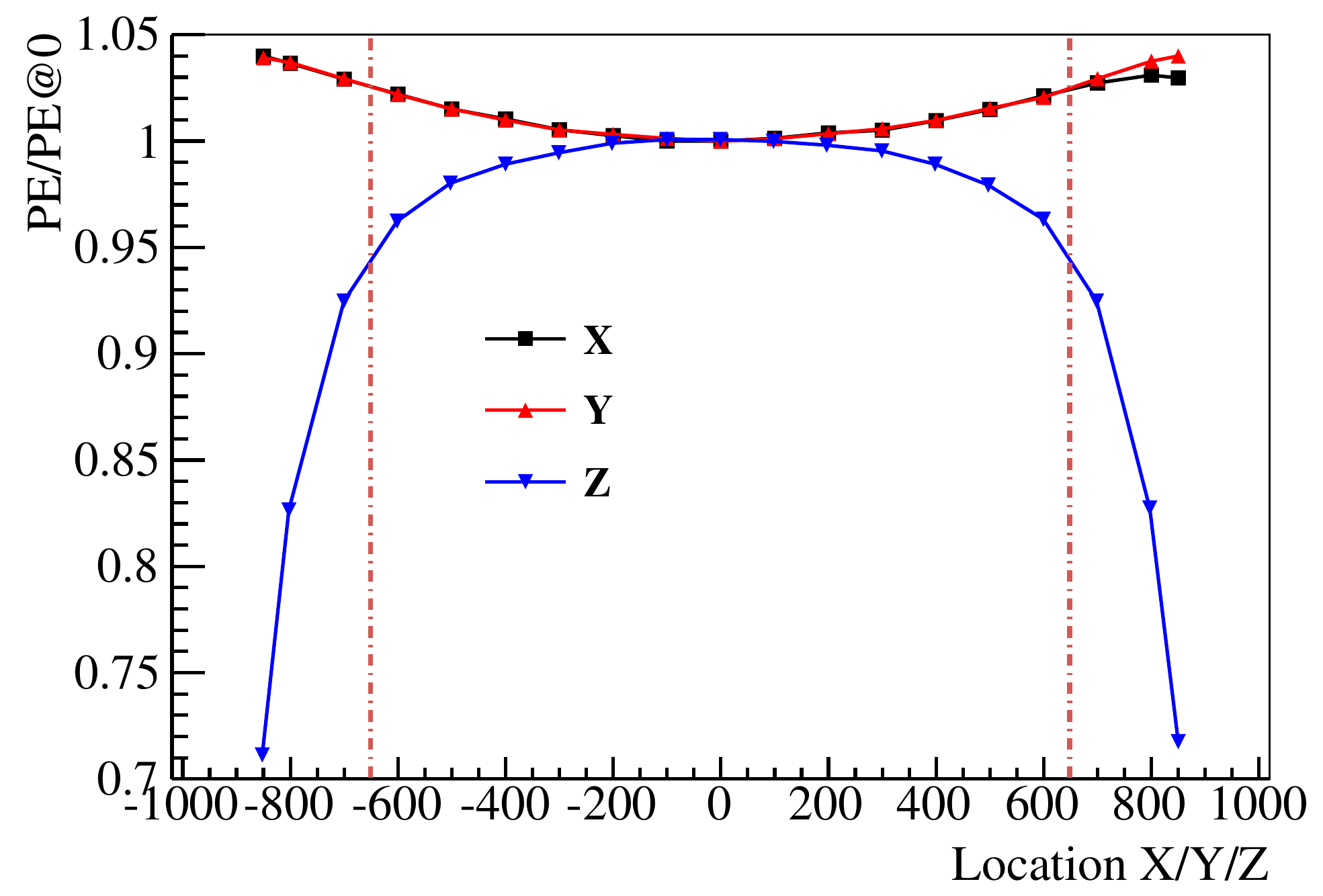}
    \caption{The simulated spatial uniformity for positions along three axes with 1 MeV electron normalized to the detector center. The vertical dashed lines show the boundary of the fiducial volume.}
    \label{fig:calib:XYZ1MeV}
\end{figure}

We simulate 1 MeV electrons along the X/Y/Z axes from the center to the edge of the GdLS volume as shown in Figure~\ref{fig:calib:XYZ1MeV}.
The events along the X and Y axes have more photoelectrons comparing to the events along the Z axis, which is caused by the calibration chimney and the supporting leg of the acrylic vessel and the sparser SiPM coverage near the poles.
The non-uniformity shown in Figure~\ref{fig:calib:XYZ1MeV} corresponds to the SiPM layout of option A in Figure~\ref{fig:coveragecom}.
Further evaluation of the layout options, possibly with non-square SiPM tiles around the poles, is under way to improve the
uniformity.

Based on the large non-uniformity of the results, it seems very demanding to deduce the spatial dependence of the energy response based solely on the source calibration points located along the detector z-axis.
The proposed solution is in two directions: 1) Try to reduce the asymmetry between X/Y and Z axes by adjusting the SiPM arrangement, especially around the calibration chimney and the supporting leg. 2) To combine the calibration results along Z axis and the uniformly distributed events (e.g. spallation neutron, alpha radioactivity from $^{212}$Bi or $^{214}$Bi in GdLS) to model the full volume response. Such a mapping could be very effective since the vertex could be reconstructed with a resolution of better than 5~cm. For the positions at the detector edge, the calibration suffers from large energy leakage.
A proper calibration source will be selected with more simulations to determine the spatial uniformity instead of using the virtual source of 1~MeV electron.
The energy leakage is important for the signal of neutron capture on Gd.
As shown in Figure~\ref{fig:delayE}, the detection efficiency of IBD delayed signal is only 59\% due to the large tail below 8~MeV.
The energy leakage is studied for the IBD positron and it will bring a systematical error described in Section~\ref{sec:Expoverview}.
Calibration at the detector edge is essential to understand the energy leakage effect and spatial uniformity.

\subsubsection{Energy scale}
\begin{figure}[htb]
    \centering
    \subfigure[LS energy nonlinearity measured with $\gamma$s.]{\includegraphics[width=0.45\columnwidth]{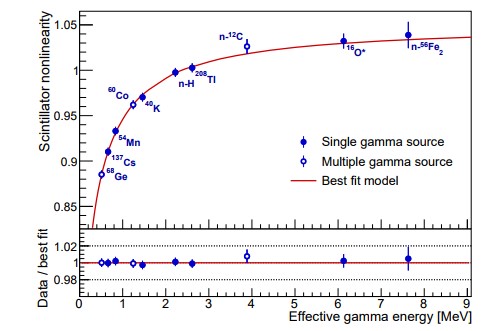}}
    \subfigure[Reconstructed electron spectrum of $^{12}B$]{\includegraphics[width=0.45\columnwidth]{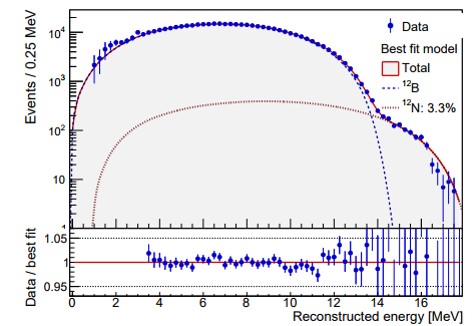}}
    \caption{Energy non-linearity curve of Daya Bay~\cite{Adey:2019zfo}.}
    \label{fig:calib:dayabay-nonlinearity}
\end{figure}

The energy scale is nonlinear due to scintillator quenching effect of liquid scintillator.
The energy linearity response for different energy and particle types is essential to reach a precise neutrino spectrum measurement.
Figure~\ref{fig:calib:dayabay-nonlinearity} shows the energy nonlinearity models, including the gamma calibration curve and electron spectrum from $^{12}B$~\cite{Adey:2019zfo}.
TAO will use the similar calibration procedure as the Daya Bay and is expected to achieve similar precision.

\subsubsection{Charge pattern}
The charge pattern calibration is crucial for the event vertex calibration to produce the expected charge distribution for all channels, which is a coupled effect of the PDE and position dependent acceptance of the SiPMs.
During the spatial uniformity calibration, the charge pattern can be obtained simultaneously by relative measurement of p.e.\ output in each channel.
The charge pattern can also be obtained from Geant4 simulation after the matching of the simulated data with calibration data placed at different positions.

\subsection{Quality assurance}
The assembled automated system has been fully tested at Daya Bay. Positioning accuracy of 2~mm at room temperature, reliability, and fail-safety of interlocks have been well established.
Radioactive sources have been tested to certify that they are leak-tight. Activity of each source will be measured and documented. The ACU system and the selected sources will be tested in a cryogenic box and then in a prototype detector.

\subsection{ES\&H}
The calibration system does not involve flammable materials or gases, high voltage, or other hazards. The radioactive sources are of very low activity, typically 1000~Bq or less, and will be operated in a shielded environment so that they do not represent a hazard to humans. Personnel involved in the installation and testing of the sources will need to be properly trained and monitored, but the dose rates will be extremely low, of an order of $\mu$rem/hr.

\subsection{Risk assessment}
The primary risk associated with the calibration systems is the interface with the detector and the reliability when working at -50$^\circ$C. Interlocks must insure that the pressure in the calibration system is equalized with the detector before deploying a source. The sources and materials must be tested to be compatible with the liquid scintillator to avoid contamination of the detector. A rigorous safety measurement in the Daya Bay experiment shows the reliability of the ACU and source design~\cite{Liu:2013ava}.

\subsection{R\&D}
The LED timing precision still needs to be further investigated to satisfy the $\sim$1~ns requirement. The laser and optical fiber are considered as an alternative light source to the LED source, which has a better timing resolution of $<0.5$~ns.
While considering the limited number of sources which can be included in the ACU in regular calibration, we need further study to finalize which sources, what activities, and how to combine the sources. At the same time, considering the unprecedented energy precision requirement, we need further understanding about the source package effect to the energy calibration as well as the source-related backgrounds. The spatial non-uniformity is large with the current detector design. Solutions should be investigated carefully. The strategy of the regular and special calibration should be designed to balance the physical data taking time the calibration performance.

%% file: VetoShielding/section.tex
\section{Shielding and Veto System}
\label{sec:veto}
\blfootnote{Editors: Oleg Smirnov (oleg.smirnov@lngs.infn.it) and Zhimin Wang (wangzhm@ihep.ac.cn)}
\blfootnote{Major contributor: Bed\v{r}ich Roskovec}

\subsection{Requirements and baseline design} 

The main backgrounds for the TAO experiment are induced by cosmic-ray muons' spallation products and by accidental coincidences mostly due to the natural radioactivity. In order to reduce the portion of these backgrounds originating outside the central detector (CD), a shielding system surrounding the detector will be used. The system should provide enough shielding while coping with the limitations of the space in the experimental hall. Moreover, it has to allow an easy access to the CD area for the installation and later for eventual maintenance.

The cosmogenic background produced in the CD or in its closest vicinity can be further reduced by a detection of passing muons and application of a short veto time in the CD. The veto system for muon detection will be therefore used in TAO. Only a short veto of $\sim$~20~$\mu$s after muon is envisaged due to the relatively high muon rate of 70~Hz/m$^2$ in the underground hall, which still rejects most of the muon associated background, including fast neutrons, without introducing excessive dead time. The veto will be applied offline providing an opportunity for careful studies and optimization of its performance. The veto system should provide $\gtrsim$ 90\% muon tagging efficiency with an uncertainty $<$ 0.5\%  and some muon track information to sufficiently reduce the muon induced background to the level determined by physics studies. The background from untagged muons (due to tagging inefficiency) can be estimated and subtracted statistically with a small uncertainty by measuring the spectrum of tagged background events and together with a precise knowledge of the tagging efficiency.  Reaching this performance, the veto system design still has to meet the limited dimensions of the experimental site.

\begin{figure}[htb]
    \centering
    \subfigure[Octagon option bird view]{\includegraphics[width=0.3\columnwidth]{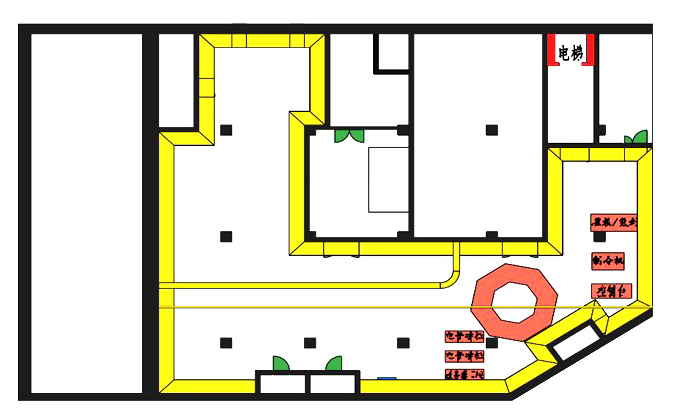}}
    \subfigure[Octagon option side view]{\includegraphics[width=0.31\columnwidth]{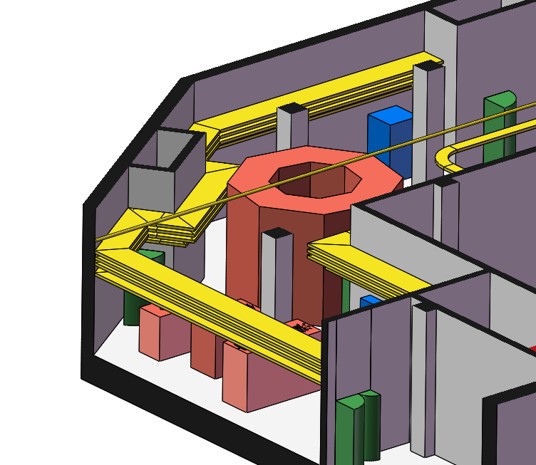}}
    \subfigure[Cut rectangle option side view]{\includegraphics[width=0.29\columnwidth]{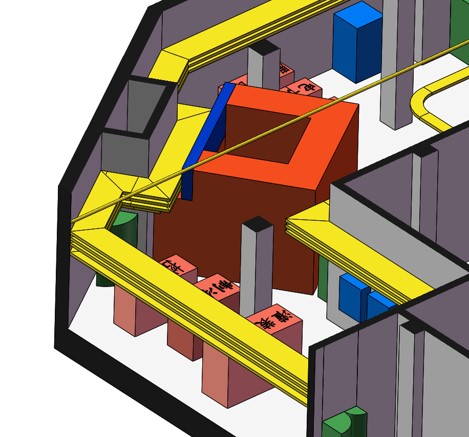}}
    \caption{Consideration on the shape arrangement of the water tanks.}
    \label{fig:veto:watertank-outer-shape}
\end{figure}

The side view of the veto detector has been shown schematically in Figure~\ref{fig:cdscheme} in the Executive Summary section. The CD is shielded by 1.2 m water in the surrounding water tanks, 1 m High Density Polyethylene on the top, and 10 cm lead at the bottom. The water tanks, instrumented with Photomultipliers (shown by red circles), and the Plastic Scintillator (PS) on the top comprise the active muon veto system.  At least one side of the tanks is movable to have access for assembly and maintenance. The exact shape of the water tanks is under optimization, keeping in mind that the maximum dimension should be $<$~5.1~m to fit in the laboratory. Figure~\ref{fig:veto:watertank-outer-shape} shows the octagon option which is less spacious with respect to the rectangle option. The top and bottom shielding is solid material instead of water due to space and installation considerations. A minimum thickness of 1.2~m of water or equivalent thickness of other materials is required to sufficiently suppress the gammas and neutrons from the natural radioactivity.

To serve as an active tagging system, the water tanks will be instrumented with 3-inch Photomultiplier Tubes (PMTs) to detect Cherenkov photons from muons going through the water. It can achieve a $\ge$~90\% efficiency for muons passing through the water tank volume with a 0.8\% photocoverage. In addition, few layers of an instrumented plastic scintillator will be placed above the CD and water tanks to detect the muon from top and further improve the muon detection and tracking, providing an independent $\ge$~90\% muon detection efficiency. Exact design of the layers is still under investigation. The combined efficiency of the veto systems has to exceed 90\%, with an uncertainty $<$~0.5\%, to satisfy the experimental requirements.

The muon rate in the veto detectors is estimated to be about 4000~Hz. Tab.~\ref{tab:veto:signal-bkg-rate} shows the expected event rates with and without shielding and muon veto time application. Background rates will be reduced to $\sim$~10\% of the IBD rate with the shielding and veto systems, which is sufficient to satisfy the TAO physical requirements.

\begin{table}[htb]
\setlength{\belowcaptionskip}{5pt}
\begin{center}
\caption{Expected signal and background rates in the TAO CD with and without a 1.2 meters-water-equivalent shielding and the active veto system. \label{tab:veto:signal-bkg-rate}}
\begin{tabular}{c c c}
\hline\hline
    Event & Without shielding and veto & With shielding and veto  \\ \hline
    IBD signal & $\sim$ 2,000/day & $\sim$ 1,800/day  \\
    Neutrons & $\sim$ 20~Hz & $\sim$ 2.7~Hz   \\
    Accidental background & $\sim$ 20,000/day & $<$ 200/day  \\
    Fast neutron background & $\sim$ 2,000/day & $\sim$ 200/day  \\ \hline
\end{tabular}
\end{center}
\end{table}

\subsection{Water tank: shielding and Cherenkov detector}
The CD will be surrounded from sides by the water tanks with a thickness of 1.2~m as a passive shielding and an active Cherenkov muon detector. The water tanks serve several crucial purposes. (i) Fast-neutron background originating from the cosmic-ray muons will be significantly reduced by such a shielding. A Monte Carlo simulation shows that the rate of fast-neutron background produced outside the detector will be reduced by a factor of 1.5--2.0 by 50~cm of water. (ii) Pure water effectively reduces the accidental background rate associated with $\gamma$ rays since low-energy $\gamma$ ray flux is reduced by a factor of ten by 50~cm of water. (iii) The water tanks will be instrumented with PMTs for the detection of the Cherenkov light induced by passing muons.

\begin{figure}[htb]
    \centering
    \subfigure[The schematic design of the water tank]{\includegraphics[width=0.4\columnwidth]{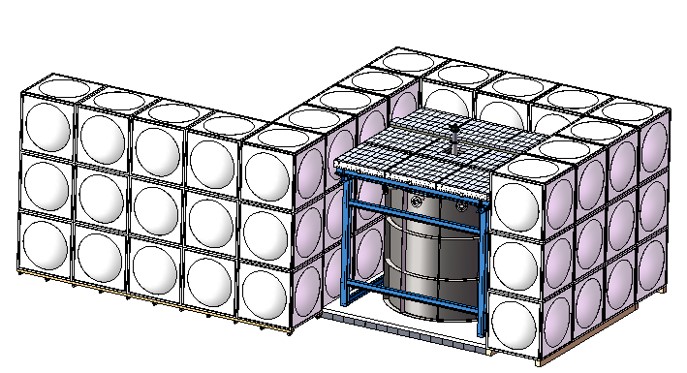}}
    \subfigure[The realized tank]{\includegraphics[width=0.3\columnwidth]{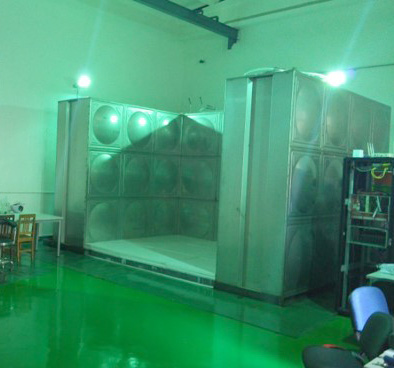}}
    \subfigure[Inside of the tank]{\includegraphics[width=0.17\columnwidth]{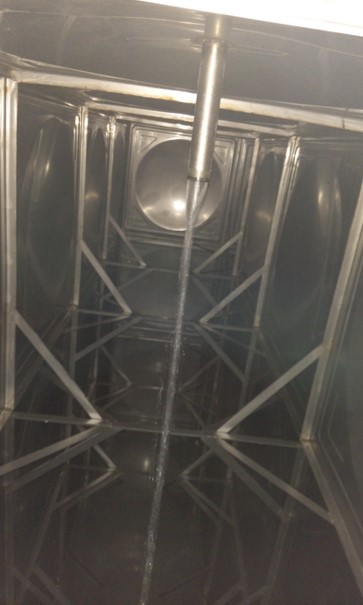}}
    \caption{The water tank design used for JUNO CD prototype shielding.}
    \label{fig:veto:JUNOprototype-water-tank}
\end{figure}

Water tanks have been used as shielding of a JUNO prototype detector of similar size as the TAO detector, as shown in Figure~\ref{fig:veto:JUNOprototype-water-tank}, and have been successfully operated for several years. They are commercially available and can be easily adapted to different geometries at low cost. To save space in the Taishan Neutrino Laboratory, rectangular, circular, and octagon shape water tanks have been considered, while keeping the requirement of 1.2-m water-equivalent shielding around the CD in all directions. The overall dimension of the TAO water tanks are limited by 5.1~m$\times$5.1~m$\times$3.8~m (high) space in the experimental hall. The octagon geometry is chosen as the baseline design for TAO, as shown in Figure~\ref{fig:veto:watertank-outer-shape}(b). With a thickness of 1.2~m, the mass of water is estimated to be $\sim$ 80~ton.

\begin{figure}[htb]
    \centering
    \subfigure[PMT assembly]{\includegraphics[width=0.45\columnwidth]{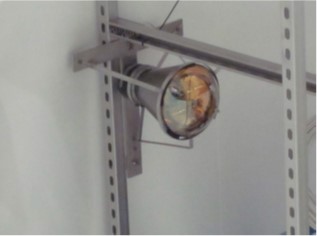}}
    \subfigure[Tyvek as a reflector]{\includegraphics[width=0.45\columnwidth]{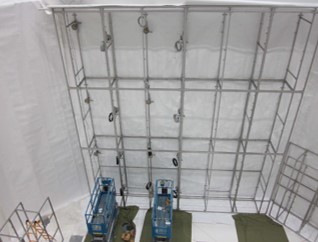}}
    \caption{8-inch PMTs and Tyvek film in the water pool of the Daya Bay experiment.}
    \label{fig:veto:watertank-3inch-pmt}
\end{figure}

The water tank will be instrumented with an array of 3-inch PMTs. Its inner surface will be covered by a reflective Tyvek film. Similar technology has been used for the water pool in the Daya Bay experiment~\cite{Dayabay:2014vka} as shown in Figure~\ref{fig:veto:watertank-3inch-pmt}, and will be used for the JUNO experiment. Inward-facing PMTs will be mounted on stainless steel frames placed on the sides and at the bottom of the tank. The frame will support PMTs before the water tank filling and hold PMTs in place afterwards to compensate the buoyant force in water. The PMTs will be approximately evenly distributed in the tank, forming a rectangular grid with a density of around 1~PMT per 0.5~m$^{2}$. This corresponds to $\sim$ 230 PMTs and a 0.8\% surface coverage. A total number of 256 PMTs might be used for redundancy. The 3-inch PMTs and their bases, potting, and electronics readout will use the same technology of the JUNO small PMT system~\cite{Djurcic:2015vqa}.

The ultra-pure water in the water tanks can be a corrosive agent. The tank material as well as the material of PMT support structure and PMTs themselves will be selected to prevent long term corrosiveness, which can cause a failure of the structural integrity or solution of impurities in the water. Furthermore, the constant clarity of the water will be maintained by a water purification system, to keep the attenuation length for the Cherenkov light significantly larger than the dimensions of the tank. For micron-sized particles, this translates to a particle number density of $<10^{10}$/m$^3$ according to the JUNO R\&D. A simple purification system with a filter stage followed by reverse osmosis will be enough to meet our specifications. The suspended particles in water could be filtered down to the size of 1~nm in diameter. Eventual radioactivity in water can be efficiently reduced to a satisfactory level. Bacterial growth in the water is also a concern for the water clarity. An ultraviolet sterilization stage or a micro-bubble stage will be integrated in the purification system. We anticipate a circulation of 50~L/min in the water system, which will allow one complete turn-around of the water in about one day. Ongoing R\&D using JUNO prototype water tanks will help to validate the design of the purification system.

The performance of the water tank Cherenkov detector, such as efficiency, position resolution, timing resolution, etc. are being further optimized with Monte Carlo simulation. The PMTs need to have a $\sim$2~ns time resolution in order to properly determine the veto window time. Their gain stability and the timing will be monitored with LED diffuser balls mounted at several locations within the water tank.

\begin{figure}[htb]
    \centering
    \includegraphics[width=0.6\columnwidth]{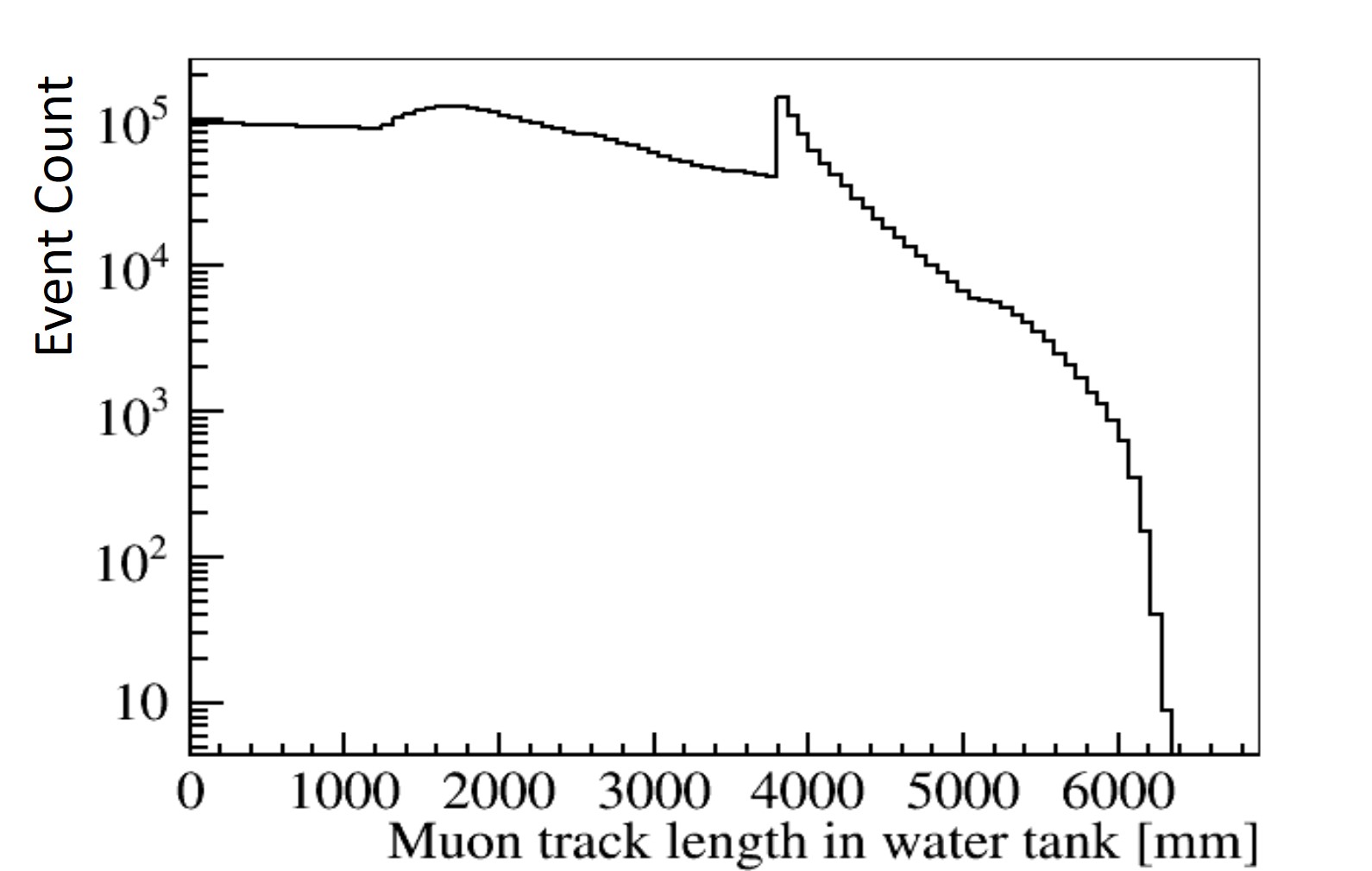}
    \caption{The distribution of muon track lengths in water tank for muons that produce fast neutron backgrounds (before veto is applied). }
    \label{fig:veto:water-tracklength}
\end{figure}

Figure~\ref{fig:veto:water-tracklength} shows the simulated distribution of track length of cosmic ray muons that produce fast neutrons in the water tank. The mean track length is about 2~m as the thickness of the water tank is 1.2~m. The mean number of photoelectrons produced by muons of such track length is $\sim$ 90, and most PMTs will see a single photoelectron.

\begin{figure}[htb]
    \centering
    \includegraphics[width=0.6\columnwidth]{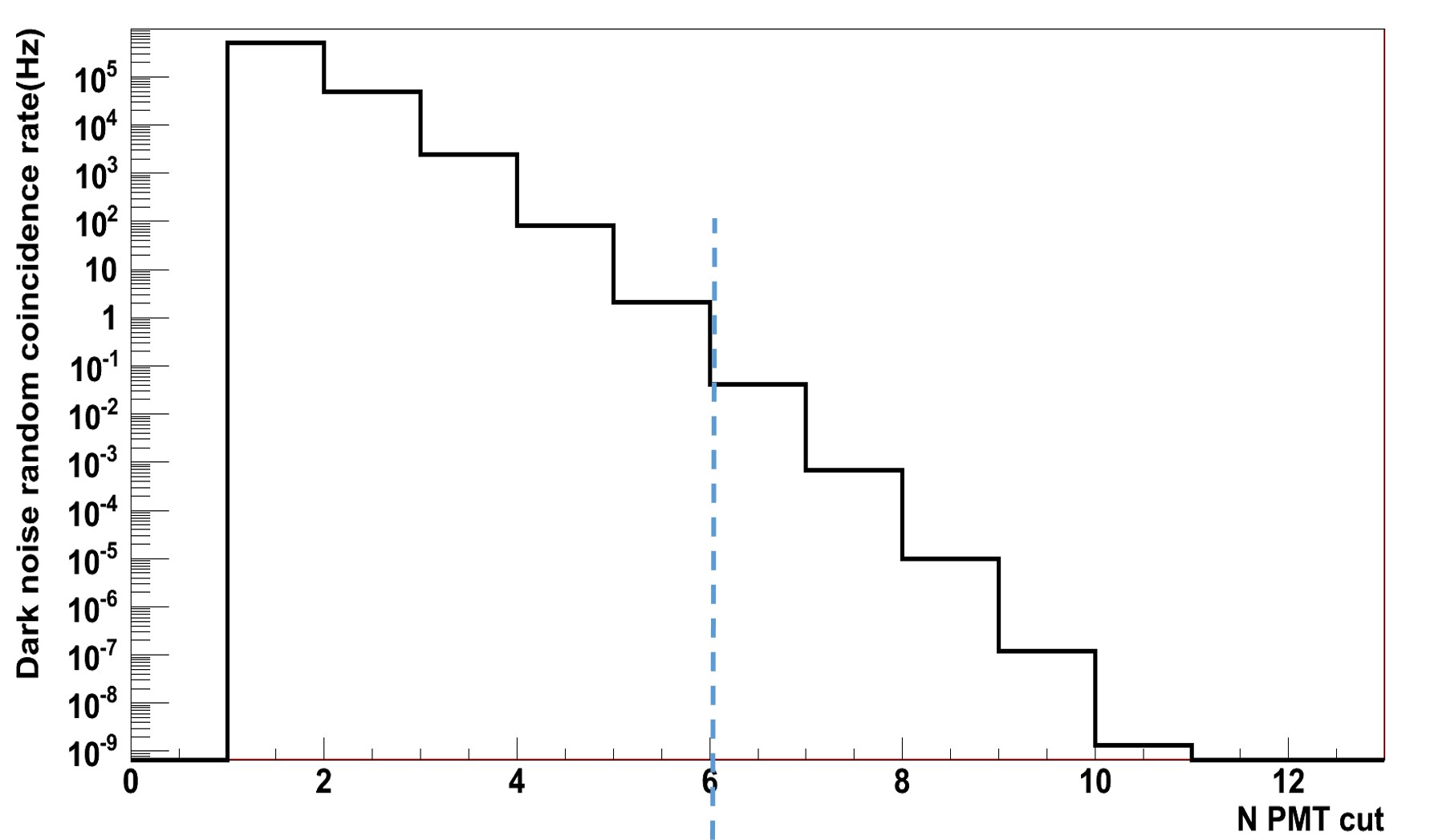}
    \caption{Dark noise random coincidence rate as a function of coincidence threshold. The threshold of $>$ 6 hit PMTs corresponds to a $<1$\% dead time due to false muons from dark noise.}
    \label{fig:veto:pmt-coincidence}
\end{figure}

Muons will be tagged when a minimum number of PMTs exceeds a certain threshold. For the baseline design of the water tank, we can achieve $\ge$ 90\% detection efficiency with a threshold of 10 hit PMTs, corresponding to the light produced by a muon of 20~cm track length in the water tank. The false rate of muons due to the random coincidence of the PMT dark noise is calculated too. Figure~\ref{fig:veto:pmt-coincidence} shows the distribution of dark noise random coincidence in a 200~ns window assuming a conservative dark noise rate of 2~kHz per PMT. A threshold of $>$ 6 hit PMTs corresponds to a $<1$\% dead time due to false muons from random coincidences. In summary, a coincidence threshold of 6--10 hit PMTs will identify muons with $\ge$ 90\% efficiency while having  $<1$\% dead time due to random dark noise coincidences.

\subsection{Top and bottom shielding and top muon detector}
The available height of the Taishan Neutrino Laboratory for the detectors is about 3.85~m due to the steel beam frames on the roof, although the height of the laboratory is about 5~m. Since the CD has a height of 2.6~m with thermal insulation layer, it is impossible to shield it with 1.2~m water both on the top and at the bottom. Simulation shows that fast neutrons coming from the side and the bottom of the CD are 50\% and 2\% of that coming from the top, respectively, as listed in Tab.~\ref{tab:veto:fndirection}. Therefore, more low-density hydrogen-rich material should be put on the top to slow down the fast neutron, while the bottom shield can be replaced with high-density high-Z material just to shield the ambient radioactivity.

\begin{table}[htb]
\setlength{\belowcaptionskip}{5pt}
\begin{center}
\caption{Fast neutrons per day in the CD from different directions, assuming 1.2~m water shielding for all directions. \label{tab:veto:fndirection}}
\begin{tabular}{r c c c c}
\hline\hline
       &  Total rate & Top & Side & Bottom \\ \hline
    Without muon veto  & 1860 & 1196 & 661 & 23 \\
    With muon Veto     &  152 &  62  &  68 & 22 \\ \hline
\end{tabular}
\end{center}
\end{table}

\begin{figure}[htb]
    \centering
    \includegraphics[width=0.4\columnwidth]{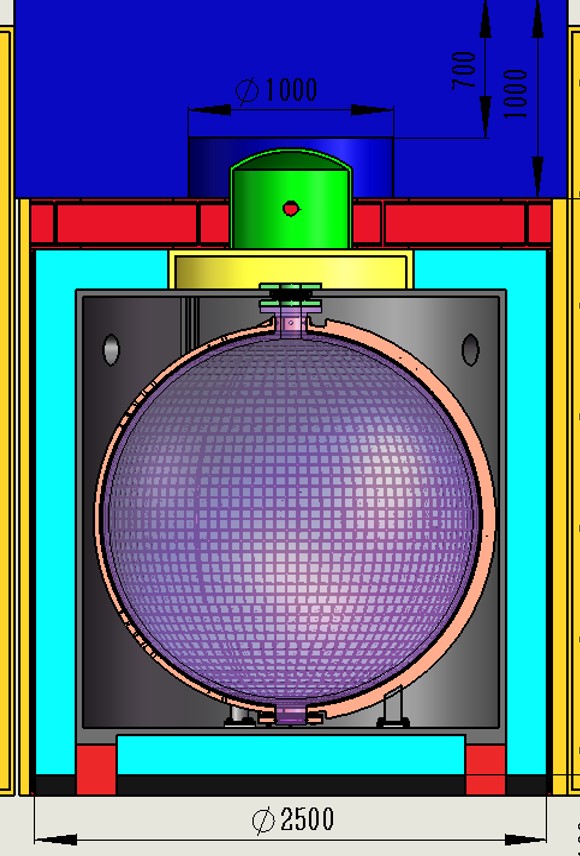}
    \caption{The interface between the ACU and the layer of HDPE top shielding.}
    \label{fig:veto:top-veto-ACU-relationship}
\end{figure}

A 10-cm layer of lead bricks will be placed at the bottom of the CD, while about 1-m of High Density Polyethylene (HDPE) will be placed on the top. Since very few fast neutrons come from the bottom, the shielding to $\gamma$s and neutrons are good in all directions. It might be possible to put more HDPE above the steel beam frame to further increase the top shielding, which will be investigated later. A HDPE ``hat" will be made for the Automated Calibration Unit (ACU) to easily implement the top shielding on a support frame, as shown in Figure~\ref{fig:veto:top-veto-ACU-relationship}. The presence of the ACU will increase the fast neutron background by $<$~10\% comparing to the full HDPE shielding case. Simulation also shows that Boron-doping in HDPE cannot apparently reduce the fast neutron background, while the single neutron events in the CD, originating from low energy neutrons, do have certain reduction.

There will be few layers of an instrumented plastic scintillator above the CD to serve as a muon detector. A multilayer coincidence would reduce the false muon rate due to the coincidence of natural radioactivity events. All plastic scintillator strips will be made from extruded polystyrene with a dimension of 2~m$\times$0.1~m$\times$0.01~m or 2~m$\times$0.1~m$\times$0.02~m, co-extruded with a coating of TiO$_2$-doped PVC. 1-inch PMTs such as Hamamatsu R6095 or Electron Tubes 9128B, running at positive high voltage, will be used to read out on each end of the scintillator. The readout electronics of the 3-inch PMT system of JUNO should to be adequate for it. The design and assembly of the plastic scintillator muon detector will be investigated in detail later.

\subsection{Summary and R\&D}
The proposed design of the shielding and veto systems can reduce the accidental and fast neutron backgrounds to less than 200 events per day each, which is less than 10\% of the antineutrino signals and has negligible impacts on the precision measurement of the reactor antineutrino spectrum after background subtraction. The exact water tank geometry, the top and bottom shielding design, and their fabrication and installation at the experimental site are still under consideration. More details on the PMT arrangement in the water tank, the muon efficiency, and the plastic scintillator design need to be elaborated too.

%% file: SiPM/section.tex
\section{Silicon Photomultiplier and Readout}
\label{sec:sipm}
\blfootnote{Editors: Nikolay Anphimov (anphimov@gmail.com), Guofu Cao
(caogf@ihep.ac.cn), Andrea Fabbri (andrea.fabbri@uniroma3.it), and Stefano Mari
(smari@os.uniroma3.it)}

The TAO detector is intended for precise measurements of the reactor antineutrino energy spectrum. With a yield of 4500 photoelectrons (p.e.) per MeV, the stochastic term of the energy resolution will be $1.5\%/\sqrt{E[{\rm (MeV)}]}$. In most regions of interest, the expected energy resolution will be sub-percent. Photodetectors more efficient than conventional Photomultiplier Tubes (PMTs) are required. Silicon Photomultiplier (SiPM) could have a Photon Detection Efficiency (PDE) twice higher than PMT and will be used for TAO. A SiPM is a silicon-based solid-state device constructed as an array of many small Single Photon Avalanche Diodes (SPADs) of dimensions from 10 to 100~$\mu$m. Each SPAD works in Geiger mode and is integrated with its passive quenching resistor. All SPADs are connected in parallel. The output charge of the SiPM is the sum of all the charges generated by the fired SPADs. It is proportional to the number of detected photons. Compared to PMTs, the SiPMs have a high PDE, but also have a very large thermal noise at room temperature. The TAO detector will operate at -50$^\circ$C to reduce the thermal noise by about three orders of magnitude compared to the room temperature.

\subsection{Requirements on the SiPM parameters}
\label{sec: param_SiPM}

The requirements on the SiPM parameters should satisfy the detector requirements to reach an energy resolution as high as possible for neutrino events, and should also be reasonable in terms of contemporary SiPM technology. The requirements for the TAO experiment are summarized in Table~\ref{tab:sipm} and described in the following.
\begin{table}[htb]
\setlength{\belowcaptionskip}{5pt}
\begin{center}
\caption{\label{tab:sipm} Requirements on the SiPM parameters.}
\begin{tabular}{ c c c }
		\hline\hline
		Parameters & Specification & Comments\\
		\hline	
		PDE  & $\geq 50$\% & at 400~nm, not including correlated noise\\
		Dark count rate  & $\leq 100$~Hz/mm$^{2}$ & at -50$^\circ$C \\
		Probability of correlated noise   & $\leq 10$\%	 & including cross talk and afterpulsing\\
		Uniformity of V$_{bd}$  & $\leq 10$\% & to avoid bias voltage tuning\\
		Size of the SiPM device  &  $\geq 6 \times 6$~mm$^2$	& for easy handling\\
		SiPM coverage within tiles &  $\geq 94$\%  & not included in SiPM's PDE\\
		\hline
\end{tabular}
\end{center}
\end{table}

Contemporary SiPMs could have a PDE higher than 50\%. However, the PDE of the SiPM usually correlates with its dark counts provoked by thermal noise, and its correlated noise, mainly including the optical cross talk and afterpulsing. We require the PDE of the SiPM to be $\geq 50$\%, and evaluate the effects of the three parameters jointly to optimize the detector energy resolution.

The uniformity of the break down voltage $V_{bd}$ is required to be $\leq 10$\% to avoid bias voltage tuning on a SiPM tile, as explained in Sec.~\ref{sec:Readout}.

Commercial SiPM devices have various sizes, e.g.\ $3\times3$, $4\times4$, $5\times5$, $6\times6$, and $10\times10$~mm$^2$. The baseline choice for TAO is $6\times6$~mm$^2$ for ease of handling, large PDE, and flexible commercial availability. The total area of SiPMs used for TAO is  $\sim 10$~m$^2$, resulting in $\sim2.7\times10^5$ SiPMs. For installation and readout considerations, the SiPMs will be assembled in $\sim4100$ SiPM tiles. Each tile consists of $8\times8$ SiPMs and has dimensions $\sim 50\times50$~mm$^2$. Technical details will be further elaborated with prototype testing. Considering the spacing between SiPMs, we require the SiPM coverage within the tile to be $\geq 94$\%.

To demonstrate the effects of the Dark Counts (DC) and the Correlated Noise (CN), we express the statistical fluctuations of the detector response under simple assumptions as
\begin{equation}
\label{eq:energy_res}
   \frac{\sigma_E}{E} = \sqrt{\frac{f_{EN}}{N_{pe}}} \,,
\end{equation}
where $N_{pe}$ is the average p.e.\ yield for an event of energy $E$, and $f_{EN}$ is the excess noise factor, defined as
\begin{equation}
\label{eq:ENF}
    f_{EN} = \frac{N_{pe}+N_{EN}}{N_{pe}} \,,
\end{equation}
with $N_{EN}$ the average excessive p.e.\ number from noise sources, mainly from the dark counts $N_{DC}$ and the correlated noise $N_{CN}$, i.e.\ $N_{EN}=N_{DC} + N_{CN}$. We have $f_{EN}=1$ in the absence of noise.\footnote{In the real case, the electronic noise and pick-up noise can also contribute to the excess noise factor.}

In the TAO case, we expect $N_{pe} \sim 4500$ for 1~MeV of deposit energy as described in Sec.~\ref{sec:energyresolution}. In the absence of any noise,
\begin{equation}
    \label{eq:energy_res_eval_TAO_wonoise}
    \frac{\sigma_E}{E} = \frac{1}{\sqrt{4500}} \approx 1.5 \% \,.
\end{equation}

Suppose we allow a moderate degradation of the energy resolution (at 1 MeV) by noise to $\sigma_E/E < 1.7 \%$, then according to Eq.~\ref{eq:energy_res} and Eq.~\ref{eq:ENF}, we have
\begin{equation}
    \label{eq:noise_level}
    f_{EN} = \left(\frac{1.7\%}{1.5\%}\right)^2 \approx 1.28
\end{equation}
and the allowed excessive p.e.\ number is
\begin{equation}
    \label{eq:noise_level_2}
    N_{EN} = (f_{EN} - 1) \cdot N_{pe} \lesssim 1260 \,.
\end{equation}

To find a compromise between the dark count contribution $N_{DC}$ and the correlated noise contribution $N_{CN}$, let us consider a realistic Dark Count Rate (DCR) of 100~Hz/mm$^2$ at a temperature of -50$^\circ$C. Given the total area of SiPMs $\sim 10$~m$^2$, the dark count rate in the TAO detector is $R_{DC}= 1.0 \times 10^9$~Hz.  Random coincidence of dark counts $N_{DC}$ in a readout time window $\tau\sim 600$~ns (determined by the slowest scintillation light component of the liquid scintillator, $\sim 200$~ns) is
\begin{equation}
    \label{eq:rndm_coinc}
    N_{DC} = R_{DC} \cdot \tau = 600 \,,
\end{equation}
and we have $f_{EN, DC} = 1.13$ for the excess noise factor driven by the dark counts. Then, the allowed correlated noise $N_{CN}$  and its excess noise factor $f_{EN, CN}$ are
\begin{equation}
    \label{eq:enf_ctdcr_2}
    N_{CN} = 1260 - 600 = 660 \,,  \quad {\rm and} \quad  f_{EN, CN} = 1.15 \,.
\end{equation}

Using the model with a branching process with a generalized Poisson distribution in Ref.~\cite{Vinogradov:2011vr}, we have
\begin{equation}
    \label{eq:CN_ENF}
  f_{EN, CN} = \frac{1}{1-\lambda} \,,
\end{equation}
where $\lambda$ is the probability of correlated noise. Using Eq.~\ref{eq:CN_ENF} and the value of $f_{EN, CN}$ in Eq.~\ref{eq:enf_ctdcr_2}, we estimate $\lambda \sim 13\%$. We therefore require the probability of correlated noise of the SiPMs for TAO, including the cross talk and afterpulsing, to be $<10$\%.

\begin{figure}[htb]
   \centering
  \includegraphics[width=0.8\linewidth]{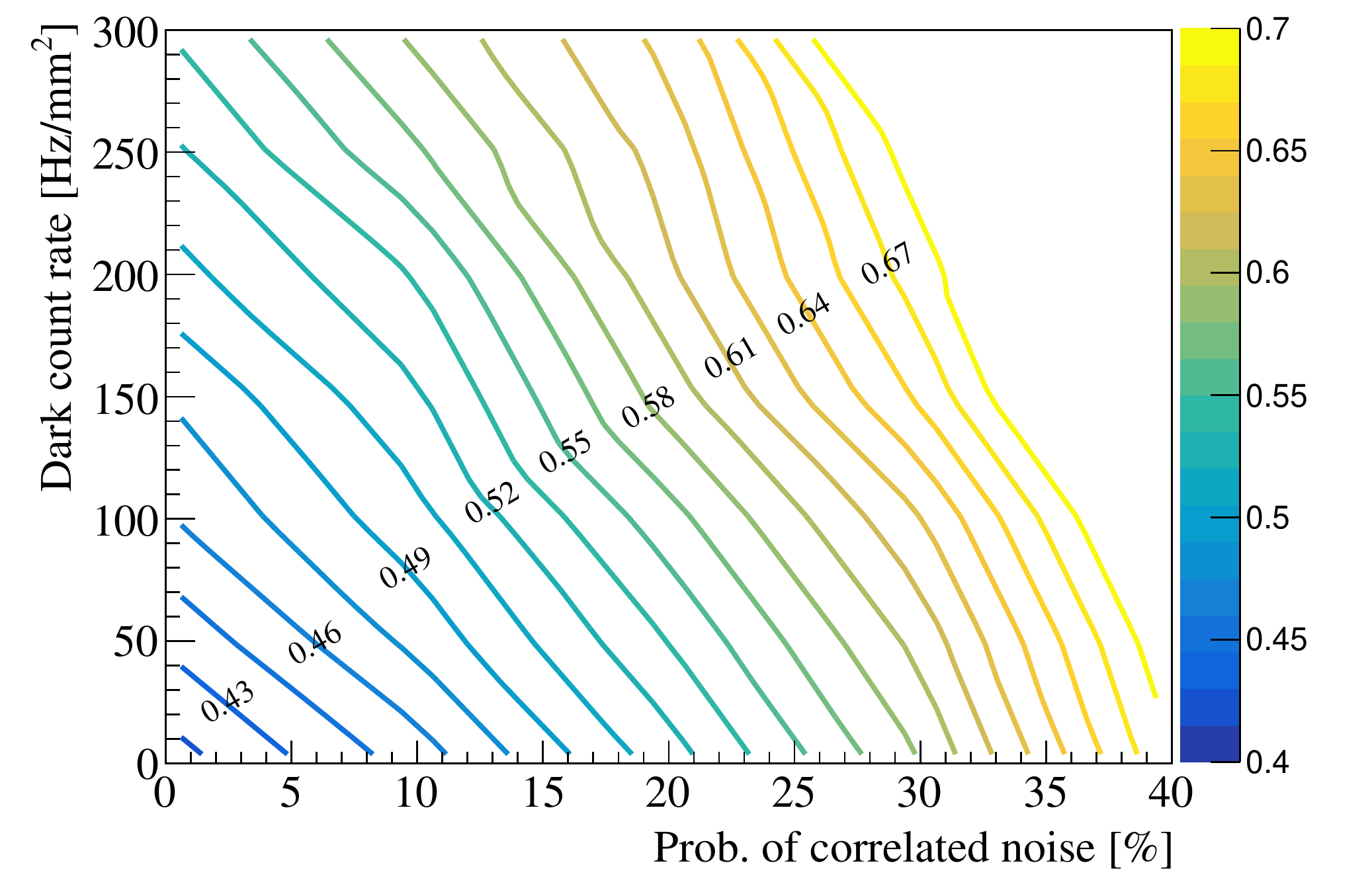}
   \caption{The required photon detection efficiency of SiPMs, showing as numbers labelled on the contours, to reach
the reference energy resolution under assumptions of different dark count rates
(y axis) and probabilities of correlated noise (x axis). \label{fig:fom}}
\end{figure}

A toy Monte Carlo simulation has been used to evaluate the joint effects of the PDE, dark count rate, and correlated noise. For given dark count rate and probability of correlated noise, the required PDE is shown as contour lines in Figure~\ref{fig:fom} by requiring the reference energy resolution of $\sigma_E/E=1.7$\% at 1~MeV from pure statistics. For example, when the dark count rate is 100~Hz/mm$^2$ and the probability of  correlated noise is 10\%, the required PDE to reach the reference energy resolution is 50\%, consistent with the analytic calculation above.

\subsection{SiPM tile}
\label{sec:SiPM_tile}
\subsubsection{Requirements}
A SiPM tile is used to support SiPMs and provide the connections to the readout electronics.
It holds multiple SiPMs and is the basic unit during detector installation. The SiPM coverage within a single SiPM tile should be larger than 94\% to ensure sufficient overall PDE. The materials of the tiles must have low radioactivity since they are close to the GdLS. The radioactivity of the SiPM tiles, including the readout electronics, should be less than 4.4~Bq/kg, 6.3~Bq/kg and 1~Bq/kg for uranium, thorium and potassium, respectively. Moreover, all materials must be compatible with the buffer liquid (LAB as the baseline design). The standard FR4 Printed Circuit Board (PCB) cannot be used to fabricate the tiles due to its high radioactivity. Some low-background materials suitable for tile fabrication are under investigation, like Pyralux, Cuflon, and others. A similar material will be used for the front-end electronics PCB.

\subsubsection{Tile design}
There are two factors related to the tile design that affect the overall PDE. The first factor is the SiPM coverage within an individual tile. It depends on the bonding of the SiPMs to the tile. If SiPMs with Through Silicon Vias (TSVs) are
available, a coverage close to 100\% can be expected. However, if wire bonding has to be used to connect the front sides of SiPMs to the tile, then the coverage will be reduced to leave space for bonding pads on the tile. After investigations into the existing technologies used by SiPM manufactures, we choose wire bonding as the baseline considering the technology maturity and the cost. A tile with an area of about 25 cm$^{2}$ consists of an $8\times8$ array of SiPMs with dimensions of $6\times 6$~mm$^{2}$, or a $6\times 4$ array of SiPMs with dimensions of $10\times 10$~mm$^{2}$. Large SiPMs have certain advantages. They provide a slightly higher coverage, but the production yield might be significantly reduced, which results in a higher cost. With $6\times 6$~mm$^{2}$ SiPMs, if we leave a 200 $\mu$m gap for bonding pads and a 100 $\mu$m gap between SiPMs, a coverage of about 95\% can be achieved. More than 97\% coverage can be achieved with $10\times 10$~mm$^{2}$ SiPMs. The second factor is the gap between SiPM tiles. After investigation of various arrangements of SiPM tiles on the supporting copper shell, we found a maximum coverage of about 95.5\% with tiles of rectangle shape and of dimensions of $5\times 5$~cm$^{2}$. Regions at the poles of the spherical copper shell have smaller tile coverage. If we use some irregular shape tiles, such as trapezoid, the coverage can be improved, and more important, the uniformity of the coverage improves. More investigations are needed to make the final decision.

To reduce the number of readout channels, it is essential to group the SiPMs either within the SiPM tile or on the front-end electronics board. For both cases, the connections to the SiPMs are routed to the back of the tiles. In the former case, the connectors need less pins.

\subsubsection{Tile packaging}
The SiPMs will be packaged on the tiles for protection and easy handling. The window material could be epoxy resin if LAB will be used as the buffer. If a liquid different from LAB will be used, silicone could be considered as the window material. A good match of refractive indices between the window material and the buffer liquid is required to reduce the light reflection on the surface. R\&D efforts are needed to choose the final window material.

\subsection{Mass characterization of SiPM}
\label{sec: MC_SiPM}

Careful characterization of all SiPMs is needed to control the quality of SiPMs used in the TAO detector. SiPMs should be tested at two levels, the wafer level and the tile level. On the wafer level, we can obtain the breakdown voltage $V_{bd}$ of each SiPM by measuring its Current-Voltage (IV) curve and therefore probe the variation of this parameter. A good uniformity of $V_{bd}$ will avoid sorting of SiPMs during assembly according to their $V_{bd}$. On the tile level, we will test every tile to ensure that the assembly was performed well and the tile has the desired performance.

\subsubsection{Characterization on wafer level}
Modern semiconductor technologies are well developed, allowing to produce SiPMs with high yield and consistent parameters. The variation of their characteristics within the wafer should be rather small. Testing a few dies on the wafer at different positions may allow us to assess the entire wafer quality. However, since the SiPM yield on the wafer is not 100\%, we still need to characterize each SiPM to remove bad ones.

A strong correlation of parameters with the bias voltage can be maintained within the same charge.
A simple and robust way to characterize SiPMs is to measure their IV curves with an automatic probe station. The method will be applied for characterization of individual SiPMs at the wafer level. The selection of SiPMs with similar breakdown voltage would allow to bias the SiPMs from a single voltage supply. This can help to reduce the number of voltage channels and make the SiPM bias robust and cheap.
\begin{figure}[htb]
   \centering
  \includegraphics[width=\textwidth]{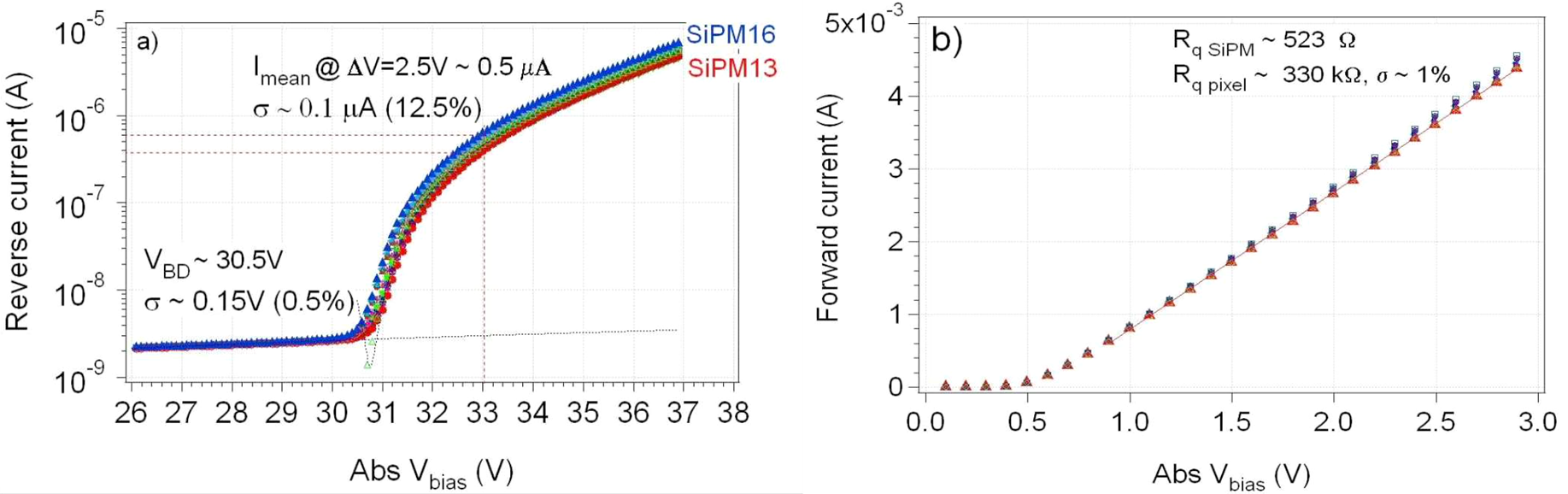} \\
   \caption{a) reverse and b) forward IV curves of 16 SiPMs~\cite{LCS:Dinu}. }
  \label{fig:lcs_fig1}
\end{figure}

We present the IV-curve measurements of 16 SiPMs from Ref.~\cite{LCS:Dinu} as an illustration. Measuring the IV curve in forward biasing (Figure~\ref{fig:lcs_fig1} b)) allows to probe the average
quenching resistance, while the reverse IV curve (Figure~\ref{fig:lcs_fig1} a)) reveals the breakdown voltage
and hence shows the operating voltage range. To find the breakdown voltage one can use the Inverse Logarithmic Derivative (ILD)~\cite{chmill}:
\begin{equation}
    f(V) = \left(\frac{d\text{ln}[I(V)]}{dV}\right)^{-1} =
\left(\frac{1}{I}\cdot\frac{dI(V)}{dV}\right)^{-1} \,.
\end{equation}
Finding the minimum of $f(V)$ returns the breakdown voltage $V_{bd}$.
Another robust approach is to apply a quadratic fit to the reverse IV curve~\cite{italian}.

Before dicing, the IV curves of each SiPM will be measured on the wafer level. This work will be conducted either by the SiPM manufacturer(s), or by the TAO team if the wafer dicing and packaging will be done by the team. For the former case, testing data will be provided by the manufacturer to help the tile-level test.

\subsubsection{Characterization on tile level}

\begin{figure}[htb]
  \begin{minipage}[ht]{0.5\linewidth}
  \centering
{\includegraphics[width=\linewidth]{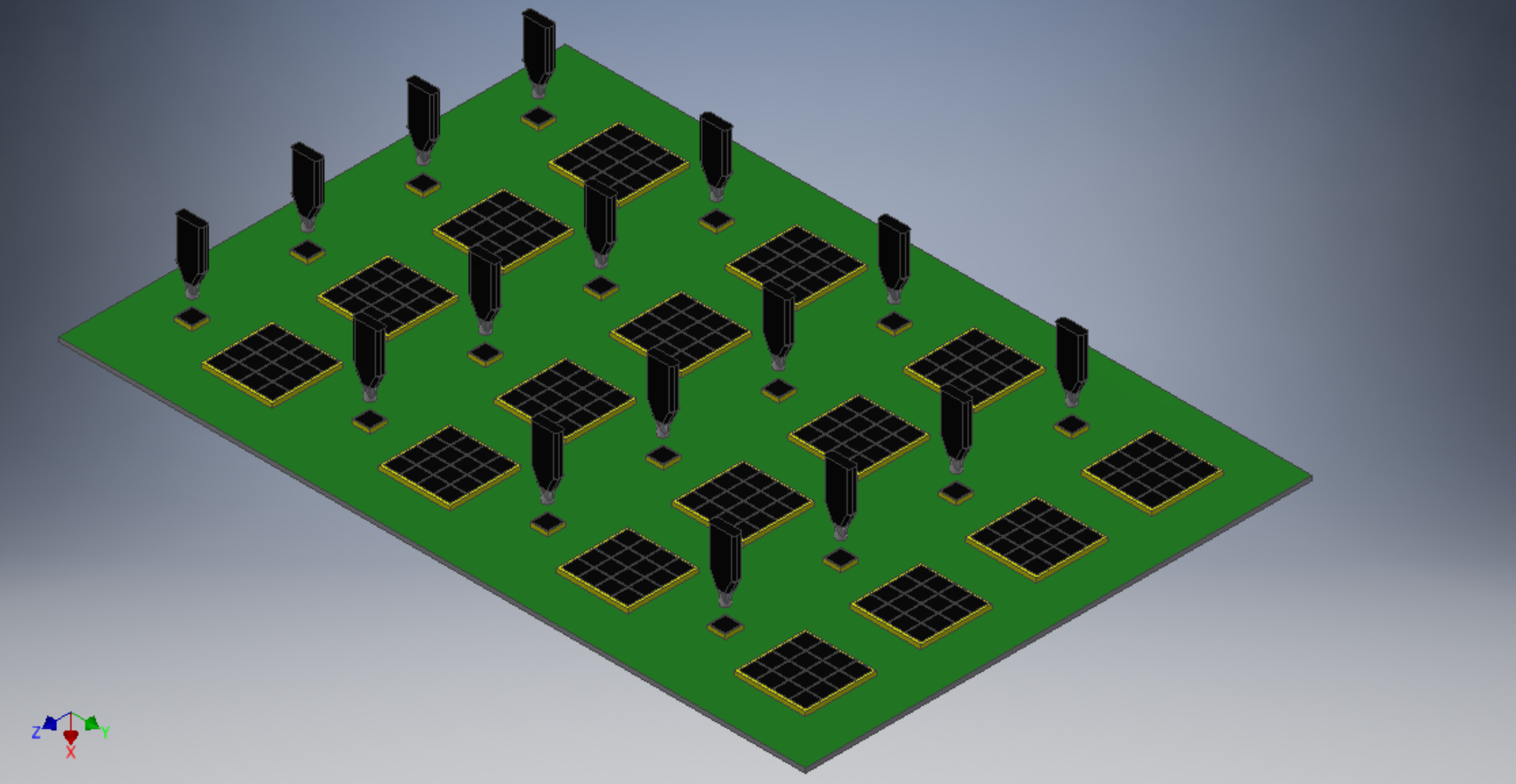}}{ (a)}
  \end{minipage}
  \hfill
  \begin{minipage}[ht]{0.5\linewidth}
  \centering
{\includegraphics[width=\linewidth]{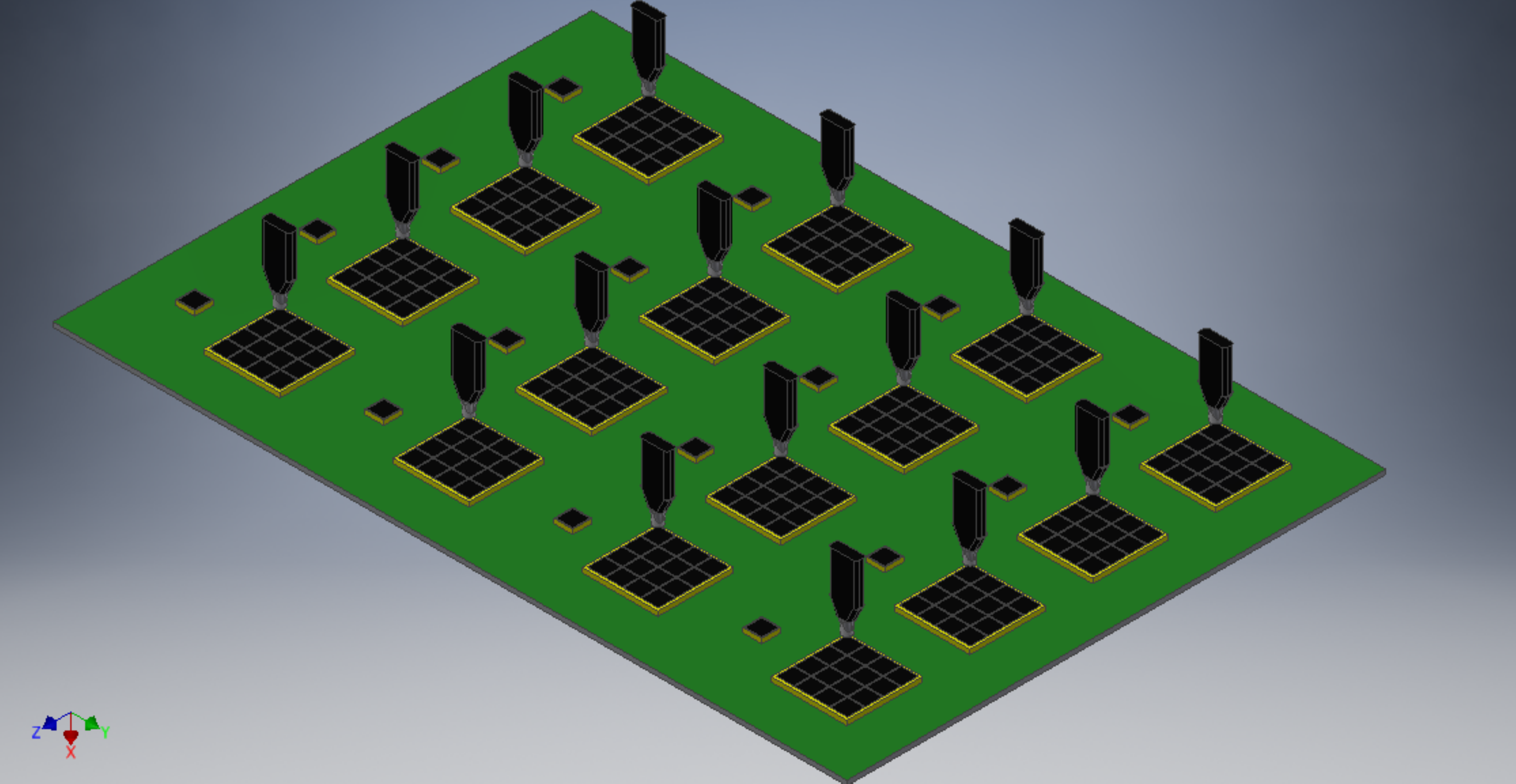}}{ (b)}
\end{minipage}
  \caption{Array tests (schematic view). (a) LEDs are in calibration mode. (b) LEDs are in scanning mode.   \label{fig:array}}
\end{figure}

In total, we will test about 4100 SiPM tiles.
The first step is a visual inspection of the window (epoxy) quality for dust and bubbles.
The second step is the simultaneous test of 16 tiles which are temporarily mounted on a large testing PCB.
Each tile is supplied from an individual voltage source that allows to precisely control the voltage of each tile.
In a dark room, we scan each SiPM on every tile with self-stabilized LEDs~\cite{hvsys1}.
Each LED is calibrated by means of a reference SiPM sitting next to each tile as shown in Figure~\ref{fig:array}~(a).
The LEDs are placed above the tiles and provide pulsed illuminations on an area of $6\times 6$~mm$^2$.
The testing PCB is moved by two step-motors positioning the LED beam precisely with respect to each SiPM as shown in Figure~\ref{fig:array}~(b).
To test all SiPMs on 16 tiles we have to provide 64 scan points for each SiPM tile.
Each scan point requires an acquisition of $\sim 10^4$ events. A full scan of 16 tiles will take about 10-20 minutes.
To test all SiPM tiles for TAO we need less than one month.
This scan allows to characterize all SiPMs in terms of PDE, gain, cross-talk, afterpulsing, and resolution of Single Photoelectron (SPE).
As cross reference to the breakdown voltage extracted from the wafer test, it could be obtained from the charge measurement of the SiPM pulses with the pulsed LED light source.\footnote{Integrating SiPM pulses from LED flashes one can obtain charge-voltage curve, from which the breakdown voltage can be extracted by performing a linear fit.}

\begin{figure}[htb]
  \begin{minipage}[ht]{0.55\linewidth}
  \centering
{\includegraphics[width=\linewidth]{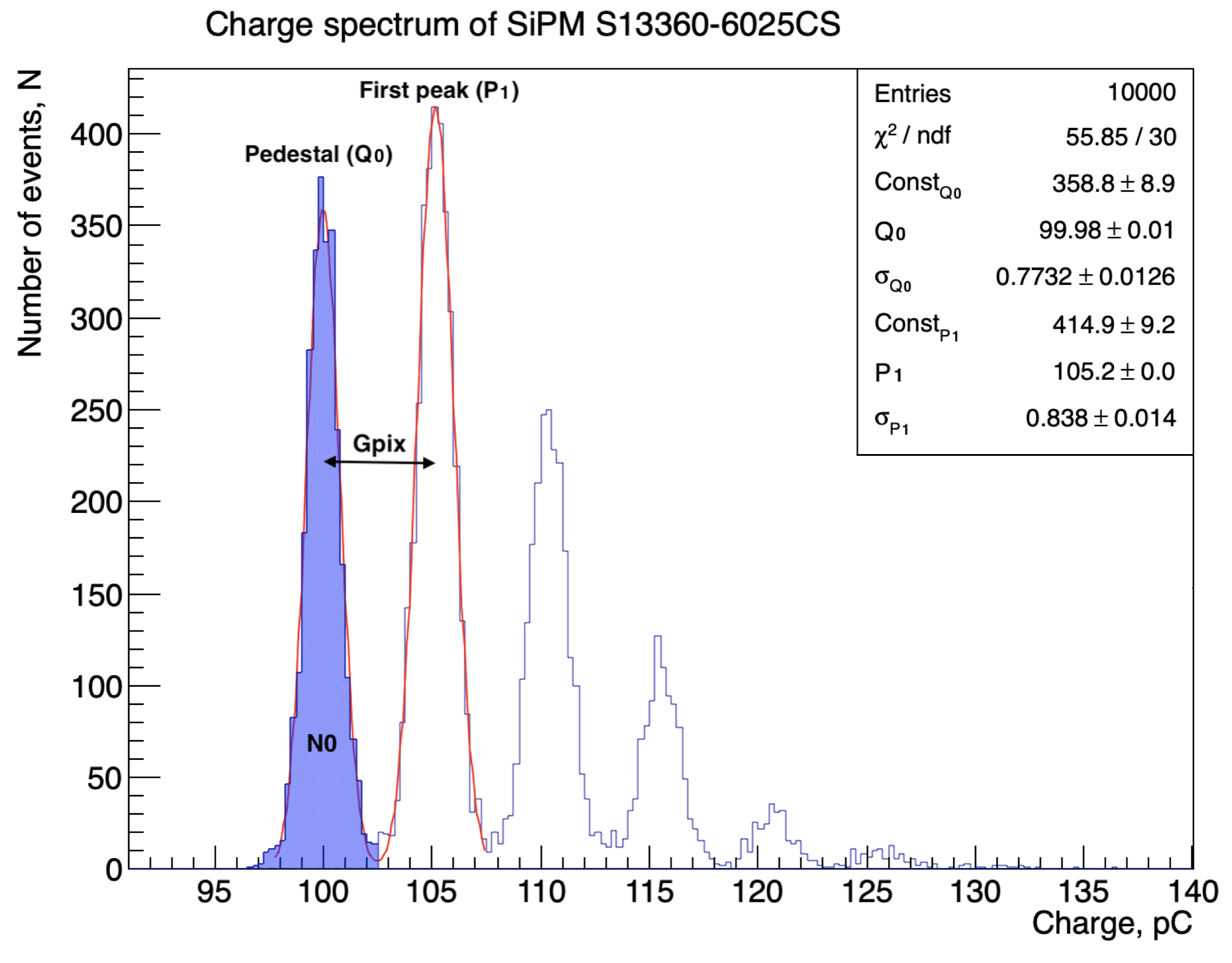}}{ a)}
  \end{minipage}
  \hfill
  \begin{minipage}[ht]{0.45\linewidth}
  \centering
{\includegraphics[width=\linewidth]{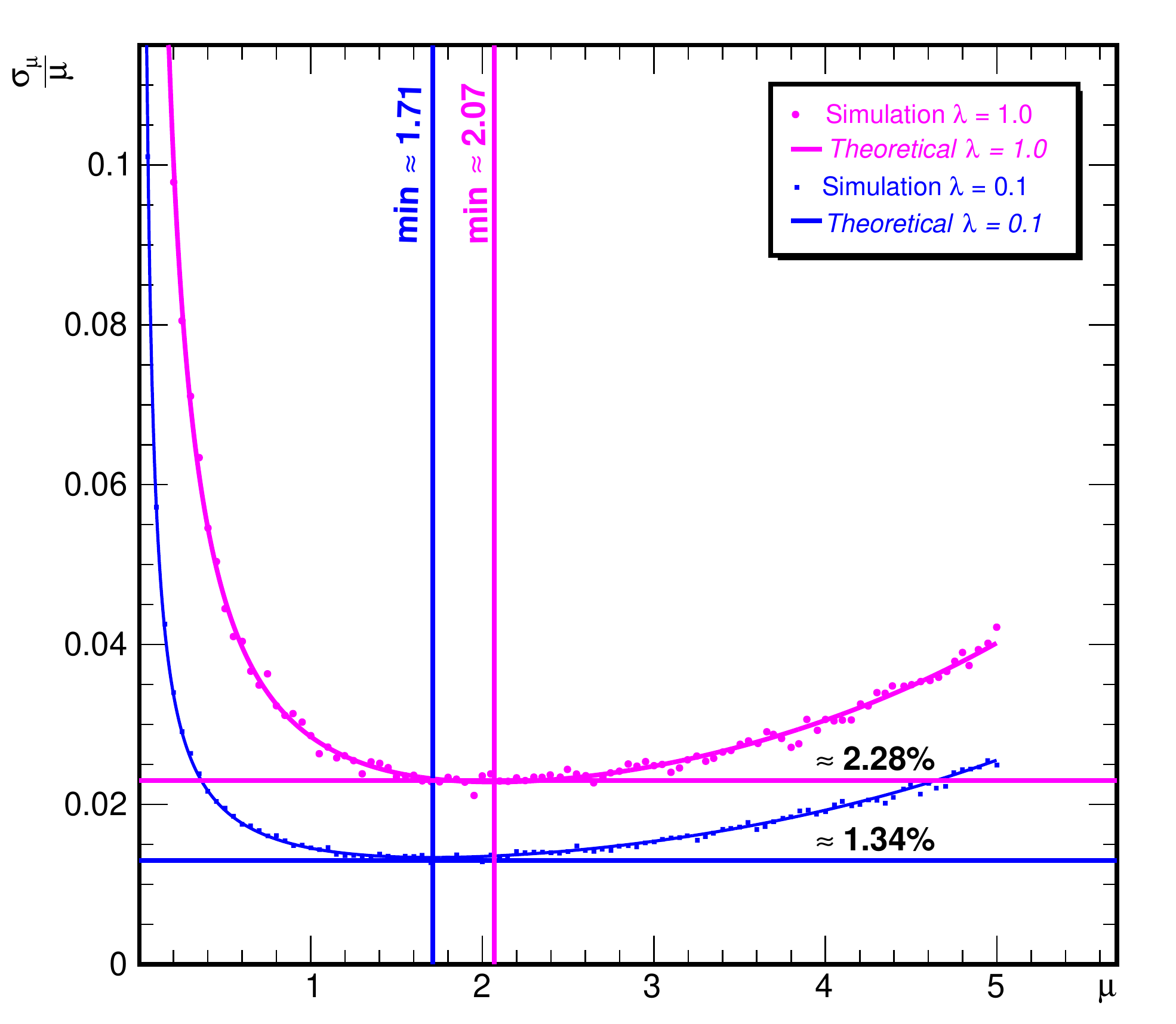}}{ b)}
         \end{minipage}
\caption{(a) Typical charge spectra of a Hamamatsu SiPM and
illustration of the pedestal method for evaluation of SiPM parameters: blue area
- pedestal events, $G_{pix}$ - pixel gain, $P_1, \sigma_1$ - single pixel's
response and its standard deviation, $Q_0, \sigma_0$ - pedestal position and
its standard deviation; (b) Relative dispersion $\sigma_{\mu}/\mu$ of $\mu$
estimation by the pedestal method for signal+noise and noise spectra  as
function of  $\mu$ for $N=10000$ triggers for two different noise levels
$\lambda=1.0$ (magenta) and $\lambda=0.1$ (blue). The points correspond to
1000 experiments. The curves correspond to Eq.~\ref{eq:dispersion_mu_00_noise}.   \label{fig:fig_spectra}}
\end{figure}

The PDE is proportional to the mean number of p.e.\ detected by the SiPM, assuming the number of detected photons follows a Poisson distribution in each LED flash. The mean number of p.e.\ can be estimated by counting the number of pedestal events~\cite{pedestal}. This method is less sensitive to the full SiPM response model, which could be quite complex~\cite{smirnov}, including the cross-talk and afterpulsing~\cite{crosstalk}. By integrating the waveforms from the SiPM recorded by a flash Analog-to-Digital Converter (ADC), its charge spectrum can be obtained, as shown in Figure~\ref{fig:fig_spectra}~(a). We acquire $N$ events for signal (LED is ON), and $D$ events for the dark spectra (LED is OFF). Using the number of events in the pedestal for the signal spectra $N_0$ (the blue area under the pedestal peak in Figure~\ref{fig:fig_spectra}~(a)) and that for the dark spectra $D_0$ in the same range, the estimation of the average number of p.e.\ $\hat{\mu}$ can be found as~\cite{anfimov}
\begin{equation}
    \label{eq:mu_00}
   \hat{\mu} = - \ln{\left(\frac{N_0}{N}\cdot\frac{D}{D_0}\right)} =
-\ln{\left(\frac{\hat{p}_{\xi_0}}{\hat{p}_{\lambda_0}}\right)} = \hat{\xi} -
\hat{\lambda} \,,
\end{equation}
where $\hat{p}_{\xi_0}$ and $\hat{p}_{\lambda_0}$ are estimators of the pedestal probability in signal ($\hat{\xi}$) and dark ($\hat{\lambda}$) spectra, respectively. The statistical error could be calculated as~\cite{anfimov}
\begin{equation}
    \label{eq:dispersion_mu_00_noise}
      \frac{\hat{\sigma}_{\hat{\mu}}}{\hat{\mu}} \approx
\frac{1}{\sqrt{N}}\sqrt{\frac{e^{\hat{\xi}}+
e^{\hat{\lambda}}-2}{(\hat{\xi}-\hat{\lambda})^2}} \,.
\end{equation}

Representation of Eq.~\ref{eq:dispersion_mu_00_noise} is show in Figure~\ref{fig:fig_spectra}~(b). The idea is to adjust the light intensity to reach the best accuracy with a limited number of acquisitions. In the case of a SiPM with a noise level of $\sim$~MHz (at room temperature) and gate window of 100~-~500~ns ($\lambda=0.1 - 0.5$), we have to use light intensity in a range of 1.5~-~2.4~photoelectrons to reach a calibration accuracy $\sim 2$\% with $10^4$ acquisitions.

The other parameters can also be easily extracted from the SiPM's charge spectrum, like the gain, cross-talk, afterpulsing, and SPE.

\subsubsection{R\&D efforts of SiPM characterization }

To study the PDE at different temperatures, we use a Dewar vessel partially filled with Liquid Nitrogen (LN). Nitrogen vapor produces a gradient of temperature at different heights.  We can use this gradient for measurements at different temperatures in a broad range, from room temperature down to LN environment. We use a stabilized LED. At the beginning, we observed a light variation through a fiber at the level of about 10\%, when temperature changed from room temperature to about -50$^\circ$C. It might be driven by changes in the optical properties of the fiber. To cancel such a behavior we decided to stabilize the temperature along the length of the fiber.

\begin{figure}[htb]
  \begin{minipage}[ht]{0.24\linewidth}
  \centering
    {\includegraphics[width=\linewidth]{SiPM/SiPM_figures/fig_setup.pdf}}{ (a)}
  \end{minipage}
  \hfill
  \begin{minipage}[ht]{0.56\linewidth}
  \centering
    {\includegraphics[width=\linewidth]{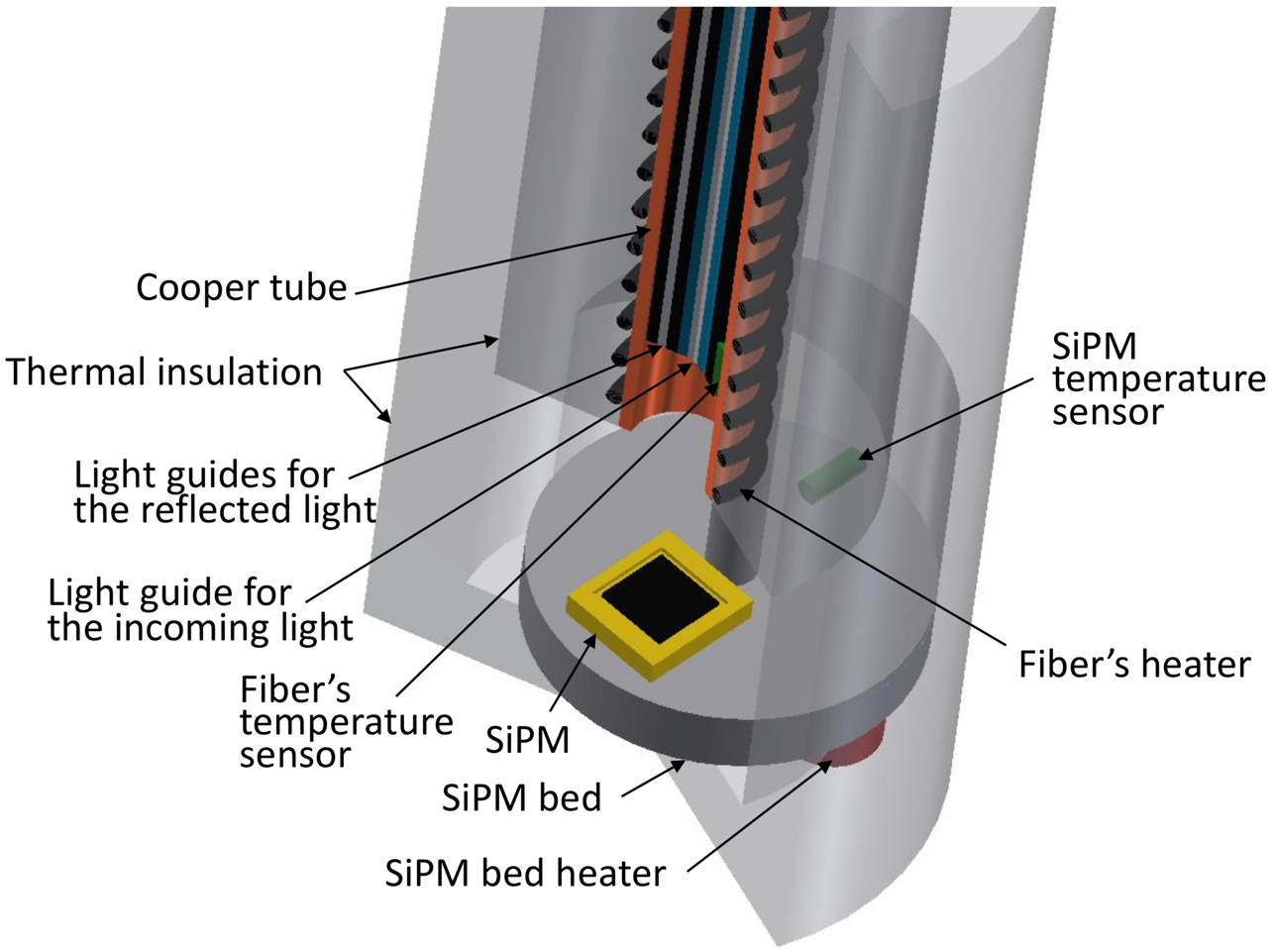}}{ (b)}
  \end{minipage}
  \caption{Setup for studying SiPM parameters at low temperatures. (a)
The assembly dip in the Dewar vessel (Nitrogen vapor); (b) Assembly of the
light delivery system with SiPM.  \label{fig:fig_setup}}
\end{figure}

The cryostat vessel of 30 liters is filled to about 1/3 with LN (see Figure~\ref{fig:fig_setup}~(a)). An assembly of the light delivery system  (see Figure~\ref{fig:fig_setup}~(b)) is moving with the SiPM along the cryostats depth. The light
delivery system is enclosed in an insulated copper pipe for screening. A bundle of optical fiber is placed inside the insulated pipe. We send light through the central fiber. The other fibers are used as a monitoring system to check the light stability when the temperature changes.  The pipe is wound by a heating cable with feedback provided by a thermal sensor inside the pipe. We place a mirror in front of the fiber bundle and check the stability with high light intensity, high enough to use a silicon PIN-photodiode to monitor the reflected light. Our measurements show that the stability is at a level of 1\%.

Then, we replace the mirror with a SiPM (S13360-6025CS from Hamamatsu) and couple it with thermal grease to an aluminum carrier (bed) with an embedded thermosensor on its backside. Temperature stability is guaranteed with a precision of better than 1$^\circ$C.

\begin{figure}[htb]
   \begin{minipage}[ht]{0.5\textwidth}
   \centering
   {\includegraphics[width=\textwidth]{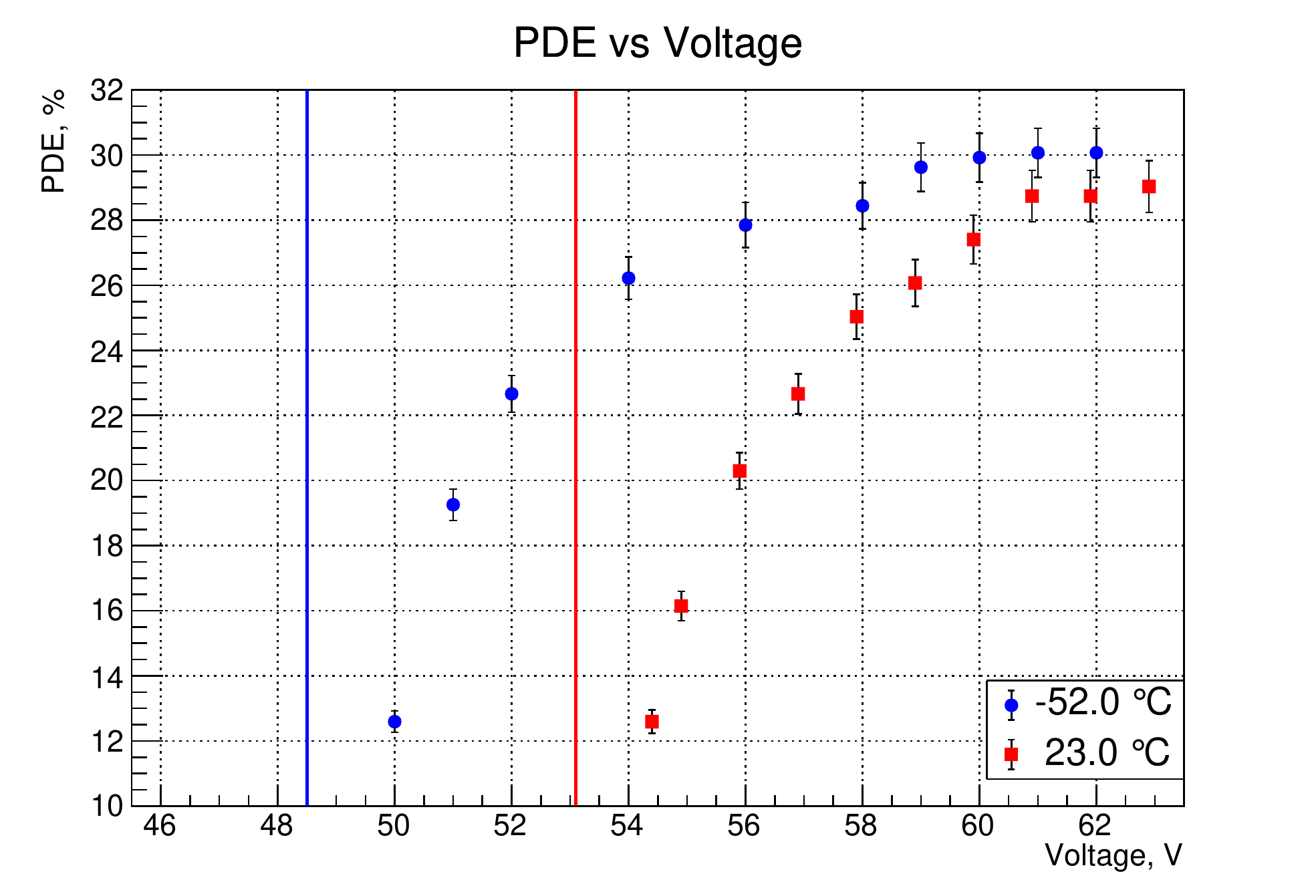}}{ (a)}
    \end{minipage}
    \hfill
    \begin{minipage}[ht]{0.5\textwidth}
    \centering
  {\includegraphics[width=\textwidth]{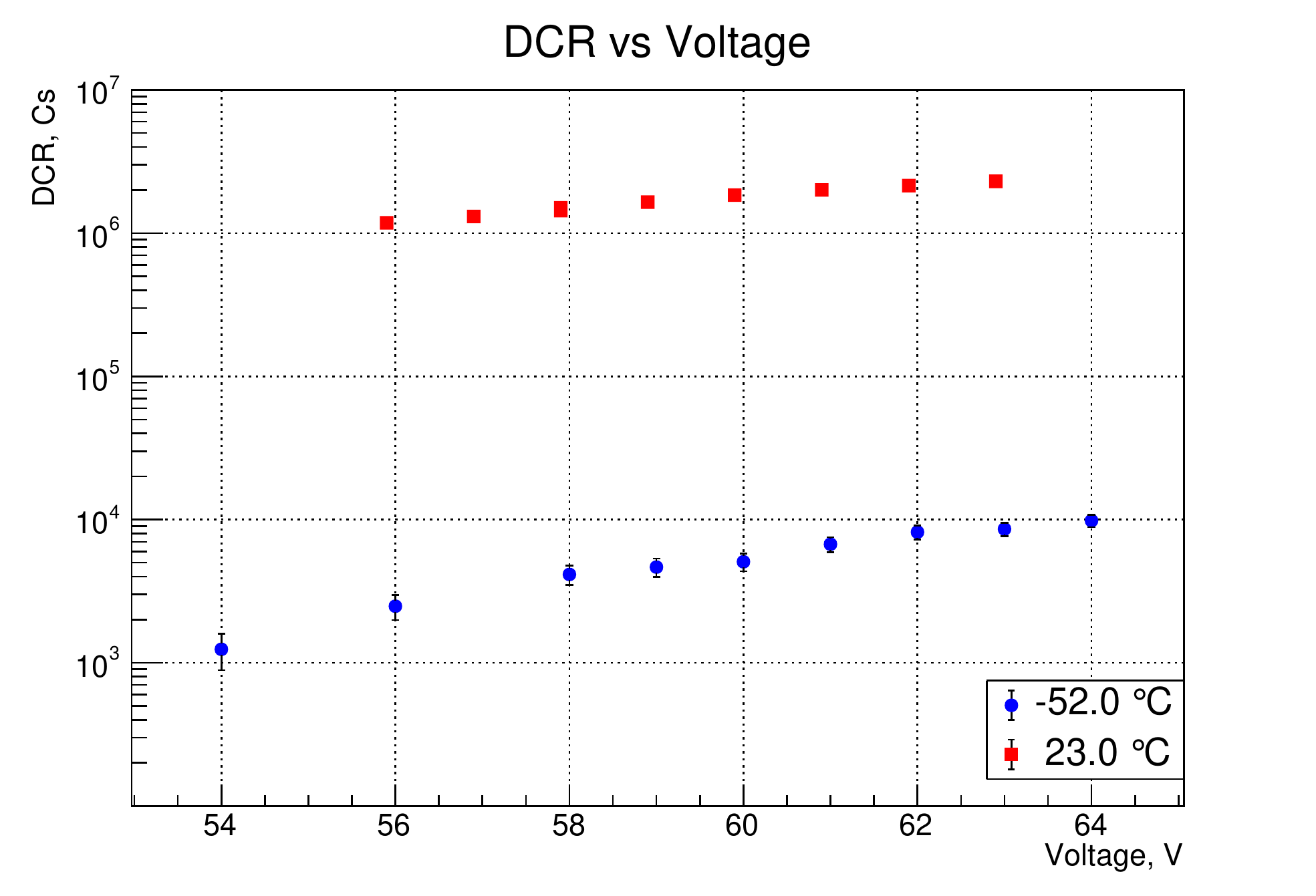}}{ (b)}
    \end{minipage}
\caption{Hamamatsu SiPM S13360-6025CS parameters at two different
temperatures, 23$^\circ$C (room temperature) and -52$^\circ$C (around the TAO working
temperature). (a) Photon detection efficiency vs voltage; (b) Dark count rate in unit of counts per second vs voltage for the 36 mm$^2$ SiPM.  \label{fig:fig_PDE}}
\end{figure}

We study the PDE (see Figure~\ref{fig:fig_PDE}~(a)) and the DCR (see Figure~\ref{fig:fig_PDE}~(b)) at two different temperatures, 23$^\circ$C (room temperature) and -52$^\circ$C (around the TAO working temperature). The breakdown voltages are $V_{bd}(23^\circ{\rm C}) \approx 53.1~V$ at room temperature and $V_{bd}(-52^\circ{\rm C}) \approx 48.5~V$ at low temperature. Lacking an absolute calibration of the PDE, we normalize our measured PDE  to 25\% at the operating voltage $V_{op}=57.9~$V at room temperature as specified by the vendor~\cite{Hamamatsu}. As one can see from Figure~\ref{fig:fig_PDE}~(a), at low temperature the PDE may reach a maximum value similar to that at the room temperature, which is $\sim 20\%$ higher than that at the reference operating voltage. A maximum PDE of about 30\% is reached at -52$^\circ$C at a bias voltage $V_{db}~=~60.0~$V, and a maximum PDE of about 29\% at 23$^\circ$C is reached at a bias voltage $V_{bd} = 61.0~$V.

\begin{figure}[htb]
   \centering
  \includegraphics[width=0.6\linewidth]{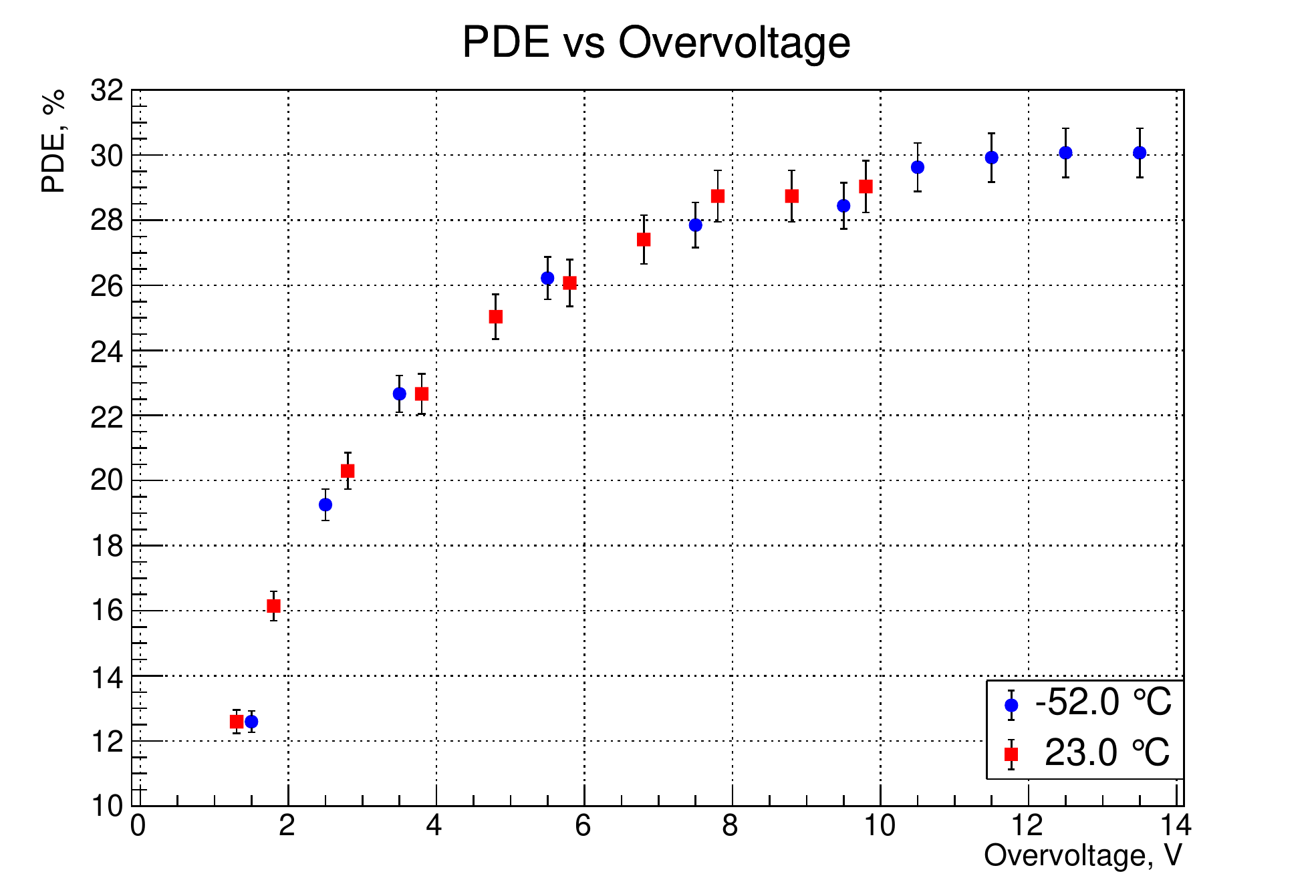}
   \caption{PDE as a function of the overvoltage at two temperatures for Hamamatsu SiPM S13360-6025CS.
    \label{fig:fig_PDE_OV}}
\end{figure}

To study the dependence of the PDE on the temperature, we plot the PDE as a function of the overvoltage
$O(V) = V-V_{br}$ (see Figure~\ref{fig:fig_PDE_OV}). One can see that there is no
any significant difference between the two curves for the room and low temperatures.

The noise level of the SiPM is shown in Figure~\ref{fig:fig_PDE}~(b). The DCR is less than 5~kHz ($\sim 140$~Hz/mm$^{2}$) for the maximum PDE at -52$^\circ$C, slightly higher than our specification.

Some of the non-detected light might be reflected from the SiPM. There is a good chance to detect this light by other SiPMs due to the high surface coverage. Therefore, it is important to understand the reflectance of the SiPMs. The active area of a SiPM has a mirror-like surface and primarily produces specular reflections. Some diffuse reflection is also expected because of the microstructures on the SiPM surface, such as the quenching resistors, trenches and traces used to connect the SPADs~\cite{Lv:2019res}. We have developed a dedicated experimental setup to measure the reflectance of SiPMs in air and in LAB at visible wavelengths. Two SiPMs have been measured with this setup. One is from FBK, model NUV-HD-lowCT~\cite{Gola}. It has pixels of 40~$\mu$m. The other one is manufactured by Hamamatsu, model number S14160-6050HS \cite{Hamamatsu2}, with a pixel size of 50~$\mu$m. The dimensions of the two SiPMs are both $6 \times 6$ mm$^2$. They are packaged with epoxy resin and silicone resin as protective layers for the FBK and Hamamatsu SiPM, respectively. Our results show that the reflectance of the FBK SiPM in air varies in the range of 14\% to 23\%, depending on the wavelength and the angle of incidence, which is twice larger than that of the Hamamatsu device. This indicates that the two manufacturers are using different anti-reflective coating on the SiPM's surface. The reflectance is reduced by about 10\% when the SiPMs are immersed in LAB, see Ref.~\cite{wangwei} for more details.

\subsection{SiPM power supply}
\label{sec:bias_SiPM}

\begin{figure}[htb]
   \centering
  \includegraphics[width=0.9\linewidth]{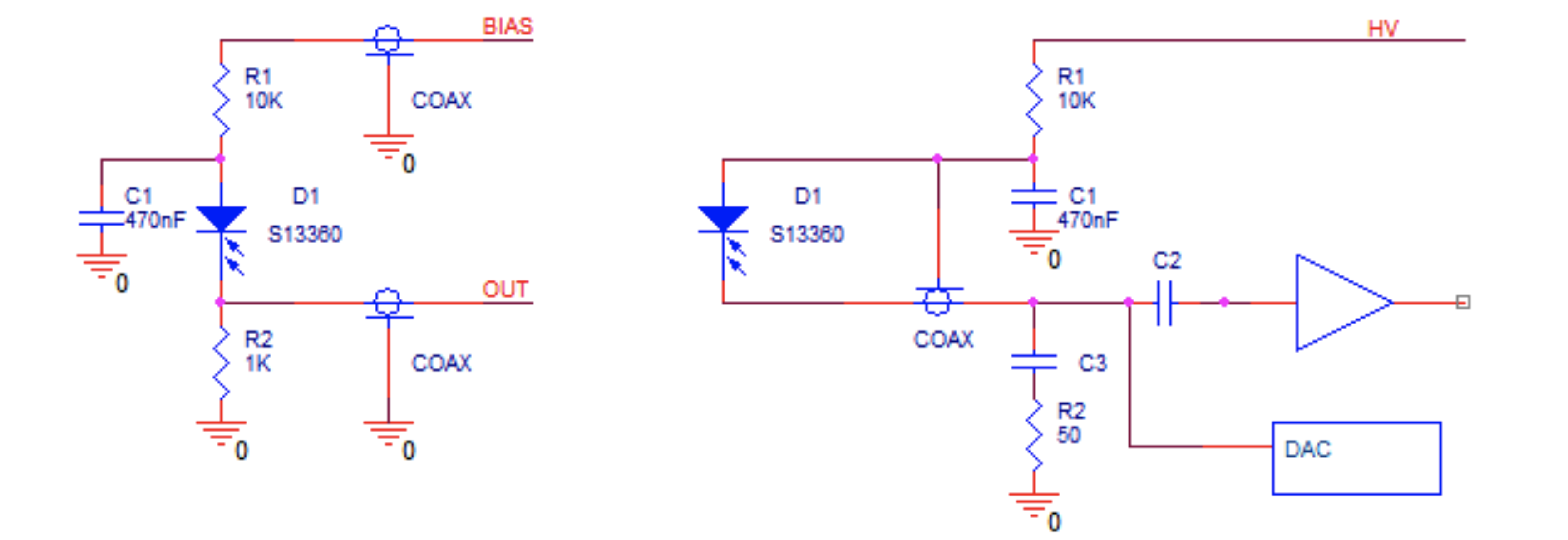}
   \caption{Schematics of the SiPM biasing. The values of the components are indicative. Left:  biasing from one side with a unipolar voltage source. Right: biasing from two sides with two sources.  \label{fig:bias} }
\end{figure}

There are two possibilities to bias the SiPMs with reverse voltage.
They are shown schematically in Figure~\ref{fig:bias}. All
values of the component are used just for demonstration. The first possibility uses unipolar power and applies the voltage from one side only. It is shown in
Figure~\ref{fig:bias}~(left). The second possibility applies a bipolar voltage from both sides, as shown in Figure~\ref{fig:bias}~(right).

The first approach has a great advantage: The readout circuit is Direct-Current (DC) coupled to the SiPM. The second one needs Alternative-Current (AC) coupling, which brings additional nuisances for high loads and long pulses, and might be problematic at high rates.

\begin{figure}[htb]
\begin{center}
  \begin{minipage}[ht]{0.45\linewidth}
  \centering
          {\includegraphics[width=\linewidth]{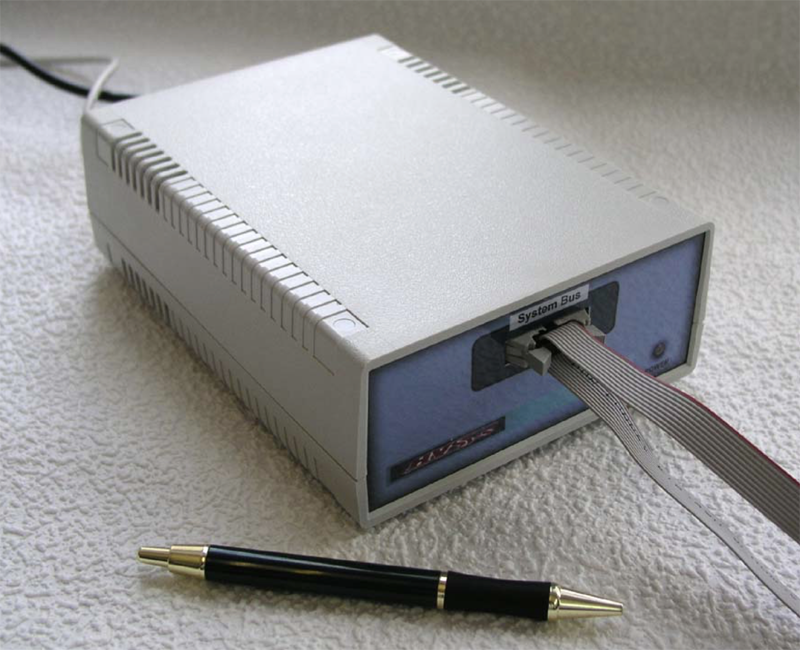} \\a) HV unit}
  \end{minipage}
  \hskip 2cm
  \begin{minipage}[ht]{0.35\linewidth}
  \centering
    {\includegraphics[width=\linewidth]{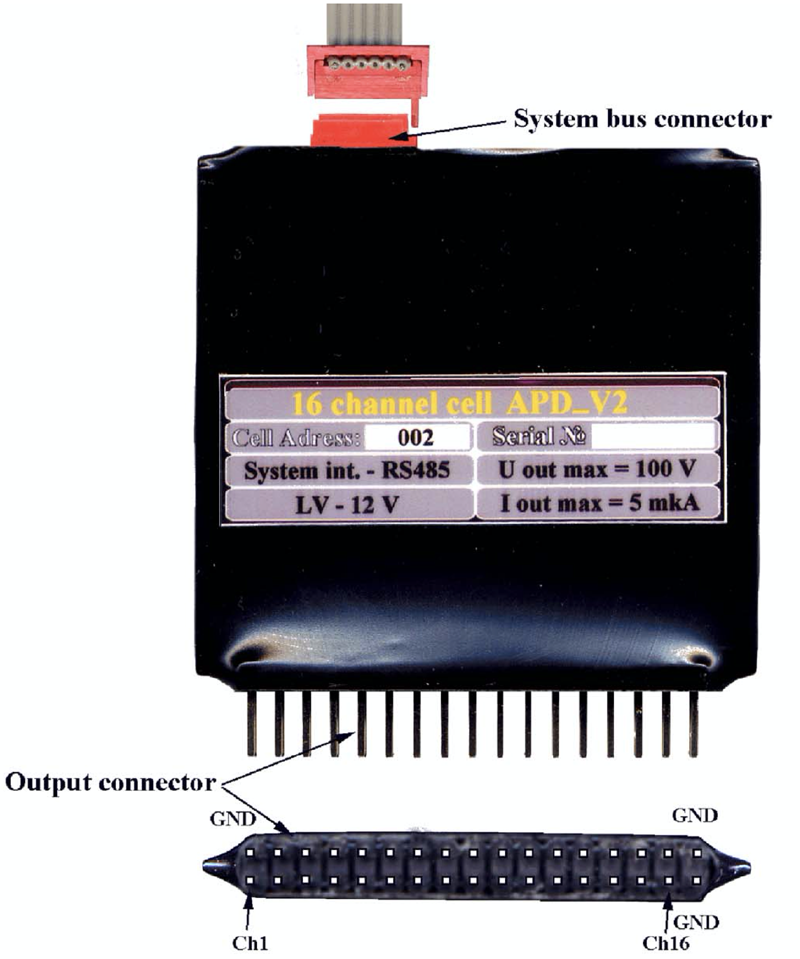} \\b) HV cell}
  \end{minipage}
  \caption{The HVsys SiPM power supply.  \label{fig:HV_unit} }
\end{center}
\end{figure}

Following the first approach, we have tested a candidate system from the company HVSys~\cite{hvsys2}. The architecture aims at low cost along with sufficiency of their characteristics, high stability of the output power voltage, and low level of fluctuation (ripples).  It provides voltage and current remote monitoring and control of all channels. The system consists of a system module, shown in Figure~\ref{fig:HV_unit}~(a), a bus, and a multichannel cell, shown in Figure~\ref{fig:HV_unit}~(b).

\begin{enumerate}
\item The system module serves up to 127 biasing cells. It includes a power supply and a microcontroller. It is connected to the mains, and through the communication line to the host computer. The system module has two designs, one desktop type where the modules are housed in a plastic case, and another type as the standard Euromechanics 6U units with 20~mm width.
\item A system bus connects the system modules to multichannel cells. The system bus is made of a 10-way ribbon cable with a spacing of 1.27~mm and Insulation Displacement Connectors (IDC).
\item Multichannel cells generate the high voltage for biasing of the SiPMs. They are made as a small-size box or printed circuit board incorporating a connector for connection to the system bus and an output connector to the detector.
\end{enumerate}

\begin{figure}[htb]
\begin{center}
  \begin{minipage}[ht]{0.35\linewidth}
  \centering
          {\includegraphics[width=\linewidth]{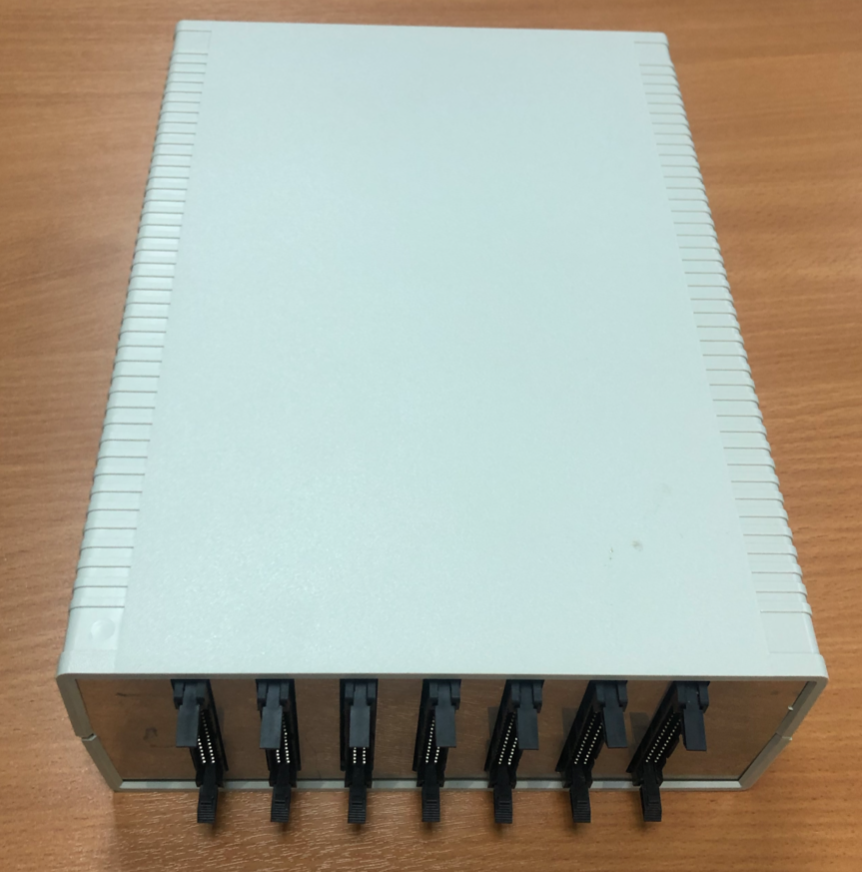} \\a) Custom made HV unit by JINR}
  \end{minipage}
\hskip 1cm
  \begin{minipage}[ht]{0.46\linewidth}
  \centering
    {\includegraphics[width=\linewidth]{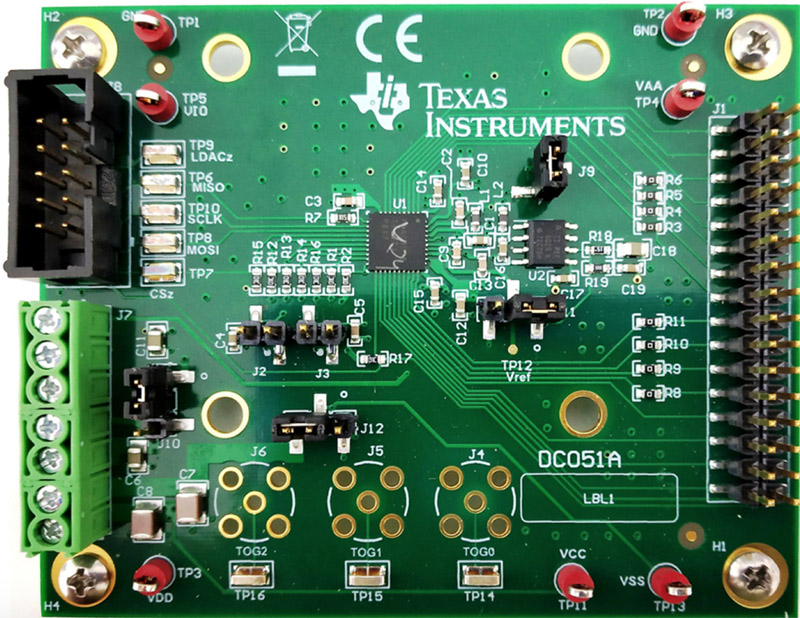} \\b) DAC81416EVM by TI}
  \end{minipage}
  \caption{Prototype of the 7x16 module of SiPM power supply based on Texas Instruments DAC.  \label{fig:HV_system}}
\end{center}
\end{figure}

Following the second approach shown in Figure~\ref{fig:bias}~(right), we have developed a HV system using a 16 channel Digital-to-Analog Converters (DAC) (See Figure~\ref{fig:HV_system}) which can produce voltages up to 40V with a Texas Instruments DAC81416~\cite{ti}, or up to 20V with a Linear Technology LTC2668-16~\cite{lt}. All the 16 channels share a common current of 25 mA. Each channel supplies a single SiPM tile with adjustable voltage within 16 (or 12) bits of dynamic range ($\pm 1$V, $\pm 2.5$V, $\pm 5$V, $\pm 10$V,$ \pm 20$V, 40V). From the other side we supply 4100 tiles with a single bias voltage. Each DAC is controlled by a micro-PC through a Serial Peripheral Interface (SPI). Using this scheme we are able to control the common current, but cannot monitor each tile individually. The main advantage of this scheme is its low cost $\approx (5 - 10)\$$/channel. By selecting SiPMs with similar operating voltage and biasing them from a common source, the cost could be further reduced.

\subsection{Readout electronics}
\label{sec:Readout}

\subsubsection{Requirements}

The total number of SiPMs is about $2.7\times10^{5}$, assuming dimensions of $6\times6$ mm$^{2}$ for a single SiPM. Simulations show that about 4500 photoelectrons will be collected by the SiPM tiles with 1~MeV of energy deposited in the GdLS. We are interested in IBD events, with energies from 1~MeV to 10~MeV. Most SiPMs will be empty and only a few photons will be captured by each fired SiPM, depending on the position of the event. The number may go up to hundreds of photons for events with large energy deposits, such as muon showers or muon bundles. The electronics system is required to precisely measure the event energy in the IBD energy region, while it is not essential to measure high energy events with good energy precision. The noise level is a challenge due to the high capacitance of the large-area SiPMs, especially with the restrictions from the very high number of channels. To reduce the number of channels, we would like to combine multiple SiPMs in one readout channel. With this method, the SiPMs connected in one readout channel are required to have good uniformity of the breakdown voltages (better than 10\% as listed in Table~\ref{tab:sipm}), in order to reduce the gain variations and improve the charge resolution. But the combination of SiPMs needs a more careful investigation.

A good timing resolution of electronics is also preferable, driven by the advantage of detecting Cerenkov photons.
The degradation on the neutrino energy resolution due to the kinetic energy spread of the neutron recoil can be significantly suppressed with information on the positron direction, which can only be reconstructed by Cerenkov photons. In
addition, Cerenkov photons might help to improve the ability of particle discrimination and suppress radioactive backgrounds.

The operation of the electronics at -50$^\circ$C puts additional constraints on the design, especially the power consumption. The light yield of the GdLS depends on the temperature. A good
temperature uniformity is necessary to reduce the light yield variation and improve the energy resolution.
The number of cables from the front-end electronics in the cryostat to the back-end of the detector at room temperature
needs to be minimized for practical reasons. The radioactivity of the materials of the electronics is also a concern since they are close to the detector target. The requirements are summarized below:
\begin{enumerate}
\item Noise: The equivalent noise charge contributed by the electronics should be less than 0.1 p.e. At this level, its contribution to the energy resolution becomes negligible compared to other factors.
\item Resolution: The resolution is required to be better than 15\%. This results in a degradation of the energy resolution of less than 0.5\%.
\item Timing: Since the fast time constant of the GdLS is about 1$\sim$3 ns, the timing resolution of the electronics is required to be better than 1~ns for the purpose of Cerenkov photon detection. \item Dynamic range: It is related to the number of SiPM devices combined in one readout channel. Dedicated simulations show that the number of photoelectrons in SiPMs per square centimeter is in a range from 0 to 15 p.e.\ for IBD events uniformly distributed in the fiducial volume (a 25~cm standoff cut). So, if we take one SiPM tile (8$\times$8 SiPMs) as one channel, the dynamic range is from 1 to 375~p.e. If the SiPM area is 1~cm$^{2}$ in one readout channel, then the dynamic range is reduced to 15~p.e.
\item Power: Power is distributed in the cryostat by the Front-End Electronics (FEE) close to the tiles. The Front-End Controller (FEC) boards might be placed at the top region in the cryostat (i.e.\ the Stainless Steel Tank) or outside of the tank at room temperature. We choose the latter case as our baseline option. The power consumption of FEE and FEC (for the case in the cryostat) is required to be less than 1~kW each. Then temperature variations of less than $\pm0.5^\circ$C in the GdLS can be guaranteed.
\item Radio-purity: The requirements for the readout boards and tiles are less than 4.4~Bq/kg, 6.3~Bq/kg and 1~Bq/kg for uranium, thorium and potassium, respectively. It is dominated by the contribution from the readout boards.
\end{enumerate}

There are two readout options considered for the $\sim 4100$ SiPM tiles. One is based on an ASIC and the other is based on commercially available discrete components. The ASIC-based readout option allows us to have a high readout granularity at the level of 1~cm$^{2}$ per channel. The discrete component readout option connects all SiPMs in one tile to a  single channel to save cost and space. The two readout options will be discussed separately in the following sections.

\subsubsection{ASIC readout option}
\label{sec:ASIC}

Several ASICs designed for SiPM readout are available. Only a few of them are suitable for TAO, which requires single photon detection, 1~ns level time resolution, high signal-to-noise ratio, and low power consumption. Thanks to the high integration, the SiPM area of each readout channel can be reduced to the
level of $\sim 1$ cm$^{2}$. Preliminary simulations show that the high readout granularity can yield more
powerful pulse shape discrimination and Cerenkov light detection, because more information is available. A small readout area also leads to a small input capacitance for the electronics, so that a much simpler passive SiPM grouping method can be used, such as parallel connections or series connections.

The KLauS ASIC~\cite{klaus_2016}, developed by Heidelberg University, is a candidate chip for TAO. It has 36 input channels and is designed for an Analog Hadron CALorimeter (AHCAL) in the CALICE collaboration~\cite{calice}.

\paragraph{KLauS chip} \mbox{}\\

\begin{figure}[htb]
  \centering
  \includegraphics[width=0.8\textwidth]{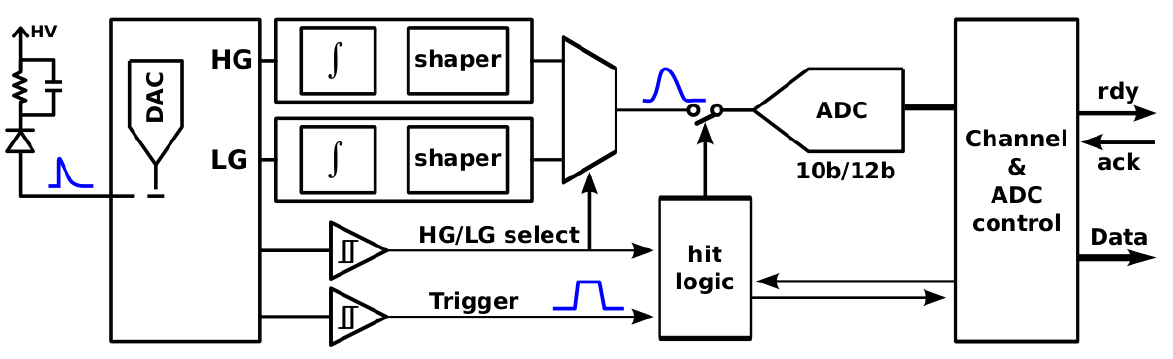}
   \caption{Block diagram of a KLauS channel~\cite{Briggl:2017kyj}.}
  \label{klaus}
\end{figure}

The KLauS chip is a 36-channel ASIC fabricated in the UMC 180~nm CMOS technology. A block level schematic diagram of one KLauS channel is shown in Figure~\ref{klaus}. It is comprised of a front-end shaping the input signal with two different gains and circuitry which detects pulses. Selected pulses are routed to an ADC which digitizes the analog information. After detection, a digital control circuit initiates and controls the analog-to-digital conversion and passes the digitized data to the following digital part, which is responsible for combining, buffering, and sending the data to the Data Acquisition (DAQ). Currently the fifth version of the KLauS chip (KLauS5) is available for testing, and the next version is under fabrication with a TDC with 200~ps steps. The main features of KLauS5 are the following.
\begin{enumerate}
\item Fine tuning of the SiPM bias voltage in a range of 2~V.
\item Two automatically switching branches with low gain and high gain.
\item The 10-bit successive-approximation-register (SAR) ADC for normal data taking and a 12-bit pipelined ADC for the readout of SiPMs with ultra-low gains.
\item Small equivalent noise charge of 6~fC with input capacitances of less than 100~pF.
\item Large dynamic range up to 450~pC.
\item Less than 3.6~mW/ch power consumption at full operation.
\item Maximum 20 events/channel for each data acquisition.
\item 25~ns event timestamps.
\end{enumerate}

\paragraph{Architecture} \mbox{}\\

\begin{figure}[htb]
  \centering
  \includegraphics[width=0.7\textwidth]{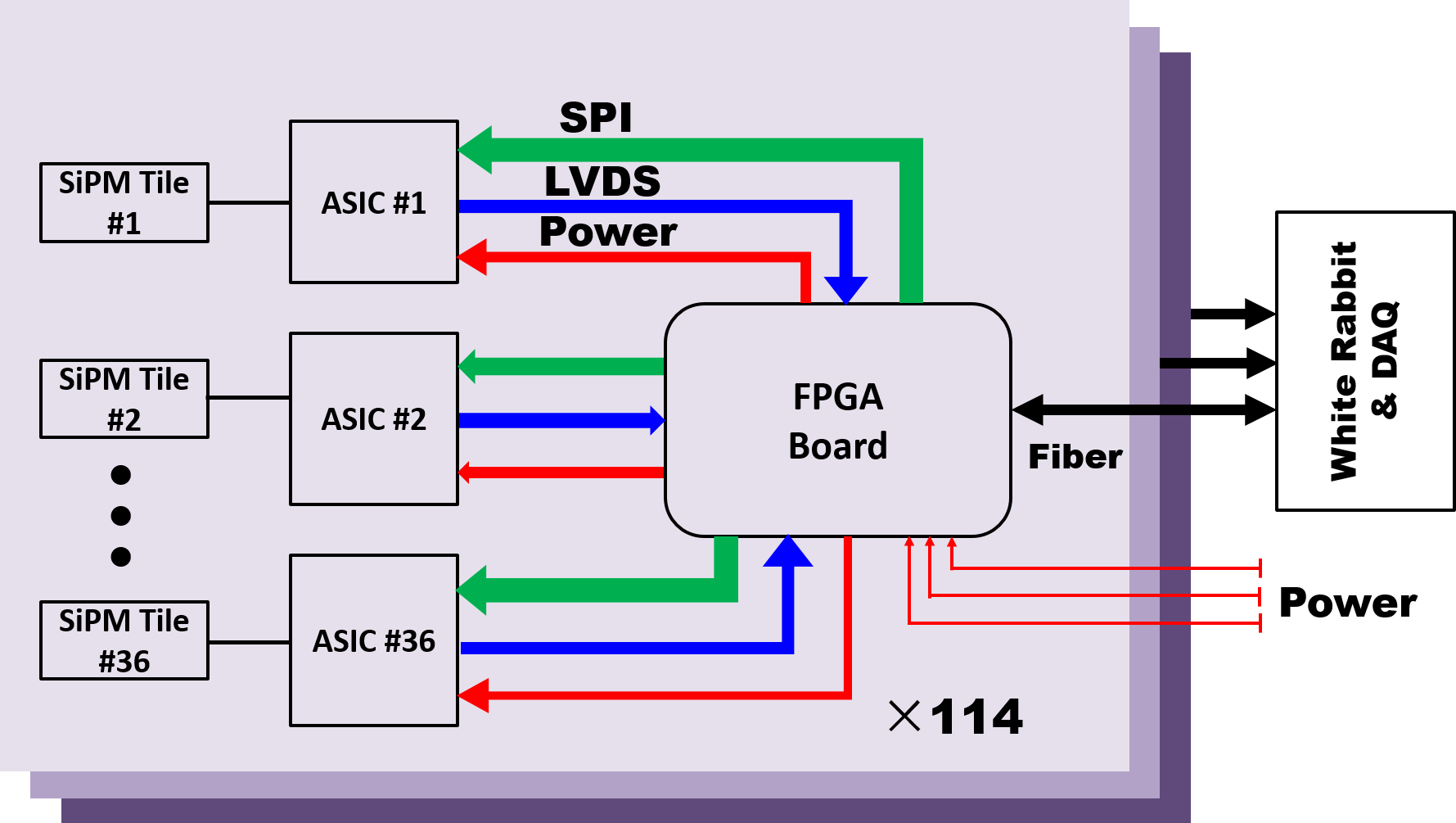}
   \caption{Architecture of ASIC-based readout electronics system.}
  \label{arch_asic}
\end{figure}

If we connect two SiPMs into one channel of the KLauS, one chip is sufficient to read a whole tile of $8 \times 8$ SiPMs. The two SiPMs can be ganged with either parallel or series connections. The architecture of the readout is shown in Figure~\ref{arch_asic}. One ASIC board will be connected to one SiPM tile by connectors or short flat cables.
If larger tiles are used eventually, several chips will be deployed on one ASIC board. The configuration, data transfer and power supply of the KLauS chips will be managed by FPGAs, which will be arranged on separated boards. Let's assume each FPGA board can handle 36 KLauS chips, then there will be 114 FPGA boards in total.

The Serial Peripheral Interface (SPI) is used to configure KLauS and provide the clock. A total of five connections is needed to each chip. The data collected by KLauS will be stored in its local FIFO. There are 3 L1 FIFO and 1 L2 FIFO in each chip. The L1 FIFO can store 64 events and the L2 FIFO can store 512 events. If the L1 and L2 FIFOs are full, the analog-to-digital conversion will be inhibited until data is read out and extra space in the FIFO becomes available. A pair of differential cables with a nominal bandwidth of 160~Mbit/s is used to transmit the data from the FIFOs to the FPGA boards via Low Voltage Differential Signaling (LVDS). The maximum data rate is expected at 10~Mbit/s for a sum of 32 channels in one chip, which is dominated by the dark noise of the SiPMs.

The FEC will also distribute the power to different ASIC boards, including the bias voltage required by the SiPMs. The FPGA boards will either be located at the top of the central detector flushed with nitrogen gas, or located outside of the Stainless Steel Tank. We can tolerate more power consumption there. The lengths of cables between the ASIC boards and the FPGA boards can vary from tens of centimeters to about 10 meters, depending on the location of the FPGA boards. One optical link from each FPGA board is used to transfer data to the DAQ hardware which is located outside of the Stainless Steel Tank. The same link will be used to send trigger signals and clock back to the FPGAs. The white rabbit system, developed at CERN~\cite{white_rabbit}, will  be used to synchronize the clocks among different FPGAs. The connection to the white rabbit system will share the optical link with data transfer. There are in total 114 optical links from the FPGA boards to the DAQ system, together with several cables for power supply.

\paragraph{R\&D with the KLauS chip} \mbox{}\\

\begin{figure}[htb]
  \centering
  \includegraphics[width=0.7\textwidth]{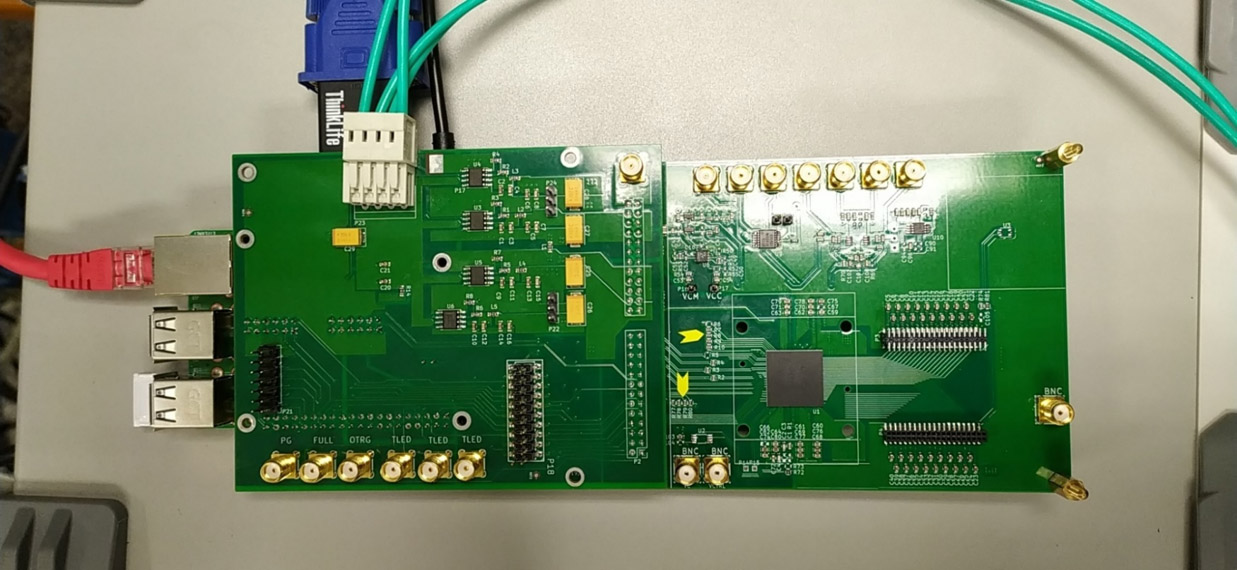}
   \caption{The KLauS5 testing boards.}
  \label{test_board}
\end{figure}

Testing boards for the KLauS5 were built, shown in Figure~\ref{test_board}. On the right, the ASIC board holds the chip. It provides interfaces for analog monitor and debugging, and connections to the SiPMs via two connectors with 72 pins in total. The left board is called interface board. It provides power and a slow control interface to the ASIC board. A Raspberry Pi (Model 3B), located under the interface board, is used to control and configure the ASIC. The Raspberry Pi is connected to a local PC via local area network for remote configuration and data taking. The DAQ software is provided by the ASIC developers from Heidelberg University. Some basic functional tests have been performed by connecting a $3 \times 3$ mm$^{2}$ SiPM (manufactured by Hamamatsu, model number S13360-3025CS) to the ASIC board. The testing boards are placed in a cryogenic box with controllable temperature from room temperature to -70$^\circ$C. A LED is positioned above the ASIC board. It is driven by a pulse generator and provides pulsed light to the SiPM.

\begin{figure}[htb]
  \centering
  \includegraphics[width=0.45\textwidth]{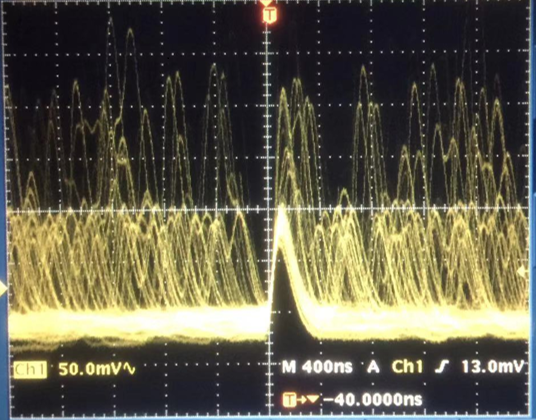}
   \caption{A snapshot of the typical waveform after the shaper in the chip
measured by an oscilloscope at -50$^\circ$C.}
  \label{waveform}
\end{figure}

\begin{figure}[htb]
  \centering
  \includegraphics[width=0.85\textwidth]{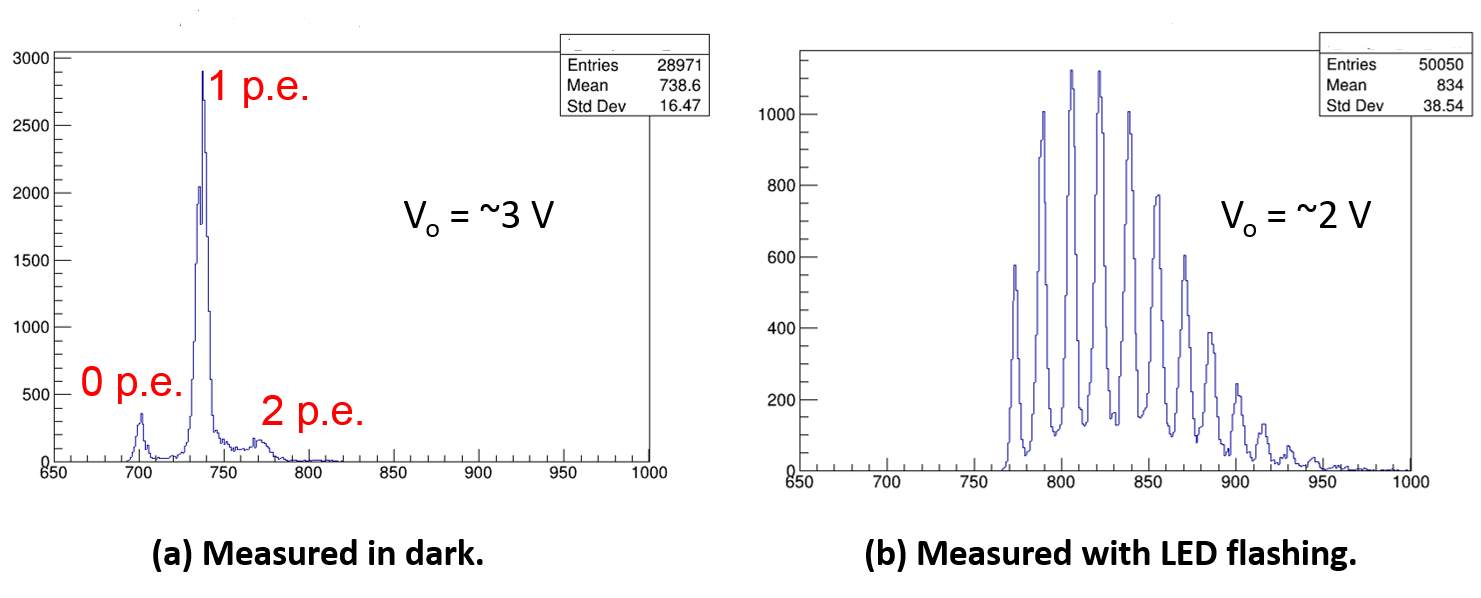}
   \caption{Charge spectrum recorded with the KLauS5 chip at -50$^\circ$C measured in the dark (a) and with pulsed
light (b).}
  \label{charge_spec}
\end{figure}

Analog pulses were monitored after the shaper circuit during the cooling process. No significant changes of the waveform were observed between room temperature and -50$^\circ$C. A snapshot of typical waveforms is shown in Figure~\ref{waveform}. A clear signal of single p.e.\ is observed. It is well separated from the baseline. The digitized output charge is shown in Figure~\ref{charge_spec}. It is measured at -50$^\circ$C. The left spectrum in Figure~\ref{charge_spec} is measured in the dark with an over voltage of about 3~V. The peak of single p.e.\ can be clearly observed, together with a fraction of two p.e.\ events caused by the optical cross-talk and a small peak of pedestal that can be eliminated by slightly increasing the trigger threshold in the chip. The right plot of Figure~\ref{charge_spec} shows the charge spectrum measured with pulsed light on the SiPM with an over voltage of about 2~V, in which the single p.e.\ signal is rejected by a relatively high trigger threshold. A good separation can be observed even for multiple p.e.

\begin{figure}[htb]
  \centering
  \includegraphics[width=0.6\textwidth]{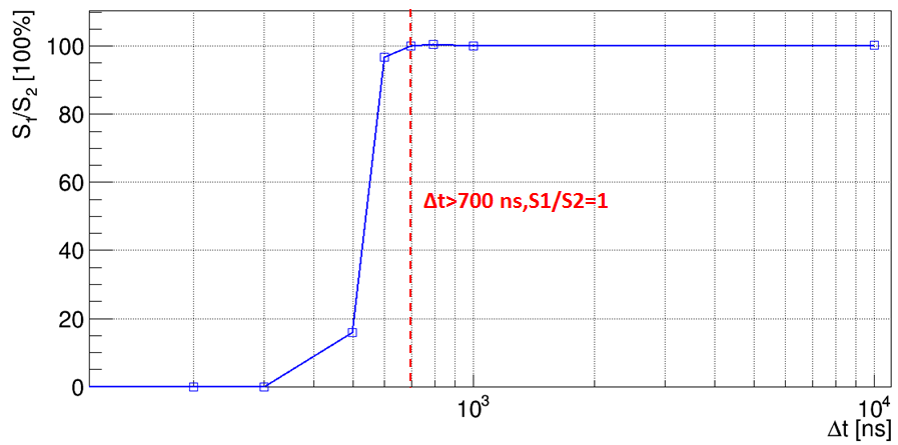}
   \caption{The fraction of the second charge detected as a function of time between the two injected charges, measured at room temperature.}
  \label{ad_time}
\end{figure}

The analog-to-digital conversion time of the KLauS5 chip is studied at room temperature by directly injecting two charges into the chip with different magnitudes. The time interval between the two injected charges is adjustable through the delay
time of the second charge in a pulse generator. During the processing of the first charge, the second charge cannot be detected.
This feature is well demonstrated in Figure~\ref{ad_time}, which shows the fraction of the second charge detected as a function of time between the two injected charges. We conclude that the KLauS5 chip fully recovers within 700~ns, which meets the requirements for IBD detection in the TAO detector.

In general, the preliminary tests show that the KLauS5 chip can work well at a temperature of -50$^\circ$C, with good performance of single photon detection and an acceptable dead time. More detailed studies are essential and will be conducted in the near future.

\subsubsection{Discrete component option}
\label{sec:discrete}

A readout system built from commercial off-the-shelf components, referred as the discrete component option here, can be adopted for TAO without a dedicated ASIC. This approach offers high flexibility on the selection of the components. The proposed solution consists of:
\begin{itemize}
\item Analog Front End Board (FEB) that amplifies and shapes the SiPM signals;
\item Front End Controller (FEC), based on an FPGA, that continuously samples and digitizes the signals coming from the FEBs. It pre-processes and formats the data. This is needed to control and eventually reduce the output bandwidth for transmitting the data to the DAQ;
\item a Central Unit (CU), a digital board based on a high-end FPGA that has the role of a global supervisor.
\end{itemize}
The FEB and the FEC are mechanically and logically separated in order to maximize flexibility and performance.

\paragraph{Architecture} \mbox{}\\
\label{sec:architecture}

\begin{figure}[htb]
  \centering
  \includegraphics[width=0.75\textwidth]{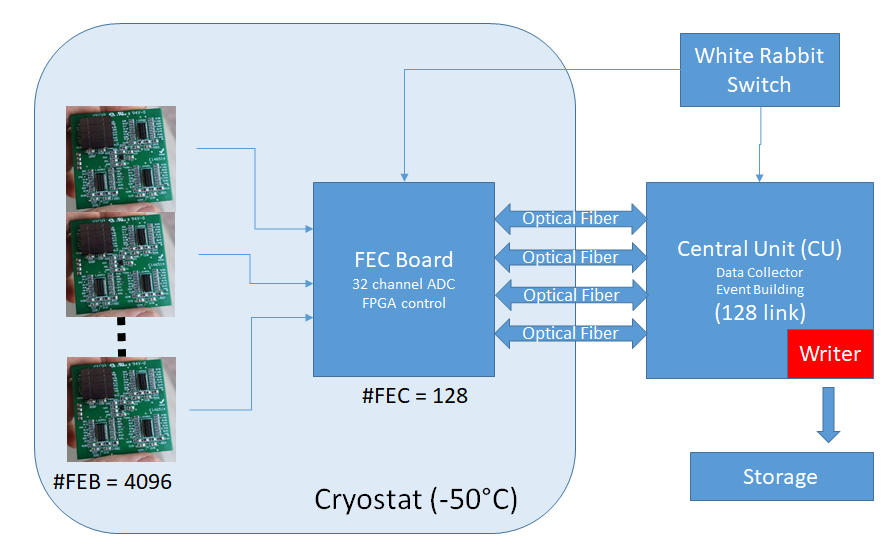}
   \caption{Architecture of the discrete component option.}
  \label{fig:architecture}
\end{figure}

The architecture of the discrete component option, shown in Figure~\ref{fig:architecture}, is based on a FEB capable reading a large-area tile ($5 \times 5$ cm$^{2}$) of SiPMs. A SiPM has an output capacitance proportional to its area, so the capacitive noise must be controlled in this approach.

\begin{figure}[htb]
  \centering
  \includegraphics[width=0.8\textwidth]{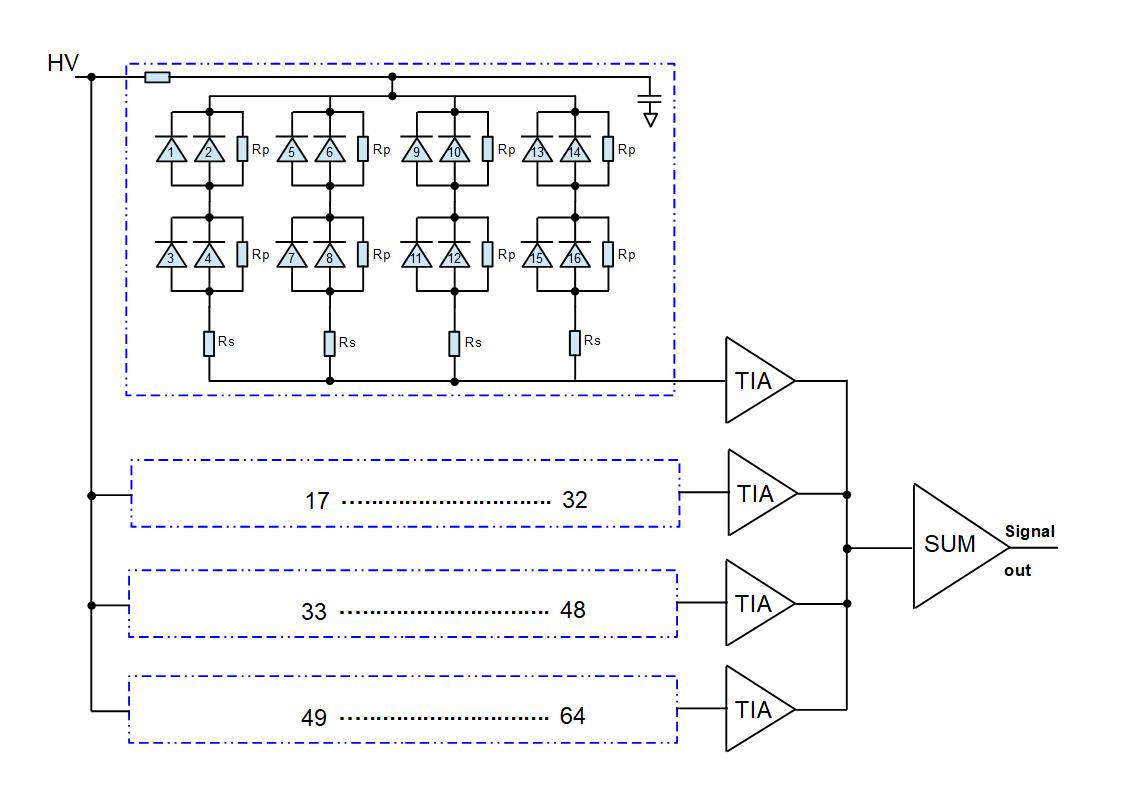}
  \caption{ Front-end board scheme of the discrete component option.}
  \label{fig:FEBscheme}
\end{figure}

On the same tile, the SiPMs are ganged together with series and parallel connections, an example of this passive ganging is shown in Figure~\ref{fig:FEBscheme}. Series connection reduces the equivalent capacitance by a factor of 2.

Each FEB reads one tile through a board-to-board connector. A first amplifier stage converts the SiPM output charge to a voltage pulse that is further amplified by a second stage. An analog signal from the whole tile is presented on the FEB output connector.

The amplified output is sent to the FEC that collects data from 32 FEBs. 128 FECs are needed for $\sim 4100$ tiles. All the FPGA boards will act as white rabbit nodes to ensure sub-nanosecond synchronization.

On the FEC, each input from the FEB is digitized by a dedicated ADC. A fast comparator ensures a good time resolution independent of the ADC's sampling rate. Digitized data collected by the FPGA is sent outside the cryostat via an optical fiber link.

The CU devoted to manage the FEBs is located outside the cryostat. It collects data from the 128 optical links. It is connected to the white rabbit network. Moreover, the CU performs a first-level triggering and event building, and manages the data transfer.

\paragraph{Front End Board (FEB)} \mbox{}\\
\label{sec:feb}

\begin{figure}[htb]
  \centering
  \includegraphics[width=0.8\textwidth]{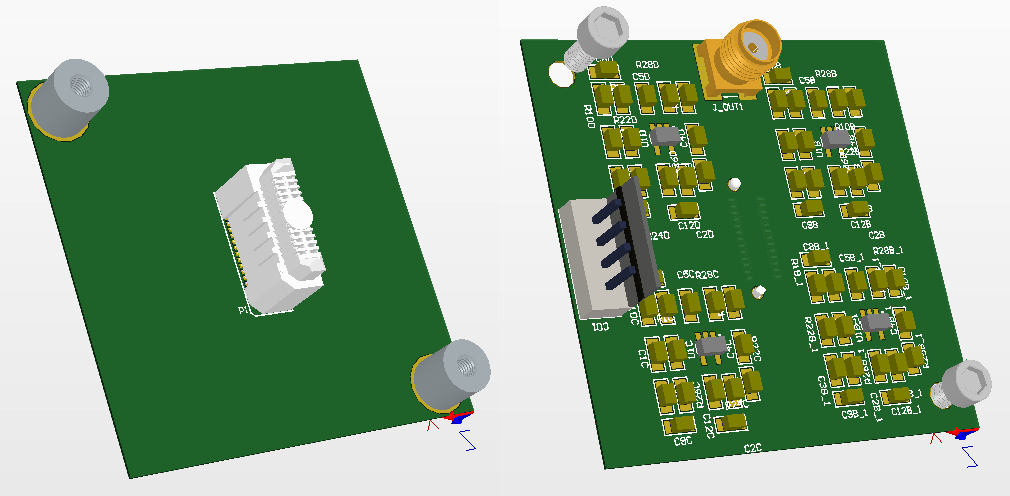}
   \caption{SiPM tile with fixing structure (left picture). Two M3 screws will bind the FEB (right picture) to the tile. The SMA connector for signal output and the supply connector are on the back side of the FEB. \label{fig:FEBfix}}
\end{figure}

The basic structure of each FEB consists of a number of Trans-Impedance Amplifier (TIA) nodes, up to 6, in order to read out of a large area SiPM tile ($\sim 5 \times 5$ cm$^2$ or more). The baseline is with 4 TIAs. Each one manages 6~cm$^2$ of SiPMs. The tile will be mechanically restrained to the FEB with a solid fixture like W{\"u}rth REDCUBE SMT series. Figure~\ref{fig:FEBfix} shows the drawing of the FEB and the connector side of a tile. This solution allows a very simple installation of FEBs. A single connector simplifies the design of the supporting copper shell and gives more space for the components. On the opposite side of the PCB a four-pin connector provides power to the FEB and the SiPM tile. All the amplifiers use the same +V/-V supply.  On the same side the analog output is routed to the FEC through an SMA connector.

\begin{figure}[htb]
  \centering
  \includegraphics[width=0.95\textwidth]{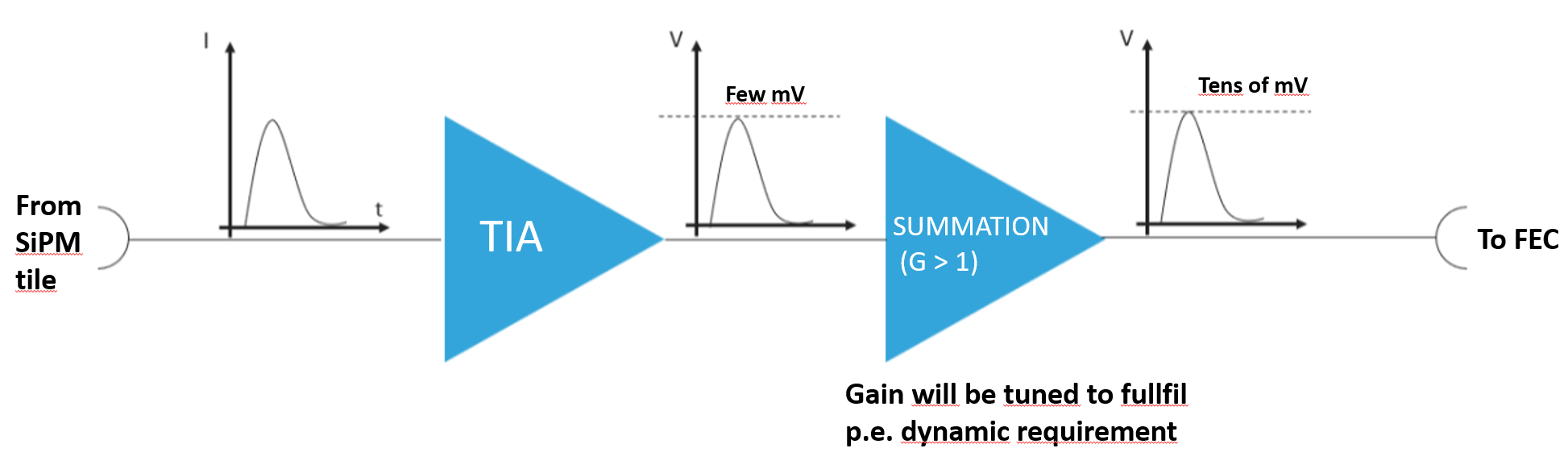}
  \caption{Block scheme about how the SiPM signal is managed by FEB.}
  \label{fig:FEBchan}
\end{figure}

The TIAs share the capacitance of the SiPM tile. A summation amplifier adds the output of the TIAs into an analog sum. This second stage can be used to enhance the total gain to ensure the best compromise between resolution of single p.e.\ and linearity for high p.e. Figure~\ref{fig:FEBscheme} shows the connection of the SiPMs to the FEB and Figure~\ref{fig:FEBchan} sketches the processing of the signal. More TIA nodes can be added on a FEB to further lower the input capacitance, especially, if we decide for larger tiles. The major advantage is the input signal bandwidth. This value is directly affected by the input capacitance. It will increase if more SiPMs are connected in parallel.

SiGe technology is a good option for the FEB operational amplifers due to the -50$^\circ$C detector working temperature and their high bandwidth. A resolution below 20\% on single photoelectron has already been achieved with this approach, while a 2~Vpp (Peak-to-Peak Voltage) output dynamic range should ensure linearity in 1-50~p.e.\ range. A power consumption of 150~mW is expected for each FEB. The board will be directly connected with the supporting copper shell, simplifying the heat removal.

\begin{figure}[htb]
  \begin{minipage}[ht]{0.38\linewidth}
  \centering
          \includegraphics[width=\linewidth]{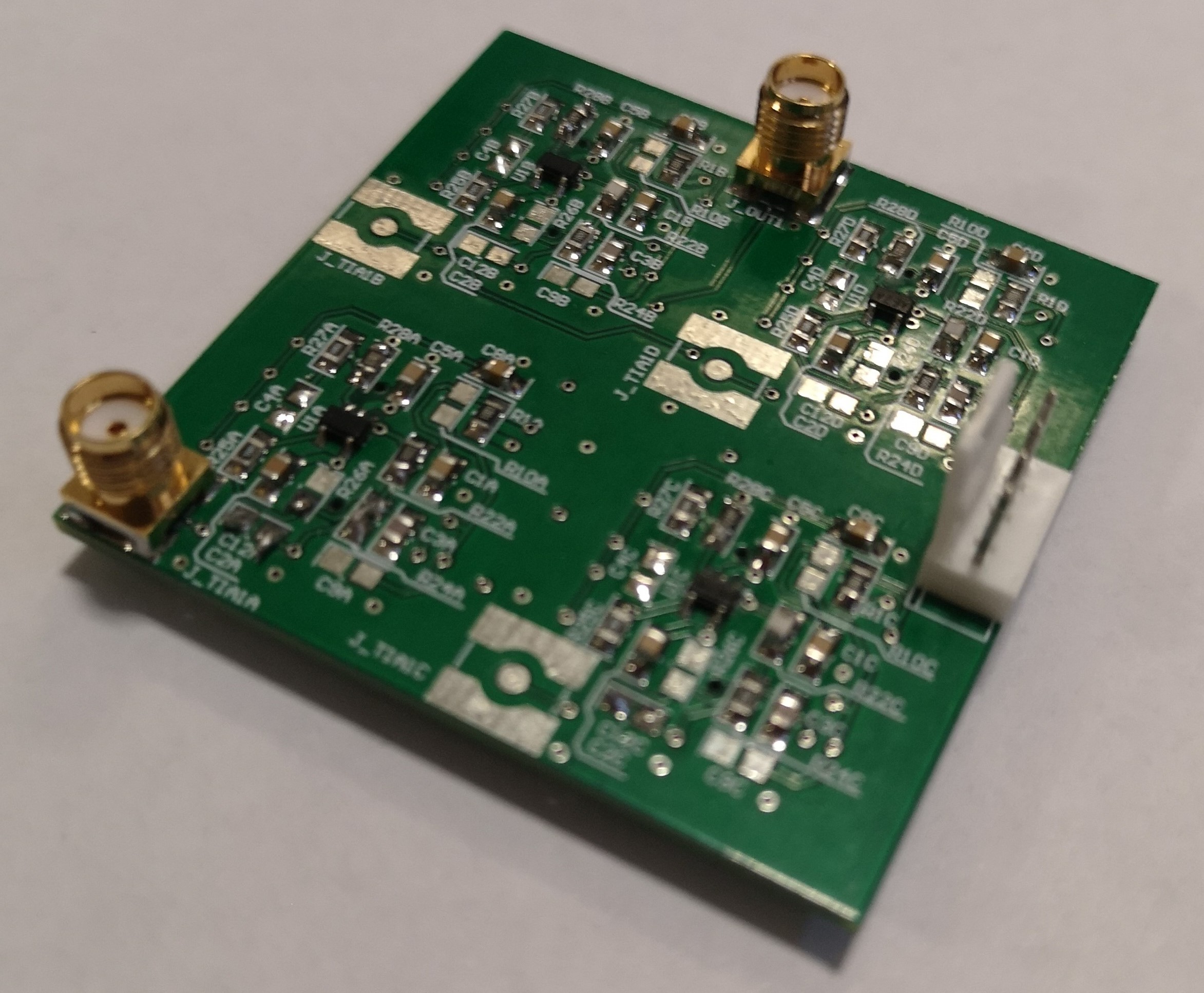}
  \end{minipage}
  \hfill
  \begin{minipage}[ht]{0.53\linewidth}
  \centering
    \includegraphics[width=\linewidth]{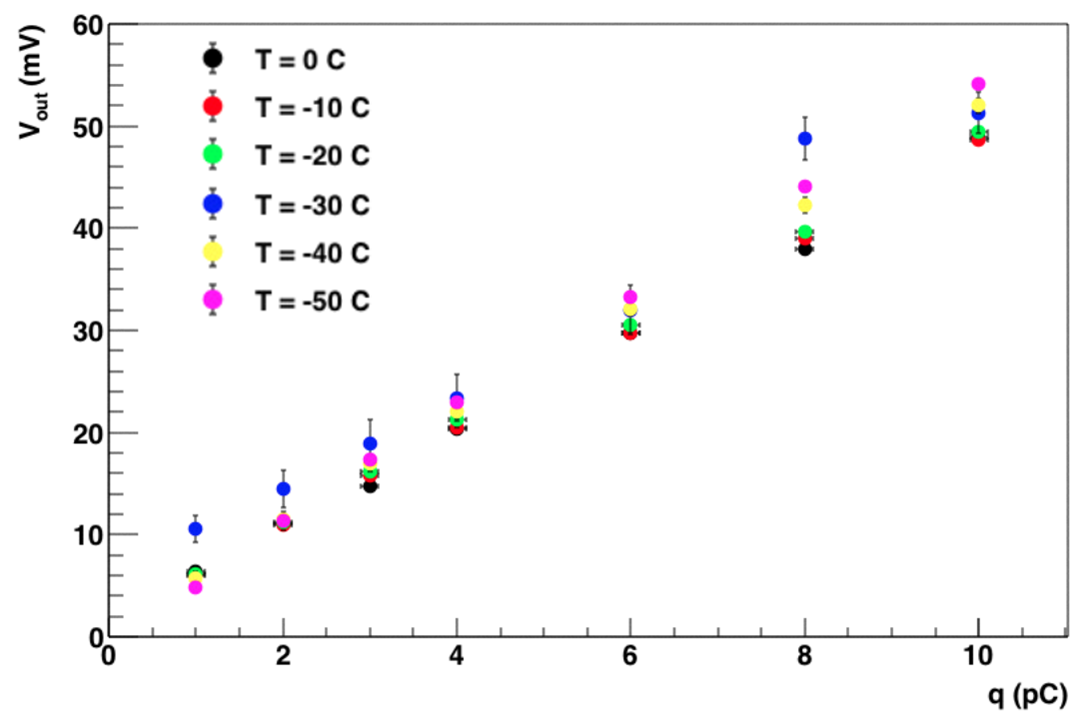}
  \end{minipage}
  \caption{Left picture shows a FEB prototype with two SMA connectors, one for the summation output, another connected to one of the TIA's output for debugging purpose. Right plot shows the TIA output linearity at different temperatures. The x-axis is the injected charge and the y-axis is the output voltage.  \label{fig:FEBboard}}
\end{figure}

\begin{figure}[htb]
  \centering
  \includegraphics[width=0.7\textwidth]{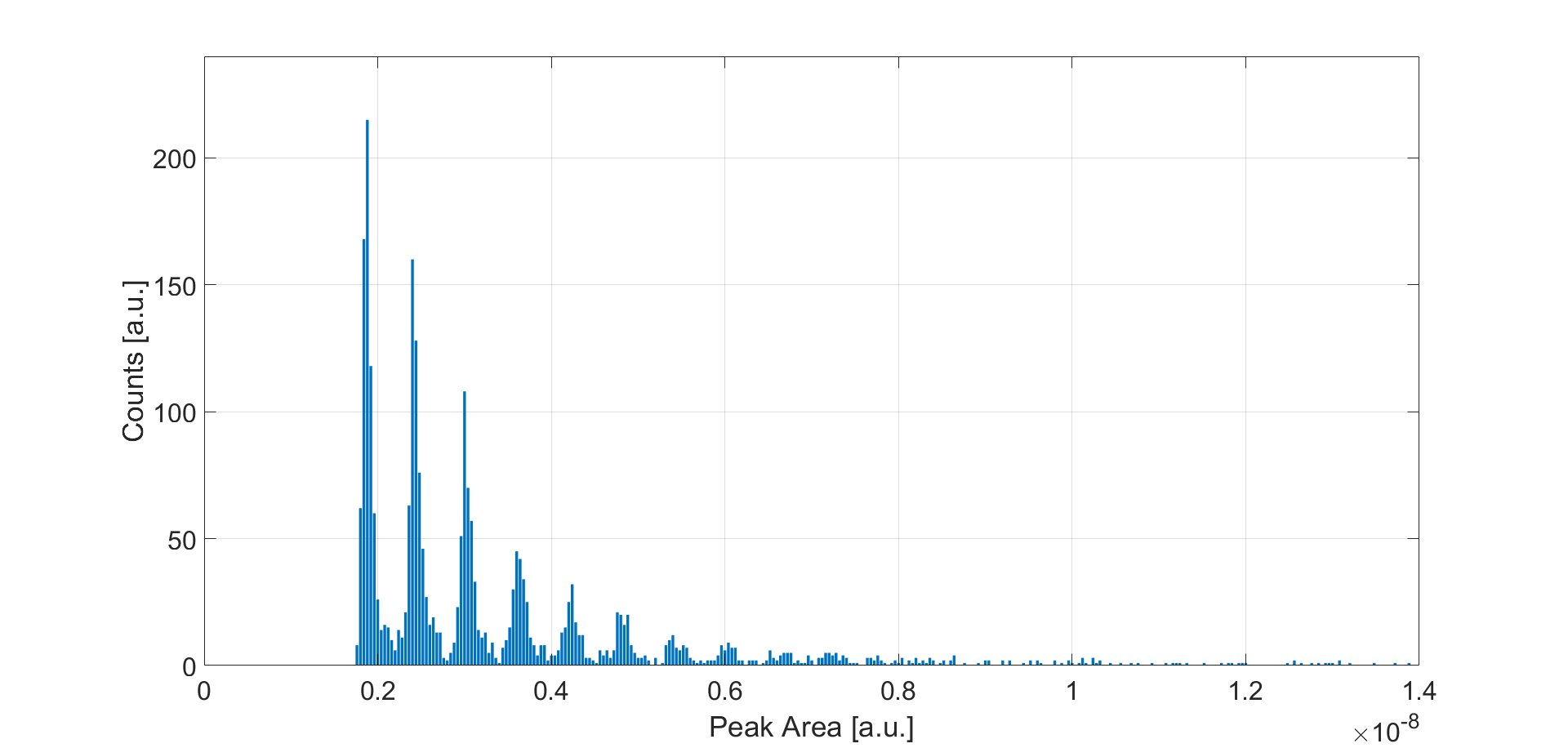}
  \caption{Charge spectrum obtained with the FEB prototype coupled to a 1-cm$^2$ SiPM.}
  \label{fig:FEBfinger}
\end{figure}

A preliminary version of the FEB board has been built and extensively tested. Following the results of these tests a new version has been produced. It is shown in Figure~\ref{fig:FEBboard}. In the same figure, linearity obtained with a current source is shown. A measured charge distribution from dark counts from an 1 cm$^2$ SiPM is shown in Figure~\ref{fig:FEBfinger}. It demonstrates the single photon counting capability and resolution of the proposed electronic.

\paragraph{Front End Controller (FEC)} \mbox{}\\
\label{sec:FEC}

\begin{figure}
    \centering
    \includegraphics[width = 0.8\textwidth]{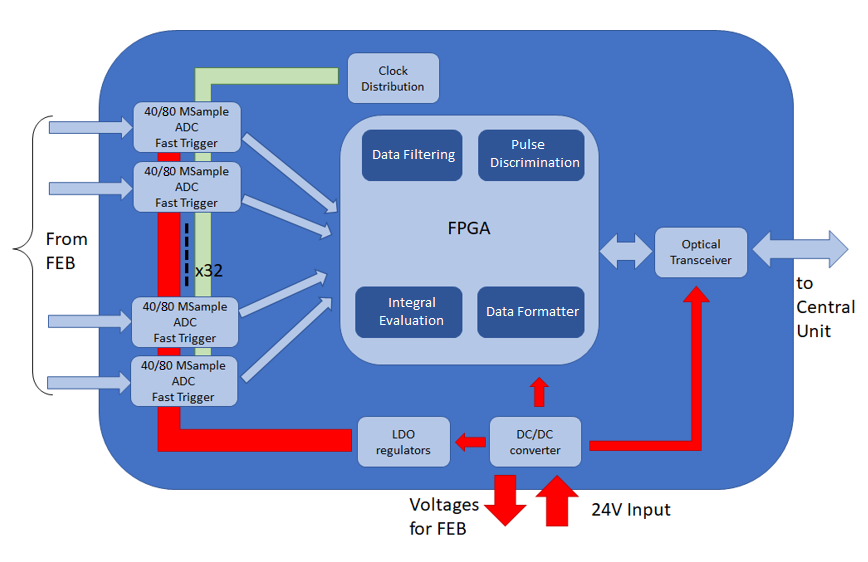}
    \caption{Block scheme of the Front-end controller in the discrete component option.}
    \label{fig:FECblock}
\end{figure}

The FEC controls the FEBs through an FPGA. Figure~\ref{fig:FECblock} shows a block diagram of the FEC. Cable assemblies like the EQRF series from SAMTEC can be used to connect the FEB's SMA connectors to the FEC. Cable lengths from 15 cm to 100 cm are possible.

The first option is to place each FEC near its FEBs. The second option might be to place all the FECs on the top of the cryostat to simplify the thermal management. A precise
evaluation of the effect of different cable lengths is necessary. A third option is to put all the FEC boards outside the cryostat.

Highly efficient DC/DC converters will be placed on the board to supply the FPGA and ADCs. We aim at least at 90\% efficiency to fulfill the power constraint. DC/DC converters for the FEBs can be located on the FEC to avoid other supply distribution boards inside the cryostat.

Each FEC collects the output of 32 FEBs. Each channel is digitized with a $40 - 80$ MSample/s ADC. Each SiPM pulse is shaped to $\sim 200$~ns. About $10 - 20$ samples are recorded for each pulse. Due to the integer nature of the SiPM signal $($the output signal represents the sum of \textit{n} SPAD signals$)$, 10 samples are sufficient to obtain a good estimate of the number of p.e.
The integral of each pulse can be computed inside FPGA, then a p.e.\ number can be associated to the obtained value. A sliding window integration can reduce the noise effect. Each channel produces a single byte of data to cover the range of 0-255 p.e. Zero suppression can be applied to reduce the bandwidth. The bandwidth becomes independent of the ADC's sampling rate, if we transfer only the computed p.e.\ number.

Most of the ADCs on the market have LVDS (Low Voltage Differential Signal) output. To manage 32 ADCs (32 differential pairs plus differential clock and control signals), an FPGA with 100 I/O ports is sufficient. The white rabbit core for synchronization can be installed on most of the seventh generation Xilinx FPGAs and on some Ultrascale models, too. The control and synchronization of ADCs can be realized on the same FPGA. The same link can be used for synchronization and data transfer. This approach can be useful to reduce power consumption of the FEC.

Multi-channel ADCs like the AD9249 consumes 58~mW per channel at a sampling rate of 65~MSample/s with 14 bit resolution. Then 2~W are needed for the 32~ADC channels. Each input link will be monitored by a fast comparator with an adjustable threshold near 0.5~p.e., the output of each comparator is directly connected to the FPGA. A timestamp will be generated each time a comparator output fires. The generated timestamp will be associated to the corresponding data from the ADCs. Another approach is to obtain a timestamp better than the sampling period with digital filters. Both approaches, fast comparator and digital filters, can be evaluated to ensure few nano-second resolution.

With a Giga-sample flash-ADC, the need for a fast comparator can be avoided. The output bandwidth is independent of the sampling rate, but the power consumption would increase.

\begin{figure}[htb]
    \centering
    \includegraphics[width = 0.55\textwidth]{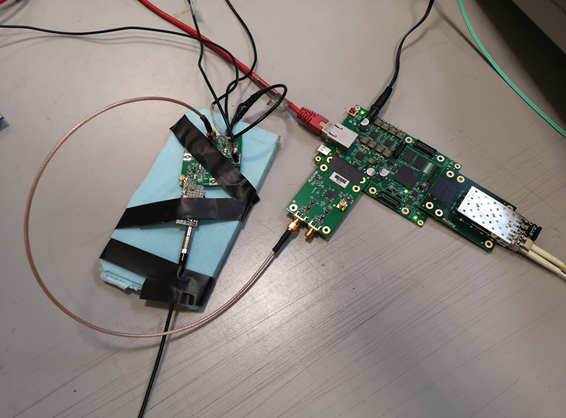}
    \caption{Test setup with the FEC board (right one) connected to a FEB (left one). The FEC prototype has two separate piggyback boards to test the ADCs and optical transceivers.  }
    \label{fig:FECboard}
\end{figure}

\begin{figure}[htb]
  \begin{minipage}[ht]{0.48\linewidth}
  \centering
\includegraphics[width=\linewidth]{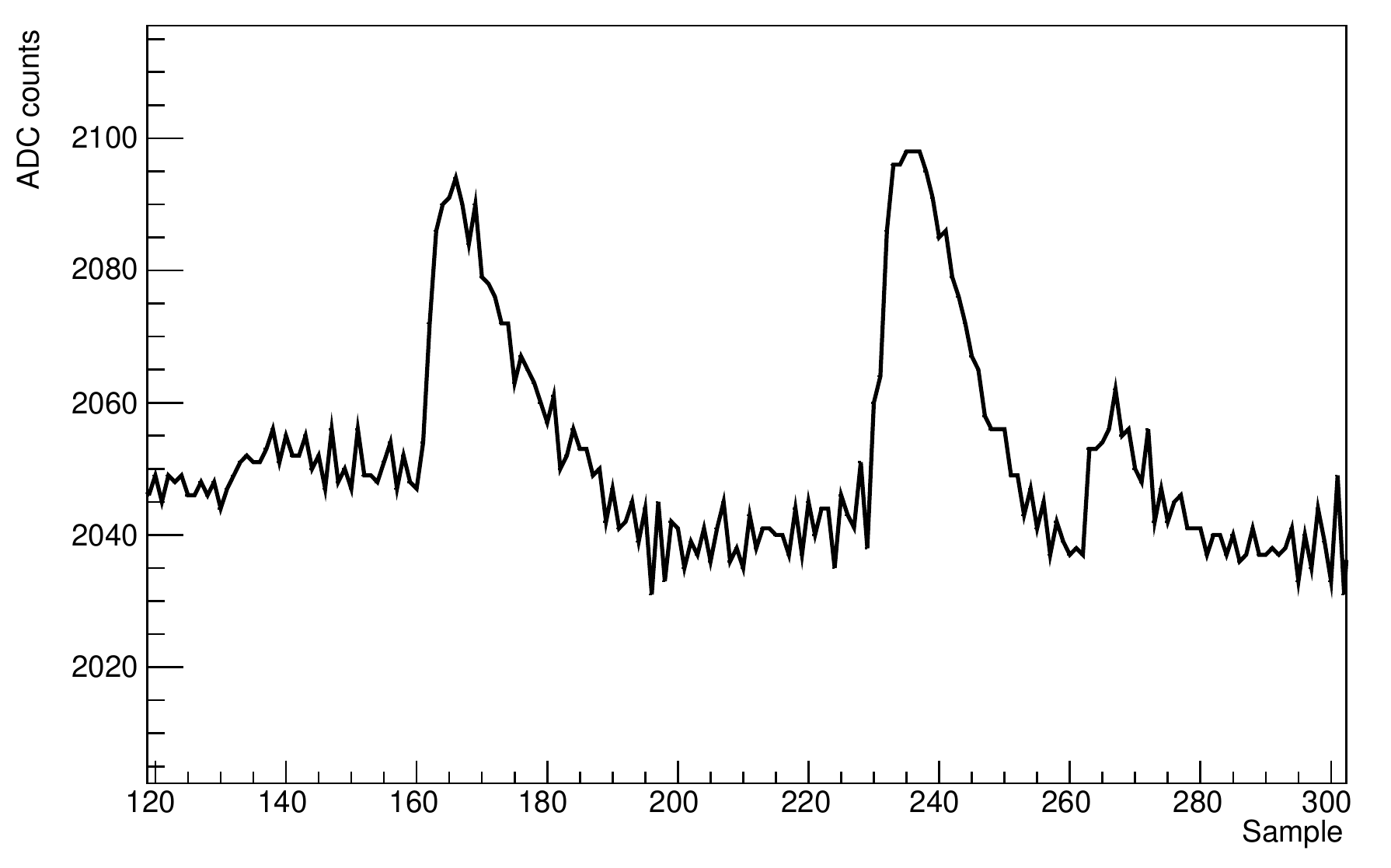}
  \end{minipage}
  \hfill
  \begin{minipage}[ht]{0.48\linewidth}
  \centering
        \includegraphics[width=\linewidth]{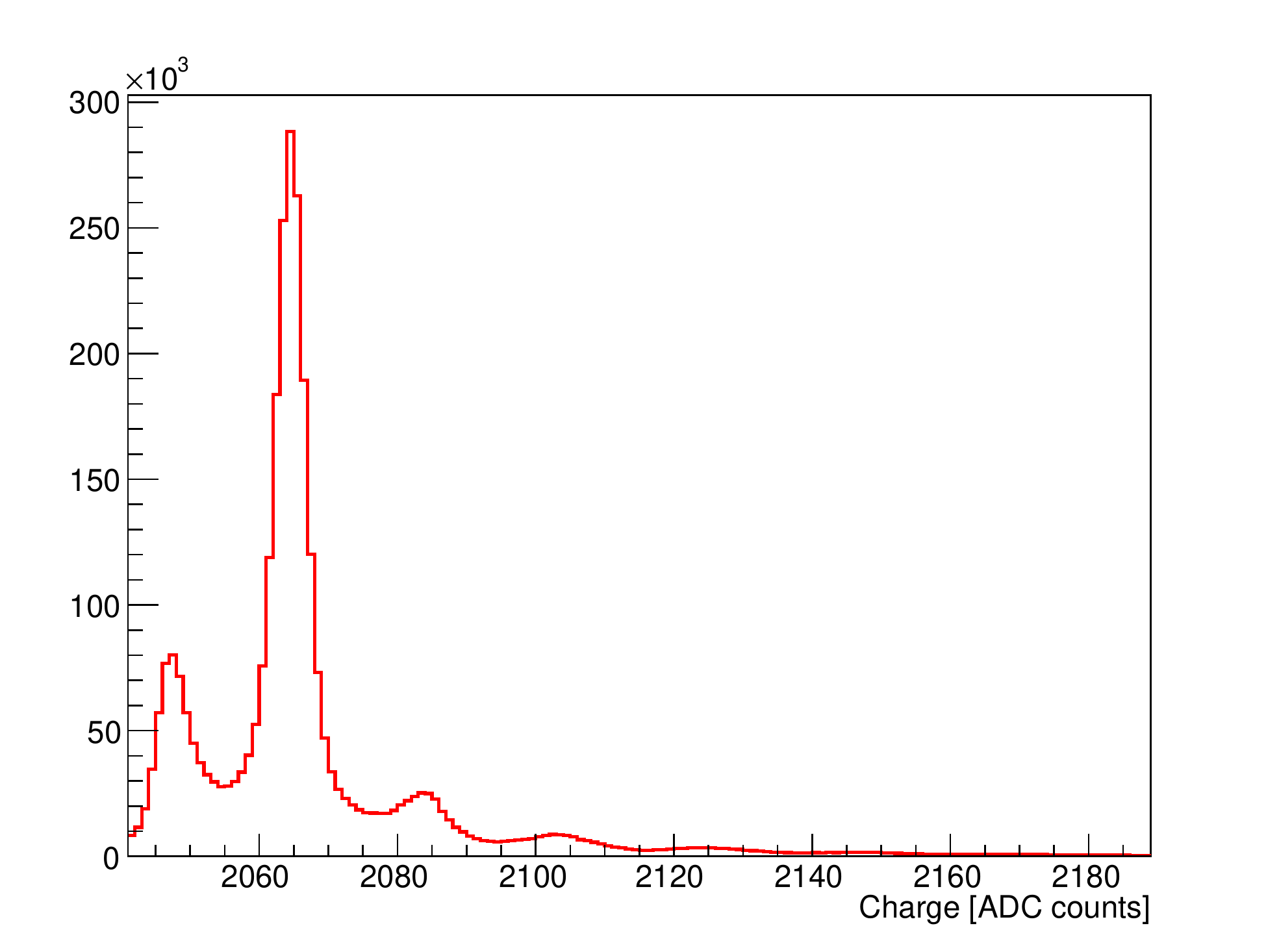}
  \end{minipage}
  \caption{Left picture shows the TIA output acquired by the FEC prototype. The x-axis is the acquired samples and the y-axis is ADC counts. Right picture shows the charge spectrum evaluated inside FPGA by computing pulse areas.}
  \label{fig:FEC_finger_plot}
\end{figure}

A preliminary version of the FEC with 2 ADC channels has been tested. It is shown in Figure~\ref{fig:FECboard}. Figure~\ref{fig:FEC_finger_plot} shows the measured charge spectrum from dark counts from a 1-cm$^2$ SiPM obtained with a FEB board connected to the FEC. The charge integral is computed inside the FPGA, thus demonstrating the single photon counting capability and resolution of the proposed electronics.

\paragraph{Central Unit (CU)} \mbox{}\\
\label{CU}

The CU will be placed outside of the cryostat with less constraints on power budget and the temperature range of the components. A high-end FPGA like the Xilinx Ultrascale+ can manage all the optical links coming from the FECs and has enough digital signal processing slices to perform online triggering.

%% file: TriggerDAQDCS/section.tex
\section{Trigger, DAQ and DCS}
\label{sec:TriggerDAQDCS}
\blfootnote{Editors: Xiaolu Ji (jixl@ihep.ac.cn), Paolo Montini (paolo.montini@uniroma3.it), and Mei Ye (yem@ihep.ac.cn)}
\blfootnote{Major contributor: Fabrizio Petrucci}

\subsection{Trigger requirements and conceptual design}
The TAO trigger will have to cope with the signals of $\sim4100$ SiPM tiles of the central detector. The average number of photoelectrons detected by each SiPM tile ranges from one for low energy events up to hundreds for high energy events with vertices located very close to the detector border. The design goals of the trigger system  are summarized below:
\begin{itemize}
    \item The efficiency for IBD events produced by reactor $\bar \nu_e$ should be close to one for an energy deposit greater than 1 MeV.
    \item The trigger system should be able to suppress the detector-related background and reduce the random coincidences due to the SiPM dark counts.
\end{itemize}
A trigger system based on a distributed FPGA architecture can be an optimal and scalable solution for either the discrete component or ASIC readout options.

\subsubsection{Discrete component readout option}

\begin{figure}[htb]
    \centering
    \includegraphics[width=0.47\textwidth]{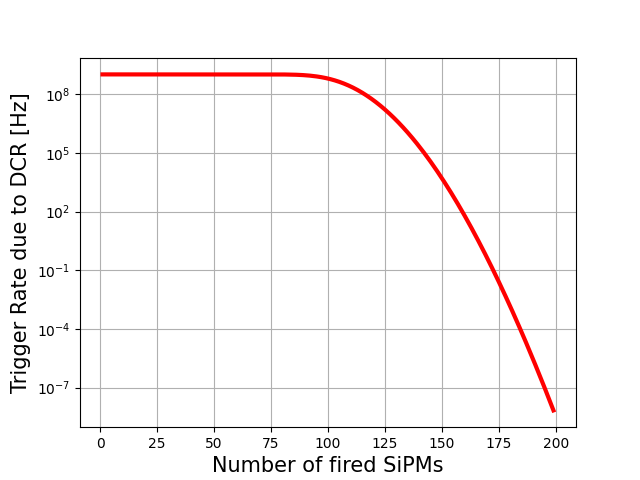}
    \includegraphics[width=0.51\textwidth]{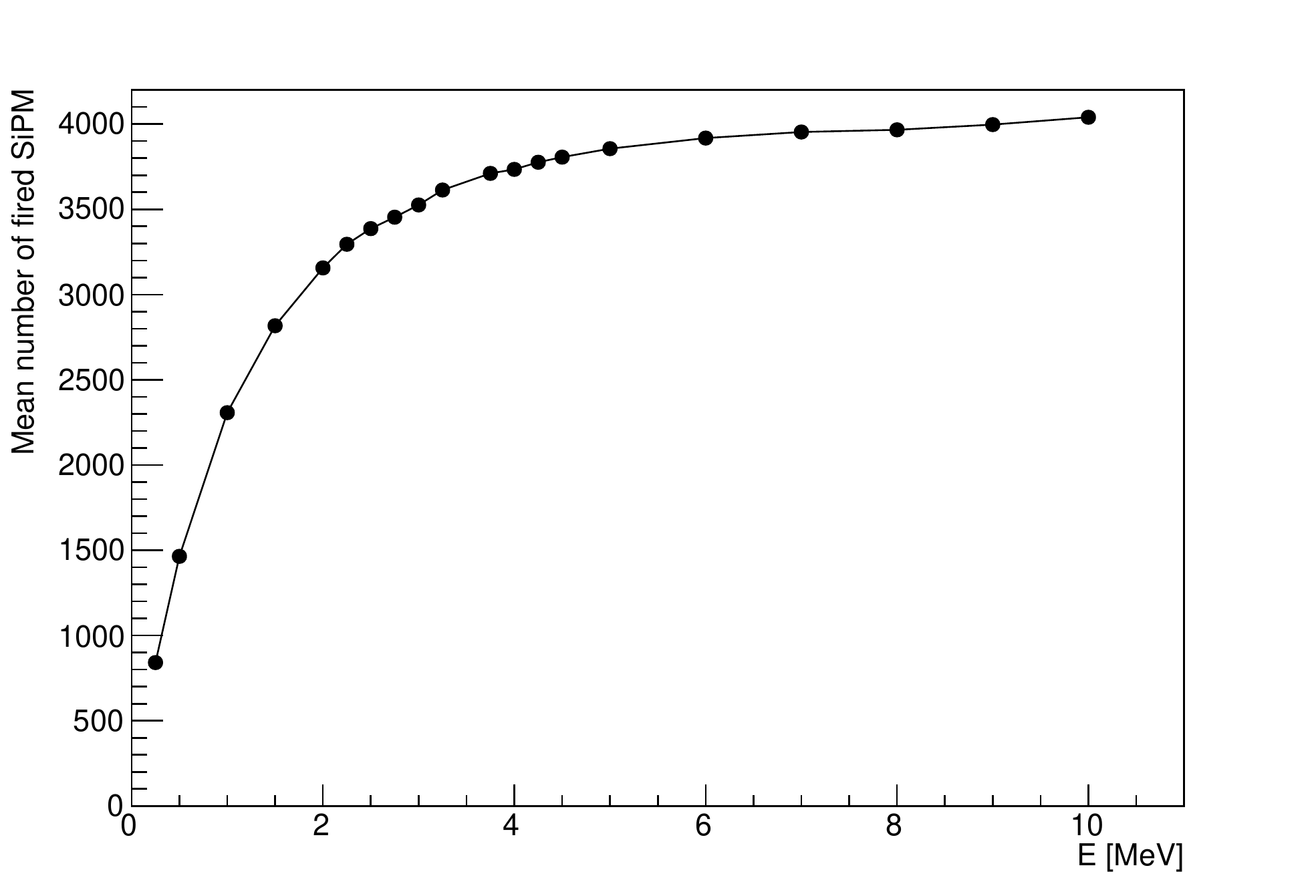}
    \caption{Left: expected trigger rate due to the dark counts as a function of the number of coincident SiPMs in a 100 ns time window assuming a dark noise of 100 Hz/mm$^2$ (250 kHz/channel). Right: Mean number of fired SiPMs within a 100 ns time window as a function of energy. Only the statistical error (not visible on this scale) is reported. The continuous line is only added to guide the eye. }
    \label{fig:darkcounts}
\end{figure}

In the discrete component readout option, described in Sec.~\ref{sec:discrete}, the Front-End Controller (FEC) board will perform a first level data management including setting of the thresholds, zero suppression and a first level (L1) trigger based on local majority and event topology. A second level (L2) trigger system will be necessary to select correlated SiPM hits produced by physical events from a sea of random dark noise hits. The L2 trigger algorithms will be implemented in the Central Unit (CU).  A simple majority logic based on the coincidence of different channels in a certain time window will suppress the dark count rate by several orders of magnitude. At -50$^\circ$C the Dark Count Rate (DCR) of a single $5\times5$ cm$^2$ tile is $\sim 250$ kHz. As shown in Figure~\ref{fig:darkcounts}, considering a coincidence time window of 100 ns, the global dark count rate will drop well below 1 kHz by requiring the coincidence among $\sim150$ channels. By requiring the coincidence of $\sim 160$ (170) channels within 100 ns the global DCR drops down to 10 Hz (1~Hz).
Additional L2 algorithms can also be implemented, for example a high-multiplicity trigger algorithm in order to tag events with a large energy deposit close to the liquid scintillator volume border. The signature of the neutron capture by gadolinium or hydrogen nuclei can be exploited in order to tag the IBD events either offline or online, thus reducing the background event rate. A preliminary estimation of the expected trigger rates is reported in Table~\ref{tab:trigger}.

\begin{table}[htb]
\setlength{\belowcaptionskip}{5pt}
\begin{center}
\caption{Expected trigger rates. \label{tab:trigger}}
\begin{tabular}{r c c}
    \hline\hline
    Signal &  Event rate & Trigger rate  \\
    \hline
     Reactor IBD & 4000/day& 4000/day\\
     Radioactivity background & $~100$ Hz & $~100$ Hz \\
     Muons & 150 Hz &  150 Hz\\
     SiPM dark counts & 1.0 GHz &  $< 1$ Hz\\
     \hline
\end{tabular}
\end{center}
\end{table}

\subsubsection{ASIC readout option}
Similar approach can be followed in the ASIC readout option, described in Sec.~\ref{sec:ASIC}. An FPGA board similar the FEC will be implemented in order to manage the signals coming from each ASIC chip. Each board will be able to control 36 ASICs. A first level local trigger can therefore be implemented in the FPGA board in order to search for locally correlated SiPM hits produced by physical events and therefore suppress the background due to the SiPM dark counts. Each FPGA board will be connected via a high-speed optical link to the Central Unit, where second level (L2) trigger algorithms can be implemented thus provide a global trigger condition.

\subsection{DAQ conceptual design scheme}
The main task of the Data Acquisition System (DAQ) is to record the antineutrino events. The DAQ has also to cope with different kind of events like muons and low-energy radioactive decays in order to understand the detector background. The DAQ has to record the data from the antineutrino and the muon veto detectors with precise timing and charge information and build the event from the data fragments coming from all the electronics devices.

\begin{figure}[htb]
    \centering
        \includegraphics[width=0.75\textwidth]{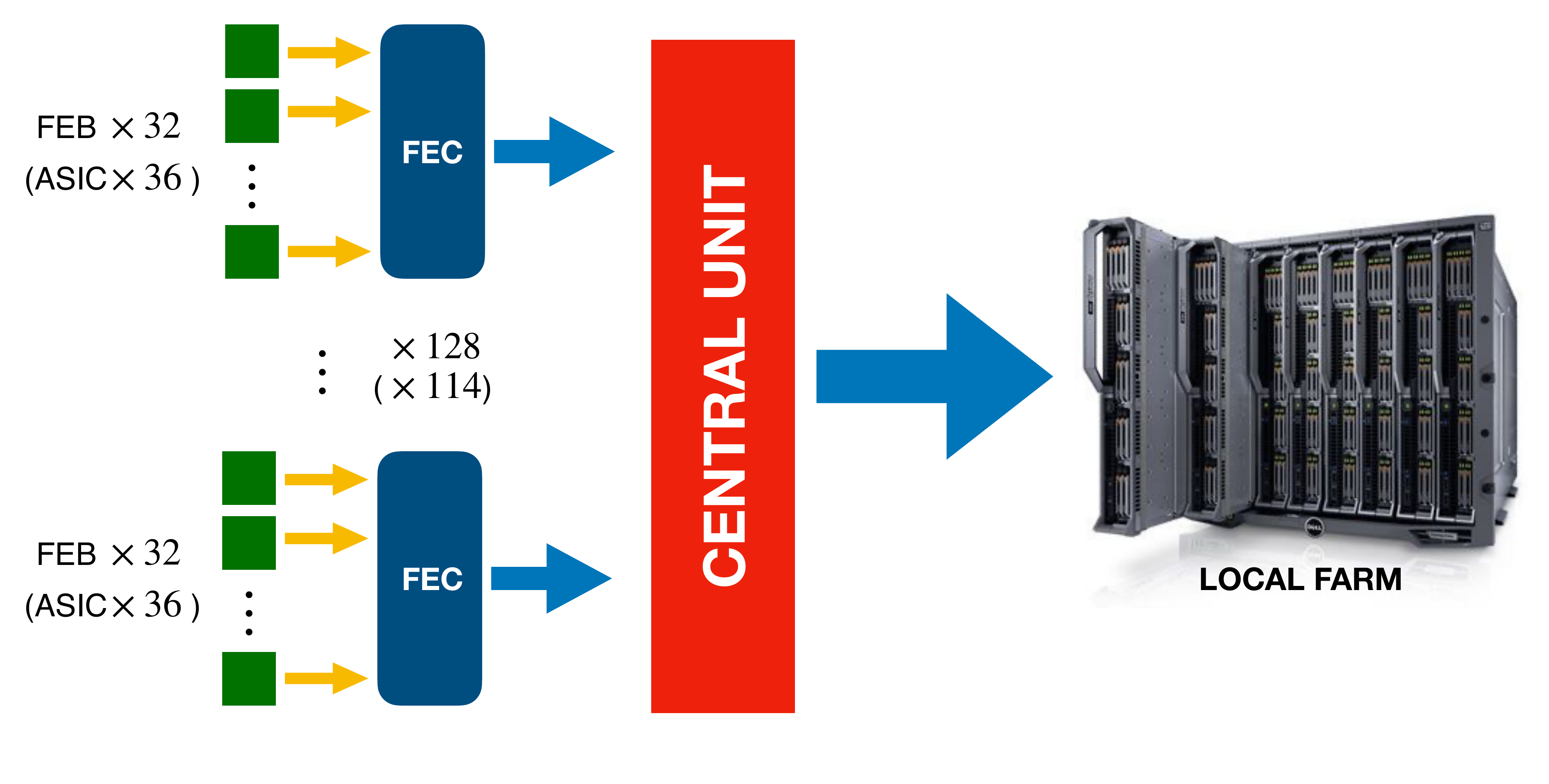}
    \caption{DAQ conceptual scheme.}
    \label{fig:daqschema}
\end{figure}

The TAO DAQ system is based on a multi-level distributed FPGA architecture.
In the discrete component readout configuration, shown in Figure~\ref{fig:daqschema}, the first level is dedicated to the management of the signals coming from the Front-end Electronics Boards (FEBs). Each FEC manages 32 FEBs and is equipped with a Xilinx Zinq/Kintex FPGA that allows continuous sampling of the signal coming from the FEBs and performs the estimation of the number of photoelectrons detected.  The trigger will be managed by the Central Unit, based on a Xilinx Ultrascale+ generation high-performance FPGA, briefly described in Sec.~\ref{CU}. The CU will receive the data from each FEC board via a high-speed optical link. Charge, timestamp and channel ID of each SiPM readout will be stored in a cyclical memory buffer. The FPGA in the CU will manage the data, look for coincidences, assert the trigger condition and build the event. In normal running conditions only the charge (i.e.\ number of photons) and time recorded by each SiPM readout will be stored. The FEC and CU firmware will also provide the possibility of saving the whole waveform which can be useful especially during the debugging, commissioning and calibration phases.

Similar architecture can also be exploited in the ASIC option, as shown in Figure~\ref{fig:daqschema}. An FPGA similar to that used in the FEC will manage the signals coming from 36 ASIC chips and deliver them to the CU that will manage the trigger and the event building. Even charge, timestamp and ID of each channel will be recorded in this approach, the ASIC does not allow to keep the whole waveform.
All the events satisfying the trigger condition will be sent to a computing farm via optical links or fast ethernet and written to disk.

\subsection{Rate and data throughput estimation}
\label{sec:rate}
The DAQ system based on the multi-level FPGA architecture provides a real-time estimation of the number of photoelectrons seen by each SiPM tile and a nanosecond timing resolution. In the ideal case the total data volume should not exceed 100~Mbps, thus allowing an easy data transfer via standard Ethernet.

\subsubsection{Discrete component readout option}
In the discrete component readout option, the electronics plans to use 16 bits for the number of photoelectrons and 16 bits for the time for each channel. Additional 32 bits will be used for control purposes. The data size of a single SiPM signal is therefore 64 bits including the header and the SiPM readout ID. The DAQ system has to deal with $\sim 4100$ channels, one for each SiPM tile. The data throughput of the TAO detector can be preliminarily evaluated for the discrete component electronics configuration. The main contributions are reported in Table~\ref{tab:tput}. The total data production rate will be less than 100~Mbps.

\begin{table}[htb]
\setlength{\belowcaptionskip}{5pt}
\begin{center}
\caption{Expected data throughput for the discrete component readout option ($\sim 4100$ channels). \label{tab:tput}}
\begin{tabular}{ccccc}
    \hline\hline
    item & average energy& event size & event rate &  data volume\\
     & [MeV] & [kbit] & [Hz] &  [kbps]\\
    \hline
     IBD & 4 & 256 & $0.05$ &  12.8 \\
     SiPM dark counts & - & $\sim 8$ ($\sim 3\%$ occupancy) & 1 & 8 \\
     Radioactivity background & 1.5 & 256  & $\sim 100$ & 25600 \\
     Muons & 223 & 256 & 153 & 38400 \\
     Cosmogenic background & 23 & 256 & 13 & 3360\\
     \hline
     Total & & & &  $< 100000$\\
     \hline
\end{tabular}
\end{center}
\end{table}

\subsubsection{ASIC readout option}
In the ASIC readout option, the DAQ system has to deal with about 140,000 channels. The event size is 52 bit/channel, 40 bits for the charge and the timestamp and 12 bits for the chip ID. The expected data throughput is reported in Table~\ref{tab:tput_asic}.

\begin{table}[htb]
\setlength{\belowcaptionskip}{5pt}
\begin{center}
\caption{Expected data throughput for the ASIC readout option ($\sim 140,000$ channels). \label{tab:tput_asic}}
\begin{tabular}{ccccc}
    \hline\hline
     &  average energy& event rate & \multicolumn{2}{c}{data rate [Mbps]}\\
    &  [MeV]  & & No global trigger & Global trigger \\
    \hline
     IBD & $\sim4$ & $0.05$ Hz & 0 & 0 \\
     SiPM dark counts & -- &  100 Hz/mm$^2$ & 52000 & 0  \\
     Radioactivity background & 1.5 & $\sim100$ Hz & 42 & 42 \\
     Muons & 223 & 153 Hz &2200  & 2200 \\
     Cosmogenic background & 23 & 13 Hz &78 & 78\\
     \hline
     Total & & & $54320$& $2320$\\
     \hline
\end{tabular}
\end{center}
\end{table}

In this case the expected data production rate is greater than 2~Gbps. It is therefore essential to apply data reduction algorithms in order to reduce the data volume to fit in 100~Mbps.

The veto data throughput has been evaluated to be only a small fraction of the total volume.

\subsection{Online system}
The online system is designed to collect the physics data from the electronics readout modules, build events, process \&  compress data, and finally save the data to disk.

\subsubsection{Requirements}

\paragraph{Performance}
The data throughput of TAO has been discussed in the Sec.~\ref{sec:rate}. The online system should support the data transfer bandwidth with negligible dead time.

\paragraph{Data Processing}
As described above, for the discrete component readout option, the input data bandwidth will be $<100$~Mbps. In this case, online system can save all the data to storage space. For the ASIC readout option, the input data rate will reach about 2.3~Gbps even after the global trigger. Therefore, to reduce the data rate to less than 100~Mbps, online data compression in real time is needed. The compression algorithm should be carefully studied to make sure the compression ratio can meet the bandwidth requirement. The performance of the algorithm should be optimized taking into account the online CPU resources, reliability and stability.

\paragraph{Other Functions}
Besides the basic dataflow related functions, the online system will also provide the common functions like run control, run monitoring, configuration, data quality check, information sharing, and so on.

\subsubsection{Conceptual design scheme}

\paragraph{Hardware Structure}
The hardware platform of the online system will be based on the advanced commercial computing servers and network equipments. The data rate produced by TAO can be managed by means of a computing infrastructure based on commercial products already available on the market thus ensuring robustness and redundancy at an affordable cost. Data will be stored in a local computing farm and subsequently transferred to IHEP data centers.
A $\sim 50$ TB onsite storage will fulfill the TAO requirements. The storage can be managed by one file server; an additional one is however foreseen for redundancy. Two online servers will be needed for the dataflow management and online functions such as computing, run monitoring, run control, data quality monitoring, and so on. The data transfer can be managed by using standard Ethernet and iSCSI interfaces. Multi-port 10~Gbit network switches will be used to manage the storage and server communications.

\begin{figure}[htb]
    \centering
        \includegraphics[width=0.75\textwidth]{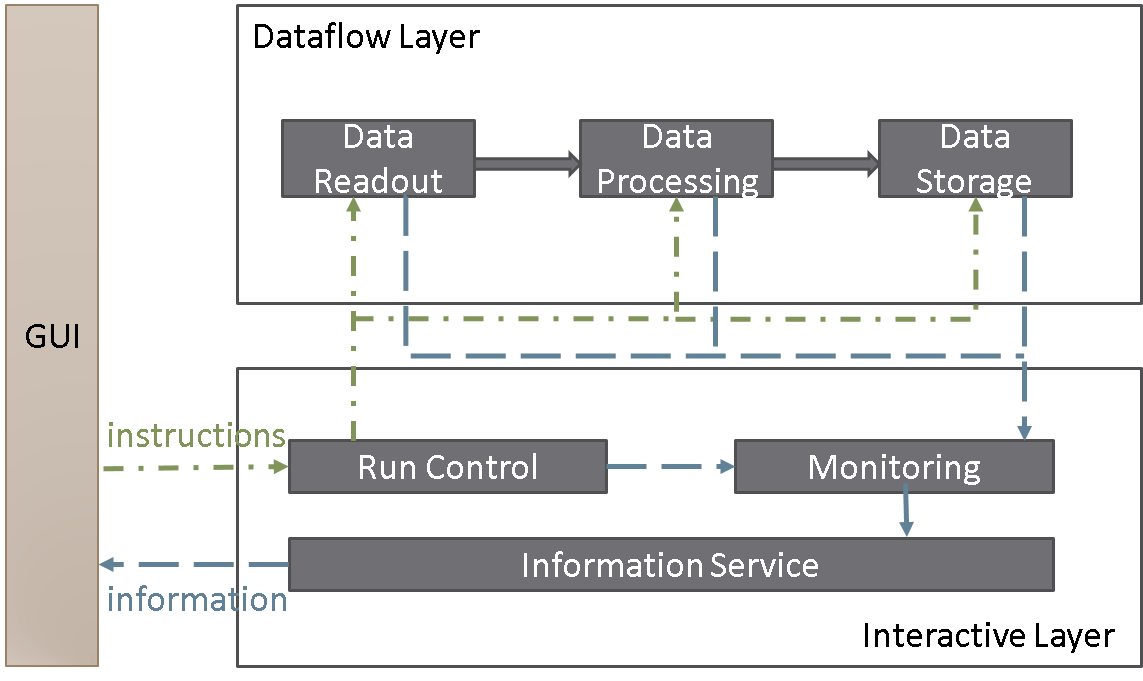}
    \caption{Online software framework.}
    \label{fig:online_soft_frame}
\end{figure}

\paragraph{Software Framework}
Based on the functional requirements of the online system, the software framework design can be divided into two layers: the dataflow layer and the interactive layer, as shown in Figure~\ref{fig:online_soft_frame}. The two layers cooperate with each other to achieve full function of the online system.

\begin{itemize}
  \item The dataflow layer is responsible for the data transfer from the Readout electronics, processing (event building) and storage.
  \item The interactive layer is responsible for all the management, controls and operations during data taking. It is the structure foundation of the online system. It is used for information transmission, real-time monitoring, online display, and so on. And it also provides the interface between users and the online system.
\end{itemize}

The two layers work independently and are connected through Ethernet, reducing interferences and thus improving the robustness of the system.

\subsection{Detector control system}

\subsubsection{Requirements}
In order to keep the TAO detector data taking stable and to guarantee the performances for a long-term operation, a standard Detector Control System (DCS) will be implemented. The main task of the DCS is to monitor and control the working conditions of the detector and to raise alarms if a specifically monitored parameter goes out of range. The DCS system is designed with different modules which are responsible for the following tasks:
\begin{itemize}
  \item Monitoring and run control
  \begin{itemize}
  \item control of the high voltage system and the SiPM low voltage power supply;
    \item monitoring of the electronics racks \& crates for the veto and central detector. The monitored parameters include: electronics temperature, power supply, fan speed, over-current protection;
the sensors include the low temperature sensors with oil-proof packaging (-70$^\circ$C to 30$^\circ$C);
    \item monitoring of the gas system, including the radon sensors, the oxygen concentration and nitrogen cover gas flow;
  \item central detector overflow tank monitoring, liquid level monitoring, video monitoring and pressure sensor monitoring from 0 to 5 meters;
    \item monitoring of the calibration system, including the monitoring of the source position and the motor control;
  \item monitoring of the experimental hall, including the temperature, humidity, pressure, video monitoring;
  \item monitoring of control room, database system and web-based remote system.
  \end{itemize}
  \item Functions
  \begin{itemize}
  \item front-end sensor digitalization \& data acquisition;
  \item GUIs for monitoring \& control and web-based remote run control;
  \item database recording, historical data query and archiver viewer;
  \item logic for alarms/errors/events;
  \item interlock logic;
  \item embedded remote-control modules;
  \item interface to DAQ \& trigger;
  \item replication of the detector settings and monitored parameters to offline software.
  \end{itemize}
\end{itemize}
Operation status of the system devices is monitored in real time and recorded in a database. Meanwhile, the devices can be protected by the safety interlock to prevent equipment damage and for personal protection.

\subsubsection{Overall goals and scheme}
\begin{figure}[htb]
    \centering
        \includegraphics[width=0.95\textwidth]{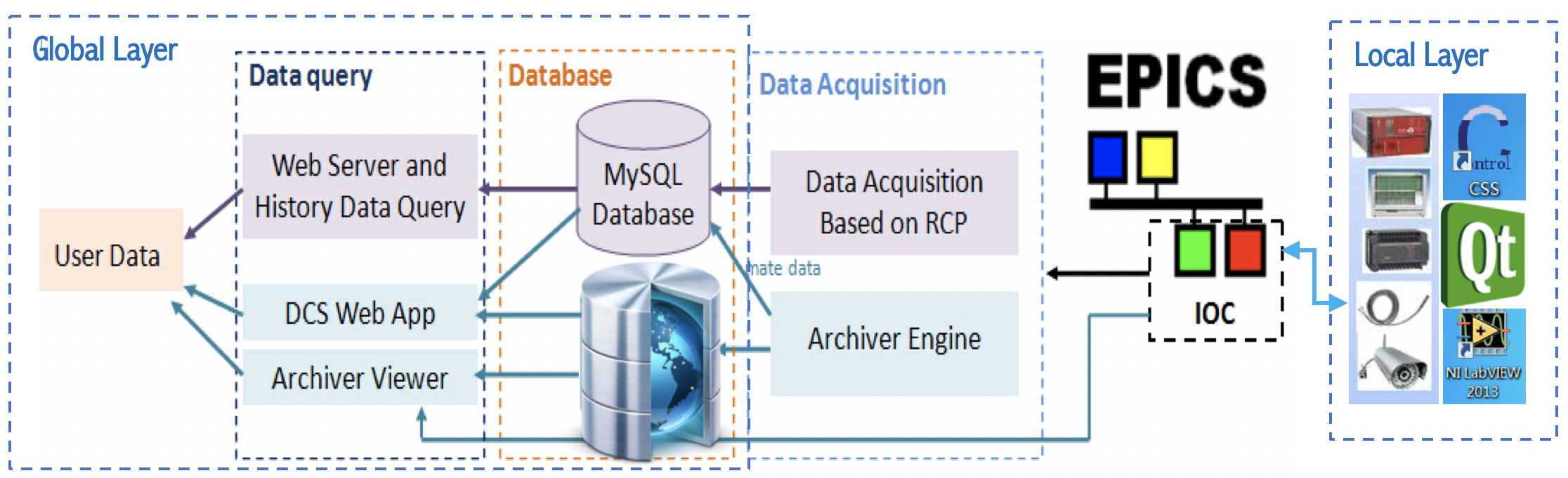}
    \caption{Framework design of the detector control system.}
    \label{fig:DCSframework}
\end{figure}

The overall goal of the design is to build a distributed system to remotely control all the equipments of the detector running industrial or self-designed devices. Following the requirements, six different subsystems as well as the common experimental infrastructure that are controlled and monitored by the DCS are foreseen, including about 60 temperature and humidity sensors and hundreds of power supply readings. The DCS framework can be divided into three parts, as shown in Figure~\ref{fig:DCSframework}. The global control layer allows for general control procedures and efficient error recognition and handling. It manages the communication among subsystems such as the central detector, the ACU, and the veto, and provides a synchronization mechanism between the DAQ and trigger system.
A database is used to store the parameters of the experiment, the configuration parameters of the systems and to replicate a subset for physics reconstruction. The local control layer provides tools for the management of local subsystems devices.
The data acquisition layer is responsible for various hardware interfaces using the Channel Access (CA) protocol. The DCS platform will be based on Experimental Physics and Industrial Control System (EPICS)~\cite{web:epics}, a set of open source software tools, libraries and applications which is widely used in experimental physics and industries to build distributed control systems.

The system adopts distributed development method. According to the experimental equipment distribution characteristics, the distributed data-exchange platform will be used for the development. Global control systems share the data and the interactively control commands by information sharing pool. Configuration files will use the text format specification.

\subsubsection{TAO DCS EPICS features}

The key technology of the development includes the remote monitoring and control for many kinds of embedded devices supported by protocols like TCP/IP, SNMP and Modbus, etc. The development method of the embedded device drivers is based on Input/Output Controller (IOC) and communicates with the up-layer software following the CA protocol. The systems can remotely monitor and control the embedded devices so that the detector can work without onsite shifter. When an exception occurs, the alarms will be raised in time and acknowledge the experts. Besides, the global layer can also manage and maintain the data in a MySQL based database.

The equipment controlled by sending and receiving strings is based on a communication module named ``StreamDevice". The IOC communication module of equipment, based on SNMP and Modbus, is developed in C language. The Control System Studio~\cite{web:controlss} will be the main platform for the development of the GUIs. Commercial systems or specific hardware/software systems will be assembled by the IOC module.

Drivers, after digitalization of the embedded devices, will be developed to collect the records related with the front-end sensors. In order to remove device-specific knowledge from the record support, each record type can have a set of device support modules. Every record support module must provide a record-processing routine to be called by the database scanners. Record processing consists of some combinations of the following functions:
\begin{itemize}
  \item input: read inputs. Inputs can be obtained, via device support routines, from hardware, from other database records via database links, or from other IOCs via CA links.
  \item conversion: conversion of the raw input to engineering units or engineering units to the raw output.
  \item output: write outputs. Output can be directed, via device support routines, to hardware, to other database records via database links, or to other IOCs via CA links.
  \item raise alarms: check for and raise alarms.
  \item monitor: trigger monitors related to CA callbacks.
  \item link: trigger processing of linked records.
\end{itemize}

The devices supported by the embedded systems can be PLC, ARM, FPGA, with standard interfaces such as USB and serial RS232, as well as network devices based on TCP/IP or UDP.

\subsubsection{R\&D plan}
A testbed will be arranged for each subsystem in order to test the hardware drivers and software systems. Each detector subsystem model will be tested in the testbed. So does the data for the pre-driver collection development. For those devices that cannot set up a testbed, simulation models should be made with the software development team to have a common definition of the interface specifications, including data format, transmission frequency, control flow, interface distribution, and physical correspondence table, etc.

A software framework will be developed with a set of integrated modules at the beginning, according to the detector hardware requirements. The interface will be defined according to the discussion with the experts of each subsystem. Data collection, visual monitoring, data recording, historical query, safety interlock of remote monitoring and control, dynamic data, functional configuration of the customization will be developed during the system assembly. During the commissioning, users' feedbacks will be collected to improve the performance. After commissioning, the formal access of the global system will be implemented.

%% file: OfflineComputing/section.tex
\section{Offline and Computing}
\label{sec:offline}
\blfootnote{Editors: Guofu Cao (caogf@ihep.ac.cn) and Paolo Montini (paolo.montini@uniroma3.it)}

\subsection{Requirements}

Offline software plays a fundamental role in the data management and in improving the efficiency of the physics analyses. The offline software system of TAO will share several structures and packages already being used in the JUNO software. A  detailed description of the software and packages shared by JUNO and TAO will be given in the next sections. The following requirements have been determined:
\begin{enumerate}
\item Reactor antineutrino events will be only a small fraction of the total number of events recorded by TAO. The offline software should provide an efficient and flexible data I/O mechanism and event buffering in order to allow high efficiency data storage and access.
\item An efficient and detailed simulation of reactor antineutrino events as well as of backgrounds is necessary to guide the detector design and evaluate detector efficiencies and systematic errors.
\item  A detailed optical model of the Gd--loaded liquid scintillator and the response model of the Silicon Photomultipliers (SiPMs) are needed in order to investigate the detector energy resolution and reconstruction performances. The reliable energy reconstruction algorithms should be developed based on charge and time information of SiPMs. An accurate vertex reconstruction is required in order to precisely determine the fiducial volume and veto natural radioactivity events coming from outside the liquid scintillator volume.
\item Event display software is needed in order to show the event structure and the reconstruction performance.
\end{enumerate}
TAO is envisaged to run for at least 3 years collecting about 1.2~PB of data during this period. A consistent part of the software infrastructure developed for JUNO will be re-used in TAO.

\subsection{Software framework}
\subsubsection{SNiPER}
\begin{figure}[!ht]
\centering
\includegraphics[width=0.8\linewidth]{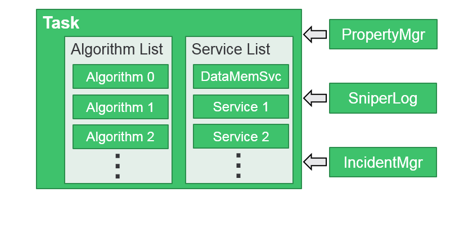}
\caption{Main components of the SNiPER framework. Description of each component can be found in the text.}
\label{sniper}
\end{figure}

It is straightforward to use the same software platform and framework in TAO as that used in JUNO, for the purposes of resource optimization, manpower integration and learning curve reduction. Therefore, we plan to develop TAO offline software based on the SNiPER framework~\cite{sniper_pro} on platform of LINUX OS and GNU compiler. Meanwhile, CMT (Configuration Management Tool)~\cite{cmt} will be selected as a tool to manage packages and generate makefiles. The SNiPER framework, standing for "Software for Non-collider Physics Experiments", has been successfully used in JUNO offline software system. It was originally designed and implemented with Object-Oriented technology and bi-language, C++ and Python for JUNO experiment. SNiPER has many innovations in multi-task processing controlling, in the handling of correlated events by introducing event buffer mechanism, and has less dependencies on the third-party software and tools. It also provides the interfaces for implementation of multi-threading computing. SNiPER requires users to implement their software as modules, and these modules can be loaded and executed dynamically at run time. Communications between modules, such as data accessing, are managed by the interfaces provided by SNiPER. According to different functionalities, the modules in SNiPER are further distinguished as the following key components, shown in Figure~\ref{sniper}.
\begin{enumerate}
 \item The Task is used to control the event loop and manage the algorithms and services.
 \item The Algorithm is implemented by users to analyze event data. It is invoked by the framework during the event loop.
 \item The Service helps users to access required parameters and data globally.
 \item The Property allows users to change parameters during the job configuration, avoiding code modification and re-compiling.
 \item The SniperLog is implemented for logs with different output levels.
 \item The Incident triggers the registered subroutines based on requirements thus making data processing more flexible.
 \item The Data Buffer stores event data over a period of time, and the length of the buffer is configurable by users.
\end{enumerate}

\begin{figure}[!ht]
\centering
\includegraphics[width=0.8\linewidth]{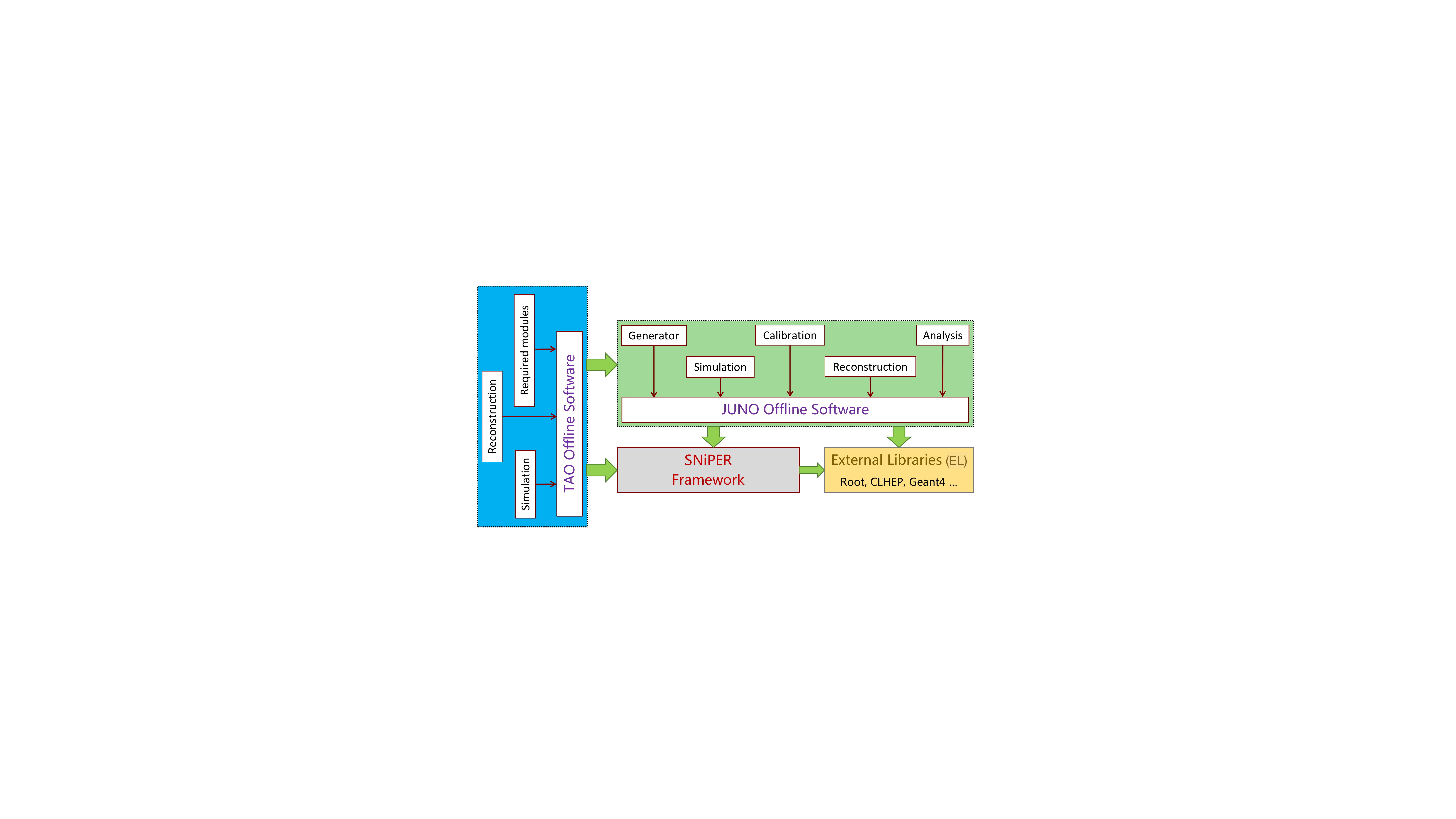}
\caption{Conceptual design of the TAO offline software, which will use the SNiPER framework and the existing JUNO offline software. Similar modules as those in JUNO will be implemented, such as simulation, reconstruction, calibration, analysis, etc.}
\label{juno_offline}
\end{figure}

JUNO's offline software system~\cite{juno_offline_pro} is designed and developed as modules within the SNiPER framework. Most of the modules in JUNO have been implemented and validated by JUNO collaborators. TAO's offline software will be  based on the JUNO offline system and the SNiPER framework, illustrated in Figure~\ref{juno_offline}. Therefore, the same external libraries, such as Geant4~\cite{Agostinelli:2002hh,Allison:2016lfl}, CLHEP~\cite{Lonnblad:1994np}, ROOT~\cite{Brun:1997pa}, etc., can be shared by JUNO and TAO. In particular, part of JUNO offline software can be re-used by TAO thus avoiding duplicated work. With this design, JUNO offline software system will not depend on TAO offline software, so that its release process and maintenance will not be affected by TAO.

\subsubsection{Event data model}
A good design on event data model for TAO is essential to achieve highly efficient data processing and easy data analysis. Due to much smaller data volume compared with JUNO, the event data model~\cite{edm_juno}, developed for JUNO, is sufficient for TAO and thus will be used. The event data model in JUNO is implemented based on ROOT TObject, in order to benefit from ROOT's features, such as schema evolution, I/O streamers, run time type identification, inspection and so on. At each data processing stage, the event data model is designed with two levels: the event header object and event object. The event header object only stores summary information of events, such as tags, timestamps, etc., which allows users to access events in a more efficient and faster way, without loading full event data. In particular, users can perform an efficient data analysis by selecting events based on information in header object and then loading the event object containing the full information of events for a deeper data analysis. The association between the header object and the event object is implemented via SmartRef, which provides control of the loading of the full event data.

\subsubsection{Geometry management}
The information of detector geometry is required in different stages of offline data processing. The consistent geometry should be used between different stages to ensure correct data handling. In JUNO, we implemented a geometry management system~\cite{geo_man} in the SNiPER framework. It is used to describe the detector properties, such as the geometrical structure, shape, dimensions, materials, positions and rotations of detector components in the coordinate system. It also provides a unified interface for applications in the offline software where geometry information is necessary, such as simulation, reconstruction, event display and data analysis thus ensuring the consistency of geometry used in different modules. In TAO offline system a similar approach of geometry management based on GDML~\cite{Chytracek:2006be} will be used. The GDML geometry will be automatically generated by an embedded G4-GDML writer in Geant4 and it will be converted to ROOT format and used by a geometry service. The geometry service will be implemented as a module in offline system to provide interfaces for applications that use geometries.

\subsection{Generators}

The Monte Carlo event generators for TAO  are quite similar with that used in Daya Bay experiment, since the both experiments are using similar target (GdLS) to detect electron anti-neutrinos emitted from reactor cores. All generators that are available in Daya Bay have been implemented in JUNO offline software. Meanwhile, some new generators have also been developed to fulfill the requirements of JUNO physics motivations, such as generators of geo-neutrinos, atmospheric neutrinos, proton decays, supernovas, etc. Since TAO offline software system will use JUNO's offline software, all generators developed for JUNO can be directly used in the TAO detector simulation. In particular radioactivity generators can be used for background simulations and calibration studies and the Inverse Beta Decay (IBD) generator can be used to study the detector response. The muon generator cannot be directly used in TAO, due to different overburden between JUNO and TAO. A much higher muon flux and smaller average muon energy are expected in TAO. In current TAO simulation we use the muon flux and energy spectrum measured at the sea level and we combine it with the implementation of geometries of the TAO's basement location and its overburden. JUNO's IBD generator is sufficient for the TAO detector R\&D studies, and it will be updated in future by taking into account the thermal power and fission fractions of the reactor cores at Taishan nuclear power plant.

Several useful tools have been developed in JUNO to help users to generate events in specific volumes or materials at specific time. However, due to different geometries in TAO, these tools have to be re-implemented in TAO's offline software.

\subsection{Simulation}

\subsubsection{Standalone simulation packages for detector R\&D}

\begin{figure}
\centering
\includegraphics[width=0.8\linewidth]{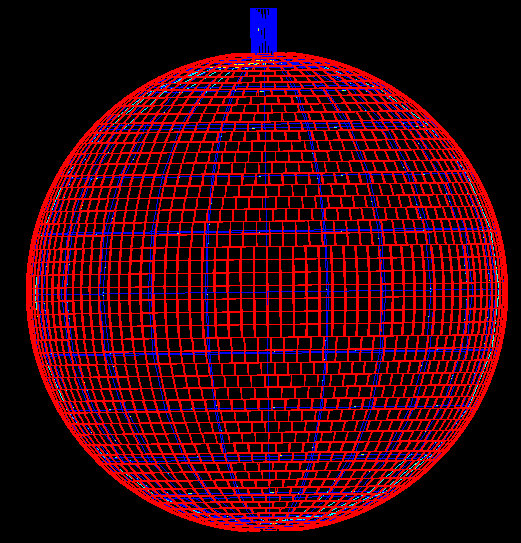}
\caption{An example of geometry of the TAO detector as implemented in a standalone Geant4 simulation package. Red lines represent the SiPM tile boundaries while blue lines represent the acrylic vessel.}
\label{sim_cd}
\end{figure}

In the beginning a dedicated standalone simulation package has been implemented based on Geant4 for the purpose of detector R\&D studies. It provides us a guidance to choose the best detector option that can fulfill the desired physics goals. Based on this package, we have conducted very detailed studies on energy resolution, natural radioactive background and fast neutron background, capability of the pulse shape discrimination, etc. In the simulation, the detector geometry is constructed with C++ codes and geometry tools provided by Geant4, including central detector, veto system and a simplified geometry of the basement and its overburden. Different detector design options have been implemented in the simulation and can be easily selected by users. The accuracy of the detector-geometry descriptions can meet the requirements of R\&D studies at the current stage. The optical properties of materials, such as GdLS, acrylic, water, etc., are managed by XML files and serves as inputs of detector geometry constructions. For simplicity, the element of SiPM sensors is constructed based on a 5~cm $\times$ 5~cm tile as a sensitive detector, instead of constructing the single cells individually. However, the coverage of SiPMs in the tile is taken as a correction factor in the Photon Detection Efficiency (PDE) of SiPMs to ensure the correct overall light detection efficiency. Figure~\ref{sim_cd} shows one of the constructed central detectors in simulation, in which the SiPM tiles are shown with red line and the acrylic sphere is indicated with blue line. The optical surfaces are also well defined during geometry construction in order to simulate optical photon propagation and its boundary processes. A complete list of the physics processes has been borrowed from the simulation code of the Daya Bay experiment, which has been extensively validated. The low energy electromagnetic processes are used for electrons and gammas, which can yield a better simulation accuracy, compared with standard electromagnetic processes. A high precision model is selected for neutrons below 20~MeV. Moreover, several bugs in Geant4 have been fixed regarding multiplicity and energy spectra of gammas emitted after neutron captured on nuclei. A customized scintillation process is implemented to handle the re-emission process of optical photons and apply multiple time constants during liquid scintillator light emissions. Besides the built-in generators in Geant4, the external generators with HEPEvt interface can also be directly used in simulation, such as the IBD generator developed for Daya Bay and JUNO experiments. A position tool is developed to help users to generate events in a specific volume or material with pre-defined position distributions. The information of the simulation output is saved in plain ROOT trees. Currently tens of variables have been saved in trees and the new variables can be easily added based on users' requirements.

\subsubsection{Integration with the SNiPER framework}
\begin{figure}
\centering
\includegraphics[width=0.9\linewidth]{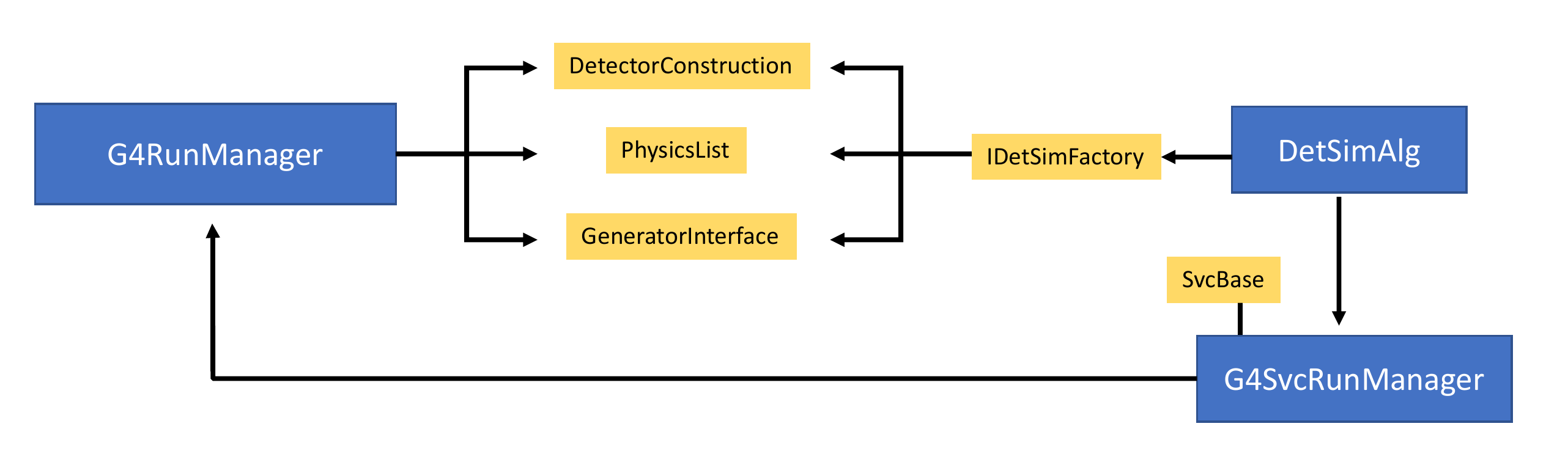}
\caption{Design schema of the simulation interface with SNiPER.}
\label{sim_framework}
\end{figure}

The detector simulation is one of the components of the TAO offline software system, so it is essential to implement the simulation within the SNiPER framework thus providing consistency with other components and maximize the benefits of the framework, like much easier data transfer between different modules. A simple way to implement the simulation in SNiPER is to integrate the developed standalone simulation package with the framework. This process, however, require some changes in order to fit the standalone package into the framework. During JUNO simulation software development, we investigated several integration solutions used by other experiments, such as BES-III~\cite{bes3}, LHC-b~\cite{Belyaev:2003be} and Daya Bay from which we worked out a final solution for JUNO experiment. The design schema is shown in Figure~\ref{sim_framework}. It can keep the integration simple, meanwhile most of parts in the standalone package can be kept unchanged, such as the most complicated detector geometry construction. The main issue arises from the fact that in the standalone package the event loop is managed by the tool G4RunManager in Geant4 while in the SNiPER framework the event looping is managed by a custom algorithm. In order to easily integrate the simulation code in the framework we will develop a SNiPER service that will deliver the G4RunManager object to a specifically designed SNiPER algorithm that will be in charge of controlling the event loop. Meanwhile, a SNiPER interface (IDetSimFactory) will be provided to users to manage the construction of necessary object instances in simulation, such as detector geometry, physics processes, primary generators and kinds of user hooks.
In the SNiPER framework, the plain ROOT trees are not suitable to store the simulation information anymore. We will use event data model to perform the data transfer between modules. For example, the generated particle information in event generators will be wrapped in the data object with a predefined format of event data model, then they are transferred to the detector simulation stage and used to accomplish simulation. The output of simulation will also be saved in data objects which will be transferred to the next stage or saved into disk. The data transfer and data I/O are automatically managed by the framework.

\subsubsection{Electronics simulation}
The main goal of the electronics simulation is to simulate the real response of the SiPM sensors, electronics system, trigger system and Data Acquisition (DAQ) system with a sufficient accuracy. It consists of SiPM and electronics modelling, trigger simulation and DAQ simulation. The output of the electronics simulation will keep the same data format as the experimental data to simplify the data analysis procedure. The electronics simulation will be implemented as a module in the SNiPER framework. To take the SiPM hit information as inputs (given by Geant4-based detector simulation), the response of SiPM sensors will be modelled first, such as photon detection efficiency, dark noise, optical cross talk, after pulse, etc. Then, we will model the full response of the electronics readout chain. For the discrete component readout option (see Sec.~\ref{sec:discrete}), signals from SiPMs will be amplified by Front-End Electronics (FEE), in which the effects of FEE will be applied, such as noise, gain, shape of waveform, etc. Then it will be followed by waveform sampling and data transfer to FPGA. In FPGA, the waveform will be integrated and converted into information of charge and time. The same integration method will be used in electronics simulation. Finally, the charge and time will be sent out to DAQ together with information of channel IDs. For the ASIC readout option, electronics simulation will simulate the response of each stage in ASIC (see Sec.~\ref{sec:ASIC}). To take KLauS ASIC as an example, it includes input stage, integrator, trigger, shaper and ADC/TDC. Trigger simulation takes the output from the electronics as its input to simulate the trigger logic and clock system, and decides whether or not to send a trigger signal if the current event passes a trigger. Currently, we have several trigger options under discussions, and the trigger simulation will be implemented for each of them to study their performance.

\subsubsection{Background mixing}
In real data, most events come from background, like natural radioactivity events and cosmic muon induced events. In order to simulate the real situation, a background mixing algorithm needs to be developed to mix signal events with background events. It is an essential module if we want to make Monte Carlo data match well with real data. There are two options for background mixing: hit level mixing and readout level mixing. For hit level mixing, the hits from both signal and background are sorted by time first, and then handled by the electronics simulation. Hit level mixing is closer to the real case, but requires a lot of computing resources. In JUNO, a hit level mixing algorithm has been implemented. Readout level mixing is much easier to implement and requires less computing resources, but it cannot accurately model the overlapping between multiple hits. Since both options have advantages and disadvantages, more studies are necessary before a final option is  selected as the official TAO background mixing algorithm.

\subsection{Reconstruction}
The major task of the reconstruction is to provide information of event vertex and energy with desired accuracy for further data analysis. The events of interest with energy deposition below 10~MeV can be treated as point-like events. The track reconstruction, even for high energy events, is not required. It is also not essential to reconstruct accurate energy for muon events, since they will be identified by setting an energy threshold, such as larger than 20~MeV used in Daya Bay analysis, then followed by a full detector veto~\cite{An:2016ses}.  The water Cherenkov veto detectors are used to tag the cosmic muons by detecting Cherenkov light and provide shielding for central detector. An additional plastic scintillator detector will be placed on top of the central detector thus improving the performance of tagging vertical muons. No track reconstruction is needed for the veto detectors. All reconstruction algorithms are required to be developed within the SNiPER framework.

\subsubsection{Vertex reconstruction}
The vertex information is critical to correct for the energy non-uniformity in the TAO detector and improve the energy resolution. A proper vertex reconstruction is needed to correctly apply a fiducial volume cut and correlate the prompt signal and delayed signal of IBD events in space. For these reasons a vertex position reconstruction is required with a resolution better than 5~cm.

\begin{figure}
\centering
\includegraphics[width=0.7\linewidth]{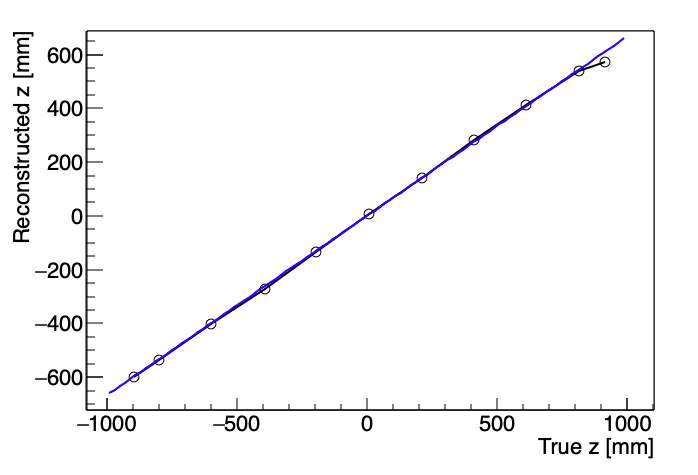}
\caption{Reconstructed vertex position according to Eq.~\ref{eq1} as a function of the true position along the z axis for positrons at rest (black). A linear dependence with a slope of $\sim0.66$ is superimposed (blue). The deviation from the linear behavior on top of the detector is due to the presence of the chimney.}
\label{rec_true_noref}
\end{figure}

\begin{figure}
\centering
\includegraphics[width=0.7\linewidth]{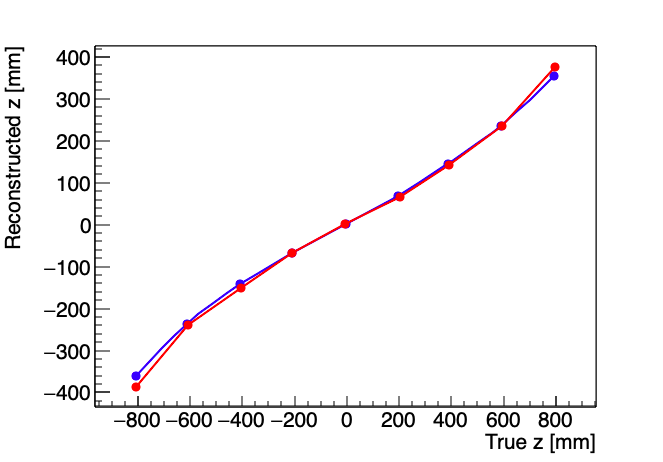}
\caption{Reconstructed vertex position according to Eq.~\ref{eq1} as a function of the true position along the z axis for positrons at rest (blue) and with 2 MeV kinetic energy (red). Reflection of optical photons on the SiPM surface is included in the simulations.}
\label{rec_true_ref}
\end{figure}

\begin{figure}
\centering
\includegraphics[width=1.\linewidth]{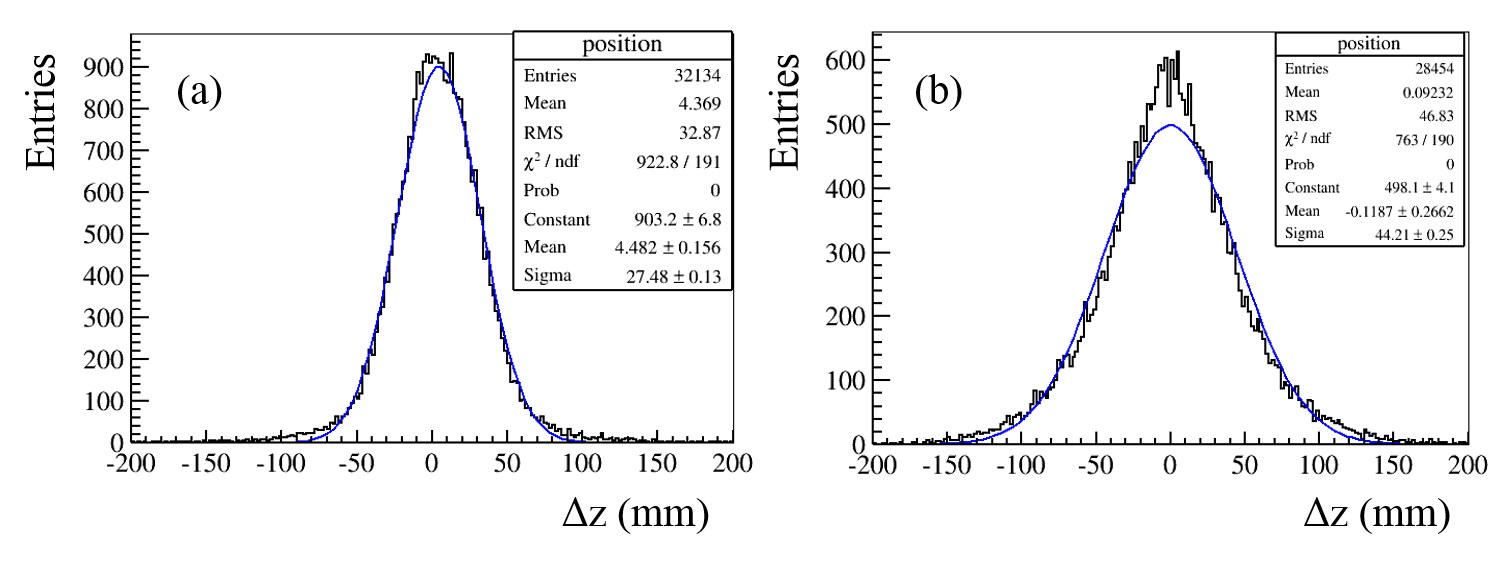}
\caption{The differences of the reconstructed z positions and the true z positions for prompt IBD events uniformly distributed in the fiducial volume (65 cm radius). (a) without reflections of the SiPMs in the IBD simulation; (b) with reflections of the SiPMs. The curves are fitted with Gaussians marked as blue lines.}
\label{ibd_rec}
\end{figure}

Both charge and time information recorded by electronics system can be used to reconstruct the event vertex. At the R\&D stage of the TAO experiment, a vertex reconstruction algorithm named center of charge and based only on the charge information has been used. This method is also successfully used in Daya Bay experiment~\cite{An:2016ses}. The charge weighted $\mathbf {\bar{x}}$ position of the vertex of one event can be simply calculated via the formula:
\begin{equation}
    \mathbf{\bar x} =\frac{\sum\limits_{i=1}^Nq_{i}\times\mathbf{r_i}}{\sum\limits_{i=1}^Nq_{i}}
    \label{eq1}
\end{equation}
in which $N$ is the total number of fired channels. $\mathbf{r_{i}}$ and $q_{i}$ denote the position and detected charge of the i-th fired readout channel, respectively.  The vertex position obtained using Eq.~\ref{eq1} has intrinsic bias caused by geometrical effects. For a spherical detector, such as TAO, this effect can be accurately predicted by performing a simple mathematical calculation as shown in Eq~\ref{eq2} (taking position in x direction as an example)
\begin{equation}
    \overline{x} = \frac{1}{4\pi}\int x\,d\Omega = \frac{1}{4\pi}\int_{0}^{2\pi} \,d\phi\int_{0}^{\pi} (x_{0}+d \cdot \cos(\theta))\sin(\theta)\,d\theta = \frac{2}{3}x_{0}
    \label{eq2}
\end{equation}
where $x_{0}$ represents the true x position of an event in the detector; $x$ indicates different x positions of photo-sensors on the sphere; $d\Omega$ denotes the open angle of photo-sensors to the true vertex position at $x$ position on the sphere; $d$ is a distance between the true vertex and the sensors. Eq.~\ref{eq2} shows that the position reconstructed using Eq.~\ref{eq1} is 2/3 of the true position. This relationship is reproduced with the TAO simulation software by simulating positron at rest deployed at different positions along the vertical z-axis and is illustrated in Figure~\ref{rec_true_noref}. The plot shows the vertex position reconstructed according to Eq.~\ref{eq1} as a function of the event true position. The linear dependence has an expected slope of 2/3 as predicted by Eq.~\ref{eq2}.
The linear relationship between the charge center and the true position, shown in Figure~\ref{rec_true_noref}, will be significantly changed after taking more effects related to optical physics processes into account, such as absorption, scattering, refraction and reflection. In particular, simulations show that the reflections on SiPM surface can make the aforementioned linear correction function no longer valid. Figure~\ref{rec_true_ref} shows the position obtained using Eq.~\ref{eq1} for positrons at rest (blue) and with 2~MeV (red) kinetic energy at different positions along the z-axis. The relation is no longer linear and a more complex relation is therefore needed.
Figure~\ref{ibd_rec} shows the differences of reconstructed positions and the true positions along z coordinate for the prompt emission of IBD events in a fiducial volume with 65~cm radius. The linear correction function is used for the case of without SiPMs' reflections, and a 3rd order polynomial function is used for the case with reflections. The reflections of SiPMs can also worsen the vertex resolution, but it still remains better than 5~cm.
In reality, the correction function can be obtained from the calibration data, like the functions used in Daya Bay experiment~\cite{DayaBay:2012aa}, since the true positions of the calibration sources are known. However, in the TAO detector, only one ACU will be installed and only the correction function along Z axis can be evaluated. The same function can be used for other axes if the good uniformity of the detector response can be guaranteed. Otherwise, we have to combine the calibration data and Monte Carlo simulation to obtain the correction functions for the whole detector volume.
More effects will be added in the simulations, such as dark noise, typically 100~Hz/mm$^{2}$, and correlated noise of SiPMs, etc., and their impacts on vertex reconstruction will be also carefully investigated. Meanwhile, a time-based reconstruction method like the one used in JUNO will also be implemented~\cite{ziyuan}, which is expected to have a better performance respect to the center of charge method, since SiPMs have much better timing performance, compared with PMTs. Besides the traditional methods, the performance of some modern vertex reconstruction methods based on machine learning will also be investigated.

\subsubsection{Energy reconstruction}
\begin{figure}
\centering
\includegraphics[width=0.8\linewidth]{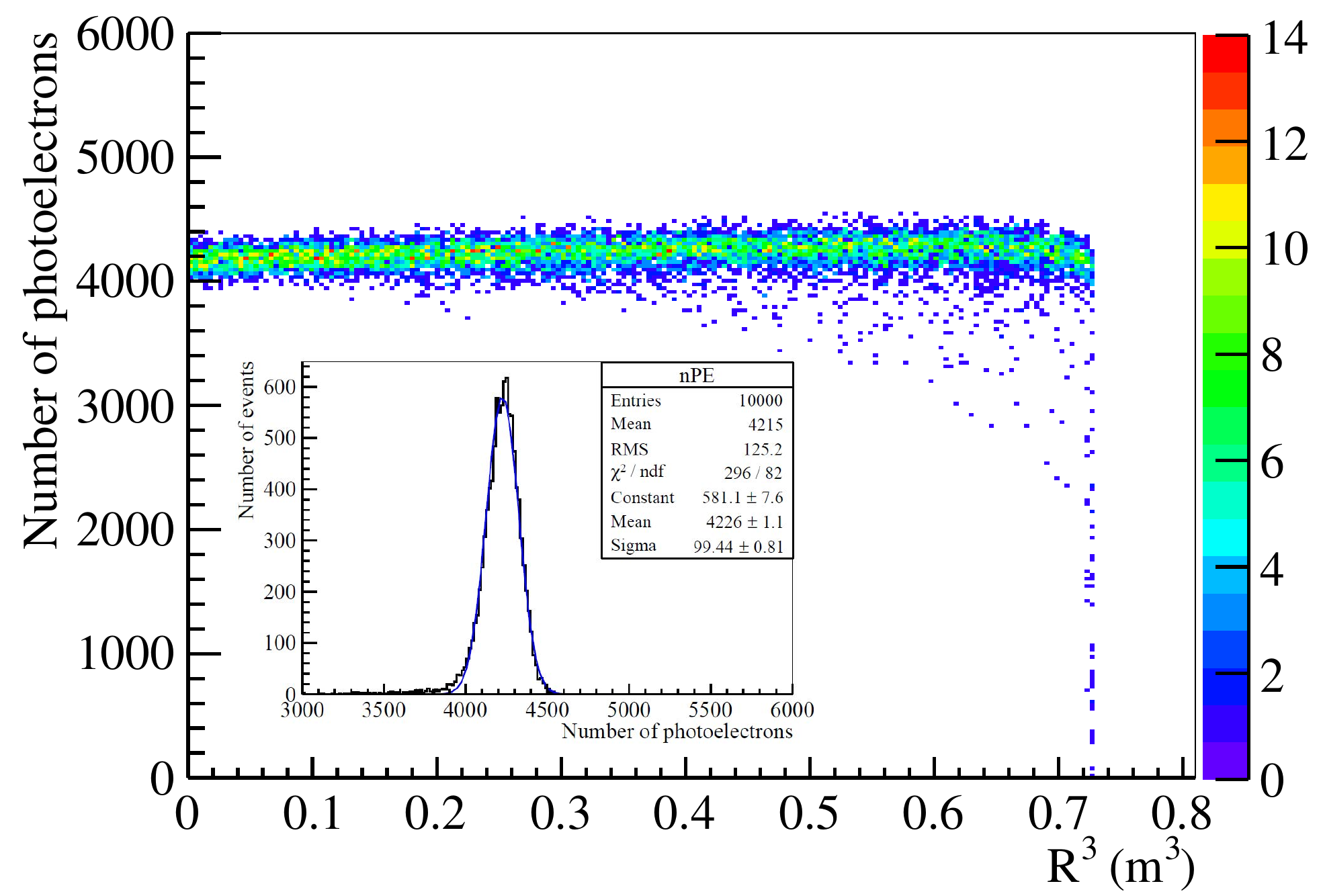}
\caption{A 2D scattering plot of the total photoelectron number versus the event radial positions for electrons with 1 MeV kinetic energy uniformly distributed in the whole detector volume. The small panel shows the photoelectron number distribution at the detector center fitted with a Gaussian.}
\label{e_rec}
\end{figure}

The charge information of each readout channel can be used to reconstruct the event energy. The gain calibration for each readout channel can be performed using a low intensity light source or a low energy calibration source. Then, based on the gain of each channel, the charge can be converted to the number of photoelectrons. The total number of photoelectrons can be therefore estimated by summing up all readout channels. A calibration source  placed at the detector center will be used to evaluate the energy scale and therefore convert the number of photoelectrons to the reconstructed energy. The non-linearity of the energy response will be reconstructed by using calibration data from different sources with known energies. Figure~\ref{e_rec} shows a scattering plot of the total number of photoelectrons as a function of the distance from the detector center obtained by simulating electrons with 1 MeV kinetic energy uniformly distributed in the whole detector volume. The plot shows a position dependence of energy response within the detector and it can be corrected with $\alpha$s from radioactivities in GdLS or with cosmogenic neutrons, since they are expected to be distributed in good quantity everywhere in the detector.
The maximum likelihood fitting is a powerful method to reconstruct the event energy and will be implemented in TAO offline software. It is based on the knowledge of the detector response model, including the light yield and the attenuation length of the liquid scintillator and the LAB buffer, the angular response of the SiPMs, and the SiPM charge resolution, as well as the reconstructed vertex as input.
Machine learning based algorithms will also be investigated and developed for TAO in order to improve the energy reconstruction and achieve the desired energy resolution better than 2\% for 1~MeV energy deposit.

\subsection{Event display}
The event display is a useful tool to show the detector structure and the event topology, in particular it can serve as an online monitor during data taking and can also help to improve the reconstruction algorithms and data analysis. The requirements of the TAO's event display system are almost same with that in JUNO. In JUNO, two event display systems have been designed and implemented. One is developed based on the ROOT Event Visualization Environment (EVE) package~\cite{dis_eve} and another one is based on the Unity engine~\cite{dis_unity}.
Thanks to the same strategies used in detector geometry management and event data model in JUNO and TAO, the developed ROOT based event display for JUNO can be transferred to TAO with limited modifications, mainly due to different variables stored in event data objects. The ROOT based event display is integrated in the offline software system as a module which usually is set up on servers with Scientific Linux system.
It can easily load the geometry file of the detector to get the structure information and uses the EVE package to generate the visual objects. Then it reads the event information from the different stages of offline data processing. It changes the visual effects of the geometry objects based on hit information and shows the detailed event information based on results of event reconstruction. Users are able to select the event that they want to display using the GUI interface.
The Unity-based event display has less dependence on the JUNO offline software, so it can be easily deployed onto different platforms after development. Besides, as a game engine, it is easier for the developer to realize fancier visual effects. However, the extra data conversion is necessary to get the event data from the ROOT format file generated by the  offline software. We expect more efforts are needed to implement the Unity-based event display for TAO, due to the data conversion. It will be decided later weather or not to use the Unity-based event display in TAO, depending on manpower.

\subsection{Database}
\begin{figure}
\centering
\includegraphics[width=0.8\linewidth]{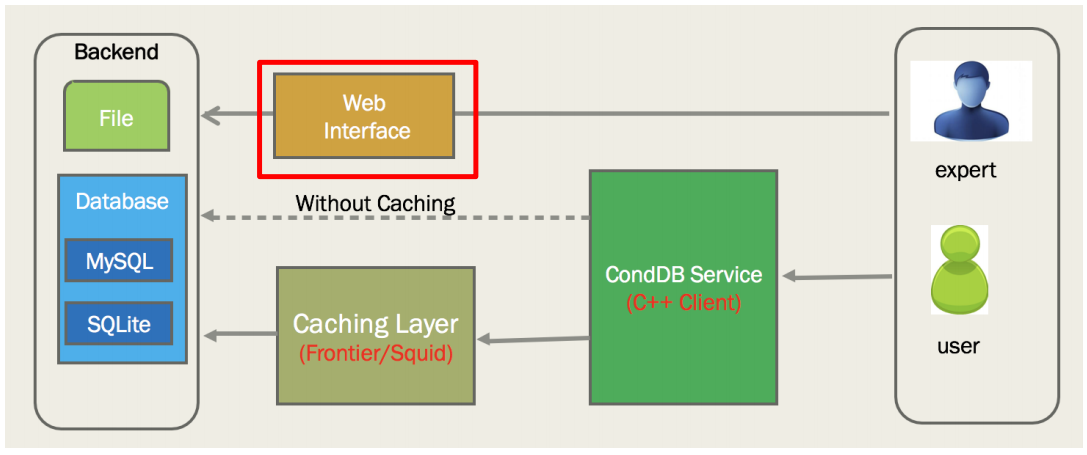}
\caption{Schema of the conditions database system designed for JUNO and envisaged to be used in TAO. Details can be found in the text.}
\label{database}
\end{figure}

Database is an indispensable component in offline data processing. It plays an important role in event data processing and data analysis. It is used to store many important information, such as detector running parameters, calibration constants, geometry parameters, optical properties of kinds of materials, schema evolution, bookkeeping and so on. Meanwhile, it also provides access to all of this information via their management services, which allow users to create, query, modify or delete stored data.
Recently, the conditions database system has been designed and a database prototype has been established for testing in JUNO offline system which in envisaged to be used in TAO as well. The design schema of the conditions database system containing three layers is shown in Figure~\ref{database}. The server layer uses the MySQL~\cite{mysql}/SQLite~\cite{sqlite} databases to store conditions data. The data model of the conditions database consists of 4 metadata tables:
\begin{enumerate}
 \item Payload, which holds conditions data or the path of the conditions data files.
 \item IOV (Interval of Valid), which describes the time information for the validity of the Payload.
 \item Tag, which collects a set of IOVs.
 \item Global Tag, which collects a set of Tags.
\end{enumerate}
Two additional auxiliary tables are used to collect Tag-IOV Map and Global Tag-Tag Map. In the client layer, the conditions database service is developed in the SNiPER framework to perform conversions from Persistent Object to Transient Object and provide database interface for different applications, such as simulation, reconstruction, data analysis, etc. An intermediate layer (Frontier/Squid) between the client and the server is adopted to provide data caching capabilities, which can efficiently decrease the heavy burden of center database when users frequently query the same conditions data at the same time. The web interface is developed for experts to manage the data in database server.

In TAO offline data processing, we will share the same database servers and conditions database framework used in JUNO. However, due to the different information stored in database for TAO, a new conditions database service will be implemented, which should inherit the existing JUNO database service with extended functionalities and fulfills the requirements of TAO.

\subsection{Data storage and computation}
\subsubsection{Requirements}
The TAO detector will produce about 0.4~PB or even less raw data every year, which will be transferred back to the Computing Center at the Institute of High Energy Physics (IHEP) in Beijing through a network connection. Meanwhile, a similar data volume of Monte Carlo (MC) data will also be produced for physics analysis. Both raw data and MC data need to be shared among collaboration members via network. JUNO will establish a computing farm with about 12,000 CPU cores, 10~PB disk storage and 30~PB archive, in which 2000 CPU cores will be deployed for TAO offline data processing. The computing nodes and storage servers will be connected to each other by a 40-Gbps backbone high-speed switching network. Moreover, the platform will also integrate the computing resources contributed by outside members via a distributed computing environment. Compared with about 2.4~PB data every year in JUNO, the data volume in TAO is not significant. The performance, security and reliability of the data transfer in JUNO have been intensively considered. Therefore, we will use the same tools developed for JUNO data transfer to transfer TAO's raw data from experimental site to IHEP's computing center.

\subsubsection{Data transfer}
\begin{figure}
\centering
\includegraphics[width=0.8\linewidth]{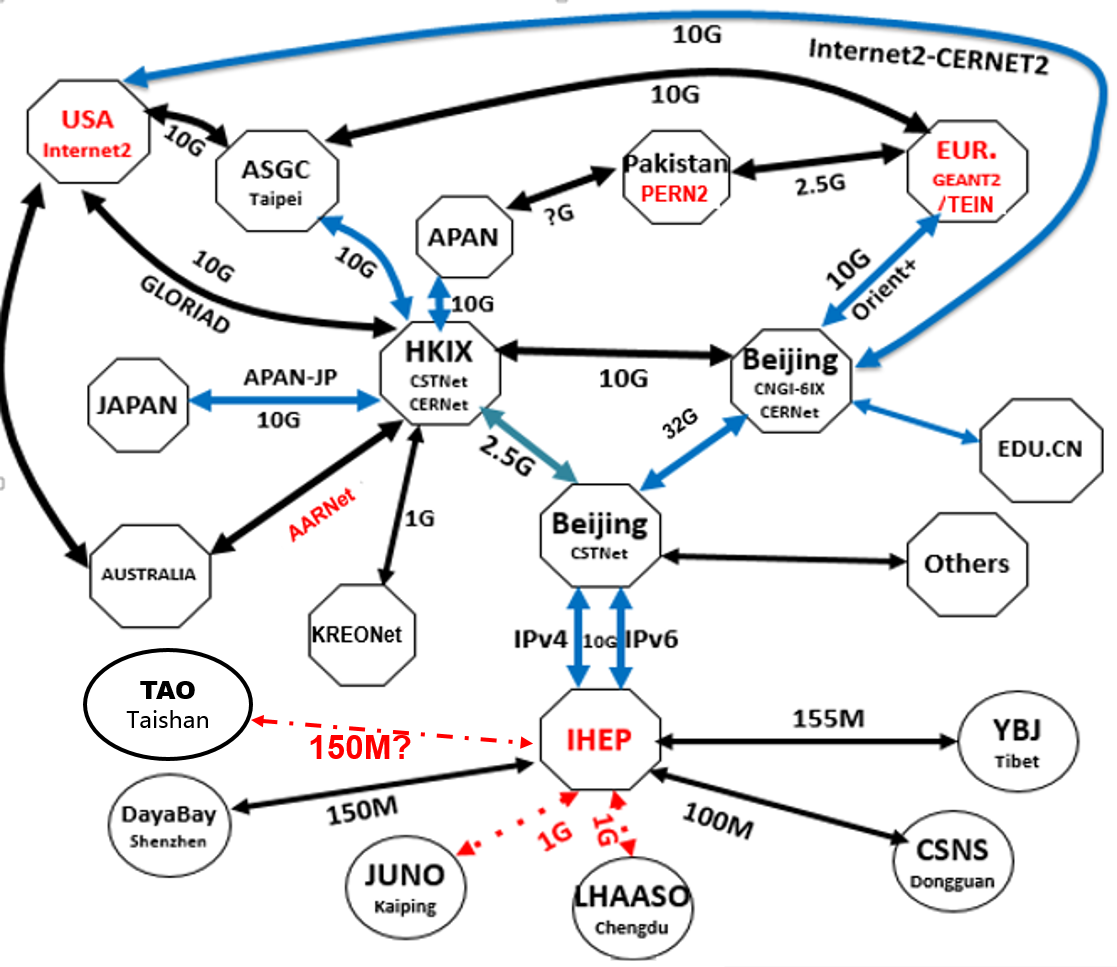}
\caption{Architecture of network connections among particular experiments including TAO, computing centers in China and other countries.}
\label{wan}
\end{figure}

Unlike JUNO planning to store the whole PMT waveforms, TAO will only store the charge and time information for each readout channel, meanwhile, the second level trigger is proposed and will be applied on onsite DAQ cluster or at read-out board FPGAs level to further reduce the data rate. By considering the event rate and event size in the TAO detector, the predicted data volume is estimated to be about 100~Mbps or less. Since only limited computing resources are expected to be deployed onsite, it is critical to transfer all raw data from experimental site to IHEP data center. By considering the raw data volume produced by the TAO detector, a link with bandwidth of about 150~Mbps is sufficient for a stable data transfer. Same with the network used by JUNO, we will also use the network provided by the Chinese Science and Technology Network (CSTNet)~\cite{cstnet}. The data will first be transferred to IHEP through the link, then relayed to collaborating sites through CSTNet. At present, IHEP is connected to the CSTNet core network through two 10~Gbps links, one of which supports IPv4 and the other IPv6. The bandwidth from IHEP to the USA is 10~Gbps and from IHEP to Europe is 5~Gbps, both of which are through CSTNet and have good network performance. The architecture of the network including TAO, other experiments in China, Chinese clusters and outside world is shown in Figure~\ref{wan}.

During the transfers, the checksum of the data will also be transferred to ensure the data integrity. If the data integrity check is failed, the raw data should be re-transferred. After the data transfer is completed and all the data integrity checks are passed, the status of the data in the DAQ local disk cache will be marked as TRANSFERRED. Moreover, a high/low water line deletion algorithm will be used to clear the outdated data in DAQ local disk cache. Most of the data will be transferred automatically by the data transfer system in order to avoid  possible human errors. To ensure the stability and robustness of the data transferring system, a monitoring system will be developed and deployed. It will monitor the data transfer and sharing status in real-time and track the efficiency of the data transfer system. Combined with the status of the IT infrastructure (including network bandwidth), the data transfer system will optimize the transfer path and recover transfer failures automatically to improve the performance and stability of the system.

\subsubsection{Data storage}
The maximum raw data volume produced by the TAO detector is about 1.2~PB with 3~years running. All of these data will be transferred from onsite DAQ to IHEP and stored in disks at the IHEP data center, meanwhile, at least one full copy of data needs to be stored on disk or tape, to ensure safety. The data will be transferred to other computing center(s) outside IHEP to share the data with collaboration members. The simulated data volume will be at the same level with the raw data.

%% file: Installation/section.tex
\section{Facility and Installation}
\label{sec:installation}
\blfootnote{Editors: Jun Cao (caoj@ihep.ac.cn), Zhimin Wang (wangzhm@ihep.ac.cn) and Yuguang Xie (ygxie@ihep.ac.cn)}
\blfootnote{Major contributor: Guofu Cao}

Assembly and installation considerations should go all through the design and R\&D of the detector system. Careful logistical coordination will be essential for the receiving, staging, assembly, installation, and testing of all detector components and subsystems, especially handling all the works around the nuclear reactor. This section discusses some of the considerations in the installation process and outlines a plan for the assembly and running of the detector system.

The working group of JUNO-TAO has a wide range of experience in the installation and operation of large detector systems, including the engineering and installation activities of Daya Bay, JUNO and Darkside-50.

The effort of WBS (Work Breakdown Structure) supports the overall planning, staging, control and execution of the final assembly and installation. It includes labor, materials and universal equipment required to perform these functions.

The assembly work of the detector will be finished in the neutrino laboratory in the campus of the Taishan Nuclear Power Plant. Due to the strict regulations on accessing the core area of the nuclear power plant, the detector will be pre-assembled at IHEP to minimize the technical risks and workload in the power plant. The assembly and installation work of JUNO-TAO includes:
\begin{enumerate}
\item Detector components will be fabricated and shipped to IHEP progressively.
\item Components and subsystem will be tested before and during assembly.
\item The central detector without, or with a small fraction of, SiPMs and electronics will be assembled and tested at IHEP first.
\item SiPMs, electronics, veto and shielding will be tested at IHEP, may or may not be integrated with the central detector test.
\item The Taishan Neutrino Laboratory will be refurbished, instrumented, and prepared for the detector installation and operation.
\item Detector components will be disassembled, shipped to the Taishan laboratory, and re-assembled again, except that the stainless steel tank (SST) and the acrylic vessel will not be re-used due to the transportation passage limitation of the laboratory. New SST and acrylic vessel components will be fabricated and welded/bonded in the Taishan laboratory.
\item GdLS will be produced at either IHEP or the Daya Bay site. All detector liquids will be transported to the Taishan laboratory and handled in clean and safe.
\end{enumerate}

The laboratory and facilities, assembly and installation of detector subsystems, and project management issues will be described in the following subsections.

\subsection{Laboratory and facility}

A neutrino laboratory will be set up in the Taishan Nuclear Power Plant. The laboratory is in a basement at 9.6~m underground outside of the concrete containment shell, about 30 m in horizontal distance to the center of the reactor core. Vertical overburden is roughly estimated to be $\sim5$ meters-water-equivalent, which comes basically from the concrete floors and the roof of the building. Muon rate and cosmogenic neutron rate were measured onsite with a plastic scintillator detector and Bonner Spheres, respectively, to be 1/3 of that on the ground. Radioactive dosage is measured with a hand-held dose meter to be $\sim0.4 \mu$Sr/h, twice higher than the ambient environment in the campus, which might largely come from the contribution of the thick concrete wall. The power supply, water supply, and ventilation are ready and satisfy the requirements of JUNO-TAO. The laboratory can be accessed via a stairway and an elevator. The elevator has a dimension of $1990~({\rm Depth)}\times 1390~({\rm Width}) \times 1990~({\rm Height})$ mm and a rated load of 2.5~ton, posing a strict limitation to the size of all detector components. The detector design and installation plan have taken this limitation into account.

\begin{figure}[htb]
    \centering
    \includegraphics[width = 0.6\columnwidth]{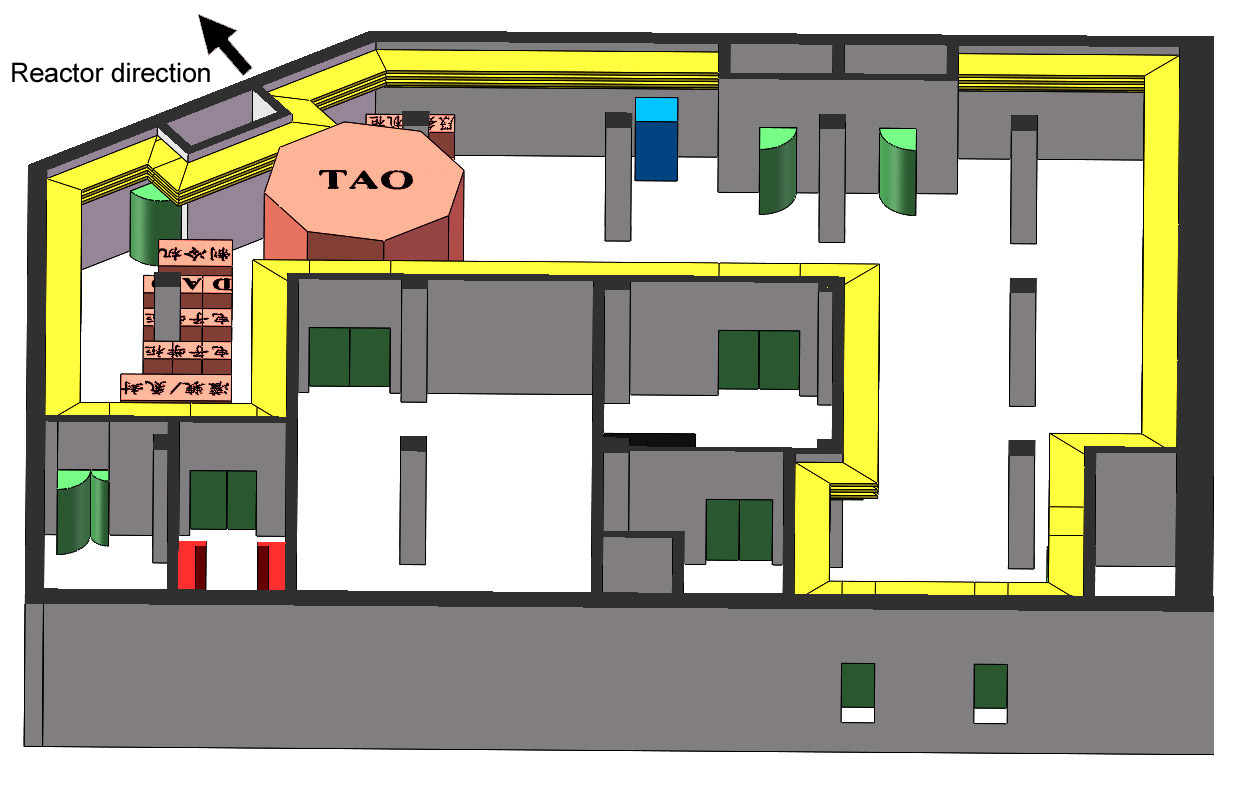}
    \caption{Taishan Neutrino Laboratory. The pink blocks show the footprint of the TAO detector and relevant facilities. The reactor core is about 30 m in the north-west direction.}
    \label{fig:TNLab}
\end{figure}

The layout of the Taishan Neutrino Laboratory is shown in Figure~\ref{fig:TNLab}. The pink blocks show the footprint of the TAO detector and relevant facilities, including the refrigerator, crates for electronics, DAQ, offline computer, network server, etc. The reactor core is about 30 m in the north-west direction. The height of the laboratory is close to 5 m, but there is a steel beam structure on the roof, resulting in an available height of 3.85 m.

The laboratory will be set up with facilities for assembly, commissioning and running, including safety related monitoring, network and management. A class 10,000 clean tent will be set up for the assembly of the central detector. Details and management rules will be elaborated with the power plant, following their requirements.

\subsection{Central detector}

\begin{figure}[htb]
    \centering
    \includegraphics[width = 0.8\columnwidth]{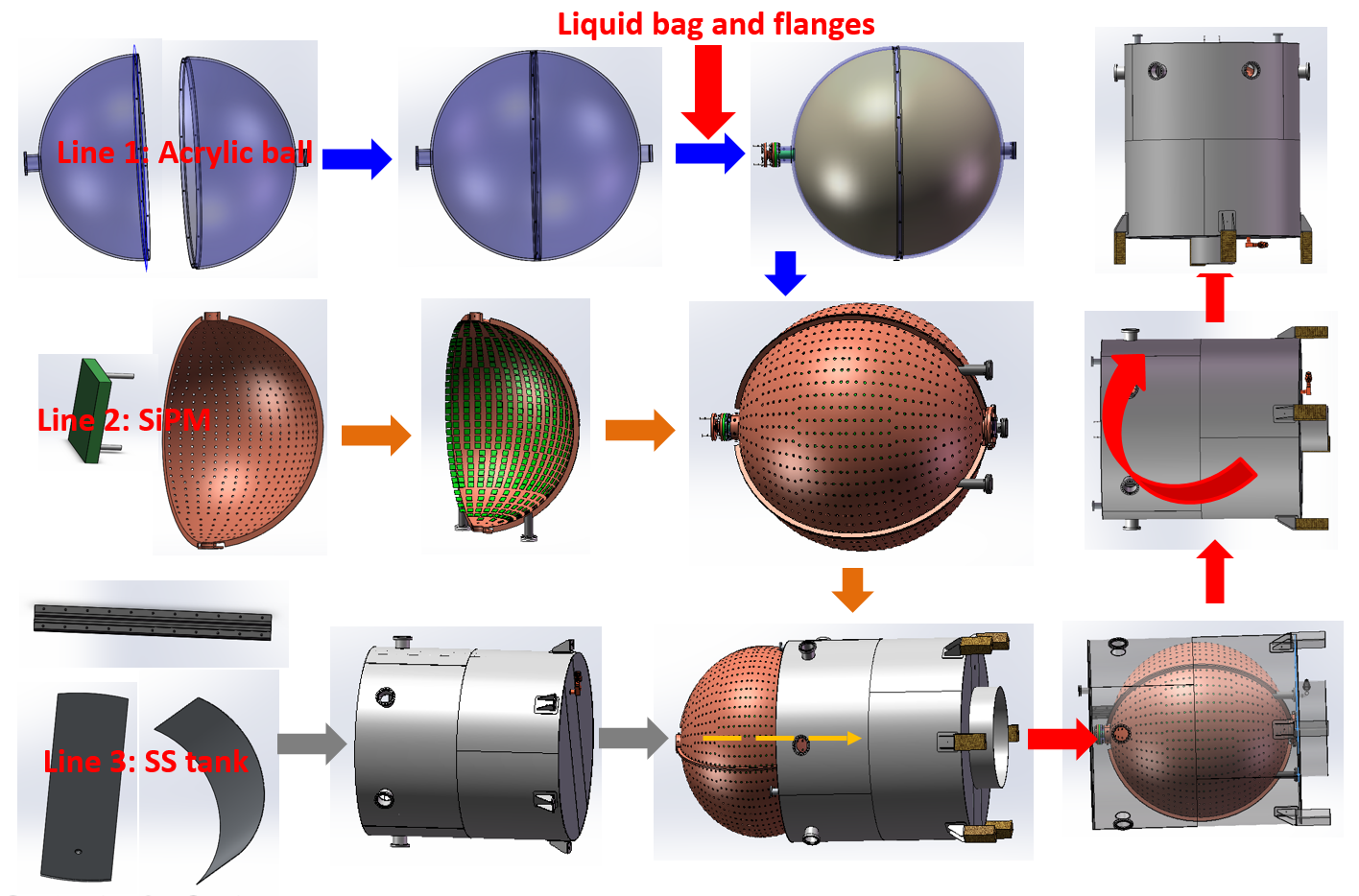}
    \caption{Central detector installation sequence.}
    \label{fig:install:cd-installation}
\end{figure}

Assembly of the central detector includes SST welding, acrylic bonding, integration of SiPM tiles and Frontend Electronics (FEE) with the copper shell, assembling all above structures, cabling, and filling liquid, etc. The general consideration of the installation sequence is shown in Figure~\ref{fig:install:cd-installation}.

The stainless steel tank will be made of 6 pieces for the barrel, 3 pieces for the bottom panel, and 3 pieces for the lid. These parts are shaped in factory and shipped to the neutrino laboratory. They will be welded together with tools to achieve required precision. The flange between the barrel and the lid need special attention since it has to be made from 3 pieces onsite due to transportation limitation. The welding lines of SST will be processed with local acid-pickling and passivation.

The acrylic vessel will be bonded via polymerization from 3 pieces. An alternative way is to clamp or latch the acrylic pieces together without bonding, and put a liquid bag inside to contain the GdLS, as described in Sec.~\ref{sec:acrylicvessel}.

The SiPM tiles and the frontend electronics boards will be mounted onto the partitions of the copper shell and tested at IHEP. After that the parts will be wrapped with plastic film and shipped to onsite.

A clean tent is needed for the assembly of the acrylic vessel and the copper shell. The parts of the copper shell with SiPM tiles mounted on its inner surface are bolted together, with the bonded (clamped) acrylic vessel inside. Then the whole copper shell will be rotated from vertical position to horizontal position. The SST will perform the same rotation. The assembled copper shell will be installed into the SST horizontally along three guide rails on the wall of the SST. The SST will then be rotated from horizontal position to vertical position with the copper shell fixed inside temporarily with tools.

For the above assembly and install procedure, available assembly space in the laboratory and possible conflicts with other subsystems have been preliminarily considered. Temporary lifting equipment will be set up in the laboratory. The height of the laboratory available for the detector is 3.85 m due to the steel beams on the roof. Space between beams is higher, which could be used to set up the lifting equipment. The assembly and rotation operation of the central detector is possible in principle while further consideration is needed. Pure water system and nitrogen system will also be set up in the laboratory. More details will be worked out.

\subsubsection{SiPM and electronics}

\begin{figure}[htb]
    \centering
    \includegraphics[width = 0.8\columnwidth]{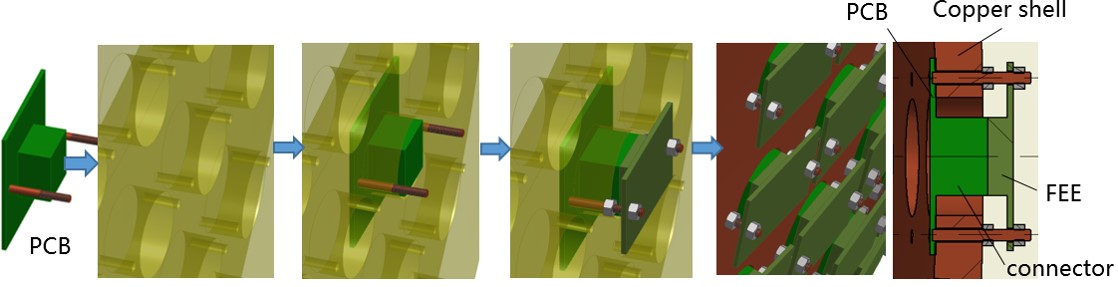}
    \caption{Assembling the SiPM tile on the copper shell and connecting with the FEE PCB.}
    \label{fig:install:SiPM-module-connection}
\end{figure}

The assembly of SiPM and electronics is a challenge since it is fragile and the clearance between SiPM tiles is tiny to achieve as high as possible photo-sensor coverage. Figure~\ref{fig:install:SiPM-module-connection} shows the preliminary consideration on how to fix a SiPM tile (on a PCB) on the copper shell, and connect to its FEE PCB. Both the distortion of the spherical copper shell and the accuracy of connecting holes are critical. Different thermal expansion of SiPM, PCB, bolt, and copper need be considered for the large range of temperature change from 25$^\circ$C during assembly to -50$^\circ$C during operation. A preliminary thermal simulation shows that the detector assembly is safe during the cooling down process. Given the risks in this assembly work, a prototype including a fraction of the copper shell and tens of the PCBs will be tested soon to practice the assembly and identify the potential design and assembly problems. Then a full-size prototype with blank PCBs as a proxy of SiPMs will be tested at IHEP.

\subsubsection{Sensors and monitoring}

The detector will be monitored with sensors for the temperature, liquid level, humidity, ethanol content, gas pressure, and power, during assembly, filling, and running. The sensors will be installed along with the assembly of main mechanical structure, and be integrated in the Detector Control System (DCS), which can be monitored both onsite and offsite via network.

\subsubsection{Liquid filling}

A lot of experiences of the liquid scintillator handling could be borrowed from Daya Bay and JUNO. Acrylic vessels or barrels with ETFE liquid bag inside would be used to ship and store the GdLS (about 2.8 t) to avoid contamination. Pumps of $\sim$~800~L/hour flux with all parts contacting the GdLS being fluorine plastic, similar to what is used in JUNO R\&D, will be used to fill the GdLS from storage container to the CD through fluorine plastic pipe. Sensors will be installed in the CD to monitor the liquid level of the GdLS and the buffer liquid. The GdLS and the buffer liquid will be filled with certain synchronization, controlling the liquid level difference being smaller than 250~mm as suggested by the analysis of the stress of the acrylic vessel. The flow will be monitored with Coriolis mass flow meter and volume flow meter.

\subsection{Calibration system}

\begin{figure}[htb]
    \centering
    \includegraphics[width = 0.8\columnwidth]{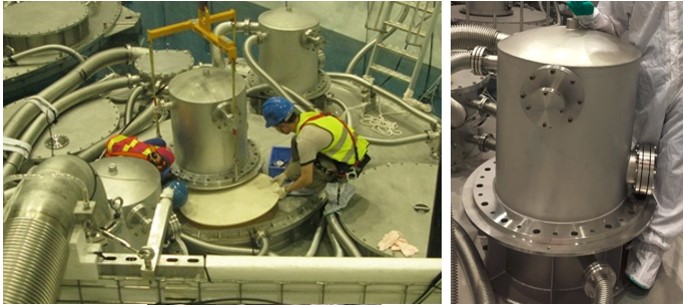}
    \caption{Automatic Calibration Unit on a Daya Bay detector.}
    \label{fig:install:ACU}
\end{figure}

The ACU will be installed on the overflow tank, which is on top of the SST lid, through a flange after the CD is installed at its location. Figure~\ref{fig:install:ACU} shows the ACU on a Daya Bay detector. One of the ACU will be used in JUNO-TAO after decommissioned from Daya Bay. It is an assembly that includes the internal source drive, the outer bell-jar, and a bottom plate to shield the radioactive sources stored in ACU. When it is rigged from the bottom plate, the ACU can be considered as a monolithic structure of $\sim$~100 kg. Its installation without an overhead crane in the laboratory of limited height needs be worked out. A movable lifting arm is an option. A specially designed thermal insulation hat made of High Density Polyethylene (HDPE) will cover the ACU.

\subsection{Veto and shielding system}

\begin{figure}[htb]
    \centering
    \includegraphics[width = 0.9\columnwidth]{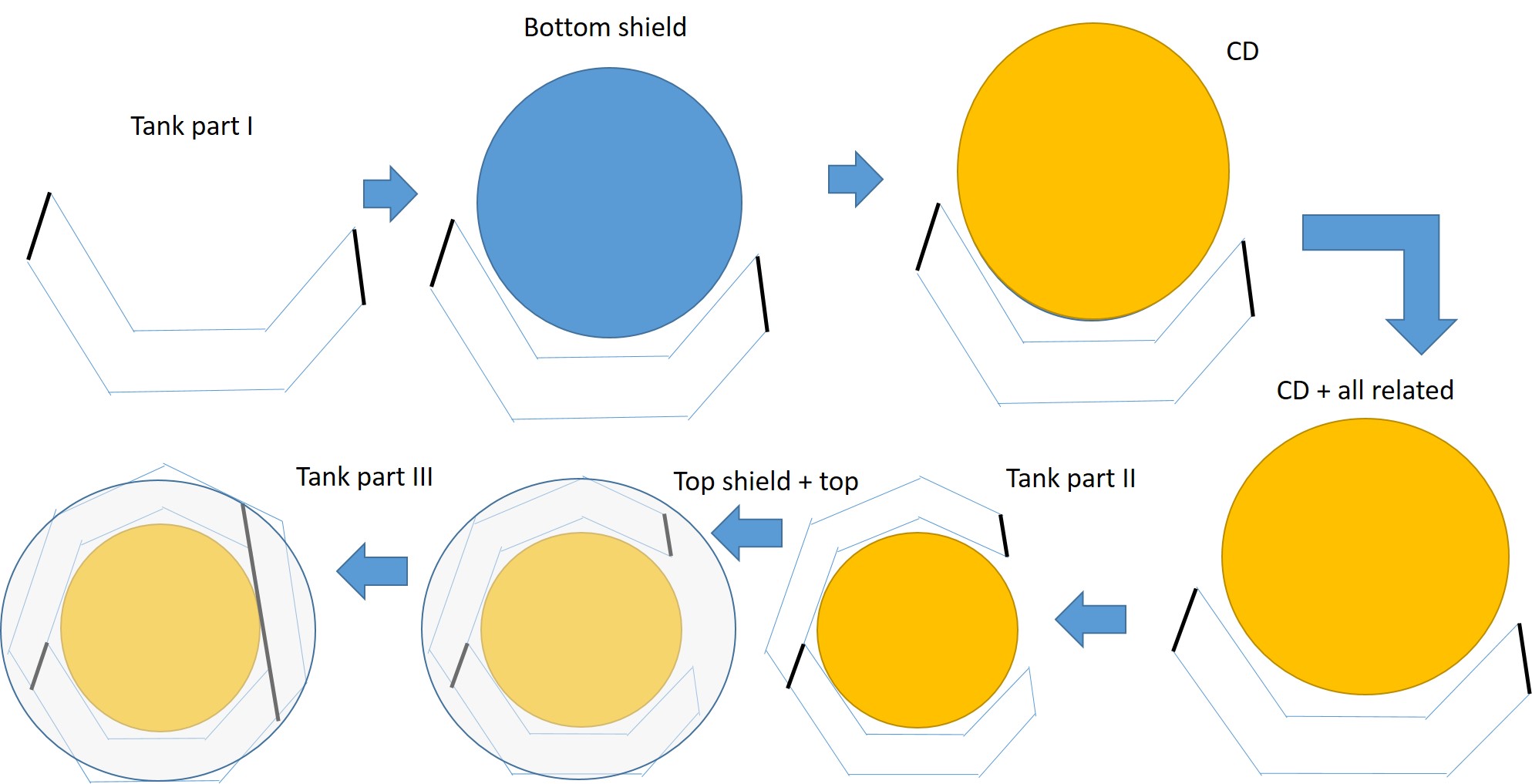}
    \caption{TAO detector installation sequence.}
    \label{fig:install:sequence}
\end{figure}

The veto and shielding system include the bottom shield of the CD made of lead, three water tanks surrounding the CD, the HDPE shielding material above the CD, and the plastic scintillator detectors on top of the HDPE shielding.
The water tanks are instrumented with 3-inch PMTs to serve as water Cherenkov muon detectors. The installation sequence of the veto and shielding system is shown in Figure~\ref{fig:install:sequence}, and is listed in the following.
\begin{enumerate}
\item Install the water circulation system.
\item Install the water tank part I.
\item Install the bottom shield of the CD.
\item Move the assembled CD to its position.
\item Install the water tank part II.
\item Install the support structure of the top shield and muon detectors.
\item Install the HDPE shielding material and the plastic scintillator detectors on the top of the CD.
\item Install the movable water tank part III.
\item Connect the supply and return water pipe.
\item Fill the tanks with pure water and run the circulation systems.
\end{enumerate}

On top of the water tanks there are manholes. PMTs and reflective film Tyvek will be installed inside the water tank in parallel with other installations.

The movable water tank part III is designed to sit on a carrier. It is the last ``door" that close the shielding of the CD. In case of inspection or repairing is needed for the CD, it can be opened to access the CD.

\subsection{Integration \& running}
The whole TAO detector will be formed through integrating all subsystems. To ensure a smooth integration process, the interfaces between subsystems should be well defined and reviewed during the detector mechanical design. A detailed detector assembling procedure has to be established and reviewed before the detector installation. To minimize the onsite workload and potential technical risks, we will process a detector pre-assembly at IHEP with most of detector components and subsystems, then the detector will be disassembled and shipped to onsite, and finally assembled again. This process can help us to verify the installation procedure and find potential problems, to train workers and improve the onsite installation efficiency, to identify the required tools and equipment, to provide more accurate estimations of the required man power and time for onsite detector installation. According to the experiences gained during the detector pre-assembly, we can make the Taishan Neutrino Laboratory to be well prepared for the detector installation and operation. Since not all of the detector components can be pre-assembled, in particular for cables, the extra onsite workload should be carefully estimated. Related installation procedure needs to be well defined before the start of onsite installation.

We will start the detector commissioning before filling the liquid to the detector. At this stage, all SiPMs and electronics will be operated and tested by using a light source or SiPMs' dark noises, in order to find potential issues of SiPM sensors, electronics and cabling. The DAQ/DCS subsystems will also be tested and tuned. The following main goals are expected to be achieved during the detector commissioning.
\begin{enumerate}
\item SiPMs and electronics can work normally both at room temperature and cryogenic temperature with affordable failure rate. The signal of single photon electron can be clearly observed at -50$^\circ$C.
\item To complete the optimization of SiPMs' operating voltage and configuration parameters of electronics at -50$^\circ$C, to obtain the best detector performance.
\item To verify the SiPM and electronics calibration.
\item To successfully cool down the central detector to -50$^\circ$C.
\item Veto detectors can run normally.
\item DAQ/DCS run smoothly.
\item A successful data transfer from onsite to IHEP.
\end{enumerate}

\subsection{Management}
In TAO, from the start of all system design, scheme argument and technology checks, we will follow the technical review system that has been established and well practiced in the JUNO collaboration. We will also follow the management system in JUNO that covers engineering drawing control, engineering change control procedure, and mechanical design standards, guidelines and reviews, etc. To well control and organize the onsite work process, a management structure will also be introduced, not only to coordinate the activities of the detector installation, commissioning and running, but also to maintain a good interface with the power plant to guarantee a smooth project process. Meanwhile, safety training and safety management structure is also required and should follow the power plant's requirements.

\subsection{Risks}
Most of the technical risks related to the detector installation can be identified and solved during the pre-assembly of the detector at IHEP, and they should not happen again in final detector assembly in the power plant. The detector components should be assembled as many as possible at the pre-assembly stage, to reduce the potential risks during the onsite detector installation. However, during the detector commissioning or running, we should also consider the following potential risks:
\begin{enumerate}
\item Compatibility issue of detector components with the buffer liquid during 3-6 years running. This can reduce the transmittance of the buffer liquid and the photon collection efficiency, thus worsen the energy resolution. To reduce this risk, we should perform a very detailed compatibility tests at R\&D stage for every material used in buffer liquid.
\item Failure of electronics components in the SST during the detector running. To replace the electronics inside the SST is rather complicated, so some efforts are required to improve the reliability of electronics components operated in cryogenic temperature and reduce its failure rate.
\end{enumerate}

\subsection{Safety}

A safety management structure will be established to ensure the safety of persons, detector components, equipment and environment during the detector installation, commissioning and running, both for pre-assembly at IHEP and final onsite assembly and installation.

As a satellite experiment of JUNO, the safety management of JUNO-TAO will follow the management of JUNO, and also follow the management of the Taishan Nuclear Power Plant, whichever stricter.

Some key points are summarized below.
\begin{itemize}
\item Design and technical review. Safety consideration need be rooted in the design of the experiment. It should be guaranteed through technical reviews. All documents including engineering drawings will be archived.
\item Hazard management. Risks and hazards should be identified and classified for the experiment, each subsystem, and the assembly and installation work. A dedicated Hazard and Operability Study (HAZOP) analysis will be performed on the whole JUNO-TAO detector and subsystems Material Safety Data Sheet (MSDS) for chemicals should be archived.
\item Safety officers. Safety officer of the experiment will approve the safety management rules and oversee the onsite activities. For each work carried out onsite, a designated local safety officer will actually oversee the whole work process.
\item Onsite work control. Every work onsite needs a procedure being approved by the safety officer and a Task Control Form being approved by onsite manager.
\item Safety training. A training program, in conjunction with the power plant, will be developed. All personnel entering the laboratory will have site training. The training will include how personnel should report and respond to emergencies.
\item Response to emergency. Documents and laboratory instructions will be prepared to indicate how personnel should react for different emergencies. Appropriate safety design and operational mitigation of safety risks associated with the use of flammable liquids will be established.
\item Environment protection. The Project is committed to protect the environment. The primary environmental concerns are with possible spills of liquid scintillator and the use of cleaning solutions. The amount of cleaning solution is small. We do not anticipate any radiological issues, but all sealed sources will be inventoried and tracked and radon will be monitored.
\item Supervision from the power plant. Usually the power plant will have safety expert to supervise the activities happening onsite.
\end{itemize}

We understand that the power plant will provide emergency response, including fire, police and emergency medical response. These forces are connected to the local government, but administratively report to the power plant.

%% file: TAOacknowledgement.tex
\section*{Acknowledgement}
\addcontentsline{toc}{section}{Acknowledgement}

We are grateful for the ongoing cooperation from the China General Nuclear Power Group.
This work was supported by
the Chinese Academy of Sciences,
the National Key R\&D Program of China,
the CAS Center for Excellence in Particle Physics,
the Joint Large-Scale Scientific Facility Funds of the NSFC and CAS,
Wuyi University,
and the Tsung-Dao Lee Institute of Shanghai Jiao Tong University in China,
the Institut National de Physique Nucl\'eaire et de Physique de Particules (IN2P3) in France,
the Istituto Nazionale di Fisica Nucleare (INFN) in Italy,
the Fond de la Recherche Scientifique (F.R.S-FNRS) and FWO under the ``Excellence of Science – EOS” in Belgium,
the Conselho Nacional de Desenvolvimento Cient\'ifico e Tecnol\`ogico in Brazil,
the Agencia Nacional de Investigaci\'on y Desarrollo in Chile,
the Charles University Research Centre and the Ministry of Education, Youth, and Sports in Czech Republic,
the Deutsche Forschungsgemeinschaft (DFG), the Helmholtz Association, and the Cluster of Excellence PRISMA+ in Germany,
the Joint Institute of Nuclear Research (JINR), Lomonosov Moscow State University, and Russian Foundation for Basic Research (RFBR) in Russia,
the MOST and MOE in Taiwan,
the Chulalongkorn University and Suranaree University of Technology in Thailand,
and the University of California at Irvine in USA.